\documentclass[aip,pof,preprint,tightenlines,longbibliography]{revtex4-2}
\usepackage{array}[=2016-10-06]
\usepackage{amsmath,amssymb,bm,graphicx,booktabs,tabularx,array,enumitem,placeins}
\usepackage{xcolor}
\definecolor{paperblue}{RGB}{0,55,153}
\usepackage{hyperref}
\graphicspath{{figures_latest/}{figures_shock/}{figures_display_fct/}{figures/}}
\newcommand{\Kn}{\mathrm{Kn}}
\newcommand{\bv}{\bm v}
\newcommand{\bu}{\bm u}
\newcommand{\bC}{\bm C}
\newcommand{\E}{\mathbb E}
\newcommand{\Var}{\operatorname{Var}}
\newcommand{\Pois}{\operatorname{Poisson}}
\AtBeginDocument{\setlength{\textfloatsep}{14pt plus 3pt minus 2pt}\setlength{\intextsep}{12pt plus 2pt minus 2pt}}
\hypersetup{colorlinks=true,linkcolor=paperblue,citecolor=paperblue,urlcolor=paperblue,pdftitle={Collision survival probabilities and transport bias in low-occupancy direct simulation Monte Carlo},pdfauthor={Ehsan Roohi, Stefan Stefanov and Ahmad Shoja-Sani}}

\newcommand{\WallFineWindow}{71.292}

\newcommand{\WallCoarseWindow}{185.915}

\newcounter{algorithm}
\begin{document}
\title{Collision survival probabilities and transport bias in low-occupancy direct simulation Monte Carlo}
\author{Ehsan Roohi}
\email{roohie@umass.edu}
\affiliation{Department of Mechanical and Industrial Engineering, University of Massachusetts Amherst, Amherst, MA 01003, USA}
\author{Stefan Stefanov}
\affiliation{Institute of Mechanics, Bulgarian Academy of Sciences, Acad. G. Bontchev Str., 1113 Sofia, Bulgaria}
\affiliation{Centre of Excellence in Informatics and Information and Communication Technologies, Sofia, Bulgaria}
\author{Ahmad Shoja-Sani}
\affiliation{Ferdowsi University of Mashhad, Mashhad, Iran}
\begin{abstract}
When a collision cell contains only two hard-sphere particles, the first isotropic collision erases their initial stress anisotropy on average. The mean anisotropy remaining after a time step is therefore set by the probability of no collision. We use this result to study momentum transport in direct simulation Monte Carlo (DSMC) with few particles per cell. Simplified Bernoulli trials (SBT) underestimate the probability of no collision. Exponential survival gives the exact mean stress for an isolated pair. For three or four particles, allowing subsequent collisions and selecting partners by their collision rates further reduces the relaxation error. In strong Couette flow at a Knudsen number of 0.05, 58\% of collisions accepted by the exponential variant involved an isolated pair. At a time step of 0.756~ns, the signed wall-shear difference from whole-cell Poisson sampling at that step changed from $+0.782$~Pa for SBT to $-0.170$~Pa with exponential survival and to about $+0.014$~Pa with multiple collisions and rate-weighted partners. In an eccentric bearing, SBT and stochastically rounded no-time-counter sampling reduced the torque magnitude by 5.5--5.8\% relative to Poisson sampling at fixed mesh, step and occupancy. In a separate fixed-mesh test with Poisson sampling, reducing the occupancy from twenty particles per cell to two raised the torque magnitude by 4.25\%. With equal total particle counts on two grids, torques agreed within 0.53\% while angular shear profiles differed by more than 4\%. Total torque can conceal local errors in momentum transfer.
\end{abstract}
\maketitle

\section{Introduction}\label{sec:intro}
Direct simulation Monte Carlo (DSMC) represents a gas by particles that move through a grid and collide in pairs within each cell \citep{Bird1963,Bird1994,Bird2013}. When a cell contains just two particles, they can collide only with each other. For isotropic hard-sphere scattering the probability that the pair finishes a time step without colliding determines how much of its initial stress anisotropy remains on average. A third particle changes the problem. A collision with one partner changes the relative velocity, and hence the collision rate, with the other \citep{RjasanowWagner1998}. Two algorithms can conserve energy and momentum and give the same mean collision rate while relaxing stress at different rates.

Bernoulli-trial methods approximate the collision operator for a finite set of particles \citep{BelotserkovskiiYanitskiy1975,Padilla2026}. In the simplified Bernoulli-trial (SBT) scheme, particles are visited in order and one eligible partner is drawn for each primary particle \citep{stefanov2011dsmc}. The generalized Bernoulli-trial (GBT) scheme changes the number of primary trials and their weights \citep{Roohi2018}. The symmetrized simplified Bernoulli-trial (SSBT) and symmetrized generalized Bernoulli-trial (SGBT) schemes change which pairs can be selected \citep{Taheri2022,Javani2024}. A particle can take part in several trials and collide more than once in a sweep.

Cell width and time step also affect momentum and heat transfer in DSMC \citep{Alexander1998CellSize,GarciaWagner2000TimeStep,Hadjiconstantinou2000Discretization,Gallis2006Normal,Rader2006Fourier}. In Fourier flow, \citet{Taheri2017} varied spatial and temporal resolution together and used wall heat flux to assess SBT at low particle counts. \citet{Roohi2026Progress} reviews how cell size and particle number affect transport. The transient adaptive subcell (TAS) method changes the region from which collision partners are drawn \citep{Amiri2014,Goshayeshi2015TAS,Stefanov2022Hybrid}. The supplement discusses these methods further \citep{RoohiAkhlaghiStefanov2025,Roohi2026Progress}.

Here we hold the ordering of SBT trials fixed and change one part of the collision rule at a time. Exponential waiting times are well established in stochastic time counters and null-collision methods \citep{RjasanowWagner1998,Koura1986}, and \citet{Shterev2021} interpreted SBT in terms of survival. Our exponential variant (SBT-exp) keeps the original partner and allows at most one collision per trial. It uses the probability of at least one collision at the selected rate during the substep. The Poisson version of SBT (P-SBT) uses the representative partner's rate to draw the number of collisions and allows further encounters during that trial. Weighted P-SBT uses the sum of the eligible pair rates and chooses each partner in proportion to its rate. Event-adaptive P-SBT (EA-P-SBT, shortened to EA in figures and tables) recalculates those rates after each collision.

Successive collisions of the same pair can also be correlated in the no-time-counter (NTC) method \citep{akhlaghi2018consequences}. Our channel implementation rounds the number of candidate collisions stochastically, draws pairs of distinct particles uniformly and accepts a collision according to the pair's current relative speed. Section~\ref{sec:fourier} gives the implementation used for Fourier flow. The whole-cell Poisson method uses the same proposals, acceptance rule and upper bound on relative speed. Drawing the number of candidates from a Poisson distribution then samples the specified collision process exactly while cell membership is fixed. Transport in a flow still depends on the grid, the time step and the particle population. Other approaches include stochastic Boltzmann schemes \citep{Nanbu1980}, virtual and pseudo-subcells \citep{LeBeau2003,Macrossan2010}, the refinements of \citet{Gallis2009Convergence} and transient subcells on unstructured grids \citep{Su2010TAS}. Convergence to the Boltzmann equation requires a suitable joint limit \citep{Wagner1992}. Table~\ref{tab:lineage} gives the collision rules most closely related to those tested here.

\begin{table}[!tbp]
\centering\small
\begin{tabularx}{\textwidth}{>{\raggedright\arraybackslash}p{.24\textwidth}>{\raggedright\arraybackslash}X>{\raggedright\arraybackslash}X}
\toprule
Earlier development & Change introduced & Relation to the present study\\
\midrule
SBT and GBT \citep{stefanov2011dsmc,Roohi2018} & Representative pair trials and probability compensation. & We keep this ordering when changing the collision probability.\\
SSBT and SGBT \citep{Taheri2022,Javani2024} & Symmetrized and generalized pair selection. & Here we vary when the rates are recalculated.\\
SBT--TAS \citep{Amiri2014,Goshayeshi2015TAS} & Adaptive collision regions and their volume/time-step treatment. & The main Couette runs use four fixed collision subcells per parent cell.\\
Time-counter and null-collision methods \citep{RjasanowWagner1998,Koura1986} & Exponential event times and current-rate acceptance. & Whole-cell Poisson sampling gives the fixed-cell reference.\\
\bottomrule
\end{tabularx}
\caption{Earlier collision methods and their relation to the present comparisons. All methods tested here use the same hard-sphere scattering law.}
\label{tab:lineage}
\end{table}

We first follow stress relaxation in a homogeneous cell containing two, three or four particles, with no molecular flight. We then measure wall shear and the momentum flux within the gas in wall-driven Couette flow. Two other channel tests examine heat transfer: body-force-driven Poiseuille flow between stationary walls (profiles in the supplement) and Fourier conduction between walls at different temperatures. Normal shocks at Mach 2 and 5 test how particle number affects shock thickness and nonequilibrium moments. At Mach 2 the thickness can be checked against an independent hard-sphere Boltzmann solution. The supplement also reports Mach 10. An impulsively started lid-driven cavity tests the transient shear in the gas just below the lid. Finally, an eccentric gas bearing lets us compare the shear around the rotor with its integral, the torque. By changing the collision rule at fixed occupancy and the occupancy at fixed collision rule, we can distinguish two effects that can shift the torque in opposite directions.

\section{Physical setting and measured moments}\label{sec:setup}
\subsection{Channel geometry and simulator populations}
In the Couette and Poiseuille calculations the gas is monatomic hard-sphere argon with molecular mass $m=6.63\times10^{-26}$ kg and diameter $d=4.17\times10^{-10}$ m. The gap is $H=10^{-3}$ m and the wall temperature $T_w=273.15$ K. The nominal Knudsen number $\Kn=\lambda_0/H=0.05$ gives a reference mean free path $\lambda_0=50\,\mu$m and number density $n_0=2.5887673\times10^{22}$ m$^{-3}$. Molecules reflect diffusely in the instantaneous wall frame. The flow is streamwise in $X$, with the centred transverse coordinate $Y/H\in[-1/2,1/2]$; the geometry is planar and one-dimensional, but the molecular velocity has three components.

In the main Couette case the walls move at $\pm U_w$ with $U_w=1000$ m s$^{-1}$ and no body force. In Poiseuille flow the walls are stationary and the acceleration is $a_X=5\times10^6$ m s$^{-2}$; the comparison with published data uses the higher acceleration $1.692936626\times10^7$ m s$^{-2}$.

The main Couette comparison uses 240 parent cells across the gap (Table~\ref{tab:runfamilies}), each divided into four independent collision subcells, so that $\Delta Y_p=H/240=4.166667\,\mu$m and $\delta Y=H/960=1.041667\,\mu$m. Collision rates use the subcell volume and partners are drawn only within the subcell, without a nearest-neighbour search; a subcell with fewer than two particles has no collisions.

We report the mean number of simulator particles per cell (PPC). For the channel flows the count refers to a parent cell, and for shocks to an upstream collision cell. A numerical suffix gives this mean, so NTC20 denotes NTC with twenty particles per cell. In the four-subcell Couette runs, PPC2 corresponds to 480 particles in the domain and an average of only 0.5 particles per collision subcell. At PPC20 the subcell average is five. The total population is fixed, but particles enter and leave the subcells as they move. In the homogeneous tests cell membership stays fixed. Table~\ref{tab:runfamilies} also identifies the order of ballistic flight (B) and collisions (C) within each step. BC denotes flight followed by collisions, CB the reverse order, and CBC a flight between two half collision steps.

\begin{table}\centering\small
\begin{tabularx}{\textwidth}{l@{\hspace{8pt}}>{\raggedright\arraybackslash}X@{\hspace{8pt}}r@{\hspace{8pt}}r@{\hspace{8pt}}l@{\hspace{8pt}}l}\toprule
Run & Flow and collision samplers & $N_Y$ & $s$ & Parent PPC & $\Delta t$ (ns)\\\midrule
C1 & Couette: SBT, SBT-exp, P-SBT, weighted P-SBT, EA, NTC, Poisson &240&4&2; NTC20 also&0.756, 0.252\\
C2 & Couette: SBT, smaller steps &240&4&2&0.126, 0.063\\
C3 & Couette: Poisson, smaller reference step &240&4&2&0.126\\
C4 & Couette: SBT with CB and CBC ordering &240&4&2&0.252\\
C5 & Couette: NTC100, EA10 and Poisson2 &240&1&100, 10, 2&3.026, 1.009\\
P1 & Poiseuille: P-SBT, weighted P-SBT, EA, Poisson &240&1&2, 4&1.009\\
P2 & Poiseuille: same four samplers, finer grid &720&1&2, 4&0.336\\
P3 & Poiseuille: NTC100, EA10, Poisson20 &240&1&100, 10, 20&3.026\\
P4 & Poiseuille: NTC100, EA10, Poisson2 &240&1&100, 10, 2&1.009\\\bottomrule
\end{tabularx}
\caption{Channel-flow calculations. $N_Y$ is the number of parent cells and $s$ is the number of collision subcells per parent ($s=1$ for an undivided cell). PPC denotes the mean number of simulator particles per parent cell. The mean per collision subcell is PPC$/s$. A number after a sampler name gives its parent PPC (e.g., NTC20). EA denotes event-adaptive P-SBT. CB denotes collision followed by flight, and CBC denotes half a collision step on either side of a full flight step. All runs use two independent seeds, $\Kn=0.05$, $T_w=273.15$ K and 102 warm-up blocks. The Couette walls move at $\pm1000$ m s$^{-1}$. The Poiseuille walls are stationary and $a_X=5\times10^6$ m s$^{-2}$. Table~\ref{tab:runwindow} and the individual captions give the sampling intervals.}\label{tab:runfamilies}
\end{table}

\subsection{Moments and the distinction between kinetic and total shear}
Let $n$ be the number density, $m$ the molecular mass, $\bv$ the molecular velocity and $\bu=\langle\bv\rangle$ the local mean velocity, where brackets denote a number-weighted average over the sampling volume. The peculiar velocity is $\bC=\bv-\bu$, with $C^2=\bC\cdot\bC$ and $C^4=(C^2)^2$. The mass density is $\rho=mn$, the thermal velocity variance $\theta=k_BT/m=\langle C^2\rangle/3$ and the scalar pressure $p=nk_BT$, with $T$ the translational temperature, $k_B$ Boltzmann's constant and $\delta_{ij}$ the Kronecker delta. Cartesian indices $i,j,k$ run over the three velocity components. The kinetic pressure tensor, its deviator and the heat-flux vector are
\begin{equation}
 P_{ij}=\rho\langle C_iC_j\rangle,\qquad
 \sigma_{ij}=P_{ij}-\rho\theta\delta_{ij},\qquad
 q_i=\frac{\rho}{2}\langle C_i C^2\rangle.
\end{equation}
With our wall convention the Couette kinetic shear $P_{XY}=\sigma_{XY}$ is negative, and we keep its sign at every sampled position. The trace-free third moment $m_{ijk}$, the deviatoric fourth moment $R_{ij}$, the scalar fourth-moment excess $\Delta$ and the contracted fifth moment $F_{5i}$ are
\begin{align}
 m_{ijk}&=\rho\langle C_{\langle i}C_jC_{k\rangle}\rangle,\\
 R_{ij}&=\rho\langle C^2 C_{\langle i}C_{j\rangle}\rangle-7\theta\sigma_{ij},\\
 \Delta&=\rho\langle C^4\rangle-15\rho\theta^2,\qquad
 F_{5i}=\rho\langle C^4 C_i\rangle .
\end{align}
Angle brackets around tensor indices denote the symmetric trace-free part; for example, $C_{\langle i}C_{j\rangle}=C_iC_j-C^2\delta_{ij}/3$ and $C_{\langle i}C_jC_{k\rangle}=C_iC_jC_k-C^2(C_i\delta_{jk}+C_j\delta_{ik}+C_k\delta_{ij})/5$, while brackets around a whole velocity product denote the local average. In the channel calculations the reference scales are $\rho_0=mn_0$ and $c_0=\sqrt{k_BT_w/m}=238.4984$ m s$^{-1}$, and moments of order $r$ are normalized by $\rho_0c_0^r$. We accumulate all 56 raw velocity monomials of total degree up to five over the stated sampling interval, then form central moments from those averages.

Because colliding particles are separated by a finite distance, collisions transfer momentum across internal sampling faces. The measured transverse flux of streamwise momentum is therefore
\begin{equation}
 J_{YX}=J_{YX}^{\mathrm{kin}}+J_{YX}^{\mathrm{coll}},
\end{equation}
where the first term counts momentum carried by particles crossing a face and the second the momentum exchanged by collisions that span it. Their sum can differ from the kinetic shear at a cell centre. Signed wall stresses come from the accumulated streamwise wall impulses, with opposite normal orientations on the two walls. Balancing the total interior flux against the wall transfer checks conservation on the grid, and comparing kinetic with total flux shows how the momentum transfer is distributed at the chosen resolution. A solver can satisfy this momentum balance while predicting a wall stress that still changes with resolution.

\section{Collision samplers and rate adaptation}\label{sec:methods}
\subsection{Common pair rate and scattering map}
Let $F_{\mathrm{num}}$ be the number of molecules represented by one simulator particle, $V_c$ the collision volume and $V=(\bv_0,\ldots,\bv_{N-1})$ the current velocities of its $N$ particles. For diameter $d$ the collision rate of the unordered pair $(i,j)$ is
\begin{equation}
 a_{ij}(V)=\frac{F_{\mathrm{num}}}{V_c}\pi d^2|\bv_i-\bv_j|.
 \label{eq:pairrate}
\end{equation}
In the ordered SBT schemes we randomly permute the local particle list before each sweep. A primary index $i$ in that ordering has the eligible partners $S_i=\{i+1,\ldots,N-1\}$, with weight $w_i=|S_i|=N-1-i$, so every unordered pair belongs to exactly one such set. The ordering and the eligible sets stay fixed during the sweep, while velocities change after accepted collisions.

The methods differ in how they choose collision times and partners. All use the same cross-section and isotropic hard-sphere scattering law. With $\bu_{ij}=(\bv_i+\bv_j)/2$ and $g=|\bv_i-\bv_j|$, a uniform unit vector $\bm e$ on the sphere gives $\bv_i'=\bu_{ij}+g\bm e/2$ and $\bv_j'=\bu_{ij}-g\bm e/2$, which conserves pair momentum and energy. We record the same quantities for all samplers in the flow comparisons and full-solver timings.

\subsection{Collision history and stress relaxation}\label{sec:physicalbasis}
All schemes use the same cross-section and scattering law. In a cell with fixed membership, differences in stress relaxation therefore arise from how the schemes sample collisions. \citet{Taheri2022} wrote the finite-cell generator and its transition exponential before symmetrizing the SBT partner proposal. In their SSBT scheme each of the $N$ particles acts as a primary and selects a partner uniformly from the other particles; the factor of one half in their Eq.~(13) accounts for the two directed proposals of a pair at first order. After provisional acceptance, a duplicate-pair check can trigger partner resampling and prevents a second accepted collision of that pair in the sweep. Homogeneous relaxation and shock moments have been examined for SBT and GBT \citep{ShojaSani2021RelaxShock} and for SSBT in normal shocks \citep{Javani2024SSBTShock}. Our variants retain the SBT ordering while varying how often collisions occur, how partners are chosen and when their rates are updated.

Let $\bC_i=\bv_i-N^{-1}\sum_j\bv_j$, $\bm g_{ij}=\bv_i-\bv_j$ and $\operatorname{dev} A=A-\tfrac13\operatorname{tr}(A)I$, with $I$ the identity tensor. The deviatoric kinetic pressure tensor of the cell is
\begin{equation}
 \bm P'=\frac{mF_{\rm num}}{V_c}\sum_i\operatorname{dev}(\bC_i\bC_i^{\mathsf T}).
\end{equation}
For equal masses and simulator weights an elastic collision leaves the cell mean velocity unchanged. The colliding pair contributes twice the outer product of its centred mean velocity plus $\bm g_{ij}\bm g_{ij}^{\mathsf T}/2$ to $\sum_i\bC_i\bC_i^{\mathsf T}$. Isotropic scattering preserves $|\bm g_{ij}|$ and makes the conditional mean of the outgoing relative dyad isotropic, so the collision generator $\mathcal L$ for a cell of fixed membership acts on the deviator as
\begin{equation}
 \mathcal L\bm P'
 =-\frac{mF_{\rm num}}{2V_c}
 \sum_{i<j}a_{ij}\operatorname{dev}(\bm g_{ij}\bm g_{ij}^{\mathsf T}).
 \label{eq:stress-generator}
\end{equation}
In Eq.~\eqref{eq:stress-generator} the contribution of each pair depends on both its rate and its relative velocity. Two samplers can conserve momentum and energy and give the same total event rate $\sum_{i<j}a_{ij}$, yet relax stress at different rates if they select different pairs.

For a cell containing just two particles, the pair is unique. Let $a$ be the pair rate and $x=ha$. For hard spheres $a$ depends only on the relative speed, which elastic scattering preserves. In the published SSBT algorithm \citep[Eq.~(13) and Appendix~A]{Taheri2022} the first primary tries the only pair with probability $x/2$; if the trial fails the reverse orientation has the same probability, and if it succeeds the second primary skips the repeated pair. For $0\leq x\leq 1$ the probabilities of no event during the substep are
\begin{equation}
 P_0^{\rm SBT}=1-x,\qquad
 P_0^{\rm SSBT}=(1-x/2)^2,\qquad
 P_0^{\rm exact}=P_0^{\text{SBT-exp}}=e^{-x}.
 \label{eq:two-pair-survival}
\end{equation}
For small $x$ the survival deficits relative to the exact value are $x^2/2+O(x^3)$ for SBT and $x^2/4+O(x^3)$ for SSBT. This comparison with SSBT is analytical. The flow simulations use the other samplers. For an isolated pair the first isotropic collision erases the mean deviator, so its expected value after a substep is $P_0\bm P'(0)$. Thus SBT relaxes stress at the effective rate $-\log(1-x)/h$ for $0<x<1$. At $x=0.4$ this is $1.277\,a$, although the mean collision count per step is $x$ for both SBT and exact pair sampling. The centred heat flux of two equal-weight particles is identically zero, since their peculiar velocities are opposite.

With several eligible partners, define the entry intensity $\lambda_i=\sum_{j\in S_i} a_{ij}$ of primary $i$. Conditional on the representative partner $J$, SBT-exp and representative-rate P-SBT have the same probability of at least one event, and both use $J$ for the first collision. For P-SBT the total number of events $K_i$ in the primary trial has the distribution
\begin{equation}
 \Pr(K_i=k\mid J)=e^{-h w_i a_{iJ}}\frac{(h w_i a_{iJ})^k}{k!},\qquad k=0,1,2,\ldots .
\end{equation}
At fixed entry velocities, convexity gives
\begin{equation}
 \E_J[e^{-h w_i a_{iJ}}]\ge e^{-h\lambda_i},
\end{equation}
where the right side is the probability that primary $i$ has no collision with any of its eligible partners when the rates stay at their initial values. Sampling one partner to estimate the total rate can therefore overestimate the probability of no collision. Weighted P-SBT removes this mixture by using $\lambda_i$ and draws partners in proportion to their frozen entry rates, so it changes both the count law and the partner law. The inequality applies to one primary trial; the sign of the stress error also depends on subsequent collisions and, in a flow, on transport.

\begin{table}
\centering\small
\begin{tabularx}{\textwidth}{@{}>{\raggedright\arraybackslash}p{.23\textwidth}@{\hspace{10pt}}>{\raggedright\arraybackslash}X@{\hspace{12pt}}>{\raggedright\arraybackslash}X@{}}
\toprule
Sampler & Event count or clock & Partner selection and rate updates\\
\midrule
SBT & Bernoulli$(h w_i a_{ij_0})$ & Uniform representative $j_0$; one event at most per primary trial.\\
SBT-exp & Bernoulli$(1-e^{-\mu_i})$, $\mu_i=h w_i a_{ij_0}$ & Same representative and one-event cap as SBT.\\
P-SBT & Poisson$(h w_i a_{ij_0})$ & First partner $j_0$; later partners uniform; count frozen per primary.\\
Weighted P-SBT & Poisson$(h\lambda_i)$ & Entry weights $a_{ij}/\lambda_i$, frozen per primary.\\
EA-P-SBT & Exponential waits at current $\lambda_i$ & Current weighted partner; rates recomputed after each event.\\
NTC & Random-rounded majorant candidates & Uniform nonself proposal; current-speed acceptance.\\
Poisson reference & Poisson majorant candidates & Same proposals and acceptance as NTC.\\
Majorant-frequency family & Upper-frequency null collisions & Historical context \citep{Ivanov1988Majorant}; no separate simulation.\\
\bottomrule
\end{tabularx}
\caption{Collision definitions used here, followed by their historical majorant-frequency context. $h$ is the duration of the collision substep, $j_0$ is uniform in $S_i$, and $\lambda_i=\sum_{j\in S_i}a_{ij}$. Each collision step in CBC uses $h=\Delta t/2$.}
\label{tab:methods}
\end{table}

\subsection{SBT, exponential survival and representative-rate P-SBT}
SBT draws a uniform $j_0\in S_i$, sets $\mu_i=h w_i a_{ij_0}$ and collides $(i,j_0)$ when a uniform random number is less than $\mu_i$. This is a probability only if $0\leq\mu_i\leq1$. In the Couette, Poiseuille, shock and bearing comparisons we stop a Bernoulli calculation when the condition fails; the separate reproduction of the 7.412 ns Fourier case accepts and counts such trials (Sec.~\ref{sec:fourier}).

SBT-exp replaces the acceptance probability by $1-e^{-\mu_i}$, evaluated as $-\operatorname{expm1}(-\mu_i)$, and keeps the representative partner, the permutation and the one-event cap of SBT. Conditional on the representative and the state at primary entry, its zero-event probability equals that of a Poisson count with mean $\mu_i$. SBT instead has zero-event probability $1-\mu_i$ and underestimates survival by $\mu_i^2/2+O(\mu_i^3)$ per step. For one hard-sphere pair the relative speed is invariant, and one isotropic collision erases the initial anisotropy of the relative direction in expectation. At $t=kh$ the exact survival is $e^{-a t}$, whereas SBT gives $(1-ha)^k=e^{-at}[1-a^2th/2+O(h^2)]$. SBT-exp therefore reproduces the exact distribution of the velocities of an isolated pair at the end of a step even though it records at most one event per step. With more particles it retains the effects of representative-rate sampling and the ordered sweep.

P-SBT replaces the count by $K_i\sim\Pois(\mu_i)$. If $K_i>0$, the first event uses $j_0$, which keeps the leading pair-rate weight that supplied the intensity. Later events draw fresh uniform partners from $S_i$ and use the current velocities. The count stays fixed during the primary trial, with no further acceptance test, and the next primary evaluates its representative rate from the updated state.

For a uniform representative $J$, let $R_i=w_i a_{iJ}$. Conditional Poisson sampling gives
\begin{equation}
 \E K_i=h\E R_i,\qquad
 \Var K_i=h\E R_i+h^2\Var R_i.
 \label{eq:mixture}
\end{equation}
Because each possible pair has its own intensity, averaging over the random partner yields a mixture of Poisson distributions. Allowing two or more events restores terms that a Bernoulli count omits, but it does not by itself produce the correct higher-order transition operator. Tying the first event to the representative gives $P_h\varphi=\varphi+h\mathcal L\varphi+O(h^2)$ for the fixed-cell generator (supplementary generator analysis); the coefficient of $h^2$ also depends on intensity mixing, event order and changing pair rates.

\begin{samepage}
\refstepcounter{algorithm}
\noindent\textbf{Algorithm \thealgorithm. Representative-rate P-SBT for one collision volume}

\begin{enumerate}[leftmargin=2em,itemsep=2pt]
\item Randomly permute the $N$ local indices and visit each primary with a nonempty $S_i$.
\item Draw $j_0$ uniformly in $S_i$, evaluate $\mu_i=h w_i a_{ij_0}$, and draw $K_i\sim\Pois(\mu_i)$.
\item If $K_i\geq1$, scatter $(i,j_0)$ with the common conservative map.
\item For events $2,\ldots,K_i$, draw a fresh uniform partner from $S_i$ and scatter using current velocities. Keep the original $K_i$.
\item Continue to the next primary, evaluating its rates from the updated state.
\end{enumerate}
\end{samepage}

\subsection{Weighted-partner P-SBT and EA-P-SBT}
The weighted-partner control evaluates every $a_{ij}$ on entry to a primary trial, sets $\lambda_i=\sum_{j\in S_i}a_{ij}$ and draws $K_i\sim\Pois(h\lambda_i)$. Each event selects $j$ with the entry probability $a_{ij}/\lambda_i$. These rates and probabilities stay frozen while collisions change the velocities and are recomputed only at the next primary.

\begin{figure}[!tp]\centering\includegraphics[width=\textwidth,height=.8\textheight,keepaspectratio]{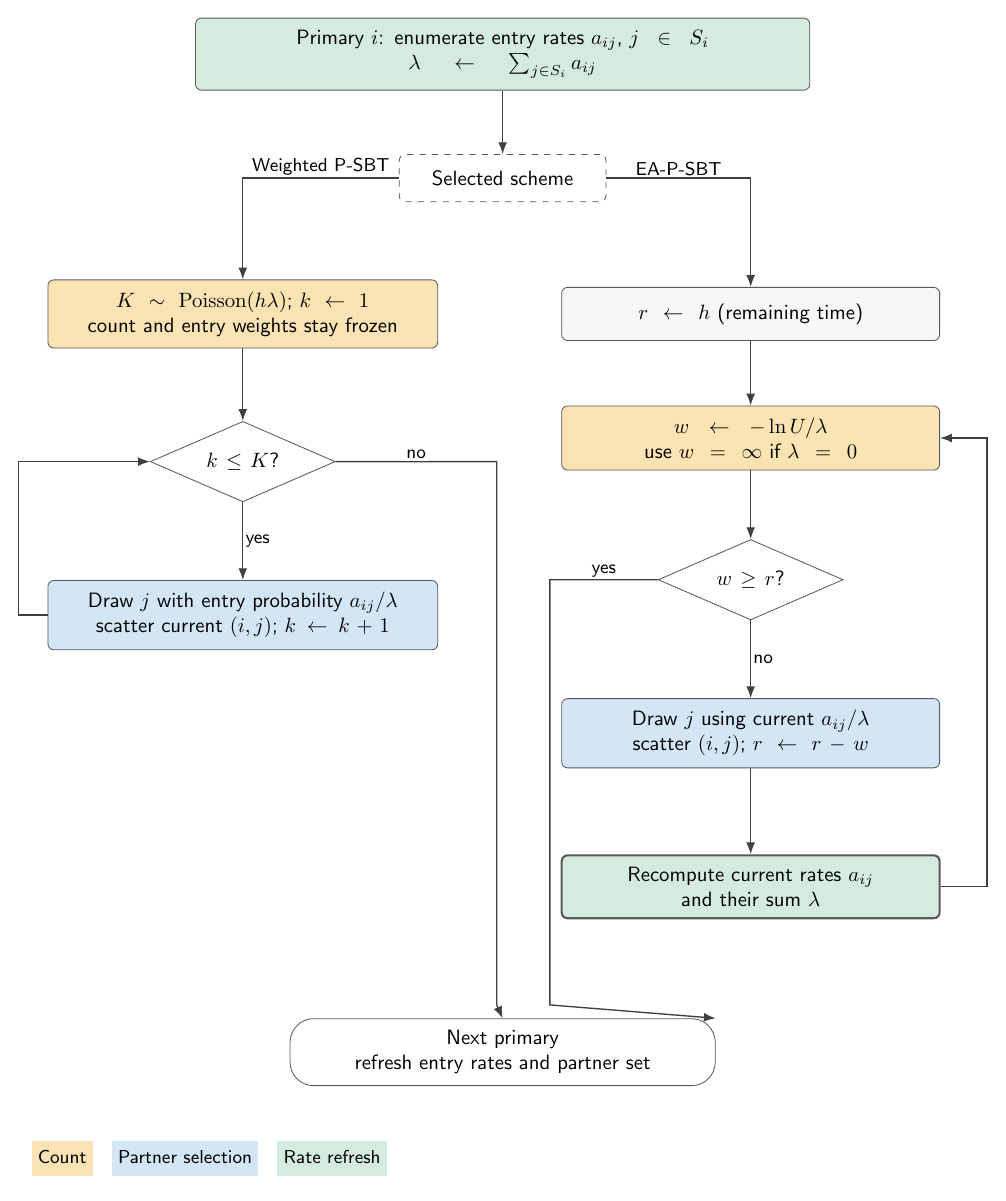}\caption{Partner selection for one primary particle. Weighted P-SBT keeps the initial rates until the next primary trial. EA recalculates them after each collision. Dashed boxes indicate which algorithm is being used.}\label{fig:flowrefresh}\end{figure}

EA-P-SBT recalculates the collision rates of the primary particle with all its eligible partners after each collision. At the start of the trial and after each event it draws the next waiting time, $-\log U/\lambda_i$, with $U$ uniform on $(0,1)$. If the waiting time exceeds the time left in the trial, the trial ends. Otherwise the method chooses a partner in proportion to its current rate and scatters the pair. Table~\ref{tab:methods}, Algorithm~\ref{alg:ea} and Fig.~\ref{fig:flowrefresh} compare the two ways of updating the rates.

\begin{samepage}
\refstepcounter{algorithm}\label{alg:ea}
\noindent\textbf{Algorithm \thealgorithm. EA-P-SBT for one primary particle}

\begin{enumerate}[leftmargin=2em,itemsep=2pt]
\item Set remaining duration $r=h$ and keep the eligible partner set $S_i$ fixed.
\item Evaluate $a_{ij}$ from current velocities and $\lambda_i=\sum_{j\in S_i}a_{ij}$. If $\lambda_i=0$, finish.
\item Draw $w=-\log U/\lambda_i$. If $w\geq r$, finish.
\item Select $j$ with probability $a_{ij}/\lambda_i$, scatter the pair, and set $r\leftarrow r-w$.
\item Return to step 2. When the remaining time has elapsed, proceed to the next primary.
\end{enumerate}
\end{samepage}

Updating pair rates after a collision differs from symmetrizing partner selection in SSBT and SGBT \citep{Taheri2022,Javani2024} and from changing the order of flight and collisions. The waiting-time rule is exact when only the pairs involving the current primary particle and its eligible partners can collide. In a full cell, however, an event changes velocities shared by different primary trials. Processing these trials in sequence still approximates the full-cell collision process. Evaluating every eligible pair costs $O(N^2)$ rate evaluations before any rates are recalculated. With a uniform partner draw of constant cost, the representative SBT and P-SBT sweeps require $O(N+\sum_i K_i)$ sampling operations, excluding sorting and diagnostics.

\subsection{NTC and the Poisson numerical reference}
For the cell-centred velocity $\bu_c$, define the invariant relative-speed bound
\begin{equation}
 G=\left[2\sum_{k=1}^N|\bv_k-\bu_c|^2\right]^{1/2},\qquad
 M=\frac{N(N-1)}{2}\frac{F_{\mathrm{num}}}{V_c}\pi d^2G.
\end{equation}
The bound remains valid through elastic collisions as long as cell membership is fixed during the substep. Our NTC implementation draws $C=\lfloor Mh+U\rfloor$ candidates; each draws two distinct particles uniformly and is accepted with its \emph{current} speed ratio $g_{ij}/G$. The reference instead draws $C\sim\Pois(Mh)$ with the same current-state acceptance, which samples the collision semigroup by uniformization (supplementary generator analysis). Majorant-frequency schemes are the historical setting for this reference \citep{Ivanov1988Majorant}, and event-driven DSMC has used exponential collision clocks before \citep{DonevGarciaAlder2008}. Two comparisons isolate individual choices within an ordered sweep. For a given entry state, SBT-exp and representative-rate P-SBT have the same probability of a first collision with each partner. A difference in their mean stress therefore measures the effect of later collisions. Weighted P-SBT and EA have the same initial rates and partner probabilities, so their mean-stress difference measures the effect of updating rates after a collision. Comparing representative-rate and weighted P-SBT changes both the count distribution and the partner probabilities. The channel solver centres the energy bound on the cell mean velocity and the shock solver uses the laboratory-frame energy; both bounds hold during elastic collision substeps.

In a two-particle channel subcell the centred bound gives $G=g_{12}$, so every candidate is accepted. For $Mh=\mu<1$, NTC with stochastic rounding then has the same survival probability $1-\mu$ as SBT. This identity holds for a single pair under the centred bound; the shock implementation uses the laboratory-frame bound.

\section{Fixed-cell stress relaxation}
\label{sec:relax}
We follow stress relaxation in a homogeneous cell containing two, three or four hard-sphere particles. The particles stay in the cell, so only collisions change their velocities. The initial velocity components are independent Gaussians with longitudinal and transverse standard deviations in the ratio $4:1$. Each realization is shifted to zero mean velocity and normalized to $\sum_i |\bm v_i|^2=N$, so that the pair rate is $|\bm v_i-\bm v_j|/N$. All methods use the same scattering law, and the primary-particle methods shuffle the particles before every sweep. We integrate to $t=3.2$ with $h=0.1$, $0.2$ and $0.4$ and follow the normal-stress anisotropy $A(t)=N^{-1}\sum_i v_{ix}^2-1/3$. Each configuration uses 16 independent groups of 12,000 realizations. The methods share initial states but use independent collision streams, and uncertainties come from the spread of the paired group means. Momentum and kinetic energy are conserved to machine precision by every method.

\begin{figure}[!tbp]
\centering
\includegraphics[width=\linewidth]{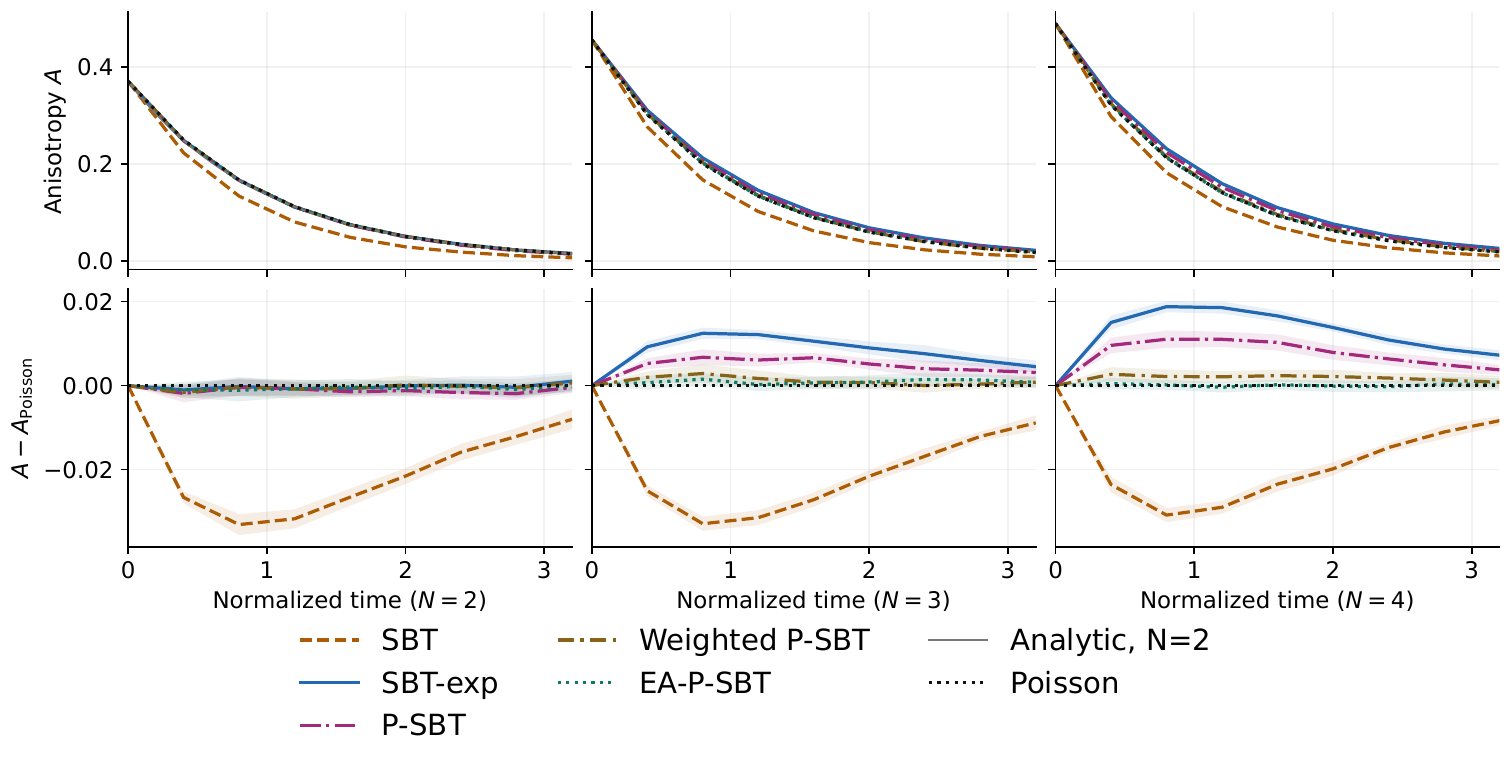}
\caption{Homogeneous normal-stress relaxation at $h=0.4$ for $N=2$, 3 and 4. The top row shows ensemble means and the analytic two-particle exponential. The bottom row shows signed differences from whole-cell Poisson. Bands show two estimated standard errors of the paired differences between group means.}
\label{fig:cellrelax}
\end{figure}

For $N=2$ the zero mean velocity and the energy normalization give a unit pair rate, and the exact mean is $\E[A(t)]=\E[A(0)]e^{-t}$ (Fig.~\ref{fig:cellrelax}). SBT-exp reproduces this curve: at $t=0.8$ its deviations range from $-3.2\times10^{-4}$ to $2.5\times10^{-4}$ over the three steps, each smaller than its standard error. At $h=0.4$ the differences from whole-cell Poisson are $-0.00019\pm0.00114$ for SBT-exp and $-0.03307\pm0.00126$ for SBT. For an isolated pair, the probability of the first collision accounts for the entire SBT error in mean stress.

With three or four particles, the one-event correction reverses the sign of the error: SBT-exp retains more anisotropy than whole-cell Poisson. At $h=0.4$ and $t=0.8$ it is $+0.01247\pm0.00067$ for $N=3$ and $+0.01881\pm0.00067$ for $N=4$. P-SBT keeps the same first-event probability and partner but allows later collisions, which reduces these differences by $0.00571\pm0.00095$ and $0.00774\pm0.00113$. For $N=4$, using the summed initial rate and choosing each partner in proportion to its own rate leaves $+0.00218\pm0.00084$, and recomputing the rates after each event leaves $+0.00018\pm0.00059$. The direct weighted-minus-adaptive difference is $+0.00199\pm0.00078$. All uncertainties are paired standard errors over the 16 groups; the supplement gives the other steps.

A later collision can involve a different partner, whose relative velocity changed in the first collision. For $N=4$ and $h=0.4$, 4.34\% of weighted trials and 4.39\% of adaptive trials contain more than one collision. In 53.0\% and 53.4\% of successive collisions within a trial, respectively, the partner changes. When at least two partners are eligible, a collision changes the pair rates by an average of about 19\% of their pre-collision sum. These updates require more rate evaluations. Selecting events takes 6.00 evaluations per step for weighted P-SBT and 8.14 for EA-P-SBT, against 1.70 candidate rates for whole-cell Poisson.

\section{Time-step criteria and transport splitting}\label{sec:step}
\subsection{Dimensionless flight and collision scales}
A time step must resolve molecular flight across a collision volume and relaxation within it, and a Bernoulli sampler additionally needs valid collision probabilities. With the nominal most probable speed $c_m=\sqrt{2k_BT_w/m}=337.2877$ m s$^{-1}$ we define
\begin{equation}
 \mathcal C_0=\frac{c_m\Delta t}{\delta Y},\qquad
 \mathcal R_0=\frac{\delta Y}{c_m\Delta t}=\mathcal C_0^{-1},\qquad
 \eta_0=\frac{\Delta t}{\tau_0}.
 \label{eq:criteria}
\end{equation}
The ratio $\delta Y/\Delta t$ alone has units of velocity; Eq.~\eqref{eq:criteria} compares flight distance with the subcell width in both orders. The same choice of length scale arises in improved SBT--TAS \citep{Goshayeshi2015TAS} and is discussed in sections~4.4.4 and~4.10 of \citet{RoohiAkhlaghiStefanov2025}. Here the subcell geometry is fixed and the Bernoulli bound is enforced directly.

For a nominal hard-sphere Maxwellian, $\overline c_0=\sqrt{8k_BT_w/(\pi m)}=380.5884$ m s$^{-1}$ and $\tau_0=\lambda_0/\overline c_0=131.3755$ ns. Table~\ref{tab:time} gives the values for the channel calculations. Quoting a step only as a small fraction of $\tau_0$ would hide the much shorter flight time across a subcell.
\begin{table}\centering\small
\begin{tabular}{rrrr}\toprule
$\Delta t$ (ns)&$c_m\Delta t/\delta Y$&$\delta Y/(c_m\Delta t)$&$100\Delta t/\tau_0$ (\%)\\\midrule
3.025965&0.979796&1.0206&2.3033\\
1.008655&0.326599&3.0619&0.7678\\
0.756491&0.244949&4.0825&0.5758\\
0.252164&0.081650&12.2474&0.1919\\
0.126082&0.040825&24.4949&0.0960\\
0.063041&0.020412&48.9898&0.0480\\
\bottomrule\end{tabular}
\caption{Nominal flight and collision scales for a parent cell divided into four collision subcells. The nominal most probable speed $c_m$ is evaluated at $T_w=273.15$ K, the subcell width is $\delta Y=1.041667\,\mu$m and the reference mean collision time is $\tau_0=131.3755$ ns. The collision cell is four times as wide in the Poiseuille comparison without subdivision.}\label{tab:time}\end{table}

At $U_w=1000$ m s$^{-1}$ the wall speed ratio is $U_w/c_m=2.9648$, and viscous heating broadens the distribution of normal molecular velocities. We therefore describe normal flight by $\Delta t\sqrt{\langle C_Y^2\rangle}/\delta Y$ and by quantiles of $|v_Y|\Delta t/\delta Y$. For the collision time we use the local Maxwellian estimate
\begin{equation}
 \tau_M(Y)=\left[n(Y)\pi d^2\,4\sqrt{\theta(Y)/\pi}\right]^{-1},
\end{equation}
which follows from density and scalar temperature and is not the exact collision time of an anisotropic nonequilibrium distribution. The local ranges are kept with the supporting data. A nominal Courant number below one does not guarantee that every particle crosses less than one subcell, because the relevant displacement is $|v_Y|\Delta t$ and viscous heating lengthens it even at fixed wall temperature.

We first reduce the baseline 3.026 and 1.009 ns steps in proportion to the fourfold decrease in collision length, giving 0.756491 and 0.252164 ns. The latter is halved to 0.126082 and 0.063041 ns for the SBT sensitivity test, with a matching 0.126082 ns Poisson calculation. For SBT at 0.252 ns the local estimate $\sqrt{2\theta}\Delta t/\delta Y$ ranges from 0.097 to 0.157, against 0.08165 at the wall temperature. We choose $\Delta t$ from the collision length, the normal velocity distribution and the local relaxation time, and check the Bernoulli probabilities where they apply. A Poisson count permits several events in a substep, but flight distance and splitting error still depend on $\Delta t$. At the exploratory 2.1024 ns step, $\mathcal R_0=1.46897$ and $\mathcal C_0=0.68075$; its separate probability-overrun test is excluded from the Couette comparisons.

\subsection{Order of flight and collisions}
Let $B_h$ denote ballistic flight with wall interactions and $C_h$ a collision update of duration $h$. The names BC, CB and CBC give the order in which the steps occur. Acting on particle states, the corresponding operators are
\begin{equation}
 \mathrm{BC}: C_h B_h,\qquad
 \mathrm{CB}: B_h C_h,\qquad
 \mathrm{CBC}: C_{h/2}B_h C_{h/2}.
\end{equation}
Each CBC half uses its own current membership and collision state. Symmetric composition cancels the leading splitting error when the suboperators are accurate enough \citep{Strang1968}, but an approximate collision kernel keeps its own discretization error. In Poiseuille flow symmetric force kicks surround the flight and collision update, and their momentum and energy contributions are recorded; they do not by themselves make the solver second order.

For particles that reach a wall during flight, we advance each particle to impact, sample the reflected velocity and complete the remaining flight time. Independent records of wall impulses, applied force and transfer across interior faces provide momentum and energy checks. Changing the order of flight and collisions can alter the kinetic shear even when the total momentum flux balances the wall stress.

\section{Sampling and uncertainty}\label{sec:protocol}
Each paired Couette or Poiseuille comparison uses two independent realizations. Matching seed indices give the methods the same initial particle state and the same segment seeds, and continuations keep the particle state and random-number stream. Particle count, force, step and collision law are fixed within a trajectory. The Fourier, shock, cavity and bearing sections state their own sampling intervals.

Each segment in the time-step comparisons covers $0.0605193$~ms of physical time. We discard the first 102 segments ($6.17336$~ms). The Couette profiles use segments $102\leq k<566$, or $28.081$~ms per realization, for every method. The longer wall-stress comparison uses 3072 segments ($185.9153$~ms per realization) at 0.252 and 0.756~ns. The supplement gives the longer sampling times used for low-occupancy Poiseuille flow. Each seed gives one independent trajectory. Subdividing a trajectory estimates temporal variation within that realization.

For each trajectory we average the raw velocity moments in time before calculating central moments. The reported profile is the mean of the two resulting profiles. Uncertainty comes from 16 contiguous temporal intervals regrouped in sets of one, two and four, keeping the largest standard error, or half the separation of the two seed means if that is larger. Shaded ranges show twice this conservative standard error pointwise; they are not simultaneous bands, and two trajectories cannot characterize the general distribution of seed-to-seed variability.

To check stationarity we compare the means of the first and second halves of the common interval with their sampling uncertainty, and we repeat the analysis starting at segment 218 instead of 102, with the same end time. These overlapping intervals test sensitivity to the starting point. All numerical comparisons and uncertainties use unfiltered data, which retain the sampled signs, peaks and asymmetries between the two sides of the channel. Over their $100.704$ and $91.021$ ms records, the changes in wall shear between halves were within one standard error for SBT at 0.063 ns and NTC20 at 0.252 ns; the shorter 0.126 ns Poisson record has a wider uncertainty.

\begin{table}\centering\small
\begin{tabular}{llrrl}\toprule
Run & Sampler & Parent PPC & $\Delta t$ (ns) & Case indices\\\midrule
M01 & SBT & 2 & 0.252164 & 2,3 \\
M02 & SBT & 2 & 0.126082 & 0,1 \\
M03 & SBT & 2 & 0.063041 & 2,3 \\
M04 & P-SBT & 2 & 0.252164 & 6,7 \\
M05 & Weighted P-SBT & 2 & 0.252164 & 10,11 \\
M06 & EA & 2 & 0.252164 & 14,15 \\
M07 & NTC2 & 2 & 0.252164 & 22,23 \\
M08 & NTC20 & 20 & 0.252164 & 26,27 \\
M09 & Poisson & 2 & 0.252164 & 18,19 \\
M10 & Poisson & 2 & 0.126082 & 0,1 \\
\bottomrule\end{tabular}
\caption{Seed pairs and time steps for the Couette comparison. The last column identifies the two realizations within each run. PPC is the mean number of simulator particles per parent cell. Each realization contributes 464 production segments ($102\leq k<566$), or 28.081 ms.}\label{tab:runwindow}
\end{table}

\subsection{Statistical noise and display filtering}\label{sec:display}
DSMC sampling uncertainty depends on the observable, the particle population and correlations between samples \citep{Hadjiconstantinou2003Statistics}. We assess it through temporal grouping and the variation between realizations, and small wall-stress differences use the uncertainty of the difference itself (Sec.~\ref{sec:cost}).

To reduce short-wavelength scatter in selected one-dimensional mean profiles we use a flux-corrected transport (FCT) filter of the type applied to microflows by \citet{Amiri2013} and described by \citet{KaplanOran2001}. Each of 32 passes applies diffusion and slope-limited antidiffusion with coefficient $1/8$; boundary fluxes vanish, the two outermost values on each side stay fixed and the discrete profile sum is conserved to roundoff. All methods in a filtered comparison use the same settings. In Fig.~\ref{fig:shear}, symbols at every 32nd cell, starting at cell 12, show twice the estimated standard error of the unfiltered mean; elsewhere the captions identify filtered profiles and their raw pointwise ranges. Spatial differences, wall stresses and reported uncertainties always use the unfiltered samples. Cavity statistics and bearing profiles use unfiltered observations. The supplement describes the bounded spatial averaging used for the cavity contours. The bearing captions state the angular sector width.

\section{Couette shear: time step, occupancy and ordering}\label{sec:couette}
\subsection{Effect of reducing the time step}
\begin{figure}[!tbp]
\centering\includegraphics[width=\textwidth]{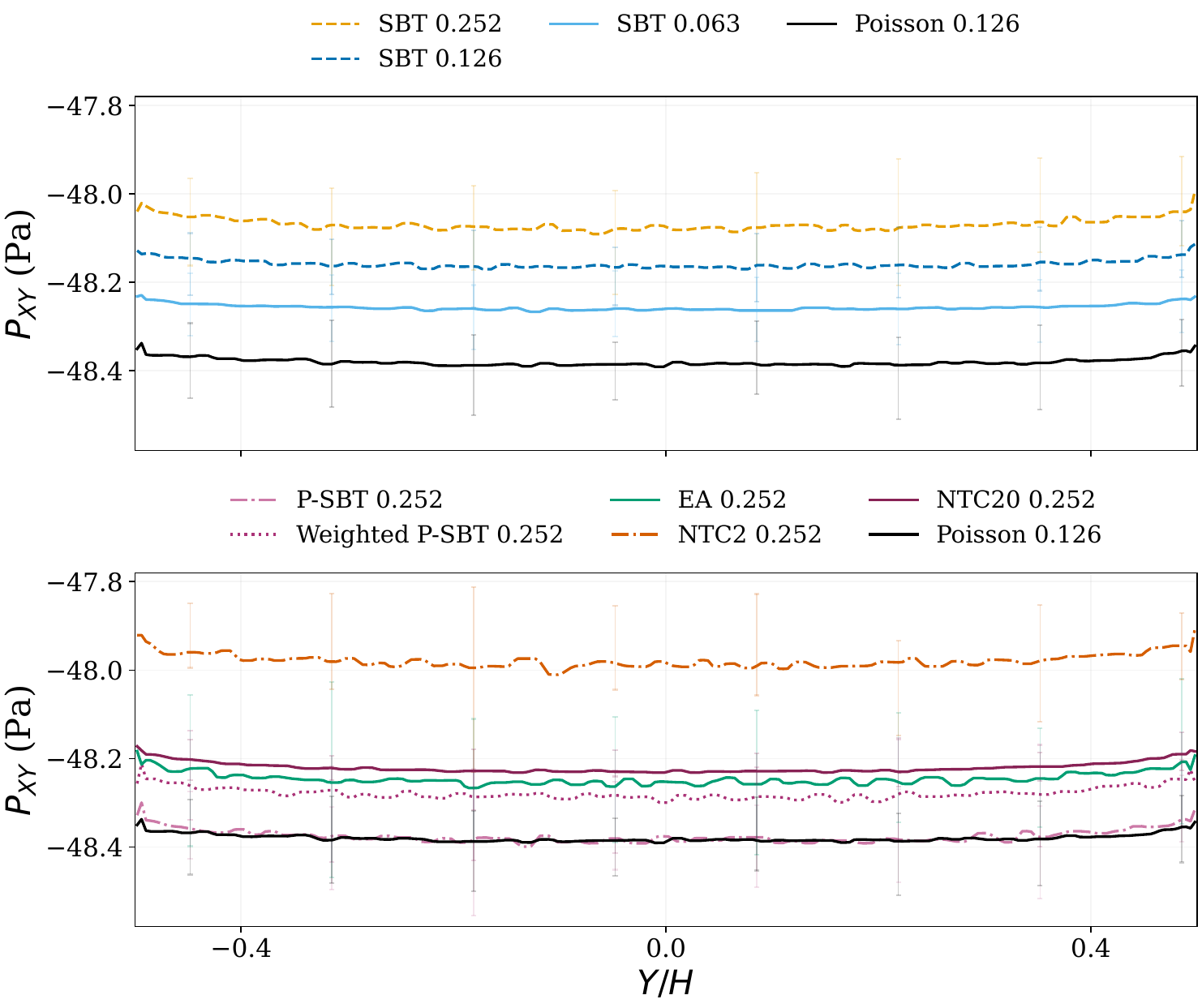}
\caption{Kinetic shear in Couette flow at $U_w=1000$ m s$^{-1}$ and $\Kn=0.05$ with 240 parent cells and four collision subcells per parent. The legend gives the time step in ns. SBT and all Poisson variants use PPC2. NTC is shown at PPC2 and PPC20. Each of two independent trajectories contributes 464 segments ($28.081$~ms) after 102 warm-up segments. The upper panel shows how SBT shear changes as the time step decreases. The lower panel compares the collision schemes and NTC occupancies. Both panels include the same 0.126~ns Poisson reference. Lines show the FCT-filtered means. Bars at every 32nd cell show twice the standard error of the unfiltered mean (Sec.~\ref{sec:display}).}
\label{fig:shear}
\end{figure}

Figure~\ref{fig:shear} compares SBT and Poisson at several steps with the other collision methods over the interval defined in Sec.~\ref{sec:protocol}. All curves share the geometry, the four collision subcells and the wall conditions. SBT and every Poisson and P-SBT variant use parent PPC2, NTC appears at PPC2 and PPC20, and SBT, P-SBT and EA use BC ordering; CBC is tested separately below.

As the SBT step is reduced, the mean kinetic shear moves steadily towards the 0.126 ns Poisson reference. Its spatial average is $-48.0687$ Pa at 0.252 ns, $-48.1586$ Pa at 0.126 ns and $-48.2565$ Pa at 0.063 ns, against $-48.3801$ Pa for the reference. The separations of 0.3114, 0.2215 and 0.1237 Pa amount to about 0.64, 0.46 and 0.26 percent of the reference magnitude. They are separations from a numerical reference over this interval, not errors relative to an exact flow solution.

Table~\ref{tab:common} lists the spatially averaged kinetic shear and the signed shear at each wall over the common interval.
\begin{table}\centering\small
\begin{tabular}{lrrrr}\toprule
Method & $\Delta t$ (ns) & $\overline P_{XY}$ (Pa)& $\tau_L$ (Pa)&$\tau_R$ (Pa)\\\midrule
SBT&0.252&-48.0687&-48.1393&-48.1398\\
SBT&0.126&-48.1586&-48.1986&-48.1982\\
SBT&0.063&-48.2565&-48.2816&-48.2810\\
P-SBT&0.252&-48.3758&-48.4457&-48.4451\\
EA-P-SBT&0.252&-48.2457&-48.3170&-48.3161\\
NTC2&0.252&-47.9788&-48.0489&-48.0488\\
NTC20&0.252&-48.2206&-48.2860&-48.2861\\
Poisson&0.252&-48.3614&-48.4325&-48.4319\\
Poisson&0.126&-48.3801&-48.4202&-48.4208\\
\bottomrule\end{tabular}
\caption{Couette flow over the same $28.081$ ms per realization. $\overline P_{XY}$ is the spatial mean kinetic shear, and $\tau_L$ and $\tau_R$ are the signed stresses at the left and right walls. The mean number of particles per parent cell is two for all runs except NTC20, which uses twenty. The accompanying data give the estimated standard errors.}\label{tab:common}\end{table}

The difference between SBT and the reference persists over these runs. The wall stresses show little drift, but the data do not determine the asymptotic order of convergence. The Poisson kinetic means at 0.252 and 0.126 ns differ by only 0.0187 Pa here, while sharing the same spatial discretization and retaining sampling uncertainty. The longer runs at 0.756 and 0.252 ns do resolve a change in the Poisson wall stress (Sec.~\ref{sec:cost}), so visual overlap at the smaller steps is weak evidence of convergence.

On the same grid and step, NTC20 gives a more negative shear than NTC2 and still differs from the fine Poisson profile. The mean subcell occupancies are five and 0.5. Changing PPC alters both the simulator weight and the distribution of particles among subcells, so this comparison measures their combined effect.

Representative-rate P-SBT stays close to the reference over this interval. Weighted P-SBT replaces the representative intensity by summed pair rates, and EA recomputes the rates after every event. Both keep the leading-order generator and change which collisions can occur within a finite step, so their accuracy has to be judged for each measured quantity at the stated step.

\subsection{Exponential survival and the wall-stress discrepancy}\label{sec:survival_couette}
To measure the effect of pair survival on wall shear, we compare SBT with SBT-exp in Couette flow at $U_w=1000$ m s$^{-1}$ and $\Kn=0.05$. The 240 parent cells each contain four collision subcells. A mean of two particles per parent gives 0.5 per subcell. SBT-exp uses the BC ordering, initial states and segment seeds of SBT and accepts a primary collision with probability $1-\exp(-\mu_i)$. The two steps correspond to $100\Delta t/\tau_0=0.192$ and $0.576$ at the wall conditions of Sec.~\ref{sec:step}. After discarding the first 102 blocks we compare SBT-exp, SBT and whole-cell Poisson in the same subcells. The sampling time is $\WallFineWindow$ ms per seed at 0.252 ns and $\WallCoarseWindow$ ms at 0.756 ns. We divide the impulse on each wall by its area and the sampling time, then average the two stresses using the same sign convention for momentum flux.

\begin{table}[!tbp]\centering\small
\setlength{\tabcolsep}{5pt}
\begin{tabular}{rrrrrrr}\toprule
& & \multicolumn{2}{c}{Difference from Poisson (Pa)}
& \multicolumn{3}{c}{SBT-exp accepted events (\%)}\\
$\Delta t$ (ns) & ms per seed & SBT & SBT-exp & $N=2$ & $N=3$ & $N\ge4$\\\midrule
0.252 & 71.292 & $+0.271\pm0.031$ & $-0.045\pm0.033$ & 58.32 & 30.75 & 10.93\\
0.756 & 185.915 & $+0.782\pm0.019$ & $-0.170\pm0.018$ & 58.47 & 30.69 & 10.85\\
\bottomrule\end{tabular}
\caption{Signed differences in mean wall stress from whole-cell Poisson at the same step, and percentage of accepted SBT-exp collisions at each instantaneous subcell population $N$. At each step the methods are compared over the same duration for both seeds. Uncertainties give one estimated standard error of each paired difference.}
\label{tab:survivalwall}\end{table}

\begin{figure}[!tbp]
\centering
\includegraphics[width=\linewidth]{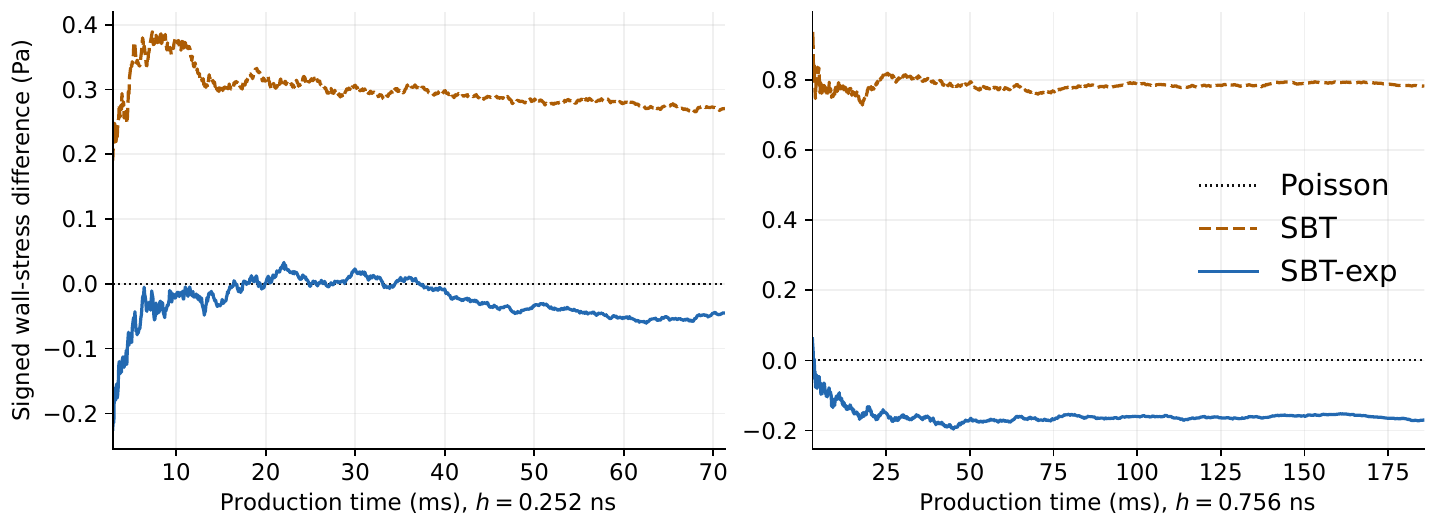}
\caption{Cumulative signed wall-stress differences in Couette flow, relative to Poisson at the same time step. Each point compares all samplers over the same elapsed time and the same two seed indices. The endpoint corresponds to Table~\ref{tab:survivalwall}. Cumulative estimates share samples.}
\label{fig:survivalwall}
\end{figure}

Exponential survival reduces the magnitude of the SBT offset at both steps (Table~\ref{tab:survivalwall}). At 0.252 ns the remaining difference is comparable to the sampling uncertainty. At 0.756 ns it reverses sign: $-0.170$ Pa with an estimated standard error of 0.018 Pa, negative in both realizations. Figure~\ref{fig:survivalwall} shows the cumulative difference as a function of sampling time. For each comparison we group matched time blocks and report the largest of the paired standard error, the zero-covariance estimate and half the absolute difference between the estimates from the two seeds.

The negative difference at 0.756 ns is consistent with the homogeneous result for $N\ge3$. There the one-event cap of SBT-exp suppresses later collisions and slows relaxation, and at a fixed local shear rate slower relaxation would mean a larger shear-stress magnitude. The wall-driven flow also adjusts its gradients and slip, however, so the cell test helps interpret the difference but predicts neither its sign nor its size.

At each collision step we record the number $N$ of particles in each subcell and the number of accepted collisions at that occupancy. About 58\% of accepted SBT-exp collisions occur at $N=2$, although such subcells make up only 7.6\% of visits, and about 31\% occur at $N=3$. These fractions depend on the available pairs, their relative speeds and the acceptance probability. They show how often an isolated pair collides. The wall impulse also depends on subsequent encounters and flight to the wall, so the 58\% cannot be interpreted as a fraction of the change in wall shear.

For a subcell containing two particles, the Bernoulli rule overestimates the probability of a first collision by $\mu-(1-\exp(-\mu))$. For small $\mu$ the relative excess over many visits is approximately $\sum_j\mu_j^2/(2\sum_j\mu_j)$, where $j$ runs over two-particle visits. Estimating this ratio requires the intensities for all two-particle visits, including those that end without a collision.

\subsection{Wall transfer and collision--flight ordering}
\begin{figure}[!tbp]
\centering\includegraphics[width=\textwidth]{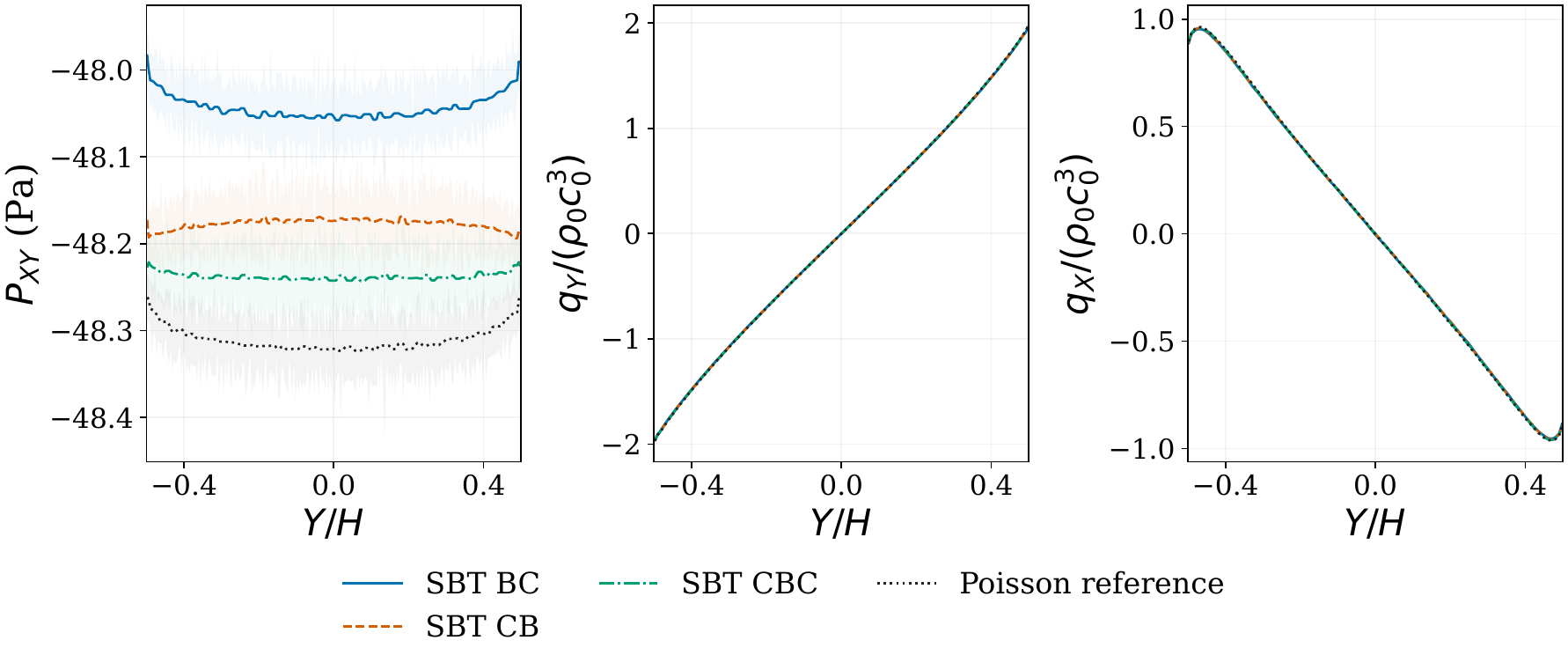}
\caption{Couette comparisons of pure SBT with BC, CB and CBC ordering and the BC Poisson numerical reference, at $\Delta t=0.252164$ ns, parent PPC2 and four collision subcells. Each of two trajectories contributes $185.915$ ms after warm-up. The panels give kinetic shear, transverse heat flux and streamwise heat flux. Bands are twice the estimated standard errors. Mean profiles use 32 FCT passes. Bands use unfiltered samples (section~\ref{sec:display}).}
\label{fig:split}
\end{figure}

The BC, CB and CBC comparison at 0.252 ns uses $185.915$ ms per independent seed. Changing the chronological order changes the kinetic-shear profile, while the heat-flux curves remain much closer at the plotted scale (Fig.~\ref{fig:split}). The separation persists as the sampling interval grows: a longer trajectory narrows the uncertainty about the value for a given ordering, whereas a different ordering changes that value.

\begin{figure}[!tbp]
\centering\includegraphics[width=.98\textwidth]{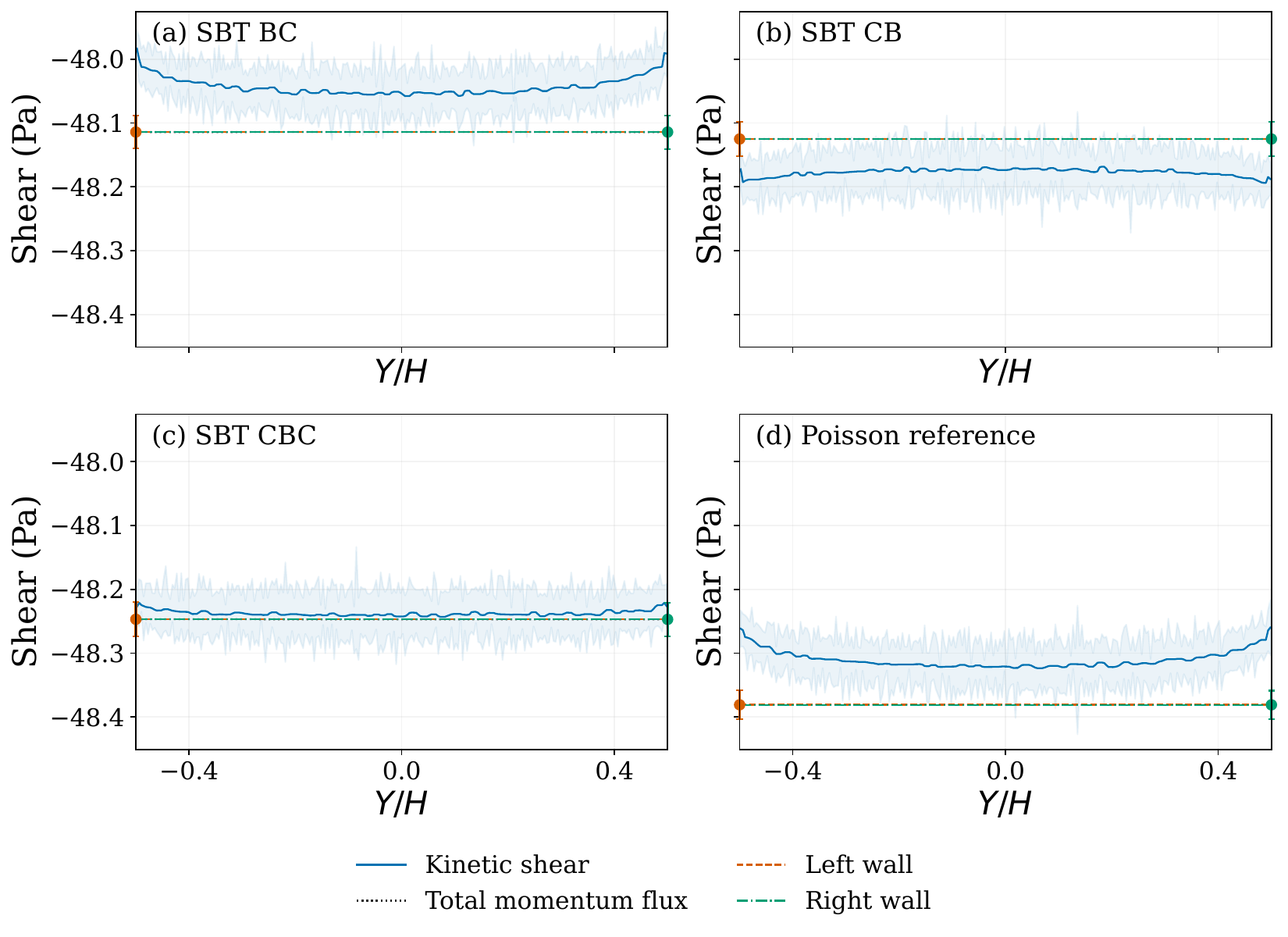}
\caption{Wall and interior shear over $185.915$ ms per realization at 0.252164 ns. The kinetic shear is sampled in parent cells; the total momentum flux includes collision transfer across interior faces. Horizontal levels and endpoint markers identify the two wall transfers, with their normal orientations mapped to the same sign convention. At a boundary, face transfer equals the recorded wall impulse. Bands and error bars are twice the estimated standard errors. Mean profiles use 32 FCT passes. Bands use unfiltered samples (section~\ref{sec:display}).}
\label{fig:wall}
\end{figure}

Figure~\ref{fig:wall} compares each kinetic profile with its own wall stresses and total interior momentum flux, and Table~\ref{tab:wall} gives the spatial averages. Among the SBT orderings, CBC brings the kinetic shear closest to the wall stress. In every case the total interior flux agrees closely with the stresses at both walls, as the discrete momentum balance requires, even though the kinetic and collisional contributions differ between methods.

\begin{table}\centering\small
\begin{tabular}{l@{\quad}r@{\quad}r@{\quad}r@{\quad}r}\toprule
Update&Kinetic (Pa)&Left wall (Pa)&Right wall (Pa)&Offset (Pa)\\\midrule
SBT BC&-48.0434&-48.1141&-48.1141&+0.0707\\
SBT CB&-48.1770&-48.1250&-48.1250&-0.0520\\
SBT CBC&-48.2376&-48.2469&-48.2471&+0.0094\\
Poisson reference&-48.3105&-48.3809&-48.3810&+0.0705\\
\bottomrule\end{tabular}
\caption{Kinetic shear averaged across the channel and signed stresses at the left and right walls over $185.915$ ms per realization at $\Delta t=0.252164$ ns. The offset is the mean kinetic shear minus the average of the two wall stresses. All rows use two particles per parent cell on average and four collision subcells per parent. Figure~\ref{fig:wall} compares the total interior momentum flux with the wall stresses.}\label{tab:wall}\end{table}

Collisions between particles on opposite sides of a sampling face carry momentum across it even when neither particle crosses the face. This contribution enters the total flux but is absent from the kinetic stress calculated from particle velocities. Changing when velocities are sampled within the step can therefore change the kinetic shear. CBC brings it closest to the wall stress on this grid. Each method can balance its own wall impulse against the interior flux and still predict a different magnitude of wall shear.

\section{Heat flux and higher moments}\label{sec:higher}
\begin{figure}[!tbp]
\centering
\includegraphics[width=\linewidth]{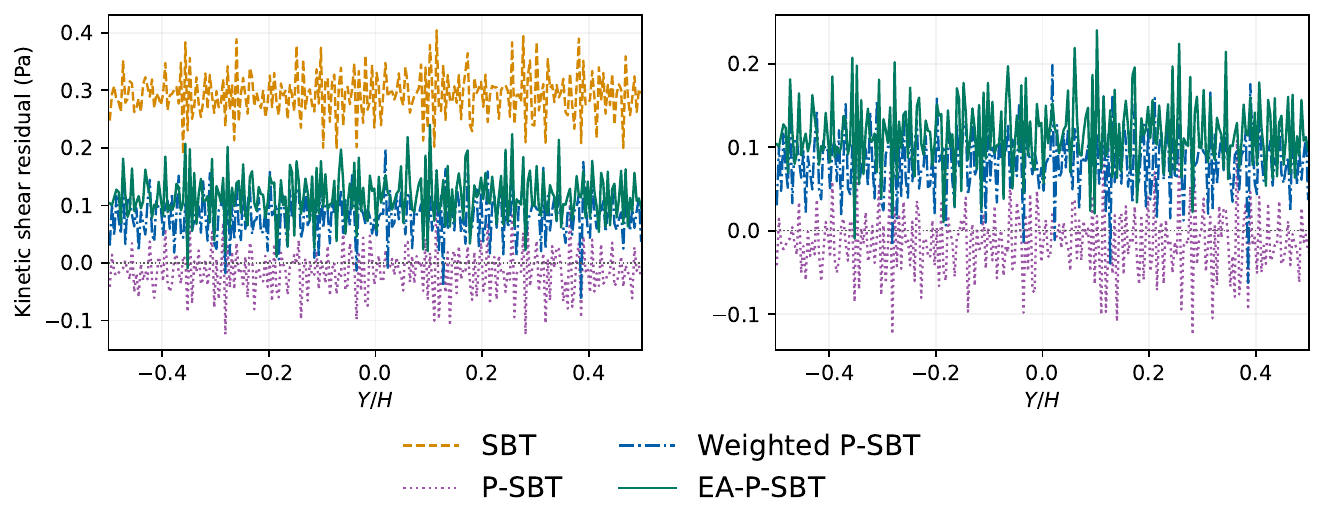}
\caption{Unfiltered differences in kinetic shear from same-step Poisson at 0.252 ns for parent PPC2. Both panels use the same 28.081 ms per realization as Fig.~\ref{fig:shear}. The right panel enlarges the differences for the Poisson primary schemes.}
\label{fig:shearresidual}
\end{figure}

Heat flux and higher moments give increasing weight to particles with large peculiar velocities. In the Couette profiles, the transverse heat flux and transverse fifth-order contraction vary little among methods at the plotted scale. The clearest differences in those moments occur in the streamwise direction, particularly near the extrema of the NTC20 heat flux. The supplementary profiles use the same $28.081$~ms per trajectory as Fig.~\ref{fig:shear}. Figure~\ref{fig:shearresidual} shows the corresponding differences in unfiltered kinetic shear.

For some methods, the largest local difference in $R_{YY}$ from the 0.126~ns Poisson result reaches several percent of the peak reference magnitude. We use this peak for normalization because the local reference can pass through zero. A pointwise maximum is sensitive to sampling noise. The accompanying data therefore give the signed profiles of all 20 quantities, standard errors from 16 intervals, and the differences between the two halves of each record and the effect of starting later. For these runs, shear distinguishes the collision methods and the order of flight and collisions more clearly than most higher moments.

\section{Fourier heat conduction}\label{sec:fourier}
In Fourier flow, a temperature difference drives a heat flux through gas with no mean flow. The stationary walls of a 1 mm gap reflect hard-sphere argon diffusely at 223.15 and 323.15 K. The gas starts at 273.15 K and 266.644 Pa. With $d=3.658\times10^{-10}$ m and $m=6.63\times10^{-26}$ kg the Knudsen number based on the gap is 0.02379. The hotter wall lies at larger $x$, so $q_x<0$; we compare magnitudes with the positive published values.

We first compare with two published SBT benchmarks of \citet{Stefanov2019Fourier}, in which the wall flux is known to be sensitive to cell width and time step \citep{Taheri2017,Stefanov2019Fourier}. With 400 cells, five particles per cell and $\Delta t=7.412$ ns, two independent SBT realizations give 1512.15 and 1501.54 W\,m$^{-2}$ over 22.26 ms each. Their mean, 1506.85 W\,m$^{-2}$, lies 0.34\% below the 1512 W\,m$^{-2}$ high-population NTC benchmark and 1.90\% below the published 1536 W\,m$^{-2}$ SBT value at the same nominal settings; the two realizations differ by 10.61 W\,m$^{-2}$. At this step a nominal SBT probability occasionally exceeds one. To reproduce the published calculation, we accept and count each such trial. Including warm-up, we recorded 57 and 70 cases among about $3.69\times10^{8}$ accepted collisions per realization. A second case in Stefanov's Table~II uses 120,000 cells and only 12,000 simulator particles, a mean occupancy of 0.1, with $\Delta t=0.024707$ ns. Here the mean SBT wall flux is 1510.03 W\,m$^{-2}$ over 0.81972 ms per realization, against the published 1514 W\,m$^{-2}$. The 3.97 W\,m$^{-2}$ difference is smaller than the 13.28 W\,m$^{-2}$ standard error estimated from the two seeds. These comparisons check the solver against published wall fluxes. The time step, total particle count and sampling time all change along with occupancy.

We compare the samplers on grids of 200, 400 or 800 collision cells using 46 pairs of realizations. For these runs NTC uses stochastic rounding of the candidate count. At 800 cells and 0.3125 ns, Poisson40 and NTC40 give $-1502.97$ and $-1505.55$ W\,m$^{-2}$ over a common 0.8192 ms per realization; their 0.17\% difference lies within the sampling uncertainty. We use Poisson40 as the numerical reference. At two particles per cell, all samplers give larger wall-flux magnitudes than the reference. Over the time steps tested, the differences are 9.4--12.5\% on 200 cells, 4.0--10.9\% on 400 cells and 0.8--5.9\% on 800 cells, using all samples. At fixed occupancy, refining the grid also increases the particle population from 400 to 1,600. Either change can affect the flux. The batch standard errors are 14--23 W\,m$^{-2}$, or about 1--1.5\%. At a fixed grid and step the samplers differ from one another by 2--7 percentage points without a consistent ranking, except that representative-rate P-SBT gives the largest magnitude in five of the seven PPC2 settings. The largest spread, at 400 cells and 1.25 ns, is comparable to the effect of the grid. The low-population samplers give similar temperatures, yet in some cases their heat-flux profiles differ by several percent (see the supplement). Temperature measures the mean peculiar kinetic energy. Heat flux also weights that energy by the signed velocity normal to the wall.

For the case with two particles per cell and 24,240 particles in total studied by \citet{Taheri2017}, we also compared the archived Fortran solver with the present implementation over the same sampling interval. The mean wall-flux magnitudes were 1533.34 and 1533.39~W\,m$^{-2}$ for the Fortran run and for the present solver with particles visited in identifier order. With random ordering the latter gave 1503.25~W\,m$^{-2}$. The supplement gives the fluxes at each wall, their sampling uncertainty and a separate Poisson comparison with eight seeds.

\section{Normal-shock structure at Mach 2 and Mach 5}\label{sec:shock}
\subsection{Gas properties, reference scales and sampling}
Normal shocks test the collision models in strongly compressive flow, and their density and temperature profiles have long served to compare shock calculations \citep{Alsmeyer1976}. All shocks use one hard-sphere model with $m=6.64\times10^{-26}$ kg, $d=4.092\times10^{-10}$ m, $T_1=293$ K and $n_1=10^{20}$ m$^{-3}$, separate from the channel-flow parameters. Dimensional quantities use $k_B=1.380622\times10^{-23}$ J K$^{-1}$, as in the solver. The Mach-2 and Mach-5 runs use 800 uniform collision cells without subdivision. The baseline runs have two particles per cell on average upstream. Additional NTC and whole-cell Poisson calculations use higher upstream populations (Table~\ref{tab:shockruns}). The mean cell occupancy rises as the gas passes through the shock.

The reference scales are $c_0=(k_BT_1/m)^{1/2}=246.824$ m s$^{-1}$, $t_0=42.659589\,\mu$s and $L_0=c_0t_0=10.5294$ mm, and the upstream hard-sphere mean free path and mean collision time are $\lambda_1=13.4420$ mm and $\tau_1=34.1277\,\mu$s. The steps $\Delta t^*=\Delta t/t_0=0.005$ and $0.015$ therefore correspond to $213.298$ and $639.894$ ns, or $0.625$ and $1.875$ percent of $\tau_1$. Table~\ref{tab:shockruns} gives the domains and cell widths, and differences are evaluated on the interval $|x/L_0|\leq12$ in the respective coordinates.

\begin{table}\centering\small
\setlength{\tabcolsep}{6pt}
\begin{tabularx}{\textwidth}{lcc>{\raggedright\arraybackslash}X}\toprule
Quantity & Mach 2 & Mach 5 & Common specification\\\midrule
Domain length $L/L_0$ & 60 & 150 & 800 full collision cells\\
Cell width $\delta x/L_0$ & 0.075 & 0.1875 & No collision subdivision\\
$\Delta t/t_0$ & 0.005, 0.015 & 0.005, 0.015 & 213.298, 639.894 ns\\
$100\Delta t/\tau_1$ & 0.625, 1.875 & 0.625, 1.875 & Upstream reference state\\
Upstream PPC & 2, 20, 40 & 2, 20, 40 & Additional populations for NTC and Poisson\\
Sampling time per run (ms) & 133.098 & 133.098 & Two seeds per curve\\
\bottomrule\end{tabularx}
\caption{Parameters for the paired Mach-2 and Mach-5 comparisons. We excluded the Mach-5 SBT run at the larger step because its Bernoulli collision probability exceeded one.}\label{tab:shockruns}\end{table}

The original Mach-2 and Mach-5 profile comparisons each use two realizations over $210\leq t/t_0\leq3330$. The solver checks the population feedback every 1,000 steps, so the physical interval between checks varies with $\Delta t^*$. All shock runs use isotropic hard-sphere scattering.

\subsection{Shock motion and variation between Poisson realizations}\label{sec:shock2}
A normal shock held in the domain by population feedback still wanders, and averaging a moving front in fixed coordinates broadens every profile. We therefore estimate the shock position $x_s$ in each time block. The density is integrated over $|x/L_0|\le 15$, and $x_s$ is the location of the step between the upstream and downstream Rankine--Hugoniot densities that has the same integral. The 56 raw moments of each block are then translated onto a common grid in $(x-x_s)/L_0$ with their integrals conserved, and we average the shifted raw moments for each realization before forming its central moments. Because the translation uses the same length $L_0$ for all methods at a given Mach number, differences in thickness remain visible. Profiles in the original coordinates are kept for comparison. Each realization contributes 133.098 ms of data on its original grid, and each method is compared with Poisson at the same Mach number and step.

\begin{figure}[!tbp]\centering
\includegraphics[width=\linewidth]{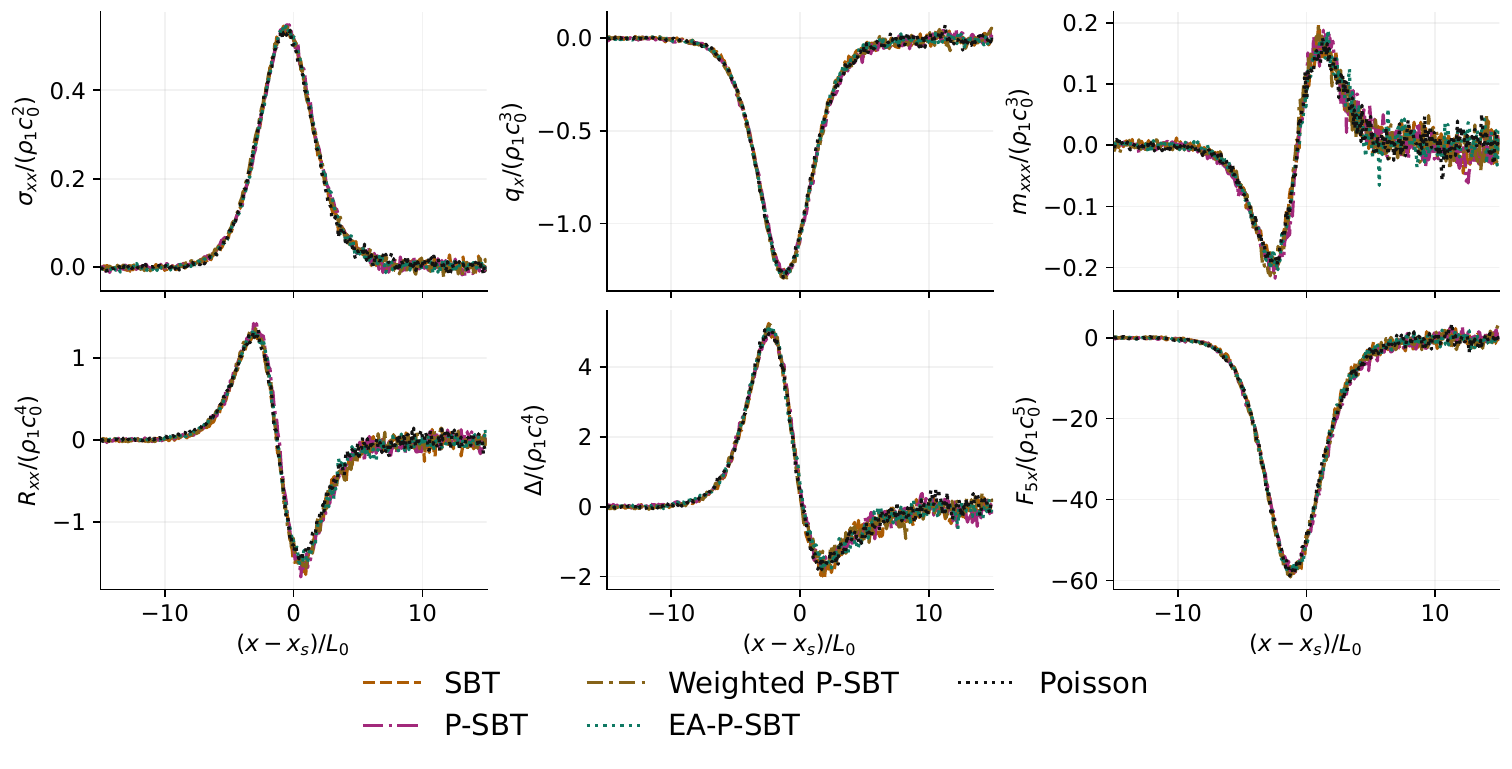}
\caption{Aligned Mach-2 moments at $\Delta t/t_0=0.005$. The panels show $\sigma_{xx}$, $q_x$, $m_{xxx}$, $R_{xx}$, $\Delta$ and $F_{5x}$ in reading order. Lines show the unfiltered mean of two realizations. The supplement also compares calculations with more particles per cell.}
\label{fig:shock2moments}
\end{figure}

\begin{figure}[!tbp]\centering
\includegraphics[width=\linewidth]{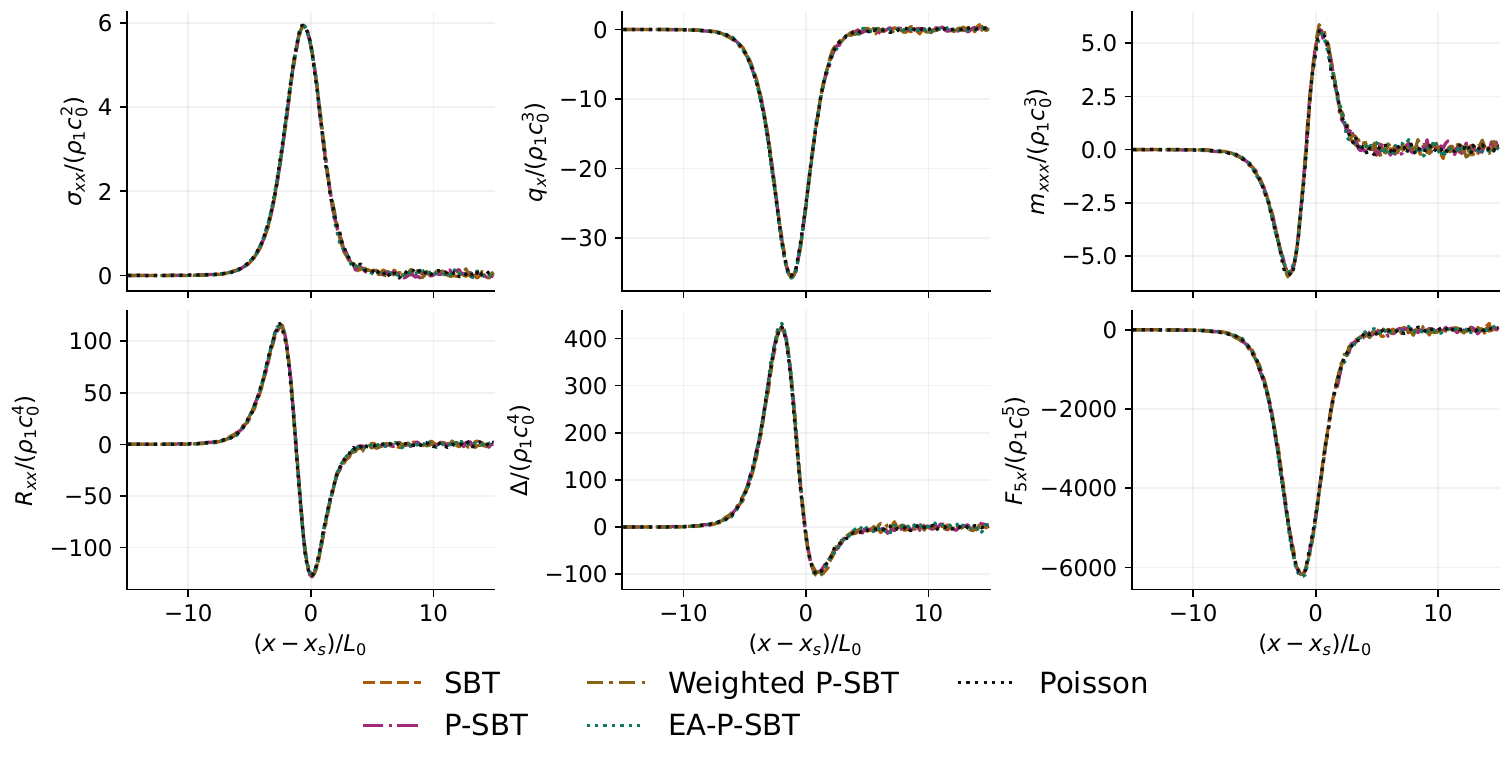}
\caption{Aligned Mach-5 moments at $\Delta t/t_0=0.005$. The panels have the same definitions as in Fig.~\ref{fig:shock2moments} and use the Mach-5 scales. Each curve averages two realizations over the same 133.098 ms.}
\label{fig:shock5moments}
\end{figure}

\begin{table}[!tbp]\centering\small
\begin{tabular}{rrrrrr}\toprule
& & \multicolumn{2}{c}{Normal stress} & \multicolumn{2}{c}{Heat flux}\\
Mach & $\Delta t/t_0$ & Fixed & Aligned & Fixed & Aligned\\\midrule
2 & 0.005 & 9.23 & 7.33 & 9.46 & 7.35\\
2 & 0.015 & 13.22 & 10.29 & 11.67 & 8.84\\
5 & 0.005 & 5.02 & 3.99 & 4.67 & 3.61\\
5 & 0.015 & 8.63 & 5.82 & 6.90 & 5.16\\
\bottomrule\end{tabular}
\caption{Relative $L_2$ differences (\%) between the two Poisson realizations at two particles per cell on $|x/L_0|\le 12$ before and after shifting the origin to follow the shock. The denominator is the norm of their mean.}
\label{tab:shockseed}\end{table}

Alignment removes part of the apparent disagreement between methods. At Mach 2 and the finer step it lowers the P-SBT stress difference relative to Poisson from 8.68\% to 6.53\%. For the four PPC2 primary-particle methods the aligned differences are 5.06--6.53\% in stress and 4.57--4.86\% in heat flux (Fig.~\ref{fig:shock2moments} and the supplement). These must be judged against the scatter of the reference itself (Table~\ref{tab:shockseed}). The two aligned Poisson realizations differ by 7.33\% in stress and 7.35\% in heat flux. For two independent estimates, each averaged over two realizations of similar variance, dividing this separation by $\sqrt{2}$ gives an estimate of their sampling variability. That scale is 5.18\% in stress and 5.20\% in heat flux at Mach 2, and 2.82\% and 2.55\% at Mach 5. Only two reference realizations are available, so these estimates are approximate.

At Mach 5 and the finer step the two Poisson realizations differ by 3.99\% in stress and 3.61\% in heat flux. Alignment lowers the SBT stress difference from 5.84\% to 2.88\%, close to the sampling scale. At the finer step, the differences in stress and heat flux among the low-occupancy methods are comparable to the variation between realizations. Alignment does not act equally on every moment. At the larger Mach-5 step it lowers the stress difference from 9.70\% to 4.58\% for EA and from 8.34\% to 4.76\% for weighted P-SBT, but raises the heat-flux difference of representative-rate P-SBT from 3.58\% to 4.35\%. Figures~\ref{fig:shock2moments} and \ref{fig:shock5moments} show the signed stress, heat flux and higher moments.

Increasing the number of particles per cell produces a larger change in the Mach-5 stress profile than changing the sampler at two particles per cell. At the finer Mach-5 step, the aligned stresses of Poisson40 and NTC20 differ from Poisson2 by 14.47\% and 14.03\%, respectively. The comparison between the two Poisson calculations keeps the sampler fixed and isolates the effect of changing the particle count, including its effect on the motion of the front. The NTC20 comparison also changes the sampler. Taking Poisson40 as the reference, the aligned stress differences are 15.32\% for EA, 15.72\% for weighted P-SBT and 0.94\% for NTC20 (supplement). The PPC2 shock profiles thus still depend on the number of particles at this resolution. On a 2,400-cell Mach-5 grid, aligned NTC20 and Poisson40 differ by only 0.96\% in normal stress and 0.68\% in heat flux over 64.42~ms per realization. The supplement gives a separate 800-to-2,400-cell comparison over 133.10~ms and the finer PPC2 grids used at Mach 2, 5 and 10. The Mach-5 SBT run at $\Delta t/t_0=0.015$ was excluded because its representative intensity exceeded one. When steps are compared, the physical interval between population corrections also changes. The accompanying data give the normalized $L_2$ differences before and after alignment for each pair of runs compared.

\subsection{Independent hard-sphere reference at Mach 2}\label{sec:ohwada}
\begin{figure}[!tbp]\centering
\includegraphics[width=\linewidth]{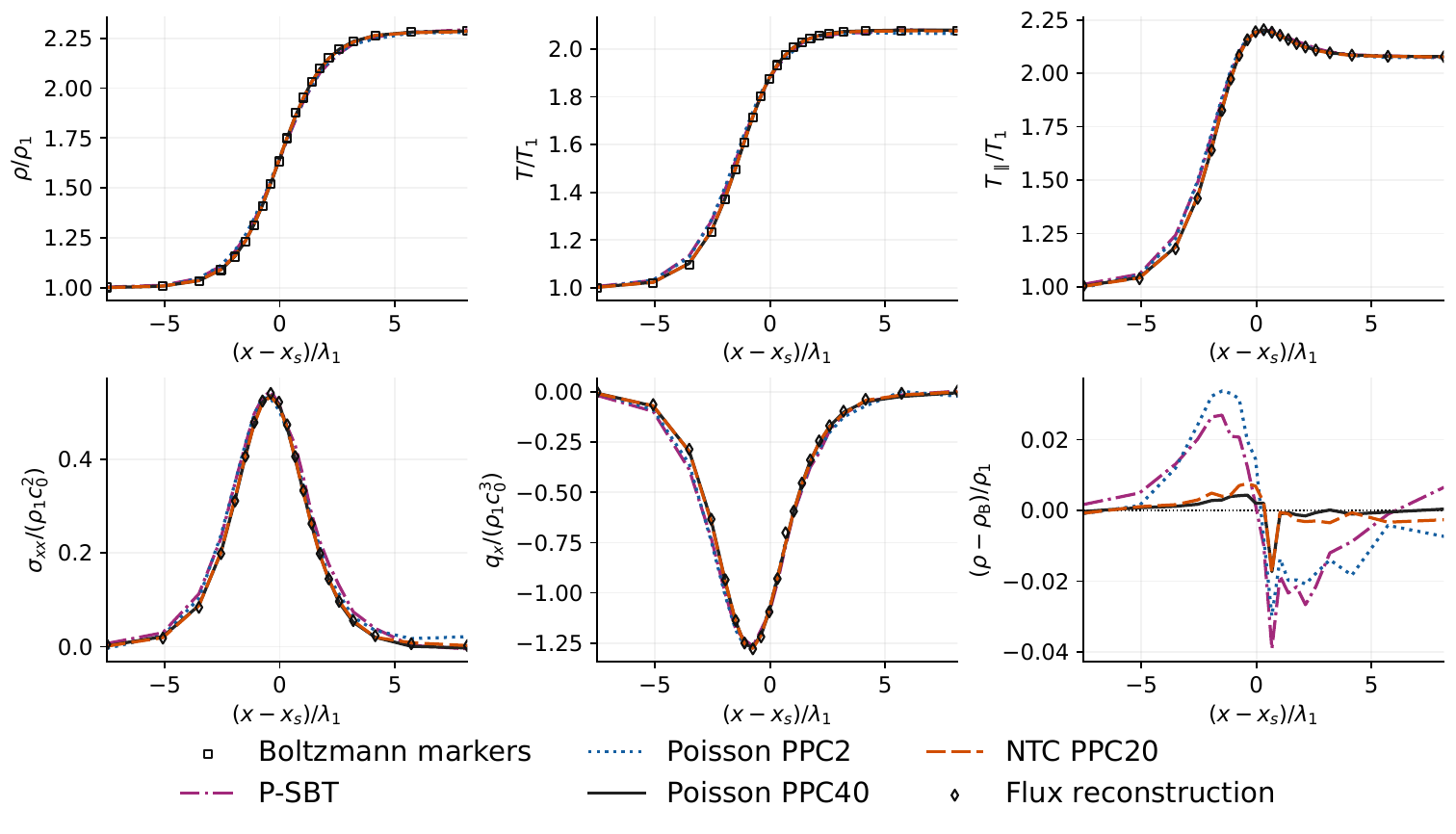}
\caption{Mach-2 profiles from selected PPC2 samplers and the two higher-occupancy calculations at the 21 published points, with distances expressed in upstream mean free paths. The DSMC collision step is $\Delta t/t_0=0.005$. Black symbols denote digitized Ohwada density and scalar temperature in the first two panels and the conservation-based reconstructions of directional temperature, stress and heat flux. The density difference in the last panel is measured from the digitized reference at the same coordinates. Table~\ref{tab:independent_mach2} includes all tested samplers.}
\label{fig:ohwada}
\end{figure}

The comparisons above use a finite-resolution Poisson calculation as the reference. At Mach 2, Ohwada's direct numerical solution of the hard-sphere Boltzmann equation provides an independent reference \citep{Ohwada1993}. \citet{DoduladTcheremissine2013} discuss the numerical accuracy of such solutions for hard-sphere and Lennard--Jones shocks. Figure~\ref{fig:ohwada} compares the density and temperature profiles and the quantities reconstructed from them. We recover 21 density and scalar-temperature values from the centres of the plotted symbols in Fig.~1 of \citet{OhwadaXu2004}. The same markers appear in Fig.~2 of \citet{XuHuang2011}, whose text defines the abscissa as $\xi=X/[(\sqrt{\pi}/2)\lambda_1]$; matching positions and values across the two figures identify this convention for both. We convert the reference positions by $X/\lambda_1=(\sqrt{\pi}/2)\xi$ and keep the DSMC coordinates physical, with $\lambda_1/L_0=16/(5\sqrt{2\pi})$ for hard spheres. Each reference or DSMC profile is centred by its density integral over the common interval $\pm7\lambda_1$, which lies within the reference coverage, and the DSMC seed means are evaluated at all 21 converted points. The width $\delta_{10\text{--}90}$ is the distance between the 10\% and 90\% levels of the equilibrium density rise.

Let $r=\rho/\rho_1$, $\Theta=T/T_1$ and $U^\dagger=u_x/c_0$, with upstream speed $U_1^\dagger=M_1\sqrt{5/3}$ for upstream Mach number $M_1$. Steady mass and momentum fluxes give
\begin{equation}
 U^\dagger=\frac{U_1^\dagger}{r},\qquad
 \frac{T_\parallel}{T_1}=\frac{1+(U_1^\dagger)^2}{r}-\frac{(U_1^\dagger)^2}{r^2}.
\end{equation}
The dimensionless stress is $s=r(T_\parallel/T_1-\Theta)$, and the energy flux gives $q_x/(\rho_1 c_0^3)=U_1^\dagger[\tfrac52(1-\Theta)+\tfrac12((U_1^\dagger)^2-(U^\dagger)^2)]-U^\dagger s$. These relations reconstruct the directional temperature, stress and heat flux from the density and temperature markers.

\begin{table}[!tbp]\centering\small
\setlength{\tabcolsep}{6pt}
\begin{tabular}{lrrrr}\toprule
Sampler & $\delta_{10\text{--}90}/\lambda_1$ &
$\rho/\rho_1$ RMSE & $T/T_1$ RMSE & $\sigma_{x_s}/L_0$\\\midrule
Ohwada (Boltzmann) & 4.38 & --- & --- & ---\\
NTC PPC2 & 4.69 & 0.0152 & 0.0180 & 0.69\\
SBT & 4.65 & 0.0165 & 0.0203 & 0.75\\
P-SBT & 4.81 & 0.0189 & 0.0194 & 0.72\\
Poisson PPC2 & 4.81 & 0.0205 & 0.0242 & 0.73\\
Poisson PPC40 & 4.41 & 0.0043 & 0.0033 & 0.17\\
Weighted P-SBT & 4.79 & 0.0204 & 0.0271 & 0.75\\
EA-P-SBT & 4.71 & 0.0192 & 0.0214 & 0.74\\
NTC PPC20 & 4.44 & 0.0053 & 0.0040 & 0.23\\
\bottomrule\end{tabular}
\caption{Mach-2 hard-sphere comparison on 800 cells with $\Delta t/t_0=0.005$ and two seeds for DSMC. The width and root-mean-square errors (RMSE) are calculated at all 21 published points after multiplying the published coordinate by $\sqrt{\pi}/2$ to obtain $X/\lambda_1$. Each density profile is centred by its integral over $\pm7\lambda_1$. The last column gives the standard deviation of the front position across time blocks, averaged over the two realizations.}
\label{tab:independent_mach2}
\end{table}

The higher-occupancy calculations reproduce the Boltzmann shock thickness within 1.6\% (Table~\ref{tab:independent_mach2}). On 800 cells at $\Delta t/t_0=0.005$ the converted reference width is $4.377\lambda_1$, and Poisson40 and NTC20 give $4.412\lambda_1$ and $4.445\lambda_1$, larger by 0.81\% and 1.56\%. Their root-mean-square deviations, normalized by the upstream values, are 0.0043 and 0.0053 in density and 0.0033 and 0.0040 in temperature. The PPC2 profiles are wider: Poisson PPC2 and representative-rate P-SBT give $4.808\lambda_1$ and $4.813\lambda_1$, about 10\% above the reference. The front also wanders more at low particle count. The standard deviation of its position across time blocks is about $0.17L_0$ for Poisson40 and $0.73L_0$ for Poisson2. Section~\ref{sec:samplecadence} tests how much of the extra width comes from motion within a block. Widths and errors use all 21 markers and the full 10--90\% rise. One published point, near $\xi=0.40$, departs from its neighbours; it controls the steepest adjacent secant, so we do not use a secant width (see the supplement).

\subsubsection{Physical scales and the reference coordinate}
The high-occupancy solvers use the hard-sphere pair collision rate $(F_{\mathrm{num}}/V_c)\pi d^2g$, a viscosity exponent of $1/2$ and isotropic scattering, so their molecular model and their definition of $\lambda_1$ match those of the Boltzmann reference. The upstream Maxwellian collision frequency and mean free path are
\begin{equation}
 \nu_1=n_1\pi d^2\frac{4}{\sqrt{\pi}}\sqrt{\frac{k_BT_1}{m}}
 =2.9301735\times10^4\ \mathrm{s}^{-1},\qquad
 \lambda_1=\frac{\sqrt{8k_BT_1/(\pi m)}}{\nu_1},
\end{equation}
which give $\tau_1=\nu_1^{-1}=34.1277\,\mu$s. This mean collision time differs from the reference time $t_0=\mu_1/(n_1k_BT_1)$ used to specify the steps.

We checked the collision frequency of the solver directly by counting accepted collisions in the upstream interval $-27<x/L_0<-18$. Each collision contributes once for each of the two particles, so
\begin{equation}
 \widehat{\nu}=\frac{2N_{\rm accepted}}{\int N_{\rm up}(t)\,dt}.
\end{equation}
We counted collisions over $3400$--$3405\,t_0$ for Poisson40 and $3330$--$3335\,t_0$ for NTC20, using copies of the saved trajectories. Pooling the two realizations gives 29,799 accepted events for Poisson40 and 15,037 for NTC20, or rates of $29237$ and $29445\ {\rm s}^{-1}$, which depart from $\nu_1$ by $-0.22\%$ and $+0.49\%$. We estimated uncertainty from the larger of the variation across time bins and half the difference between realizations. The resulting uncertainties are $1.88\%$ and $1.61\%$ of $\nu_1$. The measured frequencies are consistent with the upstream value within these uncertainties. The short samples do not resolve differences between the algorithms.

The coordinate convention matters. Ohwada's earlier account writes $X=(\sqrt{\pi}/2)\lambda_1\xi$ \citep{OhwadaRIMS1993}, and the later statement of \citet{XuHuang2011}, together with the matching markers, establishes the same convention for the republished Mach-2 data. The 10--90\% width of 4.938 read in the printed coordinate is therefore $4.938\sqrt{\pi}/2=4.377$ physical mean free paths; reading the abbreviated axis label as $X/\lambda_1$ would create a spurious difference of about 11\%. The DSMC widths in Table~\ref{tab:independent_mach2} are calculated at the converted coordinates, because rescaling a width interpolated at other sites would not give the same discrete estimate. The supplement documents the comparison of the plotted points in the two publications.

\subsection{Effect of shock motion within a sampling block}\label{sec:samplecadence}
The profiles reported above average 50 samples in each $5t_0$ block. Motion within this interval can broaden the mean density profile even after the blocks have been aligned. To measure how much this motion widens the Mach-2 profile, we continue two realizations each of Poisson PPC2, Poisson PPC40 and NTC PPC20 from saved states over $3400\le t/t_0\le3600$, keeping the 800-cell grid, $h/t_0=0.005$, the molecular model and the feedback schedule. We verified that recording density more frequently left the particle trajectories and the next block of raw moments unchanged. Each seed gives 40 blocks and 2,000 snapshots spaced $0.1t_0$ apart; we analyse these short runs separately.

\begin{figure}[!tbp]\centering
\includegraphics[width=\linewidth]{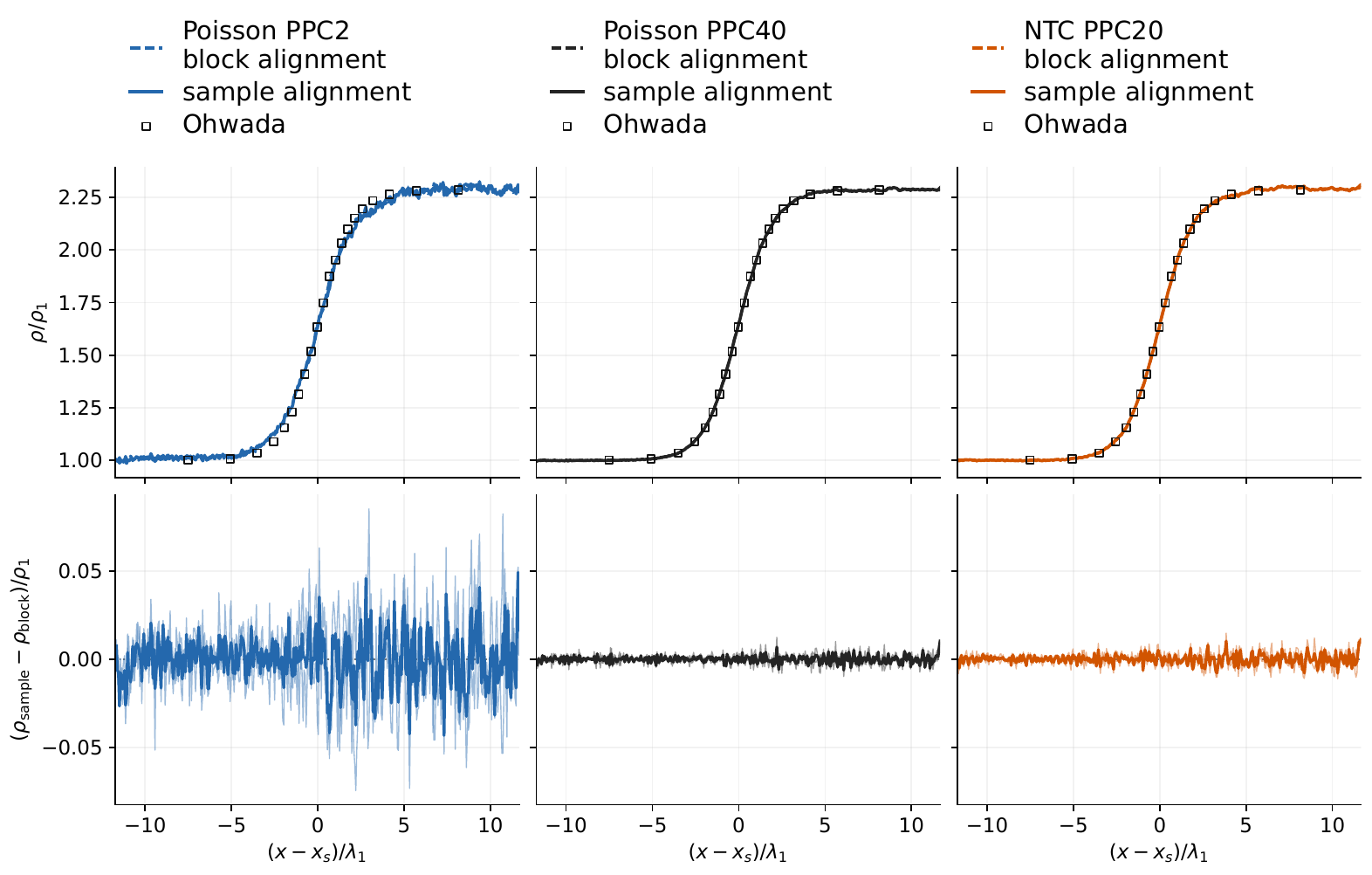}
\caption{Mach-2 density profiles over $3400\le t/t_0\le3600$. Columns show Poisson PPC2, Poisson PPC40 and NTC PPC20. Upper panels compare block-aligned means (dashed), snapshot-aligned means (solid) and the digitized Boltzmann markers. Lower panels show the difference obtained by aligning snapshots instead of blocks. Thin curves show the individual realizations and thick curves their mean. All profiles use the same physical length scale.}
\label{fig:samplecadence}
\end{figure}

We align the same density data in two ways. We either shift each snapshot by its own front position before averaging or average the snapshots within each block before shifting (Fig.~\ref{fig:samplecadence} and Table~\ref{tab:samplecadence}). The front position again comes from the density integral on $|x/L_0|\le15$, and translations conserve the cell counts. At the published coordinates, each seed's block-aligned mean is centred by its density integral over $\pm7\lambda_1$, and the same translation is applied to its snapshot-aligned mean, so that the difference isolates the alignment interval. Widths use the original density data and the physical length scale.

\begin{table}[!tbp]\centering\small
\setlength{\tabcolsep}{5pt}
\begin{tabular}{lrrrrr}\toprule
& \multicolumn{2}{c}{$\delta_{10\text{--}90}/\lambda_1$} & & \multicolumn{2}{c}{Density RMSE}\\
Sampler & Block & Snapshot & Mean width change & Block & Snapshot\\\midrule
Poisson PPC2 & 5.136 & 5.329 & $+0.289\pm 0.181$ & 0.0272 & 0.0328\\
Poisson PPC40 & 4.434 & 4.458 & $+0.028\pm 0.031$ & 0.0056 & 0.0058\\
NTC PPC20 & 4.405 & 4.414 & $+0.020\pm 0.027$ & 0.0068 & 0.0075\\
\bottomrule\end{tabular}
\caption{Mach-2 density profiles aligned by block or by snapshot over $3400\le t/t_0\le3600$. Widths and root-mean-square errors (RMSE) in density use the 21 published coordinates in Table~\ref{tab:independent_mach2}. After converting the reference coordinates by $\sqrt{\pi}/2$, the digitized shock width is $4.377\lambda_1$. At each threshold, the upward piecewise-linear crossing nearest the common origin is used, with the same rule for every profile. The first two columns are widths of the two-seed mean profiles. The next column gives the average of the width changes calculated for each realization. The number following $\pm$ is half the difference between those changes. Because threshold width is nonlinear, these two ways of averaging need not give the same change.}
\label{tab:samplecadence}
\end{table}

The root-mean-square variation in front position within a block is $0.321L_0$ for Poisson PPC2, $0.070L_0$ for Poisson PPC40 and $0.100L_0$ for NTC PPC20. Aligning individual snapshots changes the density profiles by 0.82\%, 0.11\% and 0.16\% in relative $L_2$ norm. For the higher particle counts the width hardly changes: from $4.434$ to $4.458\lambda_1$ for Poisson40 and from $4.405$ to $4.414\lambda_1$ for NTC20, against the reference $4.377\lambda_1$.

At PPC2 the result depends on how the width is measured. Snapshot alignment changes the width of the two-seed mean from $5.136$ to $5.329\lambda_1$; evaluated seed by seed at the reference sites the changes are $+0.109\lambda_1$ and $+0.470\lambda_1$, and on the native grid $-0.077\lambda_1$ and $+0.255\lambda_1$. Noisy profiles cross a threshold several times, so the width of a mean need not equal the mean of the widths, and sparse sites alter the interpolation. The root-mean-square error in density changes from 0.0272 to 0.0328, against 0.0056 to 0.0058 for Poisson40 and 0.0068 to 0.0075 for NTC20. Finite-particle noise also perturbs the estimated front position, so this test measures noise and motion together and leaves even the sign of the PPC2 width change on the native grid uncertain. With 20 or 40 particles per cell, the shock thickness remains close to the Boltzmann value under both alignment procedures. At two particles per cell, the inferred thickness is sensitive to particle number and to the estimated position of the front.

\subsection{Upstream-moving molecules in the Mach-5 shock}\label{sec:mach5-return}
Density and heat flux alone give little indication of what fraction of the local population has upstream axial velocity. We separate particles by the sign of their axial velocity $v_x$ in the shock frame. In the velocity marginal $G_x(v_x)=\int f\,\mathrm{d}v_y\,\mathrm{d}v_z$, the region $v_x<0$ contains the upstream-moving population. We use the mean velocity and temperature of \emph{all} particles in each sampling slab to form the peculiar velocity and the central moments of each group. With this convention the two contributions add to the full $q_x$ or $R_{xx}$.

We sample nine DSMC cases on 800 cells at $\Delta t/t_0=0.005$, with two trajectories each. All trajectories cover $3400\le t/t_0\le5750$, or 470 blocks and 100.25~ms per trajectory. Each $5t_0$ block contributes 50 samples. We centre each block at its density midpoint, selecting the rising crossing nearest the integral position when several occur, and average over slabs of width $0.3\lambda_1$. We compare the particle measurements with Poisson40 and with a separately computed physics-informed neural network (PINN) approximation to the hard-sphere Boltzmann equation. Roohi presented the velocity distribution at the APS Division of Fluid Dynamics meeting \citep{Roohi2026APSShock}. Section~7.2 of the preprint \citep{Roohi2026KineticPartI} describes the Mach-5 hard-sphere calculation.

\begin{figure}[!tbp]\centering
\includegraphics[width=\linewidth]{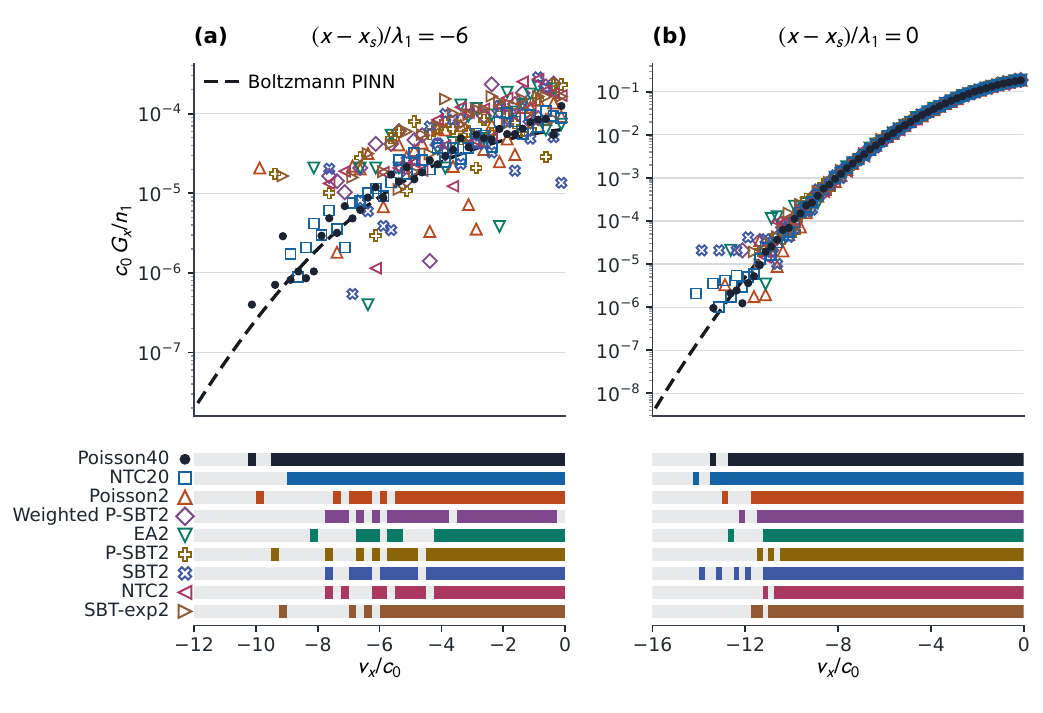}
\caption{Negative-velocity side of the Mach-5 marginal in a slab six upstream mean free paths ahead of the density midpoint and at the midpoint. Symbols give the mean bin densities from two DSMC seeds. The strips identify occupied bins for each sampler. The dashed curve is the independently calculated hard-sphere Boltzmann PINN marginal averaged over the same spatial slab. The supplement shows the full signed velocity range.}
\label{fig:m5-return-tail}
\end{figure}

At the midpoint the upstream-moving fraction is 21.07\% for Poisson40 and 21.15\% in the PINN. At this point the upstream-moving and forward-moving groups contribute $-53.14$ and $+28.44$ to $q_x/(\rho_1c_0^3)$ in Poisson40. Their sum, $-24.70$, is close to the PINN value of $-25.08$. Six mean free paths upstream, the fraction moving upstream falls to 0.0293\% in Poisson40 and 0.0226\% in the PINN. Even this small Poisson40 population carries 16.5\% of the scalar fourth central moment $\int |\boldsymbol C|^4 f\,\mathrm{d}\boldsymbol v$, where $\boldsymbol C$ is the peculiar velocity. At two particles per cell, each realization contributes only 24--62 upstream-moving particle observations to this slab after fractional bin weighting. Repeated sampling can count the same particle more than once. Figure~\ref{fig:m5-return-tail} shows which velocity bins have observations. The gaps in the far tail limit comparisons between individual schemes there.

One mean free path upstream, all nine DSMC estimates of $q_x$ are within 1.4\% of Poisson40, yet some fourth moments differ appreciably at the same position. In units of $\rho_1 c_0^4$, Poisson40 gives $R_{xx}=39.97$ and SBT at PPC2 gives $26.40$. Both terms in $R_{xx}=A_{xx}-7(k_BT/m)\sigma_{xx}$ are much larger than their difference. Here $A_{xx}$ is the traceless part of the contracted fourth central moment. The two terms are 248.96 and 209.00 for Poisson40, and 254.14 and 227.73 for SBT. Consequently, small relative changes in these terms can shift $R_{xx}$ substantially.

\begin{figure}[!tbp]\centering
\includegraphics[width=\linewidth]{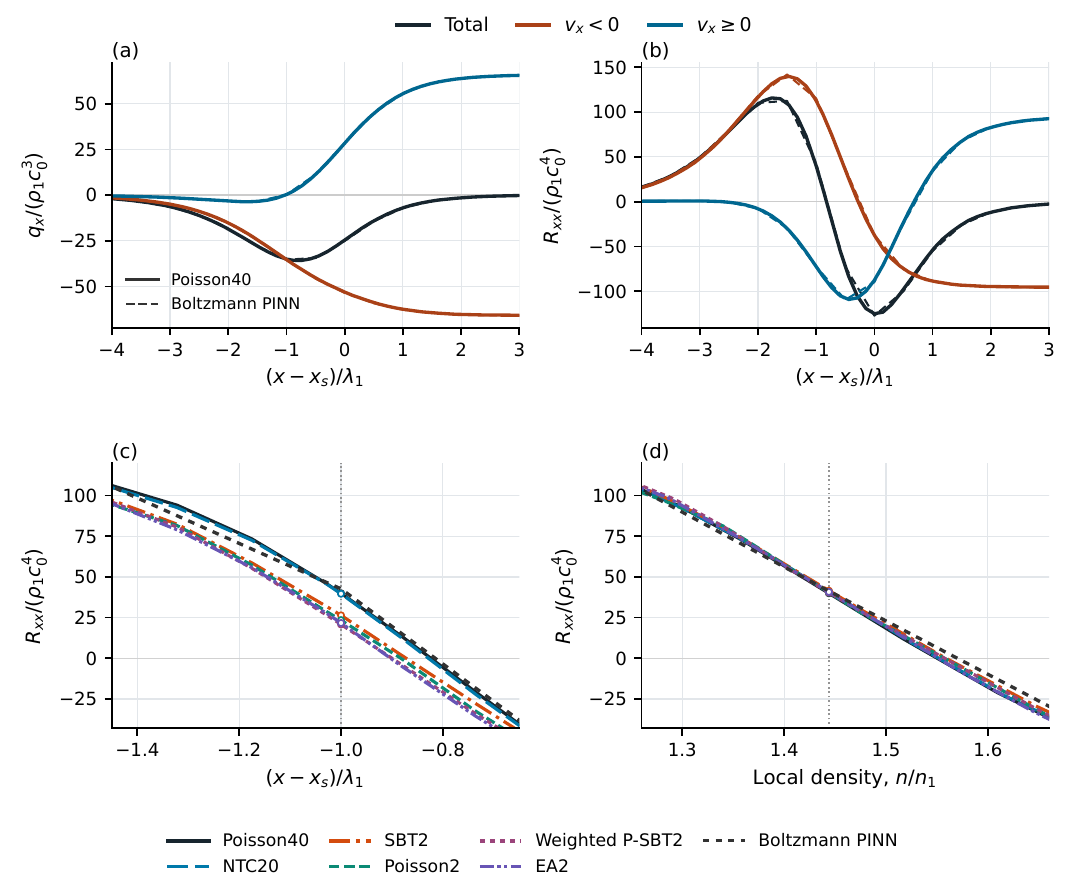}
\caption{Mach-5 transport through the shock. Panels (a,b) resolve heat flux and the contracted fourth moment into contributions from the two velocity signs. The solid curves give Poisson40 slab means and the dashed curves the independent Boltzmann PINN on its spatial grid. Panels (c,d) compare the total fourth moment on the upstream side of the shock, first at a fixed position and then at the same local density. Vertical lines mark the position and corresponding density used in the comparison. Circles denote the two-seed means and vertical segments span the two estimates. Curves in (d) use interpolation within each seed's profile before averaging.}
\label{fig:m5-return-transport}
\end{figure}

Figure~\ref{fig:m5-return-transport} compares the fourth moment at a fixed distance from the front and at equal local density. The Poisson40 density at $(x-x_s)/\lambda_1=-1$ is $1.44393n_1$. The two SBT profiles reach this density at $-1.083$ and $-1.080$; their fourth moments there are 40.92 and 42.18, compared with 39.72 and 40.22 for Poisson40. The mean SBT difference changes from $-33.9\%$ at fixed position to $+3.9\%$ at equal density. For all seven PPC2 methods the equal-density differences range from $+0.96\%$ to $+3.94\%$ relative to Poisson40. Over the full 10--90\% density rise, Poisson40, NTC20 and the slab-averaged PINN give widths of $2.727\lambda_1$, $2.736\lambda_1$ and $2.760\lambda_1$, respectively. The seven PPC2 widths span $2.897$--$2.974\lambda_1$. The broader density rise thus explains much of the local fourth-moment difference even after the density midpoints have been aligned.
We also find this dependence on the choice of coordinate at Mach 2 and Mach 10. For weighted P-SBT on the upstream flank, matching the local density reduces the relative $L_2$ difference in $R_{xx}$ from 20.0\% to 1.8\% at Mach 2 and from 8.2\% to 0.72\% on the 6000-cell Mach-10 grid. The supplement gives the corresponding heat-flux profiles and both spatial comparisons.

The supplement aligns the Mach-10 profiles with the same density integral. It compares the 6,000-cell PPC2 results with NTC20 and Poisson40 on 2,400 cells, includes the 4,000-cell runs and fixed-coordinate profiles, and holds the physical interval between population checks fixed when the step changes.

\FloatBarrier
\section{Transient lid-driven cavity}\label{sec:cavity}
When the lid starts moving, it transfers momentum to the gas and drives a recirculating flow. The lid of a 1~mm square moves at Mach~0.5 above gas initially at rest with $\Kn=0.1$. We use $64\times64$ collision cells, a common step of 11.58~ns and 21,000 trajectories per method, and define $t^*=t c_m/L$ with $L=1$ mm and $c_m=\sqrt{2k_BT_0/m}=337.29$ m\,s$^{-1}$ at the initial temperature $T_0=273.15$ K. We compare the developing flow at the same times after the lid starts.

\begin{figure}[htbp]
\centering
\includegraphics[width=.99\linewidth]{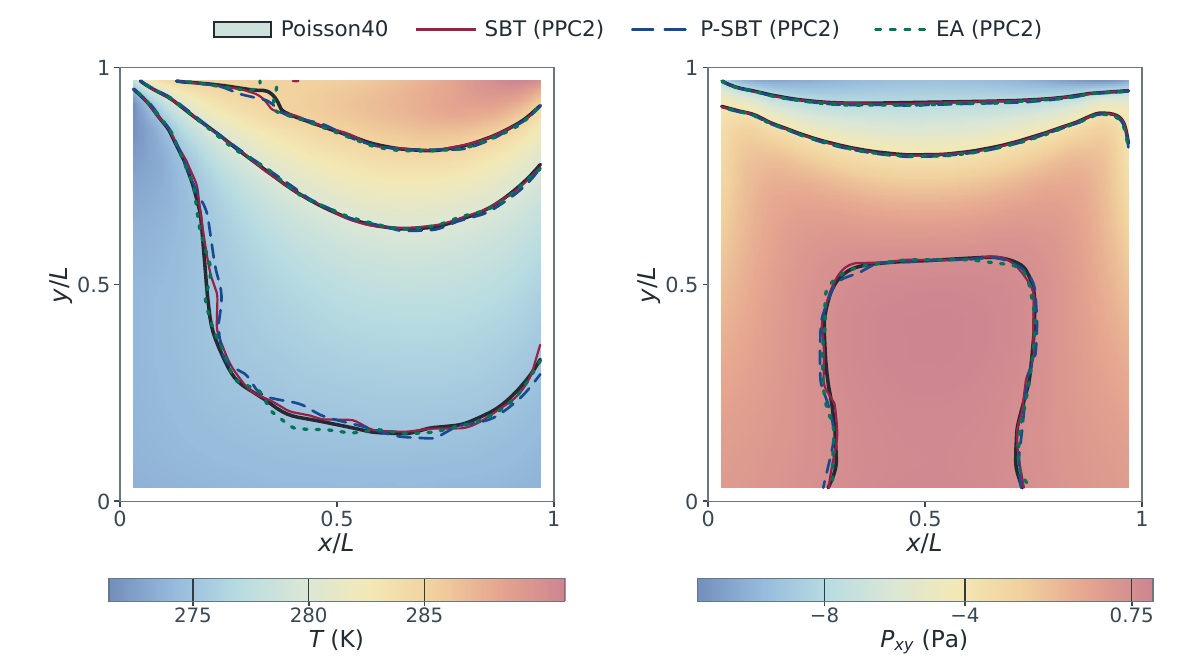}
\caption{Transient square cavity with the lid moving to the right at $y/L=1$. Each field is averaged over nine times spanning $4.5\le t^*\le5$. The color field and gray contour edges show Poisson40. The lines show the same contour levels for SBT (solid burgundy), P-SBT (dashed blue) and EA (dotted green). The isotherms are 275, 280 and 285~K. The kinetic shear levels are $-8$, $-4$ and $+0.75$~Pa. The contours join the centers of the $16\times16$ reporting bins after the bounded spatial averaging described in the supplement.}
\label{fig:cavity-fields}
\end{figure}

We average the fields over nine equally spaced times between $t^*=4.5$ and $t^*=5$. The simultaneous uncertainty intervals for the temperature differences from Poisson40 include zero at every bin. The root-mean-square differences from Poisson40 across the cavity are 0.116, 0.149 and 0.130~K for SBT, representative-rate P-SBT and EA at PPC2 (Fig.~\ref{fig:cavity-fields}). For P-SBT and EA the simultaneous intervals resolve horizontal velocity differences from Poisson40 in 13 and 8 of the 256 reporting bins, all in the upper eighth of the cavity. Their kinetic shear differs in 32 and 21 bins, including 23 and 16 in that upper region. SBT shows no resolved difference in either quantity. At $x/L=0.5$, $y/L=0.96875$, the P-SBT and EA velocities fall 0.73 and 0.63~m\,s$^{-1}$ below Poisson40, and $P_{xy}$ is more negative by 0.195 and 0.186~Pa. The SBT differences are $-0.11$~m\,s$^{-1}$ and $-0.052$~Pa. These values refer to the last sampled position inside the gas. Averaging the last five times gave shear differences of $-0.215$ and $-0.221$~Pa for P-SBT and EA. The longer average reduces the median paired standard error of the shear differences across the field by 27\%.

Each comparison with Poisson40 changes both the collision sampler and the mean population, from forty to two particles per cell. In the bearing (Sec.~\ref{sec:bearing}), reducing the mean population from ten to two particles per cell raises the mean wall-shear magnitude with whole-cell Poisson. At fixed occupancy SBT lowers it. Below the lid, the kinetic shear stress in the gas of P-SBT and EA is likewise larger in magnitude than that of Poisson40. A similar partial cancellation would explain why SBT lies closest to Poisson40 there, but without a Poisson2 ensemble the cavity data can establish neither the sign of the occupancy error nor the cancellation.

\begin{figure}[!htbp]\centering
\includegraphics[width=.92\linewidth]{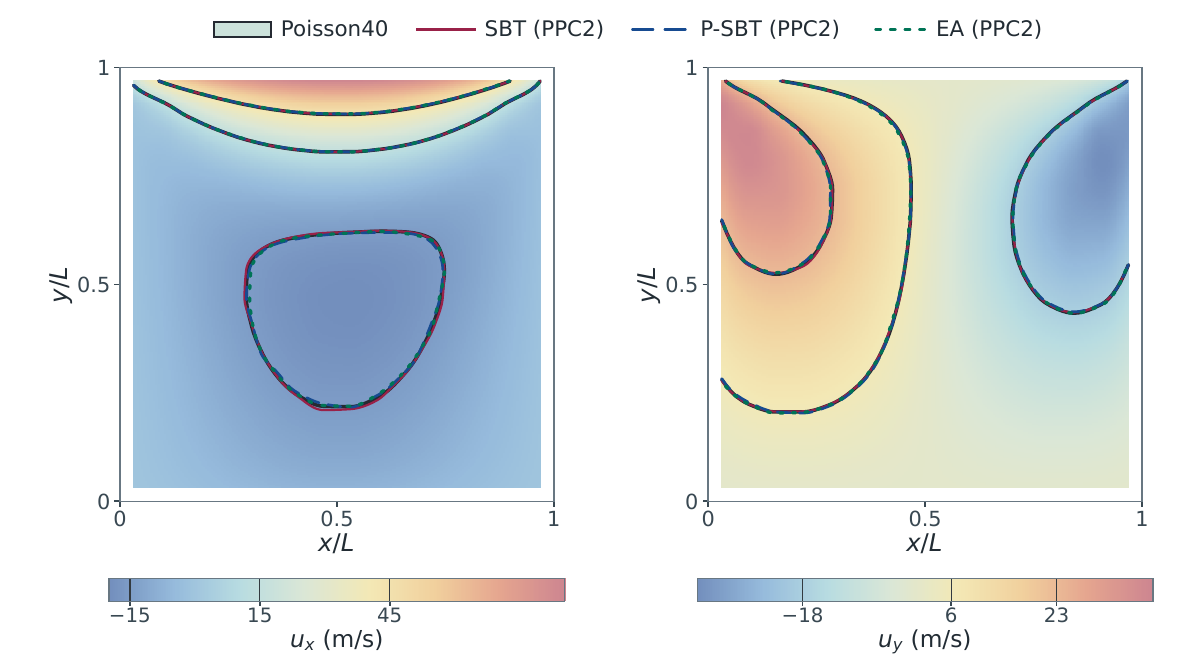}
\caption{Horizontal and vertical velocity in the transient cavity. The color field shows Poisson40. SBT, representative-rate P-SBT and EA at two particles per cell are shown as contour lines at the same velocities. Each ensemble contains 21,000 trajectories. The fields are averaged over nine times spanning $4.5\le t^*\le5$ on $16\times16$ reporting bins.}
\label{fig:cavity-velocity}
\par\medskip
\includegraphics[width=.92\linewidth]{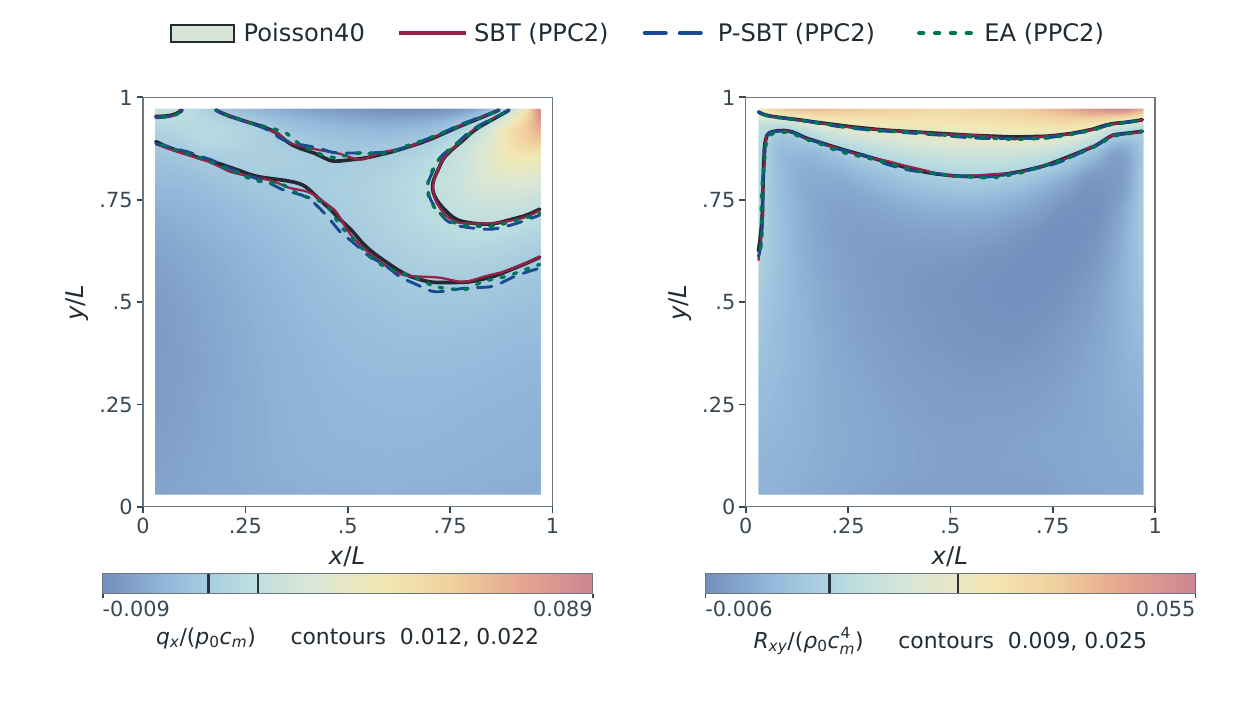}
\caption{Streamwise heat flux $q_x/(p_0c_m)$ and fourth moment $R_{xy}/(\rho_0c_m^4)$ in the same nine-phase cavity average as Fig.~\ref{fig:cavity-velocity}. Here $\rho_0$ and $p_0$ are the initial density and pressure. Colors show Poisson40. Lines compare all four methods at $0.012$ and $0.022$ for heat flux and $0.009$ and $0.025$ for the fourth moment. At these levels the estimated uncertainty in contour position is within a quarter of a reporting bin along at least 90\% of each method's contour. The supplementary figures show all contour levels.}
\label{fig:cavity-transport}\end{figure}

Figure~\ref{fig:cavity-velocity} shows the lid-driven horizontal flow and the vertical return motion in the same nine-phase average. At $y/L=0.781$ in the Poisson40 calculation, the gas rises at $37.1$ m\,s$^{-1}$ on the left ($x/L=0.094$) and descends at $35.1$ m\,s$^{-1}$ on the right ($x/L=0.906$), while the streamwise velocity near the centre is $-19.3$ m\,s$^{-1}$. Together these velocities trace a clockwise circulation. Figure~\ref{fig:cavity-transport} compares energy transport and the anisotropy of the fourth velocity moment at the same locations and times. The heat flux weights each particle by its peculiar kinetic energy, and the fourth moment weights fast particles more strongly. Neither $q_x$ nor $R_{xy}$ shows a method difference resolved by the simultaneous intervals.

We also doubled and quadrupled the step for P-SBT and EA. At 23.16 and 46.33 ns, 5,120 trajectories per PPC2 method contain the same nominal number of particle observations per phase as 256 Poisson40 trajectories, and each sampler is compared with Poisson40 at the same step and the same 81 observation times. At $t^*=5$ the root-mean-square differences on the vertical centerline range from 0.52 to 0.63 K in temperature and from 0.62 to 1.01 m\,s$^{-1}$ in horizontal velocity; the supplement gives the values by step and compares velocity and shear over $4.5\le t^*\le5$. These instantaneous centerline comparisons use fewer trajectories than the nine-time averages above.

\FloatBarrier
\section{Wall shear in an eccentric gas bearing}\label{sec:bearing}
The gap in an eccentric bearing varies around the rotor. This geometry lets us test whether calculations that agree in total torque also agree in local shear. The bearing consists of two isothermal cylinders with parallel axes. The rotor has radius $R=1$~mm and moves counterclockwise at surface speed $0.1c_m$, where $c_m=\sqrt{2k_BT_w/m}$ and $T_w=273.15$~K. The stator, of radius $1.1$~mm, is at rest. The centres are offset by half the mean clearance $H=0.1$~mm, so the gap ranges from $0.5H$ at $\theta=0$ to $1.5H$ at $\theta=\pi$. Both walls reflect diffusely in their own frames. The gas consists of hard spheres with $m=6.63\times10^{-26}$~kg, $d=3.658\times10^{-10}$~m and mean free path $\lambda/H=0.2/\sqrt{\pi}$. The 3.706~ns step is about one eighth of the mean collision time. The Couette comparison uses a 0.756~ns step and collision subcells with a mean occupancy of 0.5. \citet{Doi2025Bearing} calculated how the magnitude and direction of the force on eccentric cylinders vary with rarefaction and wall motion. Here we use whole-cell Poisson calculations to compare collision samplers.

Let $\tau(\theta)$ denote the tangential traction that the gas exerts on the rotor, positive in the direction of rotation. The torque per unit axial length, $\mathcal T=R^2\int_0^{2\pi}\tau(\theta)\,\mathrm d\theta$, depends only on the circumferential mean shear. Each nonconstant angular harmonic integrates to zero. The transverse force, however, depends on the first sine and cosine harmonics of both pressure and shear because the wall normal and tangent rotate around the cylinder. A numeral following a sampler name gives the nominal number of particles per collision cell (PPC).

\begin{figure}[!tbp]\centering
\includegraphics[width=.99\linewidth]{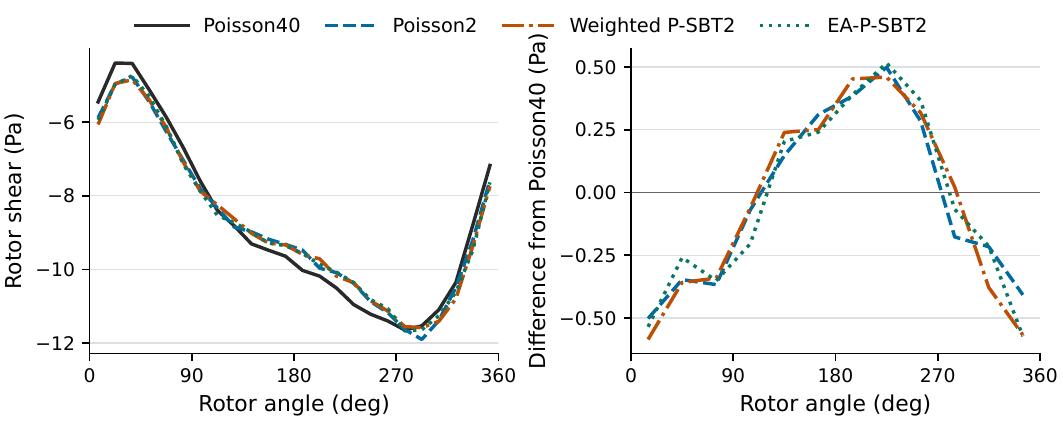}
\caption{Gas-on-rotor shear at equal total particle count (left) and its signed difference from the coarser Poisson40 reference (right). Each curve averages four trajectories over blocks 51--430, or 225.33~$\mu$s per trajectory. The curves use 24 equal angular sectors on the left and 12 on the right. Norms in the text use all 48 recorded sectors. Zero degrees is the narrowest gap.}
\label{fig:bearing-equal}
\end{figure}

\paragraph{Equal particle count.} Each calculation uses 276,480 particles and a 3.706~ns step. Poisson40 uses $576\times12$ angular and radial cells. Poisson2, weighted P-SBT and EA-P-SBT use $2560\times54$ cells at PPC2 (Fig.~\ref{fig:bearing-equal}). On the uniform fine grid, the three PPC2 torques differ from Poisson40 by 0.44--0.53\%, while their angular shear profiles differ by 4.24--4.45\% in relative $L_2$ norm. Halving the area of the first layer next to the rotor, at the same particle and cell counts, gives similar profile differences (4.46--4.69\%). Both occupancy and cell size change between the coarse and fine grids, so this comparison alone cannot isolate either effect.

\begin{figure}[!tbp]\centering
\includegraphics[width=.88\linewidth]{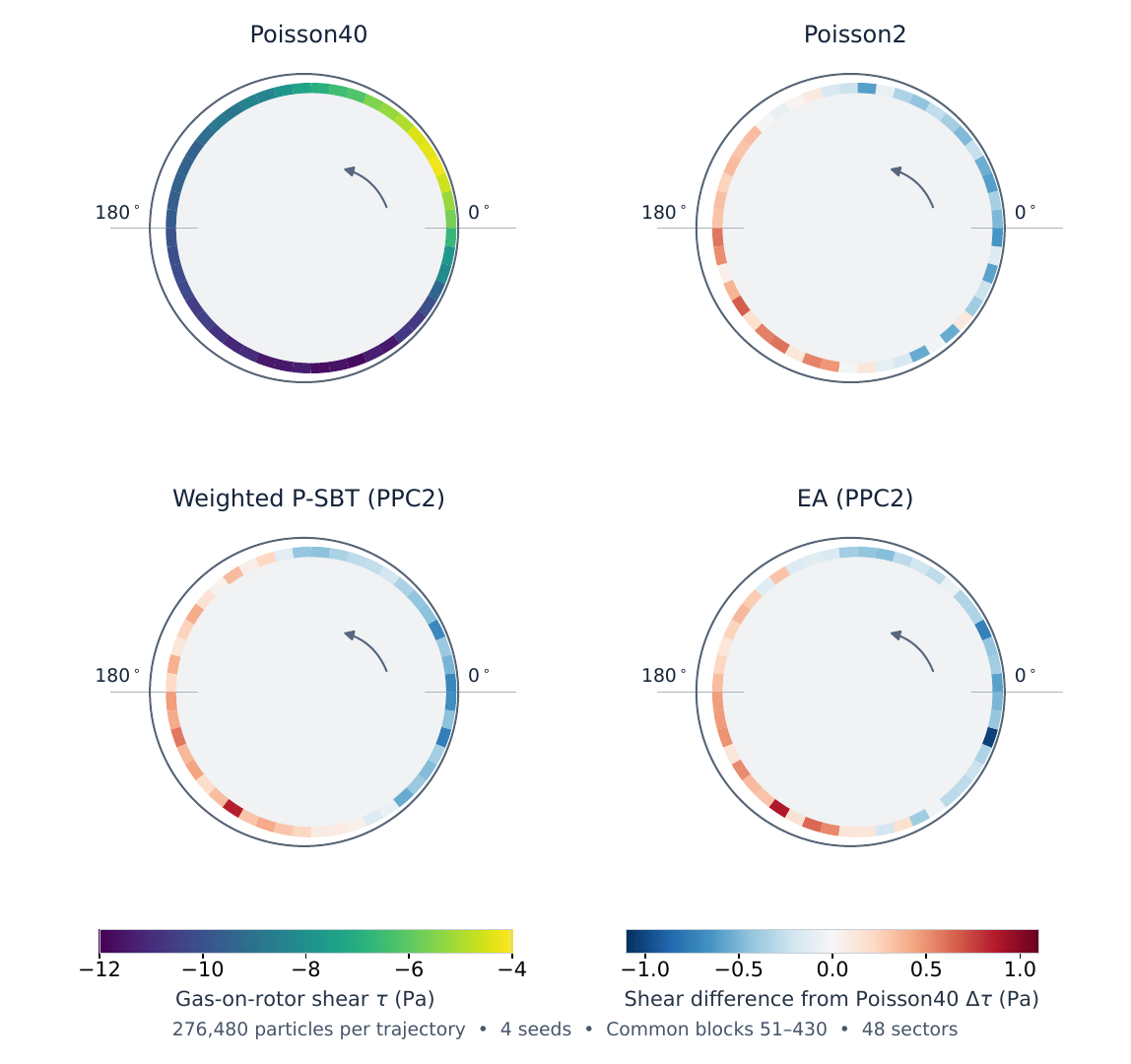}
\caption{Shear exerted by the gas on the rotor at equal total particle count. The Poisson40 panel shows the signed stress. The other three panels show the local difference, PPC2 minus Poisson40, on a common colour scale. Each segment represents one of the 48 measured sectors, averaged over blocks 51--430 for four seeds per method. The arrows mark the direction of rotation, and $0^\circ$ is the narrowest gap.}
\label{fig:bearing-rotor-color}
\end{figure}

The PPC2 calculations predict greater resistance to rotation near the narrowest gap and weaker resistance over much of the opposite side (Fig.~\ref{fig:bearing-rotor-color}). These local shear differences have opposite signs and largely cancel on integration around the rotor.

\begin{figure}[!tbp]\centering
\includegraphics[width=.99\linewidth]{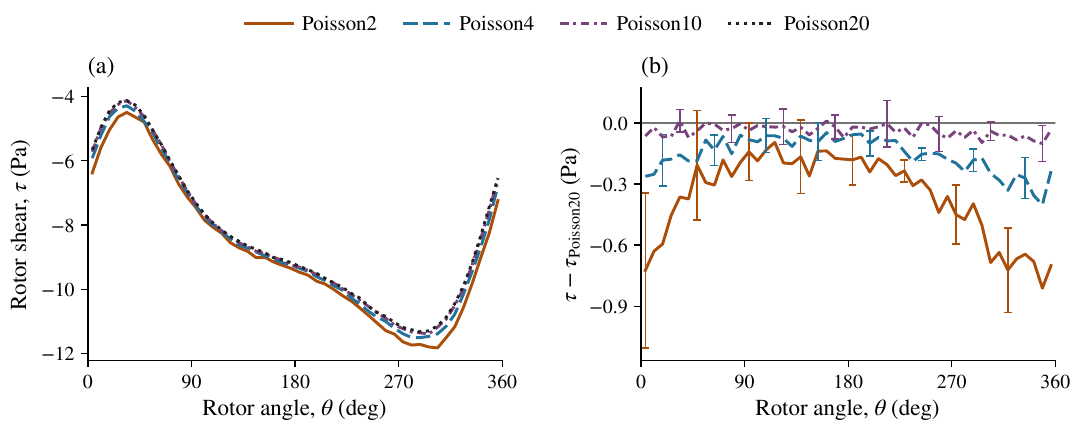}
\caption{Gas-on-rotor shear on the $5120\times108$ bearing grid at four nominal occupancies. (a) Signed mean traction in 48 angular sectors. (b) Paired differences from Poisson20. Lines join the measured sector centres. Whiskers are pointwise 95\% intervals across the four seed pairs, drawn at every sixth sector and staggered between curves for legibility. All means use blocks 151--772.}
\label{fig:bearing-ladder}
\end{figure}

\paragraph{Occupancy at fixed mesh.} On the $2560\times54$ grid, raising the nominal Poisson occupancy from 2 to 10 lowered the torque magnitude by $3.92\pm1.17$\% (blocks 151--340, four paired seeds). We then tested four occupancies on a $5120\times108$ grid. Every calculation used whole-cell Poisson sampling, a 3.706~ns step and the same 622 production blocks for averaging (Fig.~\ref{fig:bearing-ladder}). At PPC20 the mean torque per unit span was $-52.854\pm0.023~\mu\mathrm{N\,m}$ per metre and the circumferential mean shear was $-8.412\pm0.004$~Pa. We report 95\% Student $t$ intervals across four seeds, pairing runs by seed when calculating differences.

At PPC2 the torque magnitude exceeded the PPC20 value by $4.25\pm0.21$\%, at PPC4 by $1.84\pm0.11$\%, and at PPC10 by $0.47\pm0.05$\%. The corresponding relative $L_2$ differences of the angular mean profiles were 4.72, 2.04 and 0.56\%. To check the dependence on nominal occupancy $N_c$, we fitted $|\mathcal T|=a+b/N_c$ separately for each seed using PPC2, PPC4 and PPC10. The predicted PPC20 torque magnitude was $52.843~\mu\mathrm{N\,m}$ per metre. The measured magnitude exceeded the prediction by $0.021\pm0.053$\%. This inverse-occupancy relation describes the torque on the specified grid and step. An independent heat-conduction study found a leading finite-particle contribution to DSMC transport with a similar occupancy dependence \citep{Rader2006Fourier}.

\paragraph{Collision law at fixed mesh and occupancy.} On the graded $2560\times54$ grid at PPC2, we varied the sampler at fixed step and paired each run with Poisson2 by seed (Table~\ref{tab:bearingsampler}). SBT and stochastic-rounding NTC lowered the torque magnitude by $5.51\pm0.59$\% and $5.80\pm0.42$\%, respectively. The mean shear became less negative by about 0.5~Pa. For weighted P-SBT and EA-P-SBT the paired 95\% intervals for the torque difference both include zero. As in Couette flow, the two rules that predict too many collisions for an isolated pair give weaker wall shear. Here the pair collision probability per step is larger, and the fourth SBT trajectory stopped when a Bernoulli probability exceeded one. Changing the sampler on this mesh thus shifts the torque in the opposite direction to lowering the occupancy in the fixed-mesh Poisson tests. The two effects can mask one another.

\begin{table}[!tbp]\centering\small
\setlength{\tabcolsep}{1.8pt}
\begin{tabular}{@{}lrrrrrr@{}}\toprule
Sampler & Seeds & $\Delta\bar\tau$ (Pa) & Torque (\%) & Load (\%) & $\Delta a_1$ (Pa) & $\Delta b_1$ (Pa)\\\midrule
SBT & 3 & $+0.483\pm0.050$ & $-5.51\pm0.59$ & $-1.93\pm1.55$ & $+0.272\pm0.062$ & $-0.181\pm0.190$\\
NTC & 4 & $+0.509\pm0.038$ & $-5.80\pm0.42$ & $-2.33\pm1.52$ & $+0.196\pm0.083$ & $-0.120\pm0.061$\\
Weighted P-SBT & 4 & $-0.062\pm0.066$ & $+0.71\pm0.76$ & $+0.45\pm1.06$ & $-0.019\pm0.067$ & $-0.026\pm0.104$\\
EA-P-SBT & 4 & $-0.016\pm0.066$ & $+0.18\pm0.75$ & $+0.46\pm0.86$ & $-0.014\pm0.094$ & $-0.026\pm0.127$\\
\bottomrule\end{tabular}
\caption{Effect of the collision sampler on the eccentric bearing. All runs use the graded $2560\times54$ grid, two particles per cell, $\Delta t=3.706$~ns and blocks 51--430. Values are mean differences from Poisson2 with paired 95\% $t$ intervals across seeds. $\Delta\bar\tau$ is the change in mean rotor shear. Torque and load entries give percentage changes in magnitude. $\Delta a_1$ and $\Delta b_1$ are changes in the first cosine and sine coefficients of the shear profile, with $\theta=0$ at the narrowest gap. The fourth SBT realization stopped when a Bernoulli probability exceeded one.}
\label{tab:bearingsampler}
\end{table}

\begin{figure}[!tbp]\centering
\includegraphics[width=.88\linewidth]{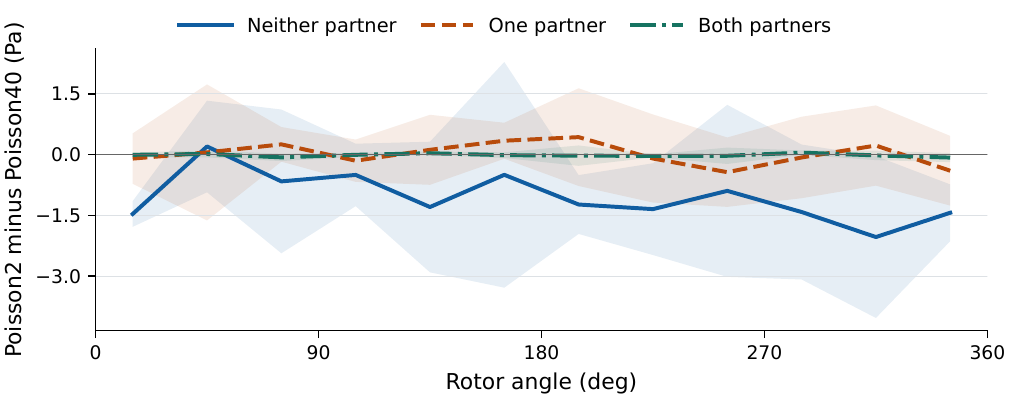}
\caption{Contributions to the rotor shear difference, Poisson2 minus Poisson40, on the common coarse grid. Curves distinguish the number of partners in the last gas collision that had recently reflected from the rotor. Each point covers $30^\circ$ at the rotor impact and blocks 153--250. Shading gives pointwise paired 95\% intervals across four realizations. Direct wall-to-wall flights are counted separately in the angular mean.}
\label{fig:bearing-hist}
\end{figure}

\paragraph{Collision histories before impact.} A molecule that has just reflected diffusely from the rotor can collide with a molecule arriving from the interior. To see whether these encounters account for the excess shear at low occupancy, we compare Poisson2 and Poisson40 on the same $576\times12$ grid. For each impact preceded by a gas collision, we count how many of the two partners had struck the rotor since their previous gas collision. We sum the signed tangential impulses in each group at the angle of impact (Fig.~\ref{fig:bearing-hist}), weighting each simulator particle by the number of molecules it represents.

The mean shear difference on this grid is $-1.072\pm0.291$~Pa. Impacts for which neither partner had recently reflected contribute $-1.049\pm0.451$~Pa. The contributions for one and two reflected partners are $+0.013\pm0.198$ and $-0.016\pm0.039$~Pa, respectively. Of the contribution for neither partner, $-0.925\pm0.510$~Pa comes from last collisions outside the first gas layer. Most of the measured excess shear thus reaches the rotor after a collision away from the wall. Earlier wall encounters could still have shaped the incoming velocities. The supplement gives the force components, occupancy comparisons, collision rates and impulse per impact.

\paragraph{Repeated partners.} On the $5120\times108$ grid we repeated the Poisson2 calculation with one change. We rejected an otherwise accepted collision if each particle's last gas collision was with the other. In ordinary Poisson2, 3.44\% of accepted gas collisions repeated the same pair within a step and 0.34\% across steps. Suppressing these repetitions changed the torque magnitude by $-0.16\pm0.23$\% over blocks 151--1190 (four paired seeds). Over blocks 151--772, the torque still exceeded the PPC20 value by $4.05\pm0.17$\% with these repetitions suppressed. Immediate recollisions therefore explain little of the excess torque at PPC2 on this grid.

\subsection{Bearing load and flow through the gap}\label{sec:bearing-doi}
Doi's kinetic solution \citep{Doi2025Bearing} provides a comparison at the same clearance ratio $H/R=0.1$, eccentricity ratio $0.5$, rotor speed $U_w/c_m=0.1$ and rarefaction parameter $k=\sqrt{\pi}\lambda/(2H)=0.1$. Both calculations use stationary outer walls and diffuse reflection at a common wall temperature. The collision models assign different viscosities to this value of $k$. With $p_0$ the mean equilibrium pressure, Doi's Bhatnagar--Gross--Krook--Welander (BGKW) model gives $\mu_0 c_m/(p_0H)=k$. For hard spheres the corresponding value is $\gamma_1 k$, where $\gamma_1=1.25$ in the first Sonine approximation and $1.27004$ for the hard-sphere Boltzmann equation \citep{Taguchi2021Rotation}. At $273.15$~K and $p_0=562.18$~Pa these values give $1.667\times10^{-5}$~Pa\,s for BGKW and $2.117\times10^{-5}$~Pa\,s for hard spheres, a 27.0\% difference.

\begin{figure}[!tbp]\centering
\includegraphics[width=\linewidth]{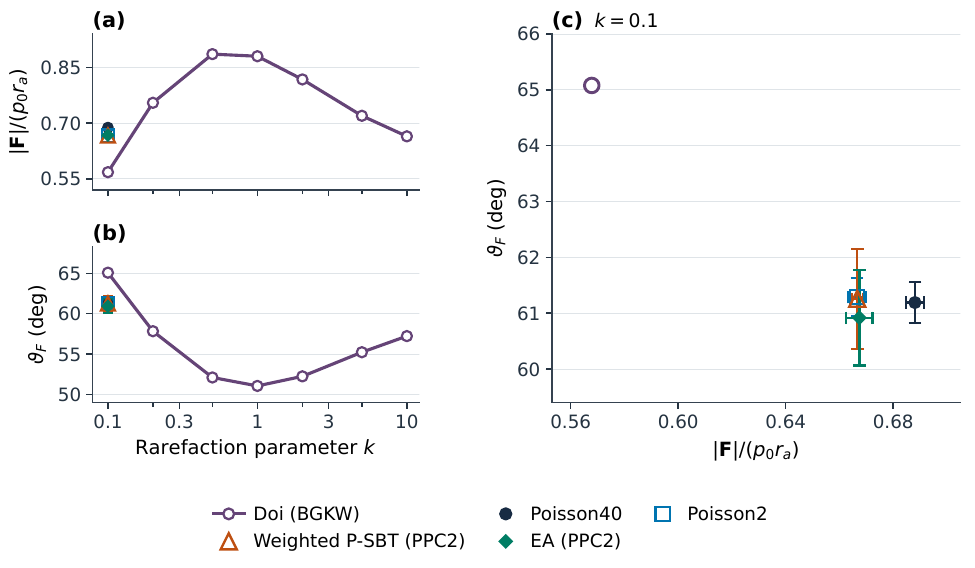}
\caption{Rotor load and force angle. Panels (a) and (b) show Doi's direct BGKW solution as $k$ varies and the four DSMC estimates at $k=0.1$. Panel (c) plots the force angle against load for both models at $k=0.1$. The radius $r_a$ in Doi's notation is $R$. In our Cartesian coordinates the force angle is $\vartheta_F=\operatorname{atan2}(F_y,-F_x)$. Error bars give separate 95\% Student intervals for load and angle from four independent trajectories per method, each averaged over $225.33\,\mu\mathrm{s}$ (blocks 1651--2030).}
\label{fig:bearing-doi-load}
\end{figure}

Figure~\ref{fig:bearing-doi-load} compares the force $\boldsymbol F$ per unit axial length exerted by the gas on the rotor. At $k=0.1$, Doi obtains $|\boldsymbol F|/(p_0R)=0.56785$ and a force angle of $65.08^\circ$. Poisson40 gives $0.6881\pm0.0033$ and $61.19^\circ\pm0.37^\circ$ (95\% intervals across four seeds). The three PPC2 calculations give loads of $0.6665$--$0.6674$ and angles of $60.9^\circ$--$61.3^\circ$. The dimensionless resisting torque $|\mathcal T|/(p_0R^2)$ is $0.07796$ for BGKW and $0.0981$--$0.0988$ in DSMC. This 26--27\% torque difference is comparable in size to the viscosity difference. When each torque is scaled by its own viscosity, the ratio $|\mathcal T|H/(\mu_0 U_w R^2)$ lies 0.26--0.96\% below Doi's value. This rescaling uses the original trajectories. For the equal-viscosity control, we increase the hard-sphere diameter from $3.658$ to $4.12243$~\AA\ for all four samplers while retaining the same density, wall temperature, rotor speed, geometry and time step. The resulting hard-sphere $k$ is $0.078738$ and $\mu_0=1.667\times10^{-5}$~Pa\,s at $T_w$, equal to the BGKW value at $k=0.1$.

\begin{figure}[!tbp]\centering
\includegraphics[width=.94\linewidth]{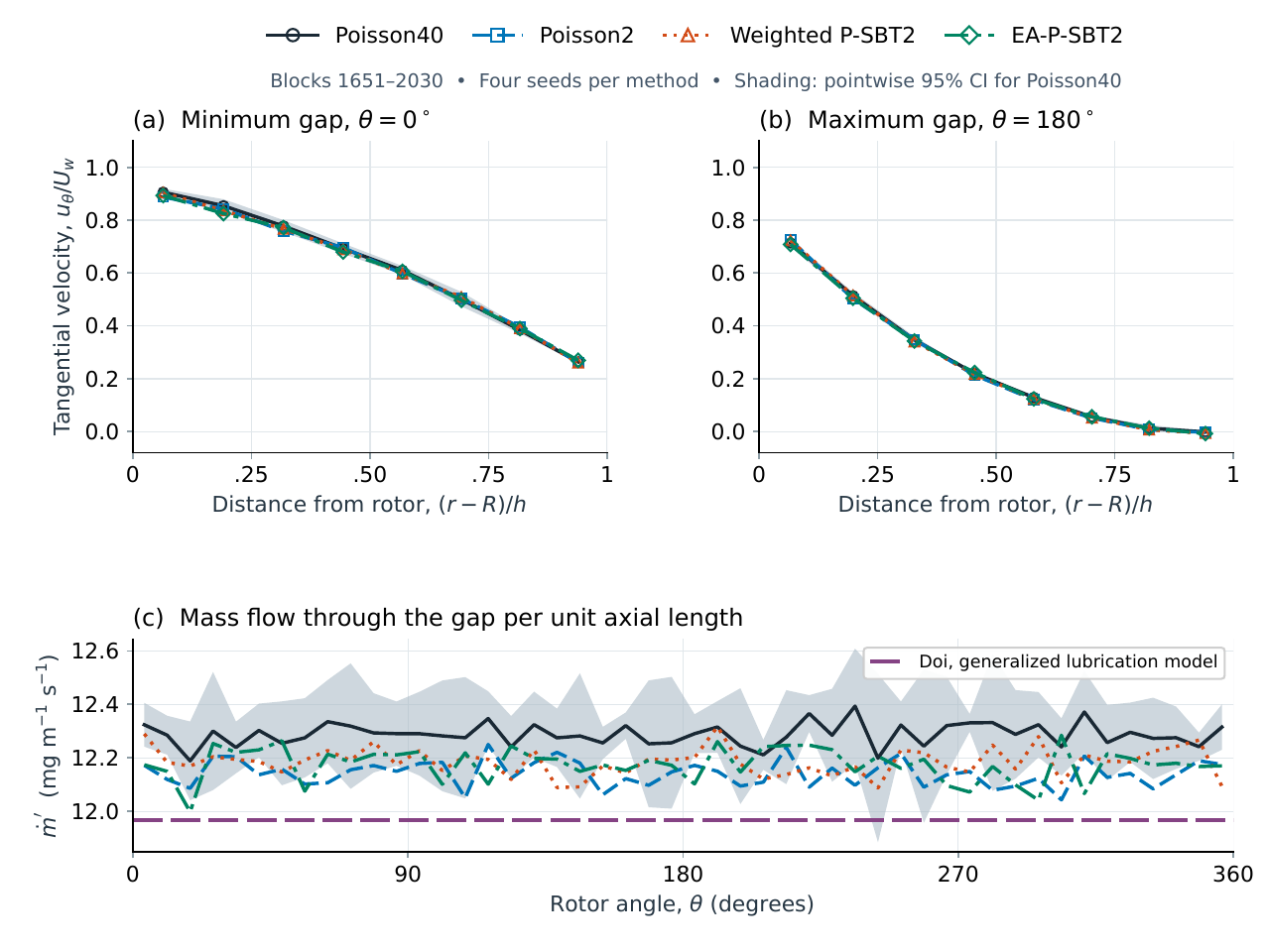}
\caption{Tangential velocity across the narrowest and widest gaps (a,b) and mass flow through each angular section (c). The radial coordinate runs from the rotor to the stator. Each velocity profile averages the two angular bins adjacent to the specified angle. Shading shows the pointwise 95\% confidence interval (CI) for Poisson40, estimated from four trajectories over blocks 1651--2030, as in Fig.~\ref{fig:bearing-doi-load}. The horizontal line in (c) gives the generalized lubrication prediction at this geometry, speed and $k$.}
\label{fig:bearing-doi-flow}
\end{figure}

The gap flow varies less between the collision schemes than the rotor load. Figure~\ref{fig:bearing-doi-flow} shows the tangential velocity at the narrowest and widest sections and the mass flow per unit axial length, $\dot m'(\theta)=\int_R^{R+h(\theta)}\rho u_\theta\,\mathrm{d}r$, where $h(\theta)$ is the local clearance. The angular mean of $\dot m'$ is $12.29\pm0.06$~mg\,m$^{-1}$\,s$^{-1}$ for Poisson40 and $12.14$--$12.19$~mg\,m$^{-1}$\,s$^{-1}$ for the three PPC2 methods. Evaluating Doi's generalized Reynolds equation with his tabulated coefficients \citep{Doi2023Lubrication} gives $\dot m'/(\rho_0 c_m H)=0.035903$, or $11.97$~mg\,m$^{-1}$\,s$^{-1}$ at our equilibrium density $\rho_0$. The DSMC means lie 1.5--2.7\% above this reduced-model prediction.

\FloatBarrier
\section{Accuracy and computational work}\label{sec:cost}
Table~\ref{tab:algorithmwork} shows how the number of operations in a collision step depends on cell occupancy.

\begin{table}\centering\small
\begin{tabularx}{\textwidth}{l>{\raggedright\arraybackslash}X@{\hspace{18pt}}l}\toprule
Sampler & Main work in a collision substep & Estimated work\\\midrule
SBT and SBT-exp & One representative rate and one Bernoulli draw per primary & $O(N)$\\
P-SBT & One representative rate, fixed count and fresh uniform partners after the first event & $O(N+K)$\\
Weighted P-SBT & Compute all entry rates and retain their cumulative distribution for this primary & $O(N^2+\sum_i K_iw_i)$\\
EA-P-SBT & Recompute pair rates and partner probabilities after each event & $O(N^2+\sum_i K_iw_i)$\\
NTC & Energy bound, stochastically rounded count and uniform pair proposals & $O(N+C)$\\
Poisson reference & Same bound and pair proposals with a Poisson candidate count & $O(N+C)$\\\bottomrule
\end{tabularx}
\caption{Estimated collision work per cell per substep, excluding flight, sorting and diagnostics. $N$ is the number of particles in the cell, $w_i=|S_i|$ is the number of eligible partners for primary $i$, and $K_i$ is its number of collisions. $K=\sum_iK_i$ counts those collisions over all primaries and $C$ counts candidate proposals. The weighted schemes select a partner by scanning the cumulative distribution of pair rates. Table~\ref{tab:cost} gives the measured times for the complete solver.}\label{tab:algorithmwork}
\end{table}

For Couette flow we compare the signed wall shear over $185.915$ ms per realization with the time needed to advance the solver at each of two time steps. After warm-up, we timed runs of equal physical duration on one Intel Xeon Gold 5220 central processing unit (CPU) in five rounds with randomized method order. Timings include transport, collisions and collection of diagnostic measurements. They exclude compilation, file output, checkpoint compression, data transfer and queue time.

\begin{table}
\centering
\small
\setlength{\tabcolsep}{5pt}
\begin{tabular}{lrrrrr}
\toprule
Scheme & PPC & $\Delta t$ (ns) & \shortstack[r]{Same-step\\difference (Pa)} & \shortstack[r]{Fine-step\\difference (Pa)} & \shortstack[r]{Relative\\time} \\
\midrule
SBT & 2 & 0.756 & $+0.7819\pm 0.0188$ & $+0.7048\pm 0.0181$ & 0.381 \\
SBT-exp & 2 & 0.756 & $-0.1703\pm 0.0184$ & $-0.2474\pm 0.0172$ & \textemdash \\
SBT & 2 & 0.252 & $+0.2668\pm 0.0167$ & $+0.2668\pm 0.0167$ & 1.036 \\
P-SBT & 2 & 0.756 & $-0.0997\pm 0.0168$ & $-0.1767\pm 0.0181$ & 0.385 \\
P-SBT & 2 & 0.252 & $-0.0273\pm 0.0172$ & $-0.0273\pm 0.0172$ & 1.047 \\
Weighted P-SBT & 2 & 0.756 & $+0.0125\pm 0.0180$ & $-0.0646\pm 0.0180$ & 0.393 \\
Weighted P-SBT & 2 & 0.252 & $+0.0133\pm 0.0168$ & $+0.0133\pm 0.0168$ & 1.065 \\
EA-P-SBT & 2 & 0.756 & $+0.0152\pm 0.0182$ & $-0.0618\pm 0.0203$ & 0.394 \\
EA-P-SBT & 2 & 0.252 & $+0.0175\pm 0.0247$ & $+0.0175\pm 0.0247$ & 1.071 \\
Poisson & 2 & 0.756 & -- & $-0.0771\pm 0.0174$ & 0.371 \\
Poisson & 2 & 0.252 & -- & -- & 1.000 \\
NTC & 2 & 0.756 & $+0.9237\pm 0.0172$ & $+0.8467\pm 0.0171$ & 0.367 \\
NTC & 2 & 0.252 & $+0.3188\pm 0.0179$ & $+0.3188\pm 0.0179$ & 0.991 \\
NTC & 20 & 0.756 & $+0.3269\pm 0.0129$ & $+0.2499\pm 0.0131$ & 3.285 \\
\bottomrule
\end{tabular}
\caption{Strong Couette flow. The two differences are the signed wall stress for each sampler minus that for whole-cell Poisson at the same step or at 0.252 ns. Uncertainties give one estimated standard error of each difference. Solver times are relative to Poisson at 0.252 ns. Dashes indicate a reference compared with itself or an unmeasured SBT-exp solver time.}
\label{tab:cost}
\end{table}

\begin{figure}[!tbp]
\centering
\includegraphics[width=\linewidth]{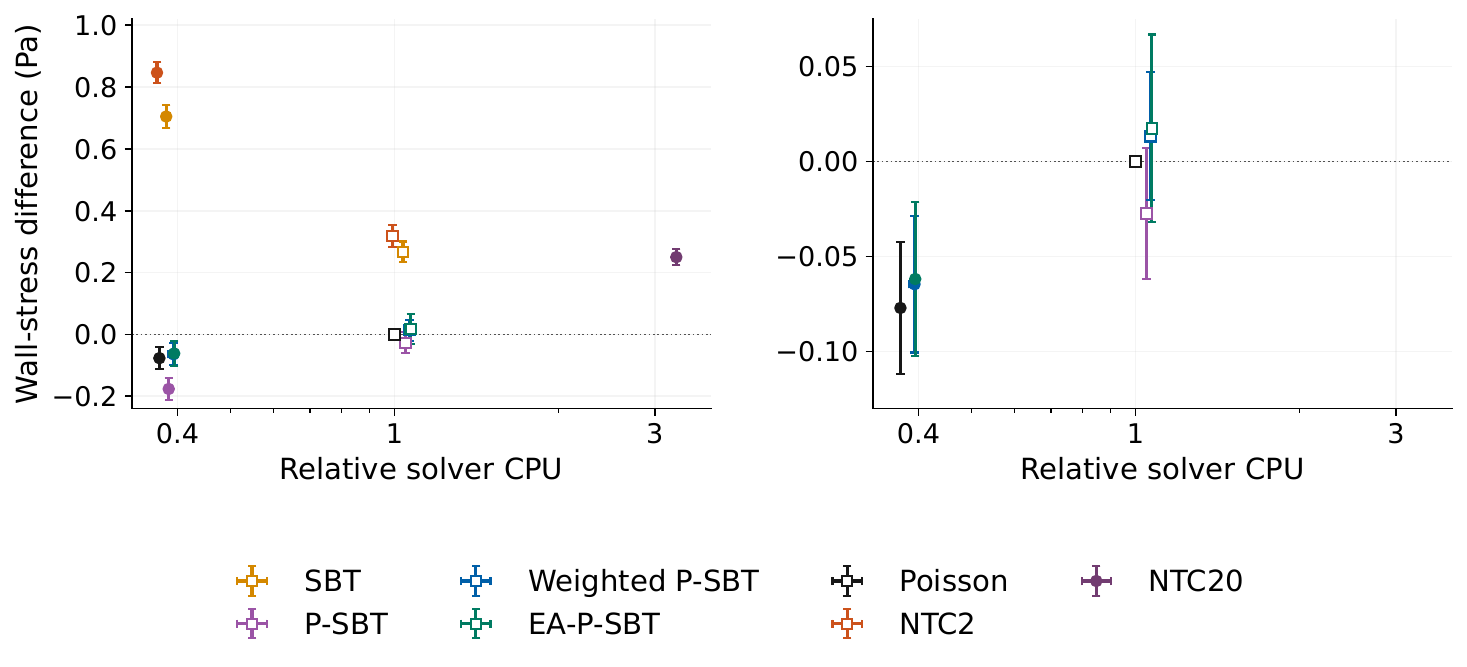}
\caption{Signed wall-stress difference versus solver time, both relative to whole-cell Poisson at 0.252 ns as in Table~\ref{tab:cost}. Circles denote 0.756 ns and squares denote 0.252 ns. The right panel enlarges the region near zero difference. Vertical bars show twice the estimated standard error and horizontal bars span the five timing rounds.}
\label{fig:costtransport}
\end{figure}

For Fig.~\ref{fig:costtransport}, we average the two wall impulses using the same sign for positive momentum transfer. We compute shear differences for matching seeds and blocks of time to account for the covariance from shared initial states and segment seeds. We group 32, 64, 128 and 256 segments and calculate standard errors both from paired observations and under a zero-covariance assumption. We report the largest estimate, increasing it to half the separation between the two seed-specific differences if that value is larger. With only two realizations, this procedure gives an approximate sampling uncertainty. Table~\ref{tab:survivalwall} lists the SBT and SBT-exp differences at each sampling duration. The 0.756 ns SBT-exp result uses the full $185.915$ ms per realization. Section~\ref{sec:occupancycost} gives timings for the collision step alone.

At 0.756 ns, the wall-stress difference from Poisson falls from $+0.7819\pm0.0188$ Pa for SBT to $+0.0125\pm0.0180$ Pa for weighted P-SBT and $+0.0152\pm0.0182$ Pa for EA-P-SBT. The sampling uncertainty prevents us from ranking the latter two methods. The Poisson wall stress at 0.756 ns differs from that at 0.252 ns by $-0.0771\pm0.0174$ Pa. Against that finer-step reference the weighted and adaptive differences are $-0.0646\pm0.0180$ Pa and $-0.0618\pm0.0203$ Pa.

At 0.756 ns the relative solver times are 0.371 for whole-cell Poisson, 0.393 for weighted P-SBT, 0.394 for EA-P-SBT and 0.367 for rounded NTC. In this Couette calculation, replacing stochastic rounding by a Poisson candidate count therefore removes the bias due to rounding at about 1\% extra solver time. The P-SBT variants retain SBT's ordered sweep and isolate the contributions of event counts and partner selection to the wall-stress error.

\subsection{Collision-kernel cost as a function of occupancy}\label{sec:occupancycost}
The full-solver timing above refers to one occupancy and two steps. To see how the cost of the collision kernels alone depends on occupancy and step, we time six kernels in a homogeneous hard-sphere cell at 13 occupancies between $N=2$ and 128 and at $h/\tau=0.01$, 0.04, 0.10, 0.20 and 0.40. The Poisson and stochastically rounded NTC kernels use the same bound derived from the conserved cell energy. Number density, cell volume and temperature are fixed, so the simulator weight scales as $1/N$, and $\tau^{-1}=n\pi d^2(4c_0/\sqrt{\pi})$ is the nominal Maxwellian collision frequency. For each setting we draw eight independent initial states, subtract the mean velocity and normalize the energy. Within each round the methods start from the same state and seed and run in random order. A timed call advances $\max(1000,\lfloor48000/N\rfloor)$ substeps on one CPU core of a single node after warm-up, with optional repeat-event diagnostics disabled. The 3,120 timing records measure collision work only, separately from the solver timings in Table~\ref{tab:cost}.

\begin{figure}[!tbp]\centering
\includegraphics[width=\linewidth]{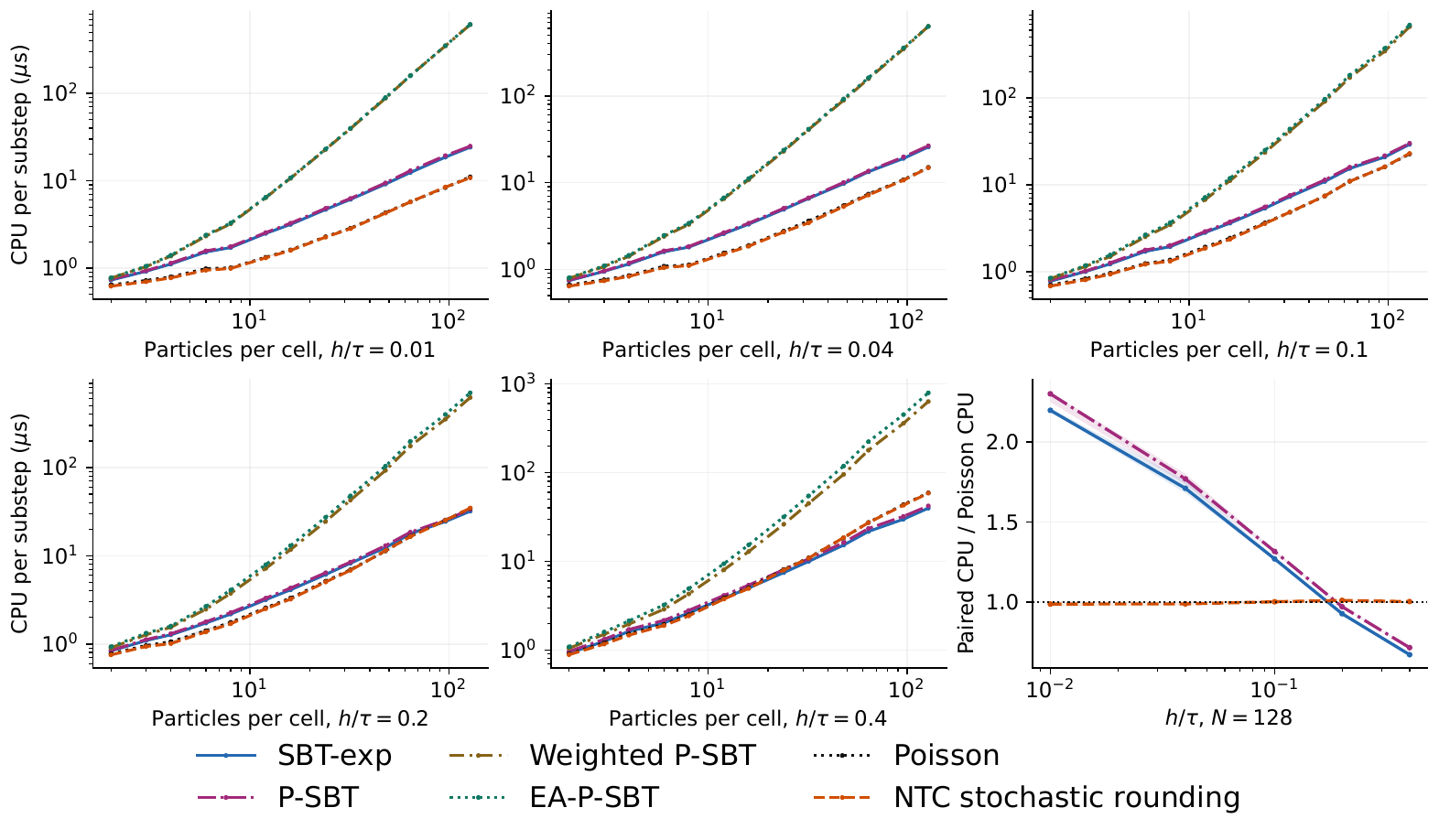}
\caption{Collision time at fixed physical density. The first five panels show the median time per substep versus particles per cell at each time step. The last panel gives time relative to whole-cell Poisson in paired runs at $N=128$ for SBT-exp, P-SBT and stochastically rounded NTC. Bands show the interquartile range of times across eight rounds.}
\label{fig:occupancycost}
\end{figure}

Figure~\ref{fig:occupancycost} shows how the cost changes with occupancy, and Table~\ref{tab:widecost} lists the timings at $N=128$.

\begin{table}[!tbp]\centering\small
\setlength{\tabcolsep}{5pt}
\begin{tabular}{rrrrrrr}\toprule
$h/\tau$ & SBT-exp & P-SBT & Weighted & EA & Poisson & Rounded NTC\\\midrule
0.01 & 24.21 & 24.87 & 611.05 & 615.48 & 11.08 & 10.80\\
0.04 & 25.79 & 26.63 & 633.20 & 632.70 & 15.02 & 14.85\\
0.10 & 29.23 & 30.06 & 658.96 & 687.33 & 22.64 & 22.96\\
0.20 & 31.94 & 33.53 & 620.91 & 702.19 & 34.41 & 34.73\\
0.40 & 39.78 & 42.18 & 633.19 & 792.31 & 59.11 & 58.67\\
\bottomrule\end{tabular}
\caption{Median collision time per substep in $\mu$s for a cell with $N=128$ particles. The step $h$ is scaled by the nominal mean collision time $\tau$. Weighted and EA denote weighted and event-adaptive P-SBT. The eight rounds use a common initial state for all methods within each round at fixed physical density. Timings exclude just-in-time compilation and file input/output.}
\label{tab:widecost}
\end{table}

The ranking reverses as the step grows. At $N=128$ and $h/\tau=0.10$, SBT-exp and P-SBT take 29.23 and 30.06 $\mu$s per substep and Poisson 22.64 $\mu$s. At $h/\tau=0.20$ the respective medians are 31.94, 33.53 and 34.41 $\mu$s. SBT-exp is faster than Poisson in all eight paired rounds and P-SBT in seven. On the tested set of steps, the order therefore changes between $h/\tau=0.10$ and 0.20 at this occupancy. At $h/\tau=0.40$ one substep takes 39.78 $\mu$s for SBT-exp, 42.18 $\mu$s for P-SBT, 59.11 $\mu$s for Poisson and 58.67 $\mu$s for rounded NTC. The median paired ratios to Poisson are 0.673 for SBT-exp and 0.717 for P-SBT, with all eight ratios below one for both. Weighted and adaptive P-SBT take 633.19 and 792.31 $\mu$s, so the advantage belongs to the methods that draw a single representative partner.

The invariant relative-speed bound explains this change in cost. At fixed energy per particle it grows as $\sqrt{N}$. With simulator weight proportional to $N^{-1}$, the mean number of whole-cell candidate pairs grows as $N^{3/2}$ at fixed density and time step. A sweep that samples one representative partner per primary takes $O(N)$ work under these conditions. Partner selection, scattering and rate evaluation add to both. These timings compare work per substep only. An equal-accuracy comparison must include the sampling error and the flight error of the chosen step, and Sec.~\ref{sec:relax} shows that SBT-exp, the fastest kernel at large steps and occupancies, retains a resolved stress error once a cell holds three or more particles.

\FloatBarrier
\section{Conclusions}\label{sec:conclusions}
All the collision schemes conserve momentum and energy in each encounter, yet they relax stress at different rates. For an isolated hard-sphere pair, the first isotropic collision erases the initial anisotropy of the relative velocity on average. SBT gives the wrong mean stress because it underestimates the probability of no collision by $\mu^2/2+O(\mu^3)$ per step. NTC with stochastic rounding has the same error for a single pair under the centred energy bound. Exponential survival corrects the isolated-pair result. With three or four particles, the first collision also changes the relative velocities available for the next one. Allowing later collisions and weighting partners by their rates reduces the error further. Updating the rates after every event brings the four-particle result within sampling uncertainty at $h=0.4$, although a small difference remains for three particles at this step.

In strong Couette flow, 58\% of the collisions accepted by SBT-exp involved the only pair present in the collision subcell. Correcting the first-event probability reversed the sign of the wall-shear difference from whole-cell Poisson sampling at 0.756~ns. The weighted and event-adaptive schemes brought the difference within one estimated standard error of the Poisson value at that step. For a solver that already selects candidates from an upper bound on the cell collision rate, replacing stochastic rounding by a Poisson draw removed the candidate-count bias at about 1\% extra cost in this Couette calculation. The simpler representative-partner schemes were faster than whole-cell Poisson only at large occupancies and time steps in the kernel tests. Their accuracy there remains to be established.

The choice of observable matters. In Fourier flow, temperature profiles that nearly coincide can give different wall heat fluxes. In the impulsively started cavity the gas shear just below the lid distinguishes methods whose temperature fields agree closely over $4.5\leq t^*\leq5$. In the shocks, averaging in coordinates that follow the front reduces broadening from its displacement. Even after alignment, Mach-2 shocks with two particles per upstream cell were 6--10\% thicker than the hard-sphere Boltzmann solution, compared with differences of at most 1.6\% for Poisson40 and NTC20. At Mach 5 the aligned normal stress from Poisson sampling differed by about 14\% between two and forty particles per upstream cell at the smaller step.

The eccentric bearing separates the effect of the collision rule from that of particle number. At fixed mesh, time step and occupancy, SBT and stochastically rounded NTC reduced the torque magnitude by 5.5--5.8\% relative to whole-cell Poisson sampling. In a separate comparison on the $5120\times108$ grid, the Poisson torque exceeded its twenty-particle value by $4.25\pm0.21$\% at two particles per cell and by $0.47\pm0.05$\% at ten. Vetoing immediate repeat collisions left most of the two-particle excess in place. An apparently accurate torque can thus result from errors of opposite sign.

Torque also averages out angular variations in shear. At equal total particle count, the fine-grid, low-occupancy runs gave more negative shear near the narrowest gap and less negative shear over much of the opposite side. Their torques agreed with the coarse-grid Poisson40 result within 0.53\%, while the relative differences between the angular shear profiles exceeded 4\%. For a calculation intended to predict wall stress, convergence with particle number should therefore be checked on a fixed mesh for the local shear as well as for its integral.

\section*{Supplementary material}
The supplement gives the Poiseuille profiles, the Fourier sampling intervals and the Mach-10 shock comparisons. It describes the Mach-2 reference coordinates and further grid and particle-count tests at each Mach number, and it contains the collision diagrams, fixed-cell statistics, moment definitions, cavity centerlines and fields, and the bearing measurements of force and wall impulse by reflection history.
\section*{Author declarations}
\subsection*{Conflict of interest}
The authors have no conflicts to disclose.
\subsection*{Author contributions}
\textbf{Ehsan Roohi:} Methodology, Project administration, Writing -- original draft.

\textbf{Stefan Stefanov:} Conceptualization, Supervision.

\textbf{Ahmad Shoja-Sani:} Investigation, Writing -- original draft.
\subsection*{Funding}
Stefan Stefanov was partially supported by the Centre of Excellence in Informatics
and Information and Communication Technologies under the Grant No BG16RFPR002-1.014-0018-C01, financed by the
Research, Innovation and Digitalization for Smart Transformation Programme
2021-2027 and co-financed by the European Union.

\subsection*{Use of computational tools}
Codex (OpenAI) assisted with analysis scripts and manuscript editing. Claude (Anthropic) assisted with the collision diagrams and editorial review. The authors checked the diagrams against the code, verified the numerical results and take responsibility for the manuscript.

\section*{Data availability}
The accompanying files contain the collision kernels, temporal and seed averages, wall-flux and timing data, shock profiles and reference coordinates, and cavity field and centerline statistics. For the bearing they include blockwise rotor data, the 48-sector profiles for each seed and occupancy, the coefficients of the inverse-occupancy fit, first-harmonic coefficients on both walls, and sector-resolved impact counts and impulses. The supplied bearing scripts reproduce the figures and numerical comparisons, including the sampler contrasts of Table~\ref{tab:bearingsampler}, from these records.
\FloatBarrier
\bibliographystyle{pof_linked}
\small
\ifdefined\bibsep\setlength{\bibsep}{1pt plus .25pt}\fi
\bibliography{references}

@book{Bird1994,
  author    = {Bird, Graeme A.},
  title     = {Molecular Gas Dynamics and the Direct Simulation of Gas Flows},
  publisher = {Clarendon Press},
  address   = {Oxford},
  year      = {1994},
  doi       = {10.1093/oso/9780198561958.001.0001}
}

@book{Bird2013,
  author    = {Bird, Graeme A.},
  title     = {The {DSMC} Method},
  publisher = {CreateSpace Independent Publishing Platform},
  year      = {2013},
  isbn      = {9781492112907},
  url       = {https://bib-pubdb1.desy.de/record/333609}
}

@article{stefanov2011dsmc,
  doi={10.1137/090751864},
  title={On {DSMC} calculations of rarefied gas flows with small number of particles in cells},
  author={Stefanov, Stefan K},
  journal={SIAM Journal on Scientific Computing},
  volume={33},
  number={2},
  pages={677--702},
  year={2011},
  publisher={SIAM}
}

@article{Alexander1998CellSize,
  author  = {Alexander, F. J. and Garcia, A. L. and Alder, B. J.},
  title   = {Cell size dependence of transport coefficients in stochastic particle algorithms},
  journal = {Physics of Fluids},
  volume  = {10},
  number  = {6},
  pages   = {1540--1542},
  year    = {1998},
  doi     = {10.1063/1.869674}
}

@article{Hadjiconstantinou2000Discretization,
  author  = {Hadjiconstantinou, N. G.},
  title   = {Analysis of discretization in the direct simulation Monte Carlo},
  journal = {Physics of Fluids},
  volume  = {12},
  number  = {10},
  pages   = {2634--2638},
  year    = {2000},
  doi     = {10.1063/1.1289393}
}

@article{GarciaWagner2000TimeStep,
  author  = {Garcia, A. L. and Wagner, W.},
  title   = {Time step truncation error in direct simulation Monte Carlo},
  journal = {Physics of Fluids},
  volume  = {12},
  number  = {10},
  pages   = {2621--2633},
  year    = {2000},
  doi     = {10.1063/1.1289691}
}

@article{Gillespie1977SSA,
  author  = {Gillespie, D. T.},
  title   = {Exact stochastic simulation of coupled chemical reactions},
  journal = {The Journal of Physical Chemistry},
  volume  = {81},
  number  = {25},
  pages   = {2340--2361},
  year    = {1977},
  doi     = {10.1021/j100540a008}
}

@article{akhlaghi2018consequences,
  doi={10.1016/j.compfluid.2017.11.005},
  title={On the consequences of successively repeated collisions in no-time-counter collision scheme in {DSMC}},
  author={Akhlaghi, Hassan and Roohi, Ehsan and Stefanov, Stefan},
  journal={Computers \& Fluids},
  volume={161},
  pages={23--32},
  year={2018},
  publisher={Elsevier}
}

@article{Roohi2018,
  doi={10.1016/j.jcp.2017.10.033},
  title={A generalized form of the {Bernoulli} Trial collision scheme in {DSMC}: Derivation and evaluation},
  author={Roohi, Ehsan and Stefanov, Stefan and Shoja-Sani, Ahmad and Ejraei, Hossein},
  journal={Journal of Computational Physics},
  volume={354},
  pages={476--492},
  year={2018},
  publisher={Elsevier}
}

@article{Taheri2022,
  doi={10.1063/5.0076025},
  author  = {Taheri, E. and Roohi, E. and Stefanov, S.},
  title   = {A symmetrized and simplified {Bernoulli} trial collision scheme in direct simulation Monte Carlo},
  journal = {Physics of Fluids},
  volume  = {34},
  pages   = {012010},
  year    = {2022}
}

@article{Javani2024,
  doi={10.1016/j.compfluid.2024.106188},
  author  = {Javani, M. and Roohi, E. and Stefanov, S.},
  title   = {Symmetrized generalized and simplified {Bernoulli}-trials collision schemes in {DSMC}},
  journal = {Computers \& Fluids},
  volume  = {272},
  pages   = {106188},
  year    = {2024}
}

@article{ShojaSani2026,
 author={Shoja-Sani, Ahmad and Roohi, Ehsan and Javani, Maryam and Akhlaghi, Hassan and Stefanov, Stefan},
 title={Efficient collision algorithms in {DSMC} for rarefied gas dynamics: Markovian {NTC}-pre-scan and {Bernoulli}-trial schemes},
 journal={Vacuum}, volume={247}, pages={115138}, year={2026},
 doi={10.1016/j.vacuum.2026.115138}
}

@article{Bird1963,
 author={Bird, G. A.},title={Approach to Translational Equilibrium in a Rigid Sphere Gas},
 journal={Physics of Fluids},volume={6},number={10},pages={1518--1519},year={1963},doi={10.1063/1.1710976}}

@article{Taheri2017,
 author={Taheri, Elmira and Roohi, Ehsan and Stefanov, Stefan},
 title={On the convergence of the simplified {Bernoulli} trial collision scheme in rarefied {Fourier} flow},
 journal={Physics of Fluids},volume={29},pages={062003},year={2017},doi={10.1063/1.4985712}}

@article{Strang1968,
 author={Strang, Gilbert},title={On the Construction and Comparison of Difference Schemes},
 journal={SIAM Journal on Numerical Analysis},volume={5},number={3},pages={506--517},year={1968},doi={10.1137/0705041}}

@article{Amiri2014,
 author={Amiri-Jaghargh, Ali and Roohi, Ehsan and Stefanov, Stefan and Nami, Hassan and Niazmand, Hamid},
 title={{DSMC} simulation of micro/nano flows using {SBT}--{TAS} technique},
 journal={Computers \& Fluids},
 year={2014},
 doi={10.1016/j.compfluid.2014.07.003},
 volume={102},
 pages={266--276}
}

@article{Roohi2016,
 author={Roohi, Ehsan and Stefanov, Stefan},
 title={Collision partner selection schemes in {DSMC}: From micro/nano flows to hypersonic flows},
 journal={Physics Reports},
 year={2016},
 doi={10.1016/j.physrep.2016.08.002},
 volume={656},
 pages={1--38}
}

@article{Goshayeshi2015ISBT,
 author={Goshayeshi, Bijan and Roohi, Ehsan and Stefanov, Stefan},
 title={A novel simplified {Bernoulli} trials collision scheme in the direct simulation Monte Carlo with intelligence over particle distances},
 journal={Physics of Fluids},
 year={2015},
 doi={10.1063/1.4933251},
 volume={27},
 number={10},
 pages={107104}
}

@article{Goshayeshi2015TAS,
 author={Goshayeshi, Bijan and Roohi, Ehsan and Stefanov, Stefan},
 title={{DSMC} simulation of hypersonic flows using an improved {SBT}-{TAS} technique},
 journal={Journal of Computational Physics},
 year={2015},
 doi={10.1016/j.jcp.2015.09.027},
 volume={303},
 pages={28--44}
}

@article{Stefanov2022Hybrid,
 author={Stefanov, Stefan and Roohi, Ehsan and Shoja-Sani, Ahmad},
 title={A novel transient-adaptive subcell algorithm with a hybrid application of different collision techniques in direct simulation Monte Carlo ({DSMC})},
 journal={Physics of Fluids},
 year={2022},
 doi={10.1063/5.0104613},
 volume={34},
 number={9},
 pages={092003}
}

@article{Shterev2021,
 author={Shterev, Kiril},
 title={The Correctness of the Simplified {Bernoulli} Trial ({SBT}) Collision Scheme of Calculations of Two-Dimensional Flows},
 journal={Micromachines},
 year={2021},
 doi={10.3390/mi12020127},
 volume={12},
 number={2},
 pages={127}
}

@article{Nanbu1980,
 author={Nanbu, Kenichi},
 title={Direct Simulation Scheme Derived from the {Boltzmann} Equation. I. Monocomponent Gases},
 journal={Journal of the Physical Society of Japan},
 year={1980},
 doi={10.1143/jpsj.49.2042},
 volume={49},
 number={5},
 pages={2042--2049}
}

@article{Koura1986,
 author={Koura, Katsuhisa},
 title={Null-collision technique in the direct-simulation Monte Carlo method},
 journal={The Physics of Fluids},
 year={1986},
 doi={10.1063/1.865826},
 volume={29},
 number={11},
 pages={3509--3511}
}

@article{Gallis2009Convergence,
 author={Gallis, M.A. and Torczynski, J.R. and Rader, D.J. and Bird, G.A.},
 title={Convergence behavior of a new {DSMC} algorithm},
 journal={Journal of Computational Physics},
 year={2009},
 doi={10.1016/j.jcp.2009.03.021},
 volume={228},
 number={12},
 pages={4532--4548}
}

@article{Hadjiconstantinou2003Statistics,
 author={Hadjiconstantinou, Nicolas G. and Garcia, Alejandro L. and Bazant, Martin Z. and He, Gang},
 title={Statistical error in particle simulations of hydrodynamic phenomena},
 journal={Journal of Computational Physics},
 year={2003},
 doi={10.1016/s0021-9991(03)00099-8},
 volume={187},
 number={1},
 pages={274--297}
}

@inproceedings{KaplanOran2001,
 author={Kaplan, Carolyn R. and Oran, Elaine S.},
 title={Nonlinear filtering for low-velocity gaseous microflows},
 year={2001},
 doi={10.1063/1.1407598},
 volume={585},
 pages={472--479},
 booktitle={AIP Conference Proceedings}
}

@inproceedings{LeBeau2003,
 author={LeBeau, Gerald and Jacikas, Katie and Lumpkin, Forrest},
 title={Virtual Sub-Cells for the Direct Simulation Monte Carlo Method},
 year={2003},
 doi={10.2514/6.2003-1031},
 booktitle={41st Aerospace Sciences Meeting and Exhibit}
}

@article{Padilla2026,
 author={Padilla, Francis and Roohi, Ehsan and Stefanov, Stefan},
 title={{Bernoulli} Trial ({BT}) Collision Operator Scheme Family in the Direct Simulation Monte-Carlo ({DSMC}): Derivation and Evaluation},
 journal={Cybernetics and Information Technologies},
 year={2026},
 doi={10.2478/cait-2026-0011},
 volume={26},
 number={2},
 pages={3--14}
}

@article{GBTTAS2020,
 author={Shoja-Sani, Ahmad and Roohi, Ehsan and Stefanov, Stefan},
 title={Evaluation of the generalized bernoulli trial-transient adaptive subcell ({GBT}-{TAS}) collision scheme in treating rarefied gas flows},
 journal={Computers \& Fluids},
 year={2020},
 doi={10.1016/j.compfluid.2020.104740},
 volume={213},
 pages={104740}
}

@article{Su2010TAS,
 author={Su, C.C. and Tseng, K.C. and Cave, H.M. and Wu, J.S. and Lian, Y.Y. and Kuo, T.C. and Jermy, M.C.},
 title={Implementation of a transient adaptive sub-cell module for the parallel-{DSMC} code using unstructured grids},
 journal={Computers \& Fluids},
 year={2010},
 doi={10.1016/j.compfluid.2010.02.003},
 volume={39},
 number={7},
 pages={1136--1145}
}

@article{Wagner1992,
 author={Wagner, Wolfgang},
 title={A convergence proof for Bird's direct simulation Monte Carlo method for the {Boltzmann} equation},
 journal={Journal of Statistical Physics},
 year={1992},
 doi={10.1007/bf01055714},
 volume={66},
 number={3-4},
 pages={1011--1044}
}

@article{Amiri2013,
 author={Amiri-Jaghargh, Ali and Roohi, Ehsan and Niazmand, Hamid and Stefanov, Stefan},
 title={{DSMC} Simulation of Low {Knudsen} Micro/Nanoflows Using Small Number of Particles per Cells},
 journal={Journal of Heat Transfer},volume={135},number={10},pages={101008},year={2013},doi={10.1115/1.4024505}}

@article{Ivanov1988Majorant,
 author={Ivanov, M. S. and Rogazinskii, S. V.},
 title={Comparative analysis of direct simulation algorithms in rarefied gas dynamics},
 journal={U.S.S.R. Computational Mathematics and Mathematical Physics},
 volume={28}, number={4}, pages={63--71}, year={1988},
 doi={10.1016/0041-5553(88)90112-7}}

@article{Alsmeyer1976,
 author={Alsmeyer, H.},
 title={Density profiles in argon and nitrogen shock waves measured by the absorption of an electron beam},
 journal={Journal of Fluid Mechanics}, volume={74}, number={3}, pages={497--513}, year={1976},
 doi={10.1017/S0022112076001912}}

@article{DonevGarciaAlder2008,
  author = {Donev, Aleksandar and Garcia, Alejandro L. and Alder, Berni J.},
  title = {Stochastic Event-Driven Molecular Dynamics},
  journal = {Journal of Computational Physics},
  volume = {227},
  pages = {2644--2665},
  year = {2008},
  doi = {10.1016/j.jcp.2007.11.010}
}

@article{Ohwada1993,
  author={Ohwada, Taku},
  title={Structure of normal shock waves: Direct numerical analysis of the Boltzmann equation for hard-sphere molecules},
  journal={Phys. Fluids A}, year={1993}, volume={5}, pages={217--234}, doi={10.1063/1.858777}
}

@article{OhwadaXu2004,
  author={Ohwada, Taku and Xu, Kun},
  title={The kinetic scheme for the full-Burnett equations},
  journal={J. Comput. Phys.}, year={2004}, volume={201}, pages={315--332}, doi={10.1016/j.jcp.2004.05.017}
}

@article{Gallis2006Normal,
  author={Gallis, M. A. and Torczynski, J. R. and Rader, D. J. and Tij, M. and Santos, A.},
  title={Normal solutions of the {Boltzmann} equation for highly nonequilibrium {Fourier} flow and {Couette} flow},
  journal={Phys. Fluids}, year={2006}, volume={18}, pages={017104}, doi={10.1063/1.2166449}
}

@article{Rader2006Fourier,
  author={Rader, D. J. and Gallis, M. A. and Torczynski, J. R. and Wagner, W.},
  title={Direct simulation {Monte Carlo} convergence behavior of the hard-sphere-gas thermal conductivity for {Fourier} heat flow},
  journal={Phys. Fluids}, year={2006}, volume={18}, pages={077102}, doi={10.1063/1.2213640}
}

@techreport{RjasanowWagner1998,
  author={Rjasanow, Sergej and Wagner, Wolfgang},
  title={On time counting procedures in the {DSMC} method for rarefied gases},
  institution={Weierstrass Institute for Applied Analysis and Stochastics}, number={349},
  year={1997}, doi={10.20347/WIAS.PREPRINT.349}
}

@book{RoohiAkhlaghiStefanov2025,
 author={Roohi, Ehsan and Akhlaghi, Hassan and Stefanov, Stefan},
 title={Advances in Direct Simulation Monte Carlo: From Micro-Scale to Rarefied Flow Phenomena},
 publisher={Springer Nature Singapore}, address={Singapore}, year={2025},
 doi={10.1007/978-981-96-8200-3}
}

@article{Roohi2026Progress,
 author={Roohi, Ehsan},
 title={From direct simulation Monte Carlo to physics-aware neural operators: A critical review of kinetic modeling for rarefied and hypersonic aerospace flows},
 journal={Progress in Aerospace Sciences}, volume={167}, pages={101261}, year={2026},
 doi={10.1016/j.paerosci.2026.101261}
}

@article{Roohi2026TAITAS,
 author={Roohi, Ehsan and Shoja-Sani, Ahmad and Goshayeshi, Bijan and Stefanov, Stefan},
 title={A Time-Averaged Inverse Transient Adaptive Sub-cell strategy for {DSMC} applied to shock tube problem},
 journal={Vacuum}, volume={244}, pages={114824}, year={2026},
 doi={10.1016/j.vacuum.2025.114824}
}

@article{BelotserkovskiiYanitskiy1975,
 author={Belotserkovskii, O. M. and Yanitskii, V. E.},
 title={The statistical particles-in-cells method for solving rarefied gas dynamics problems},
 journal={U.S.S.R. Computational Mathematics and Mathematical Physics},
 volume={15}, number={5}, pages={101--114}, year={1975},
 doi={10.1016/0041-5553(75)90108-1}
}

@article{Macrossan2010,
 author={Macrossan, M. N.},
 title={Searching for a near neighbor particle in {DSMC} cells using pseudo-subcells},
 journal={Journal of Computational Physics}, volume={229}, number={17},
 pages={5857--5861}, year={2010}, doi={10.1016/j.jcp.2010.04.042}
}

@article{DoduladTcheremissine2013,
 author={Dodulad, O. I. and Tcheremissine, F. G.},
 title={Computation of a shock wave structure in monatomic gas with accuracy control},
 journal={Computational Mathematics and Mathematical Physics},
 volume={53}, number={6}, pages={827--844}, year={2013},
 doi={10.1134/S0965542513060055}
}

@article{OhwadaRIMS1993,
 author={Ohwada, Taku},
 title={Structure of normal shock waves},
 journal={RIMS Kokyuroku},
 volume={824},
 pages={255--267},
 year={1993},
 archiveid={2433/83238},
 url={https://hdl.handle.net/2433/83238}
}

@article{XuHuang2011,
 author={Xu, Kun and Huang, Juan-Chen},
 title={An improved unified gas-kinetic scheme and the study of shock structures},
 journal={IMA Journal of Applied Mathematics},
 volume={76}, pages={698--711}, year={2011},
 doi={10.1093/imamat/hxr002}
}

@article{ShojaSani2021RelaxShock,
 author={Ahmad Shoja-Sani and Ehsan Roohi and Stefan Stefanov},
 title={Homogeneous relaxation and shock wave problems: Assessment of the simplified and generalized {Bernoulli} trial collision schemes},
 journal={Physics of Fluids},volume={33},number={3},pages={032004},year={2021},
 doi={10.1063/5.0039071}}

@article{Javani2024SSBTShock,
 author={Maryam Javani and Ehsan Roohi and Elmira Taheri},
 title={On the convergence of the symmetrized and simplified {Bernoulli} trial ({SSBT}) collision scheme in shock wave problem},
 journal={AIP Conference Proceedings},volume={3050},pages={080001},year={2024},
 doi={10.1063/5.0187366}}

@article{Stefanov2019Fourier,
  author  = {Stefanov, S. K.},
  title   = {On the basic concepts of the direct simulation {Monte Carlo} method},
  journal = {Physics of Fluids},
  volume  = {31}, number = {6}, pages = {067104}, year = {2019},
  doi     = {10.1063/1.5099042}
}

@article{Doi2025Bearing,
  author = {Doi, Toshiyuki},
  title = {Kinetic theory analysis of microscale lubrication of a gas between eccentric circular cylinders: Effect of rotation of the outer cylinder},
  journal = {Phys. Rev. Fluids},
  volume = {10},
  pages = {024201},
  year = {2025},
  doi = {10.1103/PhysRevFluids.10.024201}
}

@article{Doi2023Lubrication,
  author = {Doi, Toshiyuki},
  title = {Generalized {Reynolds} equation for microscale lubrication between eccentric circular cylinders based on kinetic theory},
  journal = {J. Fluid Mech.}, volume = {974}, pages = {A13}, year = {2023},
  doi = {10.1017/jfm.2023.732}
}

@article{Taguchi2021Rotation,
  author = {Taguchi, Satoshi and Tsuji, Tetsuro and Kotera, Masashi},
  title = {Transient behaviour of a rarefied gas around a sphere caused by impulsive rotation},
  journal = {J. Fluid Mech.}, volume = {909}, pages = {A6}, year = {2021},
  doi = {10.1017/jfm.2020.907}
}

@misc{Roohi2026APSShock,
  author = {Roohi, Ehsan},
  title = {What do {PINNs} really learn in rarefied gases? {Shocks}, scalar-excess recovery, and cold-to-hot cavities},
  howpublished = {APS Division of Fluid Dynamics Meeting, Orlando, Florida, presentation C38.2},
  url = {https://dfd-meeting.aps.org/dfd/2026/events/C38/2},
  month = nov, year = {2026}
}

@misc{Roohi2026KineticPartI,
  author = {Roohi, Ehsan},
  title = {Physics-informed neural networks for strong normal shocks in kinetic relaxation models and the hard-sphere {Boltzmann} equation},
  year = {2026},
  doi = {10.48550/arXiv.2611.31996},
  url = {https://arxiv.org/abs/2611.31996},
  eprint = {2611.31996},
  archivePrefix = {arXiv},
  primaryClass = {physics.flu-dyn}
}
\end{document}

\title{Supplementary material: Collision survival probabilities and transport bias in low-occupancy direct simulation Monte Carlo}
\author{Ehsan Roohi}
\email{roohie@umass.edu}
\affiliation{Department of Mechanical and Industrial Engineering, University of Massachusetts Amherst, Amherst, MA 01003, USA}
\author{Stefan Stefanov}
\affiliation{Institute of Mechanics, Bulgarian Academy of Sciences, Acad. G. Bontchev Str., 1113 Sofia, Bulgaria}
\affiliation{Centre of Excellence in Informatics and Information and Communication Technologies, Sofia, Bulgaria}
\author{Ahmad Shoja-Sani}
\affiliation{Ferdowsi University of Mashhad, Mashhad, Iran}
\maketitle
\FloatBarrier

\section{Related collision methods}
Earlier direct simulation Monte Carlo (DSMC) methods selected collision partners using shifted dual grids \citep{Amiri2013}, selection based on interparticle distance \citep{Goshayeshi2015ISBT}, generalized Bernoulli trials with transient adaptive subcells (GBT--TAS) \citep{GBTTAS2020}, occupancy-dependent hybrid sampling \citep{Stefanov2022Hybrid} and time-averaged inverse adaptation \citep{Roohi2026TAITAS}. The pre-scan version of the no-time-counter (NTC) method estimates an upper bound for the candidate collision rate \citep{ShojaSani2026}. Reviews \citep{Roohi2016,Roohi2026Progress} and the monograph \citep{RoohiAkhlaghiStefanov2025} discuss these developments. Our comparisons use fixed subcells to study the probability and sequence of collisions.

\FloatBarrier
\section{Fixed-cell generator, count laws and interpretation}\label{app:generator}
For a fixed set of $N$ particles and an observable $\varphi(V)$, the intended collision generator is
\begin{equation}
 \mathcal L\varphi(V)=\sum_{i<j}a_{ij}(V)
 \left[\E_\Omega\varphi(V'_{ij})-\varphi(V)\right].
\end{equation}
The bound $M$ from the main article exceeds or equals the total collision rate in every state with the same conserved energy. Uniform pair proposals and acceptance by the current speed produce $Q=I+\mathcal L/M$, hence
\begin{equation}
 e^{h\mathcal L}=e^{-Mh}\sum_{k=0}^{\infty}\frac{(Mh)^k}{k!}Q^k.
\end{equation}
This identity shows why Poisson sampling of candidate pairs gives the exact collision evolution while cell membership remains fixed. The accepted event count need not be Poisson at the initial total physical rate, since the rates can change after an event. Algorithms that draw waiting times from the current collision rates give a related construction \citep{Gillespie1977SSA}. Spatial transport still depends on the grid.

\paragraph{First-event consistency.} For a primary with $w_i>0$, choose the representative $J$ uniformly among its eligible partners $S_i$ and draw the event count from a Poisson distribution with mean $h w_i a_{iJ}$. Coupling the first collision to $J$ gives
\begin{equation}
 \E[\varphi(V_{\rm out})-\varphi(V)]
 =h\sum_{j\in S_i}a_{ij}(V)
 \left[\E_\Omega\varphi(V'_{ij})-\varphi(V)\right]+O(h^2)
 =h\mathcal L_i\varphi+O(h^2).
 \label{eq:firstevent}
\end{equation}
Here $V_{\rm out}$ is the velocity state after this primary trial. An independent, uniformly redrawn first partner would replace each pair rate in the leading term by the mean rate across partners.
Multiple events contribute at order $h^2$ on bounded energy shells. Summing over primary particles therefore gives first-order consistency with $\mathcal L=\sum_i\mathcal L_i$. Second order would also require the coefficient $\mathcal L^2/2$. Drawing a representative rate and holding the rate fixed after a collision can each change this coefficient. For example, a representative intensity equal to 2 or 6 with equal probability has mean 4 but second moment 20, not 16. The leading coefficient in the probability of two events is then 10 instead of 8. In a state chain with successive rates $a$ and $b$, the two-event coefficient is $ab/2$, while freezing the initial count produces $a^2/2$. Thus the probability of a second event depends on how the first event changes the rates.

For $N=2$, elastic collisions preserve the relative speed $g$ of the two hard spheres. Given the initial velocities, the Poisson variant (P-SBT) of simplified Bernoulli trials (SBT) samples the exact Poisson law for the number of collisions, with mean $bg$, where $b=hF_{\mathrm{num}}\pi d^2/V_c$. Independently Maxwellian initial velocities give
\begin{equation}
 P(C=k)=\int_0^\infty 4\pi g^2
 \left(\frac{m}{4\pi k_BT}\right)^{3/2}e^{-mg^2/(4k_BT)}
 e^{-bg}\frac{(bg)^k}{k!}\,\dd g.
\end{equation}
Here the variation in collision rate comes entirely from the initial relative speed. In a cell with several possible partners, collisions also change the rates of subsequent encounters. Counts measured at fixed membership therefore isolate the collision rule. Large relative differences in very rare counts may still contribute little to momentum transfer.

\FloatBarrier
\section{Collision sampling schematics}
\begin{figure}[!tb]\centering\includegraphics[width=.7\textwidth]{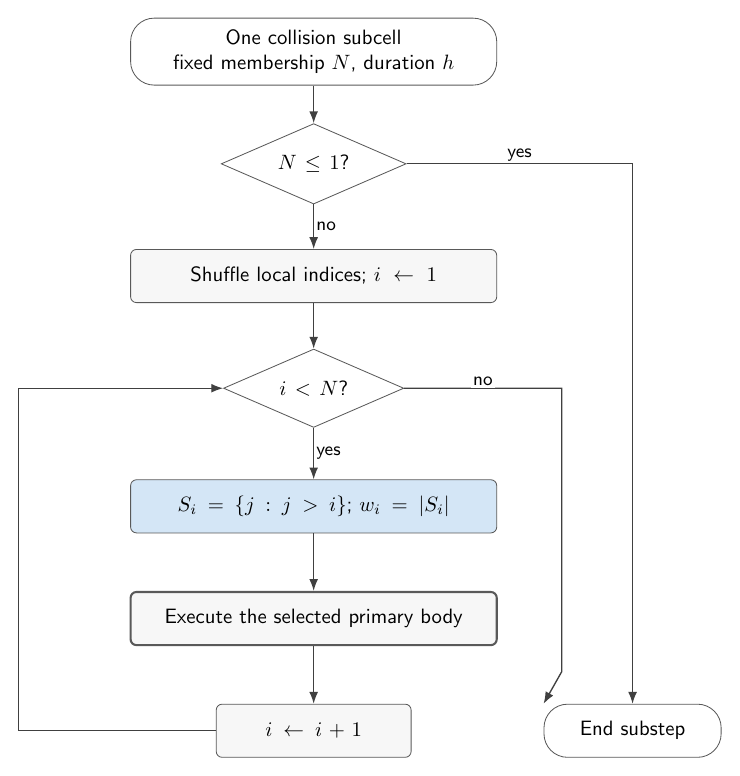}\caption{Shared primary sweep, drawn with one-based indices. The eligible set contains later shuffled indices, so a particle can participate again as another primary's partner. Membership is fixed during a collision substep.}\label{fig:flowouter}\end{figure}

\begin{figure}[!tp]\centering\includegraphics[width=\textwidth,height=.8\textheight,keepaspectratio]{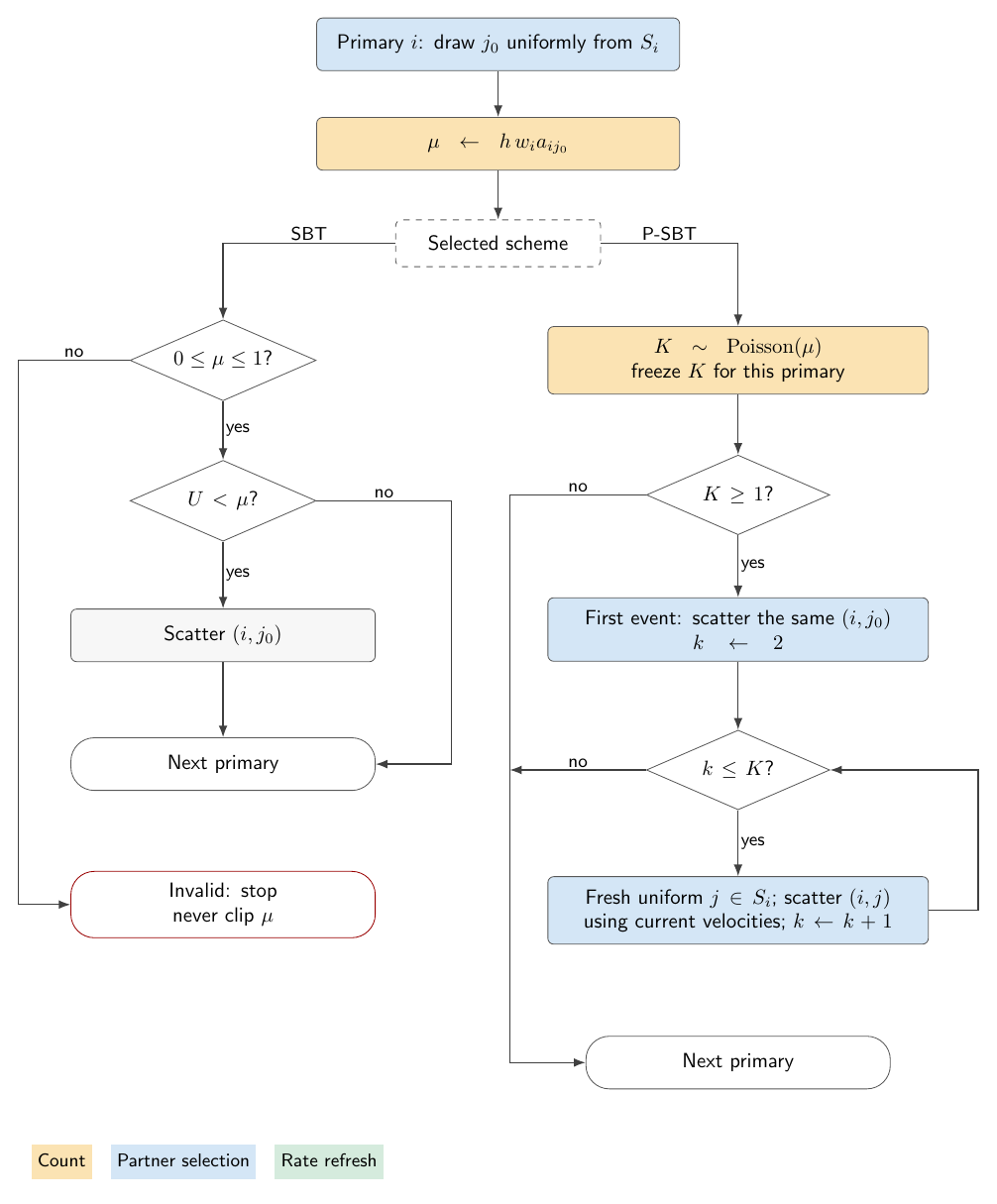}\caption{Collision counts in the sweep of Fig.~\ref{fig:flowouter} for SBT and representative-rate P-SBT. The first P-SBT event remains tied to the representative partner. Subsequent events use fresh uniform partners while keeping the sampled count fixed.}\label{fig:flowcounts}\end{figure}

\begin{figure}[!tp]\centering\includegraphics[width=\textwidth,height=.8\textheight,keepaspectratio]{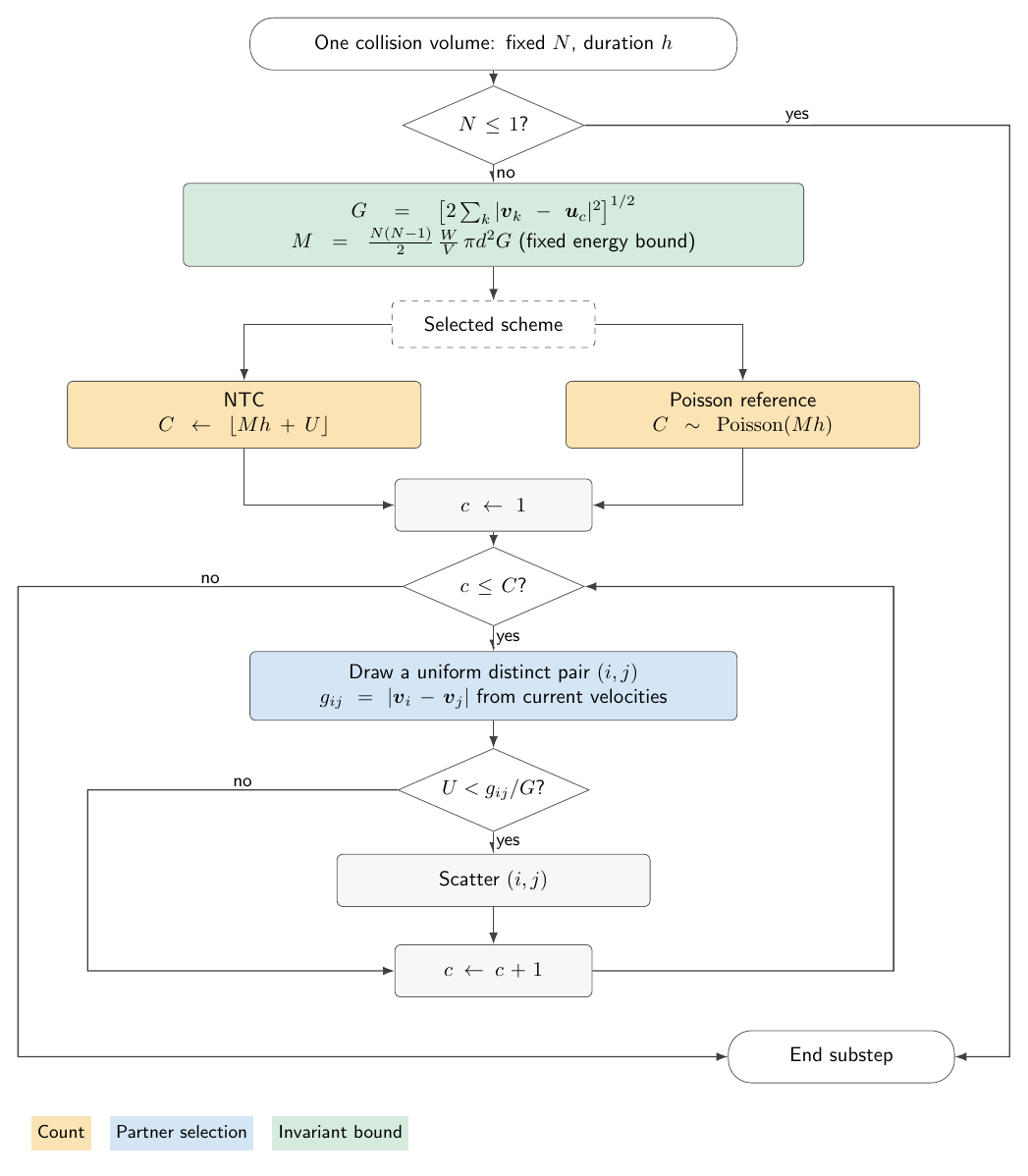}\caption{Selection of candidate pairs in NTC and whole-cell Poisson sampling. Both methods use the same energy bound, propose pairs uniformly and accept them according to their current relative speed. Only the candidate count differs. The count draws and acceptance tests use independent uniform random variates. Dashed boxes indicate the choice of method.}\label{fig:flowcandidates}\end{figure}

Figure~\ref{fig:flowouter} defines the partner list. Figure~\ref{fig:flowcounts} shows how a primary draws the number of collisions, and Fig.~\ref{fig:flowcandidates} shows how the whole-cell algorithms generate candidate pairs.

\FloatBarrier
\section{Additional homogeneous relaxation results}
Tables~\ref{tab:relax-small} and~\ref{tab:relax-four} report repeated collisions and the paired deviations of the normal-stress anisotropy $A$ from the whole-cell Poisson result. Weighted P-SBT draws partners from rates evaluated at the start of the primary trial. Event-adaptive P-SBT (EA-P-SBT, abbreviated EA in the figures) updates these rates after each event. The exponential-survival variant of SBT (SBT-exp) allows one event per trial. With only one eligible partner, repeated collisions must involve the same pair. We therefore average changes in rate only over events in trials with at least two eligible partners.
\begin{table}
\centering
\small
\begin{tabular}{rrlrr}
\toprule
$N$ & $h$ & Scheme & Multiple events (\%) & $A(0.8)-A_{\mathrm{Pois}}(0.8)$ \\
\midrule
2 & 0.1 & SBT & 0.00 & $-0.00805\pm 0.00108$ \\
2 & 0.1 & P-SBT & 0.47 & $-0.00012\pm 0.00136$ \\
2 & 0.1 & Weighted P-SBT & 0.47 & $-0.00083\pm 0.00089$ \\
2 & 0.1 & EA-P-SBT & 0.46 & $+0.00055\pm 0.00095$ \\
2 & 0.2 & SBT & 0.00 & $-0.01543\pm 0.00096$ \\
2 & 0.2 & P-SBT & 1.75 & $-0.00175\pm 0.00113$ \\
2 & 0.2 & Weighted P-SBT & 1.74 & $+0.00006\pm 0.00091$ \\
2 & 0.2 & EA-P-SBT & 1.75 & $-0.00099\pm 0.00139$ \\
2 & 0.4 & SBT & 0.00 & $-0.03307\pm 0.00126$ \\
2 & 0.4 & P-SBT & 6.20 & $-0.00041\pm 0.00097$ \\
2 & 0.4 & Weighted P-SBT & 6.16 & $-0.00033\pm 0.00105$ \\
2 & 0.4 & EA-P-SBT & 6.16 & $-0.00109\pm 0.00132$ \\
3 & 0.1 & SBT & 0.00 & $-0.00524\pm 0.00082$ \\
3 & 0.1 & P-SBT & 0.39 & $+0.00298\pm 0.00098$ \\
3 & 0.1 & Weighted P-SBT & 0.37 & $+0.00253\pm 0.00101$ \\
3 & 0.1 & EA-P-SBT & 0.37 & $+0.00103\pm 0.00109$ \\
3 & 0.2 & SBT & 0.00 & $-0.01387\pm 0.00096$ \\
3 & 0.2 & P-SBT & 1.43 & $+0.00409\pm 0.00071$ \\
3 & 0.2 & Weighted P-SBT & 1.37 & $+0.00189\pm 0.00083$ \\
3 & 0.2 & EA-P-SBT & 1.39 & $+0.00065\pm 0.00098$ \\
3 & 0.4 & SBT & 0.00 & $-0.03288\pm 0.00084$ \\
3 & 0.4 & P-SBT & 4.97 & $+0.00676\pm 0.00094$ \\
3 & 0.4 & Weighted P-SBT & 4.80 & $+0.00289\pm 0.00109$ \\
3 & 0.4 & EA-P-SBT & 4.81 & $+0.00154\pm 0.00071$ \\
\bottomrule
\end{tabular}
\caption{Relaxation of stress anisotropy in two- and three-particle cells. The percentage gives the fraction of primary trials with at least two collisions. Uncertainties are one standard error of the paired differences across 16 groups.}
\label{tab:relax-small}\end{table}
\begin{table}
\centering
\small
\begin{tabular}{rrlrr}
\toprule
$N$ & $h$ & Scheme & Multiple events (\%) & $A(0.8)-A_{\mathrm{Pois}}(0.8)$ \\
\midrule
4 & 0.1 & SBT & 0.00 & $-0.00624\pm 0.00087$ \\
4 & 0.1 & P-SBT & 0.36 & $+0.00287\pm 0.00090$ \\
4 & 0.1 & Weighted P-SBT & 0.33 & $+0.00084\pm 0.00080$ \\
4 & 0.1 & EA-P-SBT & 0.33 & $+0.00106\pm 0.00079$ \\
4 & 0.2 & SBT & 0.00 & $-0.01270\pm 0.00086$ \\
4 & 0.2 & P-SBT & 1.33 & $+0.00567\pm 0.00081$ \\
4 & 0.2 & Weighted P-SBT & 1.25 & $+0.00063\pm 0.00076$ \\
4 & 0.2 & EA-P-SBT & 1.26 & $+0.00035\pm 0.00084$ \\
4 & 0.4 & SBT & 0.00 & $-0.03082\pm 0.00085$ \\
4 & 0.4 & P-SBT & 4.58 & $+0.01106\pm 0.00103$ \\
4 & 0.4 & Weighted P-SBT & 4.34 & $+0.00218\pm 0.00084$ \\
4 & 0.4 & EA-P-SBT & 4.39 & $+0.00018\pm 0.00059$ \\
\bottomrule
\end{tabular}
\caption{Relaxation of stress anisotropy in four-particle cells. The percentage gives the fraction of primary trials with at least two collisions. Uncertainties are one standard error of the paired differences across 16 groups.}
\label{tab:relax-four}\end{table}

\FloatBarrier
\begin{table}[!tbp]\centering\small
\begin{tabular}{ccrrr}\toprule
$N$ & $h$ & SBT & SBT-exp & P-SBT\\\midrule
2 & 0.1 & $-0.00805\pm0.00108$ & $-0.00053\pm0.00106$ & $-0.00012\pm0.00136$\\
2 & 0.2 & $-0.01543\pm0.00096$ & $-0.00085\pm0.00136$ & $-0.00175\pm0.00113$\\
2 & 0.4 & $-0.03307\pm0.00126$ & $-0.00019\pm0.00114$ & $-0.00041\pm0.00097$\\
3 & 0.1 & $-0.00524\pm0.00082$ & $+0.00367\pm0.00129$ & $+0.00298\pm0.00098$\\
3 & 0.2 & $-0.01387\pm0.00096$ & $+0.00666\pm0.00087$ & $+0.00409\pm0.00071$\\
3 & 0.4 & $-0.03288\pm0.00084$ & $+0.01247\pm0.00067$ & $+0.00676\pm0.00094$\\
4 & 0.1 & $-0.00624\pm0.00087$ & $+0.00501\pm0.00086$ & $+0.00287\pm0.00090$\\
4 & 0.2 & $-0.01270\pm0.00086$ & $+0.01032\pm0.00075$ & $+0.00567\pm0.00081$\\
4 & 0.4 & $-0.03082\pm0.00085$ & $+0.01881\pm0.00067$ & $+0.01106\pm0.00103$\\
\bottomrule\end{tabular}
\caption{Normal-stress anisotropy $A$ at $t=0.8$, expressed as each sampler mean minus the whole-cell Poisson mean. $N$ is the number of particles in the fixed cell and $h$ is the dimensionless time step. Each uncertainty is one standard error of the paired differences across 16 independent groups.}
\end{table}

\FloatBarrier
\section{Calculation of the velocity moments}\label{app:moments}
In the Couette and Poiseuille figures $X$ is streamwise and $Y$ is normal to the walls. For the shock figures, $x$ is normal to the front and $y,z$ are tangential. In the cavity, $x$ points along the moving lid and $y$ upward. We map stored components into these physical coordinates before labelling the figures. The solver uses an affine offset when accumulating raw velocity moments. We remove that offset before forming local central moments.

For integer $r$, the central tensor of order $r$ has components $M^{(r)}_{i_1\ldots i_r}=\rho\langle C_{i_1}\cdots C_{i_r}\rangle$. Its contractions use $C^2=C_x^2+C_y^2+C_z^2$. Thus $q_i=M^{(3)}_{ikk}/2$, $F_{5i}=M^{(5)}_{ikkll}$ and $\Delta=M^{(4)}_{kkll}-15\rho\theta^2$, with repeated Cartesian indices summed. The fourth-order tensor $R_{ij}=M^{(4)}_{kkij}-\delta_{ij} M^{(4)}_{kkll}/3-7\theta\sigma_{ij}$ selects the anisotropic part. For an uncontracted fourth moment, the stated Cartesian indices are retained without summation. In the channel data the 56 raw monomials of degree zero through five determine density, velocity, temperature and all reported moments.

The accompanying data give pointwise means, standard errors and the variation with sampling period and seed. Local Maxwellian collision-time estimates in the strong-shear Couette comparison are approximately 56--88 ns, depending on method and position. For example, the local $\Delta t/\tau_M$ range for SBT at 0.252 ns is about 0.286--0.442 percent, larger than its nominal 0.192 percent. Its temperature-based local $\sqrt{2\theta}\Delta t/\delta Y$ range is about 0.097--0.157, rather than the nominal 0.08165. The scalar temperature alone does not specify the fraction of particles with large wall-normal speeds.

The Bernoulli channel and shock comparisons use the probability bound specified in the article. The separate Fourier reproduction below records all nominal probabilities above unity.

\FloatBarrier
\section{Additional channel moment profiles}
\begin{figure}[!tbp]
\centering\includegraphics[width=\textwidth]{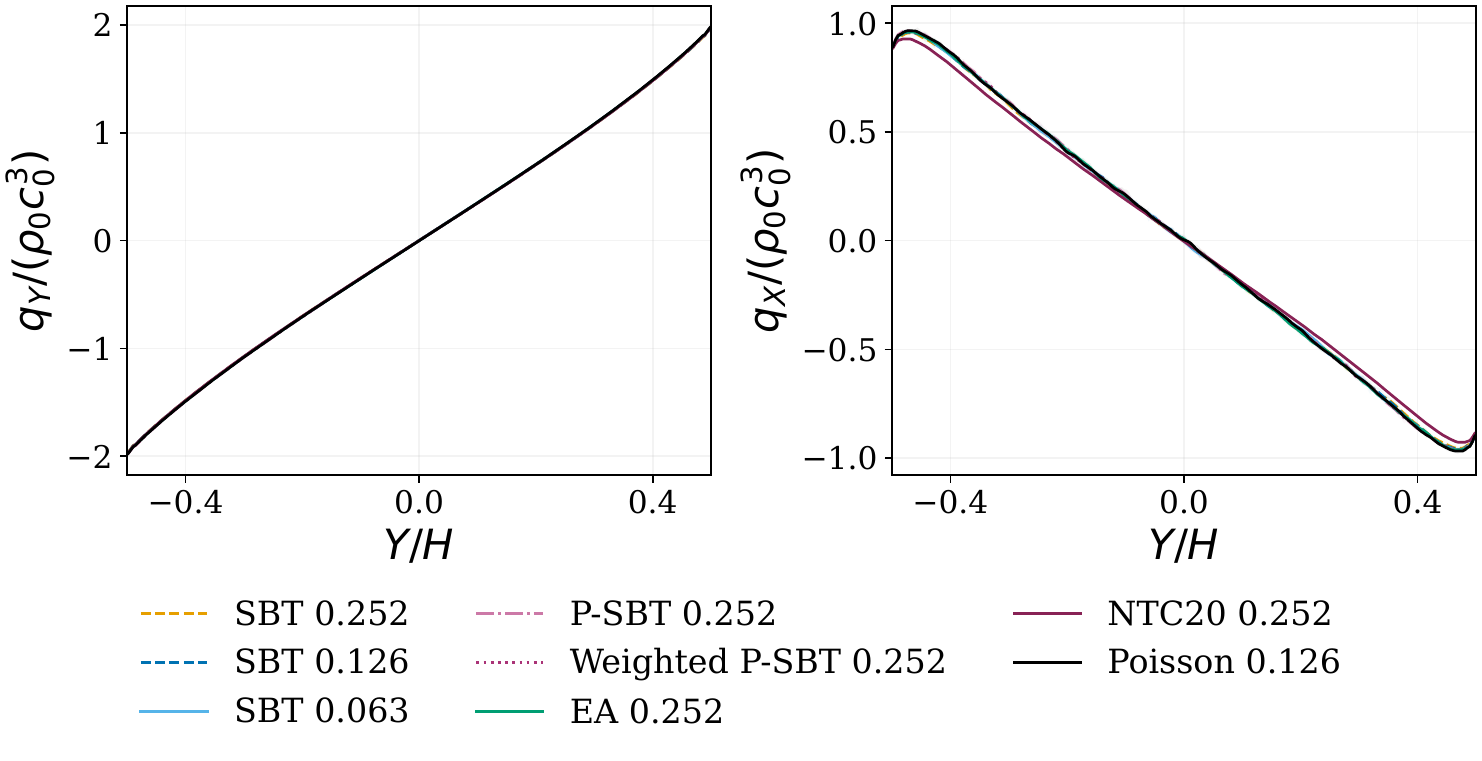}
\caption{Transverse and streamwise heat fluxes for the shared Couette sampling interval. The legend gives the time step in ns; all samplers use parent PPC2 except NTC20. The two seeds, $28.081$ ms per seed, flow conditions and uncertainty estimates match the main Couette shear comparison. For weighted P-SBT the partner probabilities are fixed during each primary trial. Means use 32 FCT passes. Bands use unfiltered samples.}
\label{fig:heat}
\end{figure}

\begin{figure}[!tbp]
\centering\includegraphics[width=\textwidth]{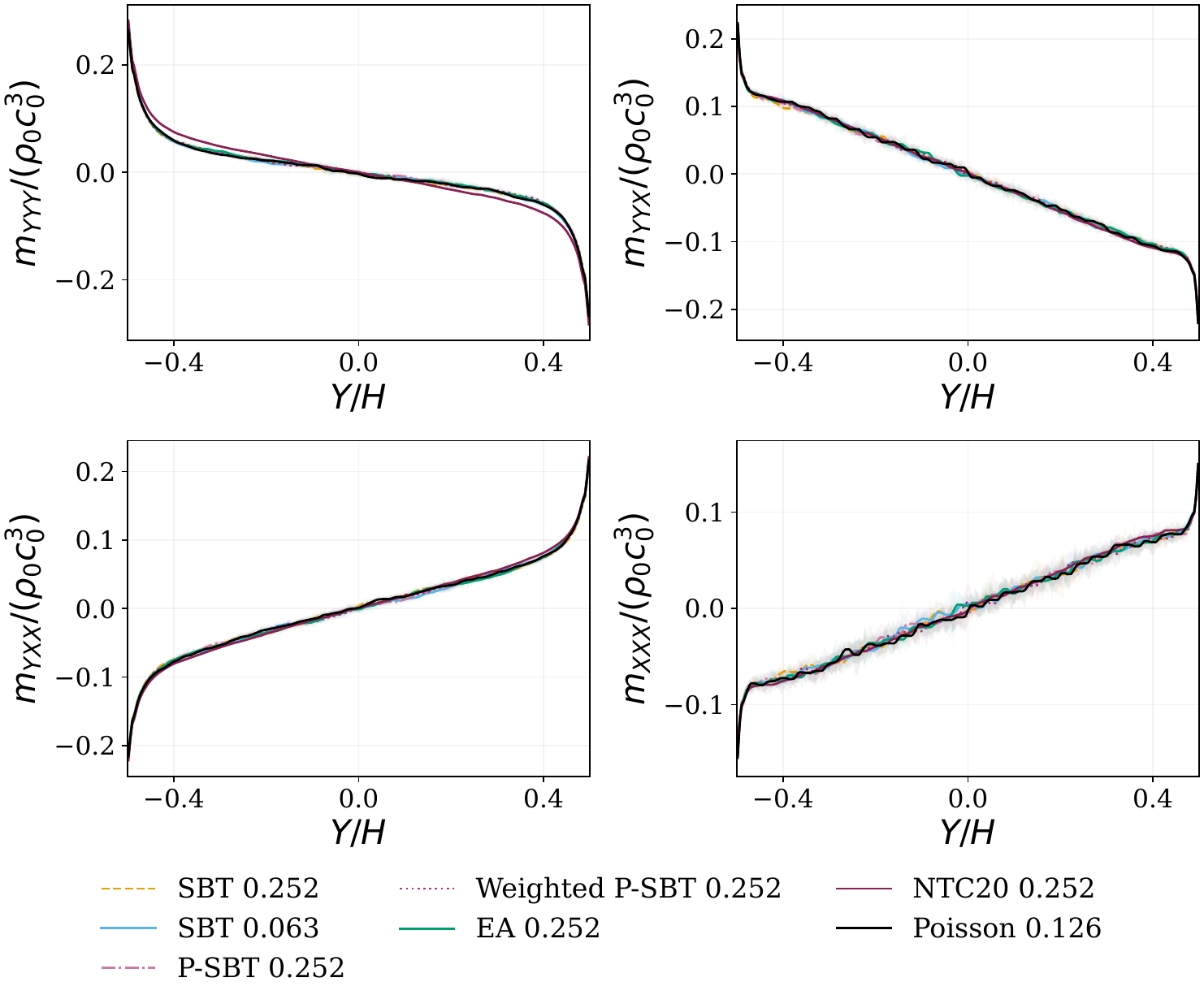}
\caption{Trace-free third moments in the common Couette comparison: $m_{YYY}$, $m_{YYX}$, $m_{YXX}$ and $m_{XXX}$. All curves use the same $28.081$ ms per seed. No symmetry constraint or mirror averaging has been applied. The legend gives the time step in ns. Means use 32 FCT passes. Bands use unfiltered samples.}
\label{fig:third}
\end{figure}

\begin{figure}[!tbp]
\centering\includegraphics[width=\textwidth]{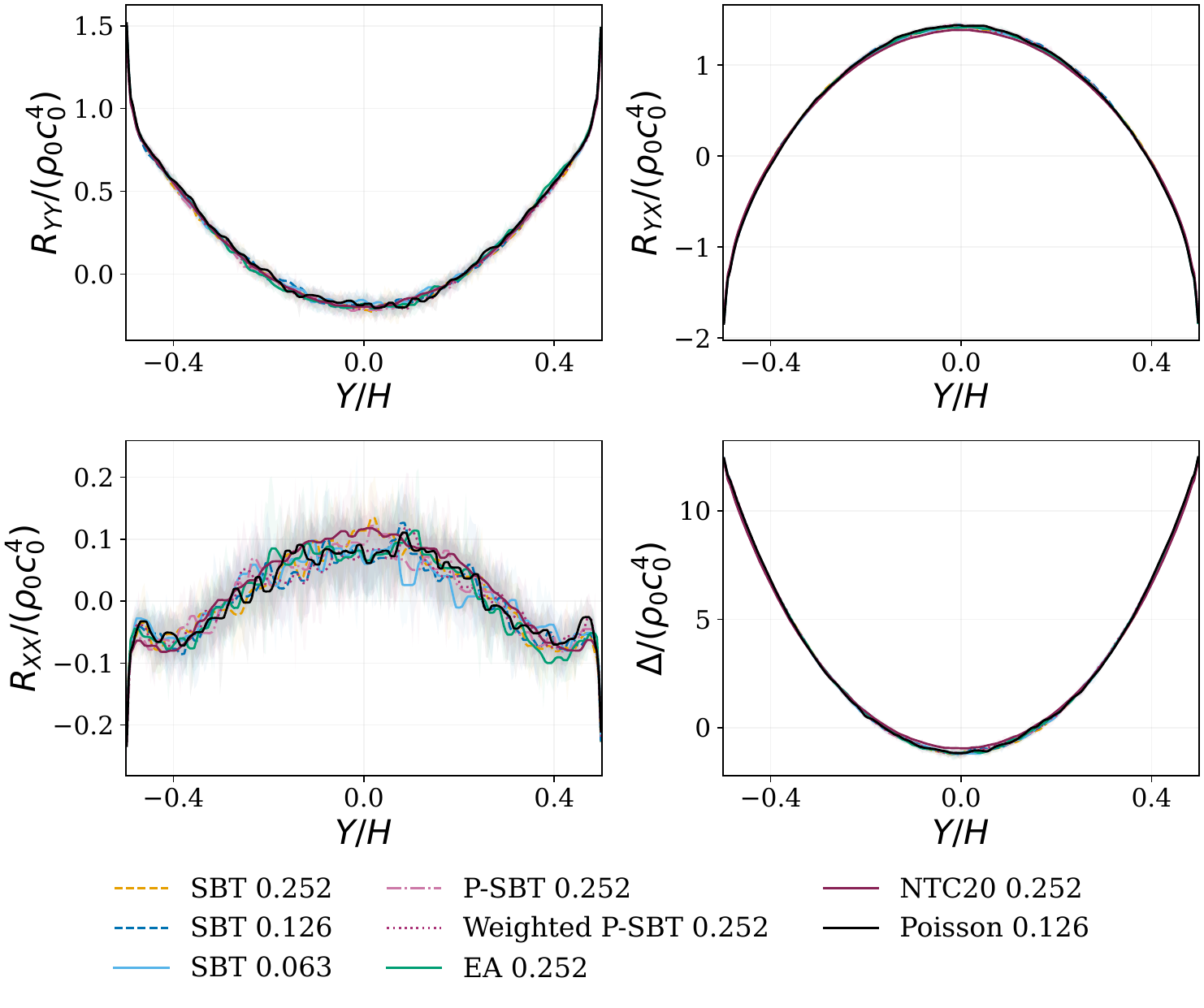}
\caption{Fourth-order contracted moments $R_{YY}$, $R_{YX}$, $R_{XX}$ and $\Delta$ over the same Couette sampling interval. The legend gives the time step in ns. The transverse coordinate is $Y$; the streamwise coordinate is $X$. The subtraction terms defining $R$ and $\Delta$ are formed from the pooled raw moments and the temperature of each trajectory. Relative differences are sensitive to small offsets near a zero crossing. Means use 32 FCT passes. Bands use unfiltered samples.}
\label{fig:fourth}
\end{figure}

\begin{figure}[!tbp]
\centering\includegraphics[width=\textwidth]{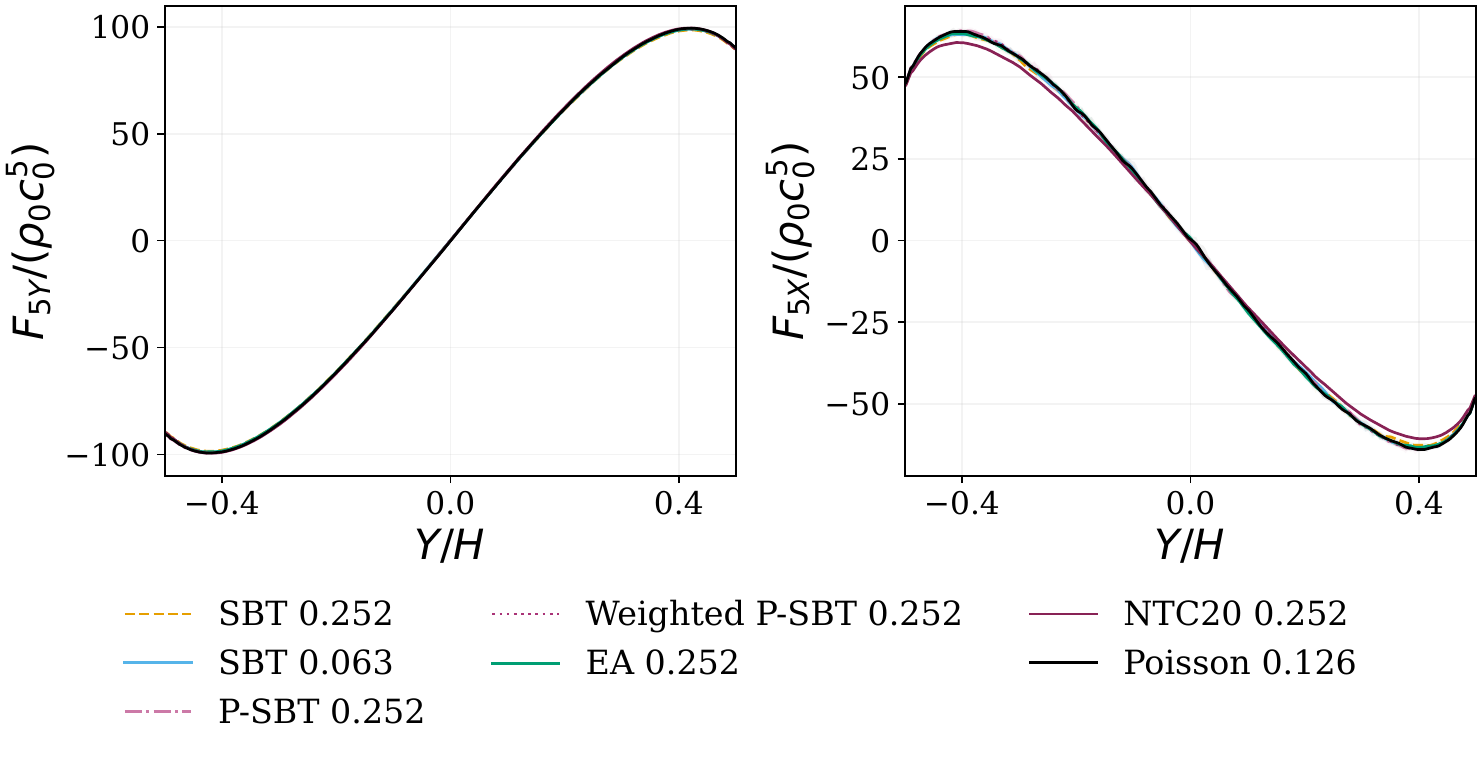}
\caption{Fifth-order contractions $F_{5Y}$ and $F_{5X}$ in the shared Couette sampling interval. The profiles show contractions of the central fifth moments, normalized by $\rho_0c_0^5$. The legend gives PPC and time step as in Fig.~\ref{fig:heat}. Means use 32 FCT passes. Bands use unfiltered samples.}
\label{fig:fifth}
\end{figure}

\begin{figure}[!tbp]
\centering\includegraphics[width=\textwidth]{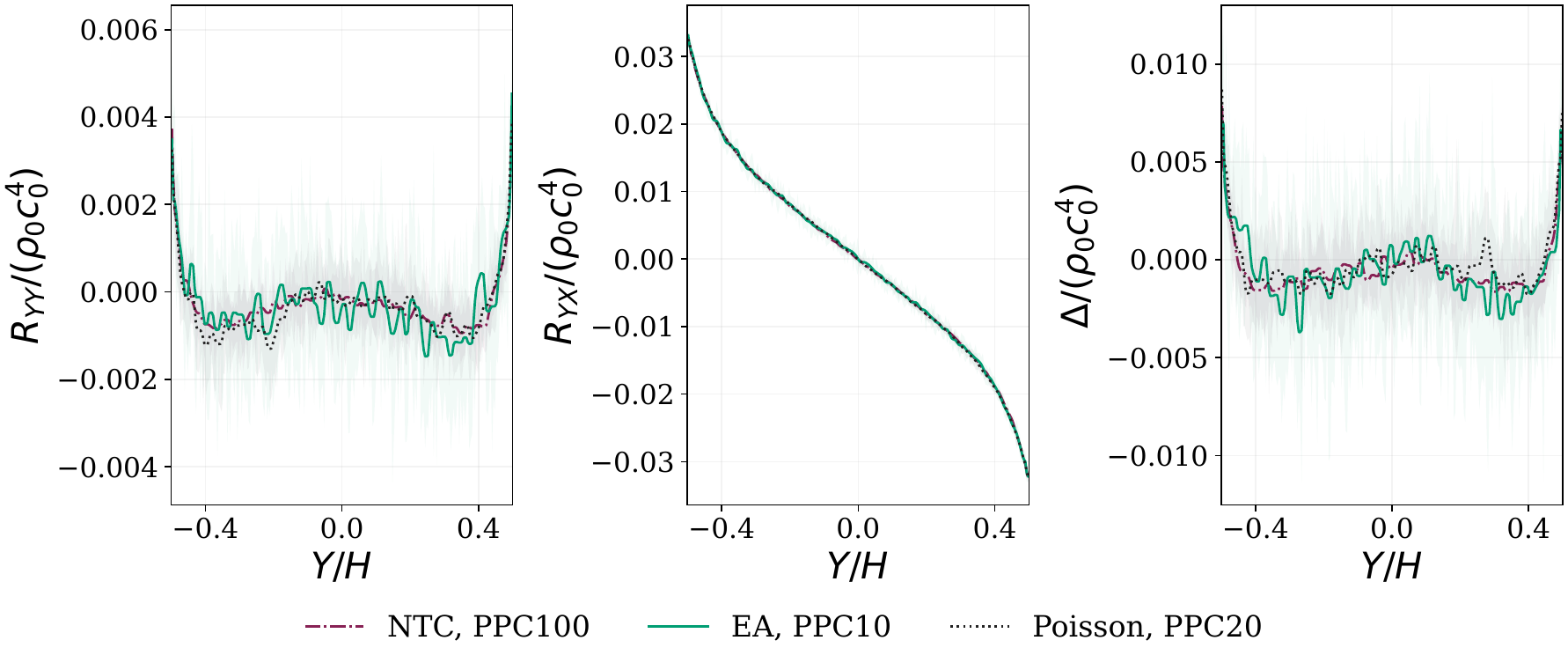}
\caption{Fourth-order Poiseuille contractions for the same trajectories and parameters as Fig.~\ref{fig:poiseuille}. The definitions and normalizations are the same as in Couette flow. Here a body force drives the gas and collisions use undivided cells. Means use 32 FCT passes. Bands use unfiltered samples.}
\label{fig:poisfourth}
\end{figure}

In the channel figures, particles per cell (PPC) refers to the mean number of simulator particles in a parent cell. A numeral after a method name gives this occupancy. We use flux-corrected transport (FCT) to smooth the channel means for display. The signed heat flux and third-, fourth- and fifth-order Couette moments appear in Figs.~\ref{fig:heat}, \ref{fig:third}, \ref{fig:fourth} and~\ref{fig:fifth}. Figure~\ref{fig:poisfourth} gives the fourth moments for force-driven Poiseuille flow.

\FloatBarrier

The Couette comparisons use matched initial states and segment seeds across methods. The accompanying data report signed wall stresses for each block and seed, and the number of particles in the subcell at each accepted collision. At each time step, the reported fractions use the total number of accepted events for that run.

\FloatBarrier
\section{Channel comparisons with 100 particles per cell in NTC}\label{app:ntc100}
\begin{figure}[!tbp]
\centering\includegraphics[width=\textwidth]{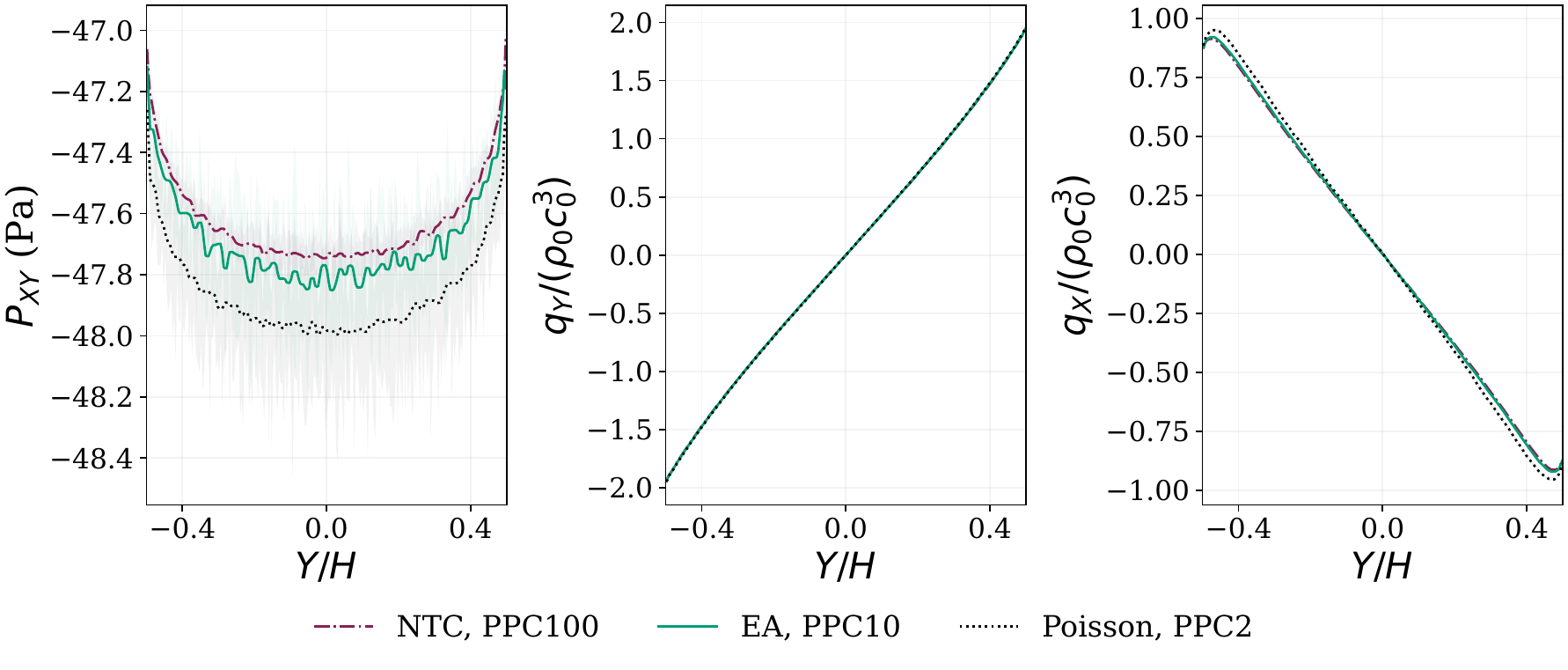}
\caption{Couette comparison without collision subcells at $\Delta t=3.025965$ ns. NTC100, EA10 and Poisson2 use 240 cells, Knudsen number $\Kn=0.05$, $U_w=1000$ m s$^{-1}$ and a common $65.845$ ms interval per seed after the first 102 segments. Bands span twice the estimated standard error from two seeds. Means use 32 FCT passes. Bands use unfiltered samples.}\label{fig:ntc100c3}
\end{figure}

\begin{figure}[!tbp]
\centering\includegraphics[width=\textwidth]{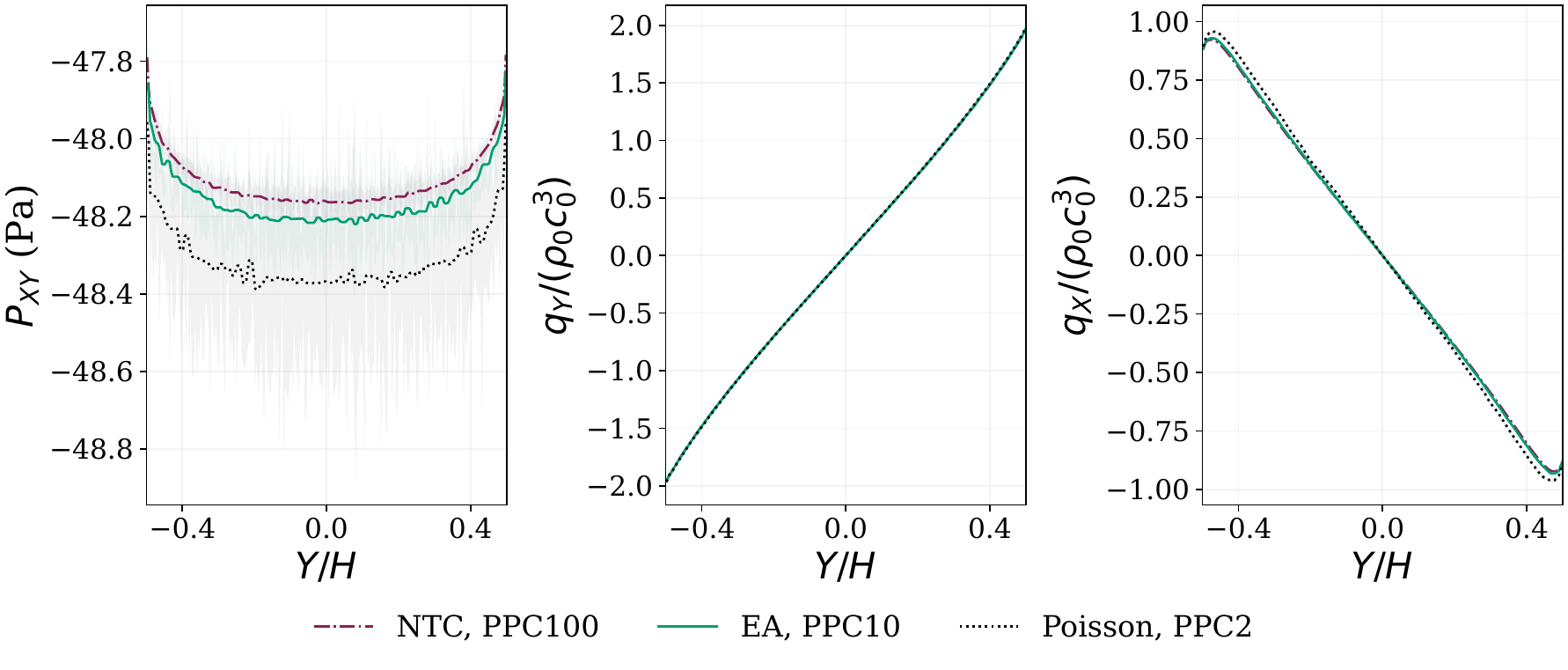}
\caption{Couette flow without subcells at $\Delta t=1.008655$ ns, using the same physical interval, geometry and particle counts as Fig.~\ref{fig:ntc100c3}. Means use 32 FCT passes. Bands use unfiltered samples.}\label{fig:ntc100c1}
\end{figure}

Here collision pairs are selected from the entire cell. Figures~\ref{fig:ntc100c3} and~\ref{fig:ntc100c1} use a $65.845$ ms sampling interval per seed for both Couette time steps, after 102 warm-up segments. At each step, NTC100 is compared with EA10 and the same-step Poisson sampler at PPC2. All cases have 240 cells, walls at $\pm1000$ m s$^{-1}$ and two trajectories. These are separate from the four-subcell comparisons in the main text.

The NTC100 mean kinetic shear changes from $-47.6245$ Pa at 3.026 ns to $-48.1106$ Pa at 1.009 ns. EA10 changes from $-47.6904$ to $-48.1565$ Pa and Poisson2 from $-47.8607$ to $-48.3136$ Pa. The kinetic shear therefore remains sensitive to the time step even at 100 particles per cell. The difference between NTC100 and Poisson2 includes the finite-occupancy and spatial errors of the Poisson reference.

\begin{figure}[!tbp]
\centering\includegraphics[width=\textwidth]{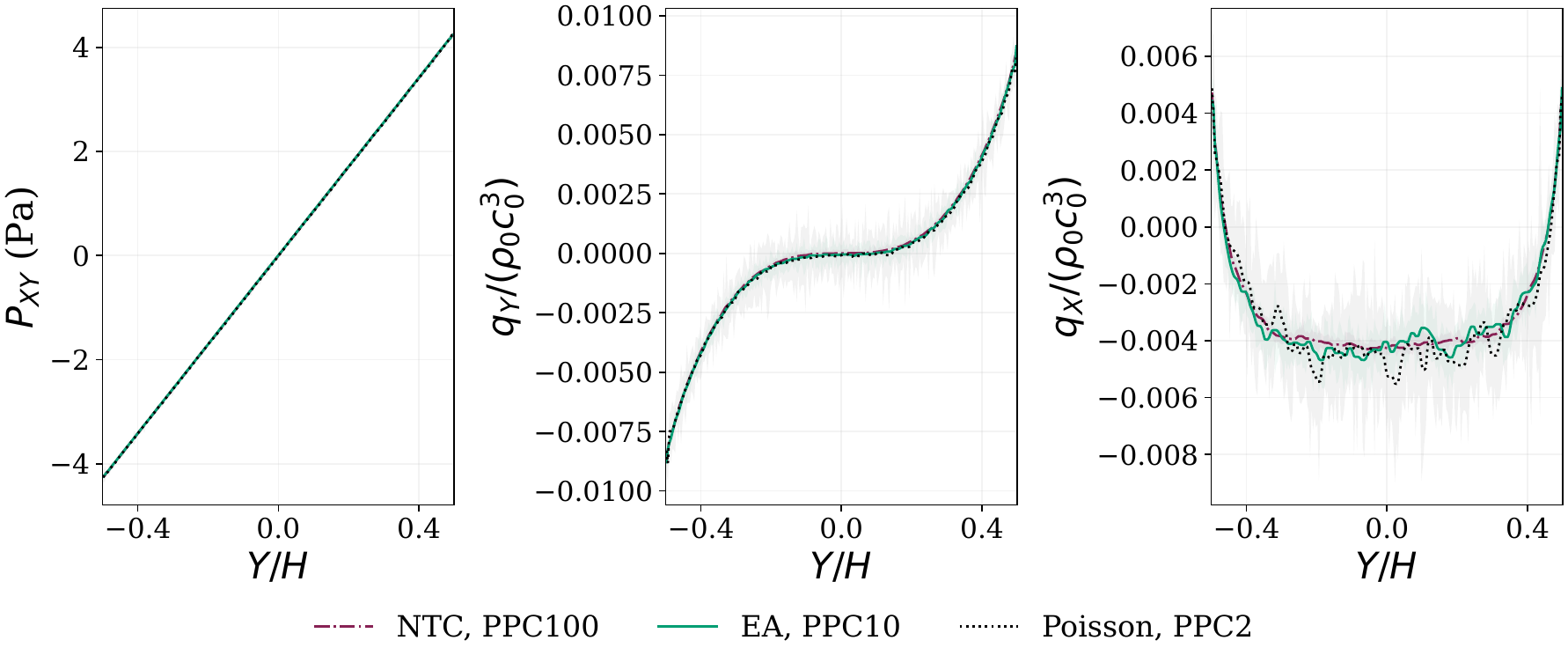}
\caption{Poiseuille kinetic shear and heat fluxes at 1.008655 ns, $\Kn=0.05$, $a_X=5\times10^6$ m s$^{-2}$, 240 cells and no collision subcells. NTC100, EA10 and Poisson2 each contribute the same $71.655$ ms per seed after warm-up, with two seeds. The Poisson reference uses PPC2 here and PPC20 in the 3.026 ns comparison. Means use 32 FCT passes. Bands use unfiltered samples.}\label{fig:ntc100p1}
\end{figure}

At 1.009 ns we compare NTC100, EA10 and same-step Poisson2 over $71.655$ ms per seed after warm-up (Fig.~\ref{fig:ntc100p1}). The acceleration is $5\times10^6$ m s$^{-2}$ and collisions use whole cells. Compared with the 3.026 ns Poisson20 result in Sec.~\ref{app:poiseuille}, both the time step and particle count change. The shear is nearly antisymmetric and can have a small spatial mean even when the wall stresses are nonzero. We assess stationarity from the signed profile and the two wall stresses separately.
Table~\ref{tab:stopped} gives the sampling duration for each trajectory that ended before its planned production interval.
\begin{table}\centering\small
\begin{tabular}{llrrr}\toprule
Flow&Sampler&$\Delta t$ (ns)&Production segments, seed 0&Seed 1\\\midrule
Couette&NTC100&3.025965&2984&2900\\
Couette&NTC100&1.008655&1203&1096\\
Poiseuille&NTC100&1.008655&1190&1288\\
Couette&NTC20&0.252164&1519&1597\\
Couette&SBT&0.063041&1753&1699\\
Couette&Poisson&0.126082&476&464\\
\bottomrule\end{tabular}
\caption{Number of production segments from each of the twelve trajectories that stopped before the planned 3072 segments. Counts exclude the first 102 segments. Each segment lasts $0.0605193$ ms. Each figure compares methods over the same physical interval.}\label{tab:stopped}\end{table}

\FloatBarrier
\section{Poiseuille flow with different particle counts}\label{app:poiseuille}
\begin{figure}[!tbp]
\centering\includegraphics[width=\textwidth]{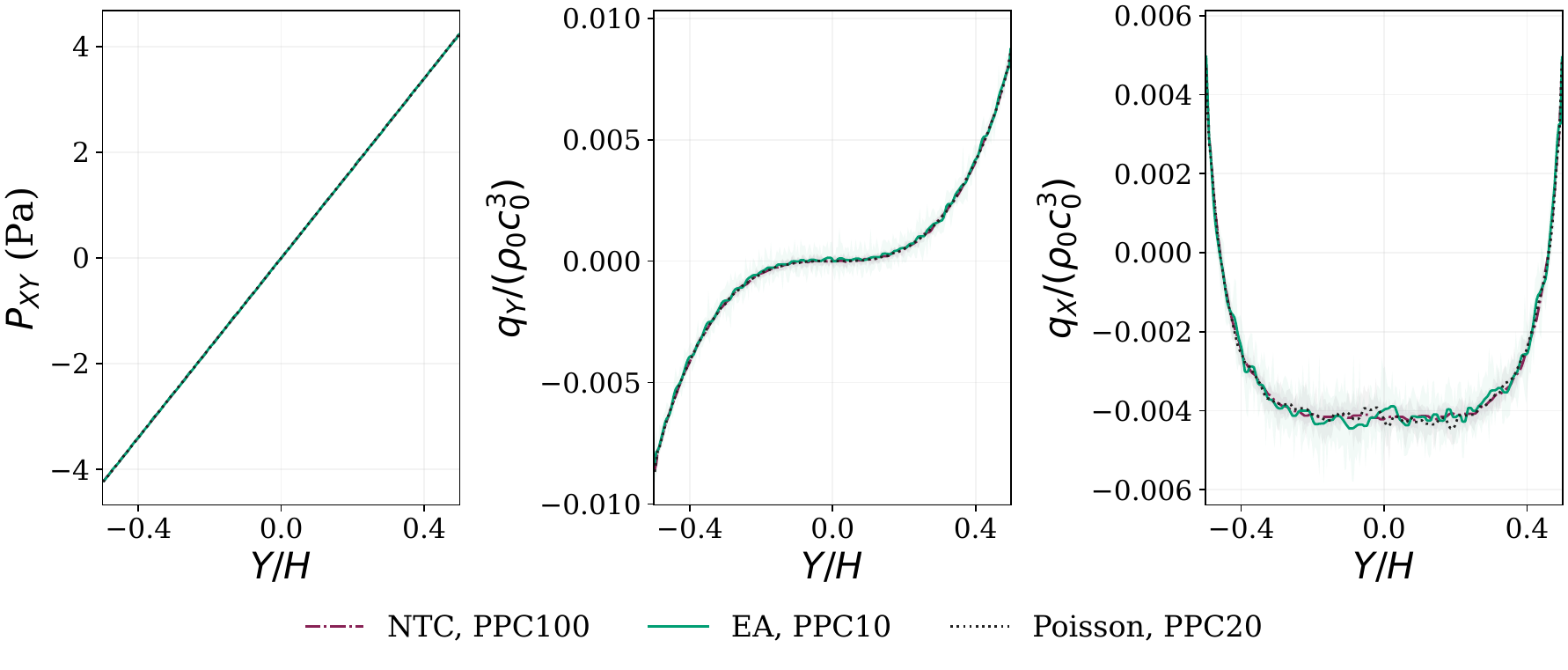}
\caption{Poiseuille kinetic shear and heat fluxes at $a_X=5\times10^6$ m s$^{-2}$, $\Kn=0.05$, 240 cells, no collision subcells and $\Delta t=3.025965$ ns. The particle counts differ: NTC100, EA10 and Poisson20. Each method uses two independent trajectories and $185.915$ ms per trajectory. Bands are twice the estimated standard errors. Means use 32 FCT passes. Bands use unfiltered samples.}
\label{fig:poiseuille}
\end{figure}

The larger-population Poiseuille comparison uses 240 collision cells without subdivision, stationary diffuse walls, $a_X=5\times10^6$ m s$^{-2}$ and $\Delta t=3.025965$ ns. NTC, EA and Poisson use PPC100, PPC10 and PPC20, respectively. Each method contributes two realizations of $185.915$ ms. Figure~\ref{fig:poiseuille} shows the kinetic shear and heat flux. Figure~\ref{fig:poisfourth} shows the higher contractions.

Signed kinetic shear and both heat-flux components nearly coincide, although simulator weight and collision method both vary. The forcing in Poiseuille flow also gives a different temperature field, distribution of flight speeds and sampling phase from those in strong Couette flow.

On the 240- and 720-cell grids, the samplers agree on the pressure curvature at lower occupancy. Temperature differences remain sensitive to sampling.

\FloatBarrier
\section{Poiseuille profiles at low occupancy on two grids}\label{app:longpois}
The 240-cell Poiseuille profiles compare representative-rate P-SBT, weighted-partner P-SBT, EA and the Poisson numerical reference at parent PPC2 and PPC4. They use $\Delta t=1.008655$ ns, an acceleration of $5\times10^6$ m s$^{-2}$ and no collision subcells. The sampling intervals are 23040 production segments for PPC2 ($1394.365$ ms per seed) and 9216 for PPC4 ($557.746$ ms per seed). Within each occupancy group, every method uses two independent trajectories and the same sampling interval after 102 discarded segments.

We form the scalar pressure $p=\rho\theta$ from the pooled density and temperature for each trajectory and normalize by $p_0=\rho_0c_0^2$. The time-interval estimates retain the joint variation of density and temperature. The long averages resolve the shallow pressure curvature and small differences between methods. Fluctuations in the nearly uniform centreline temperature obscure any double peak. Figures~\ref{fig:long2mom} and~\ref{fig:long4mom} give the fourth moments at two and four particles per cell.
\begin{figure}[!tbp]
\centering\includegraphics[width=\textwidth]{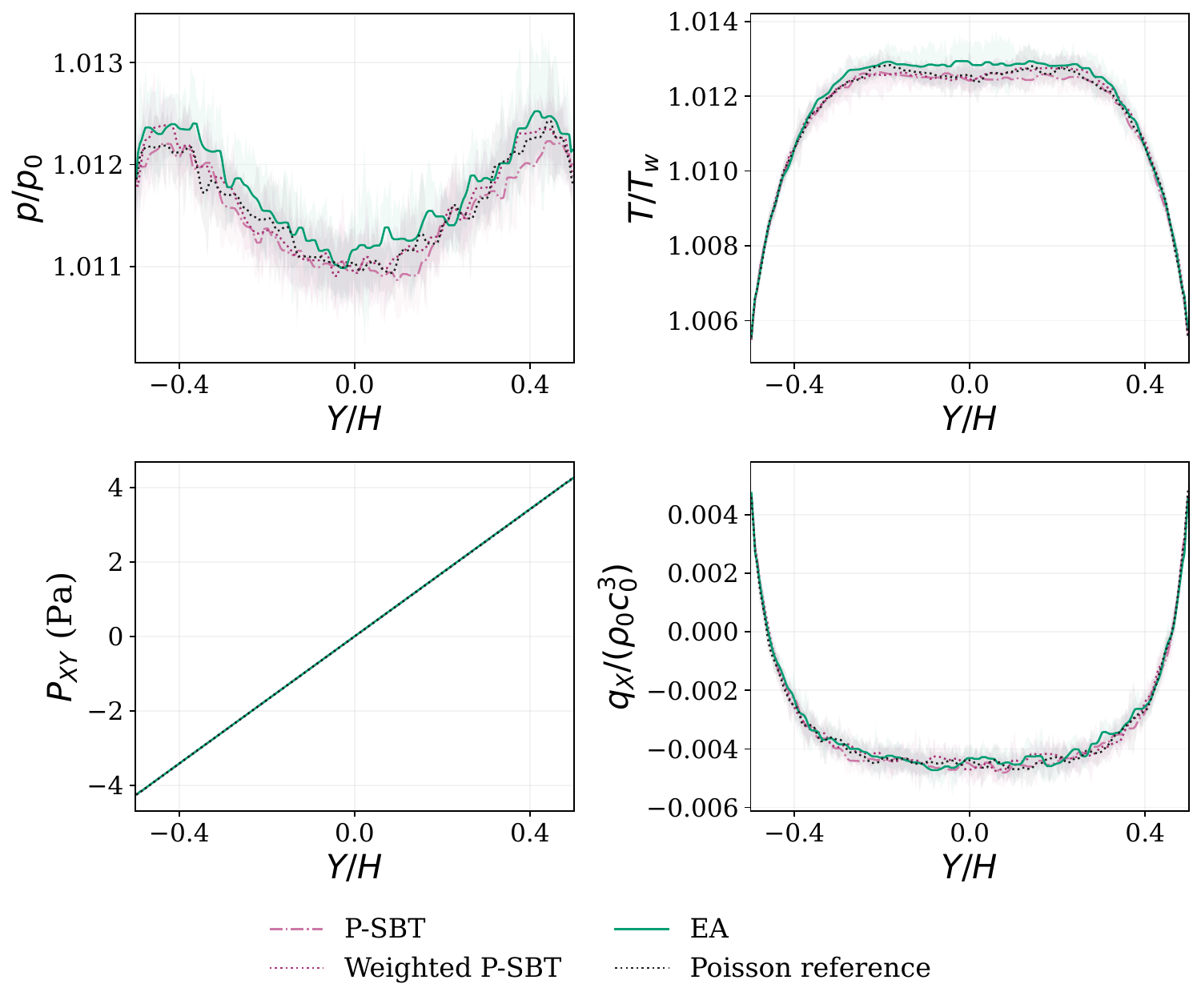}
\caption{240-cell Poiseuille PPC2 comparison: pressure, temperature, signed kinetic shear and streamwise heat flux. All four methods use $1394.365$ ms per seed, two seeds, $\Delta t=1.008655$ ns, $\Kn=0.05$, $a_X=5\times10^6$ m s$^{-2}$ and no collision subcells. Bands span twice the estimated standard error. The two sides of the channel are sampled separately. Means use 32 FCT passes. Bands use unfiltered samples.}\label{fig:long2}
\end{figure}
\begin{figure}[!tbp]
\centering\includegraphics[width=\textwidth]{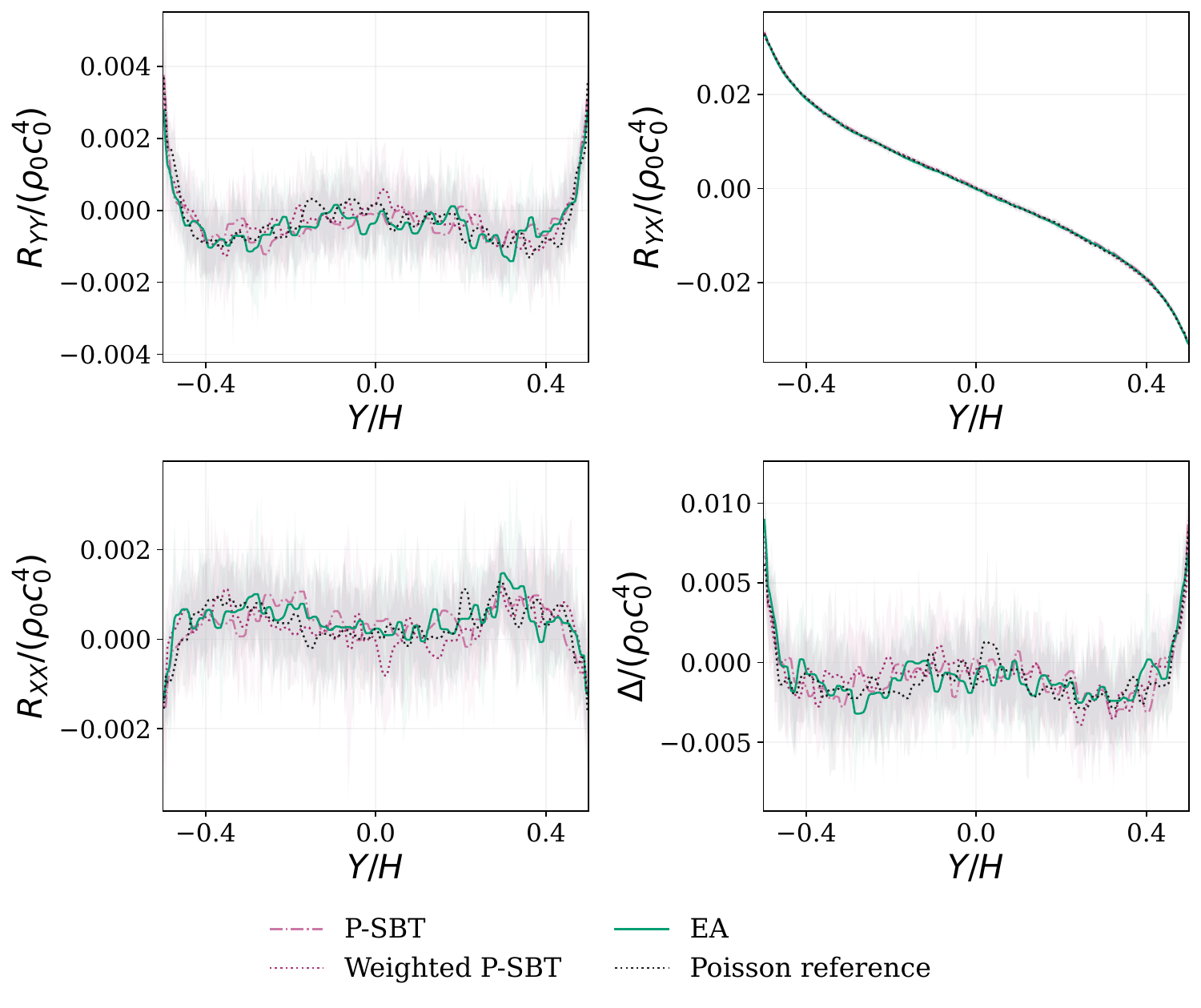}
\caption{Fourth-order contractions for the same PPC2 trajectories and $1394.365$ ms per-seed interval as Fig.~\ref{fig:long2}. Zero crossings are retained without imposing symmetry. Means use 32 FCT passes. Bands use unfiltered samples.}\label{fig:long2mom}
\end{figure}
\begin{figure}[!tbp]
\centering\includegraphics[width=\textwidth]{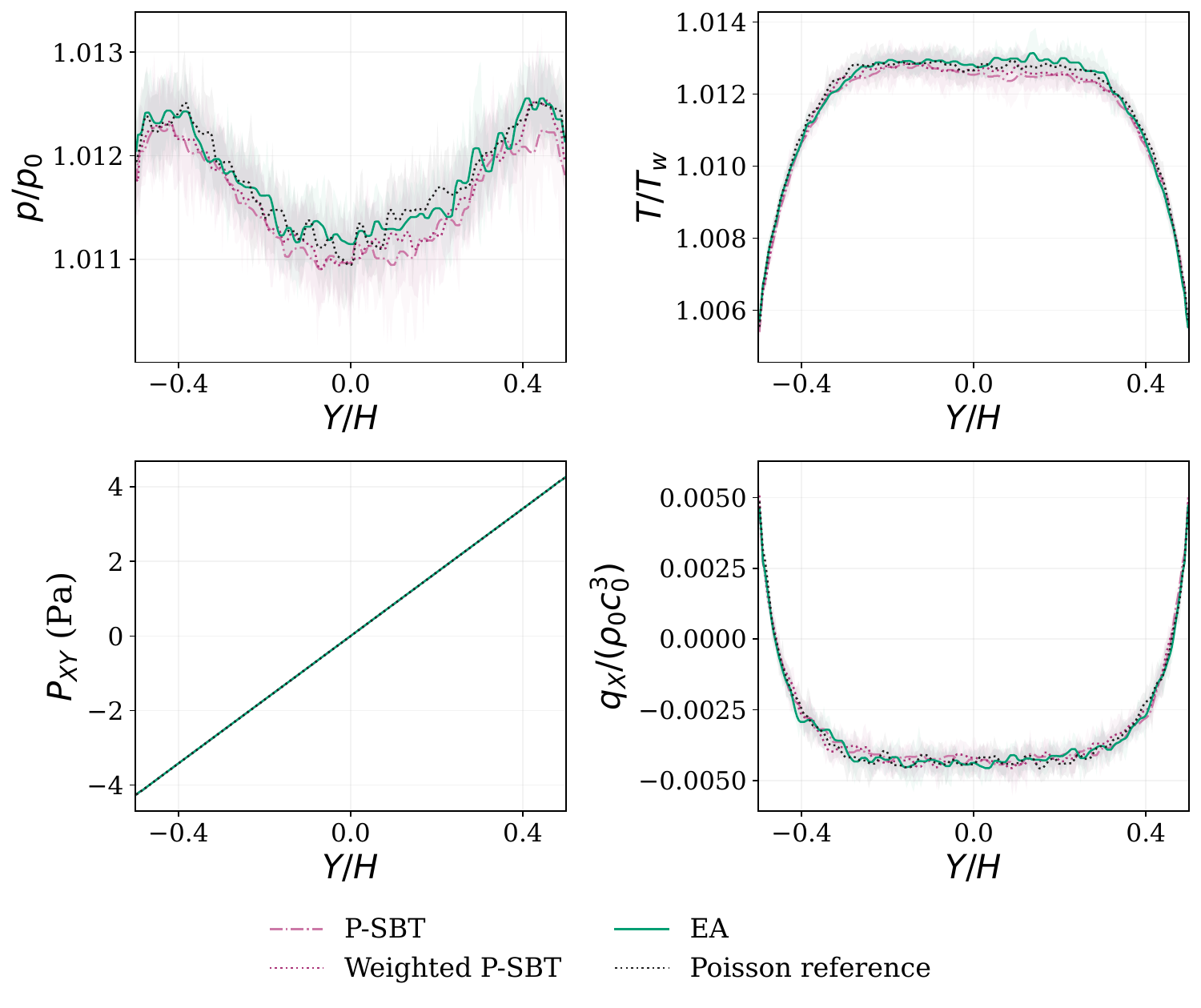}
\caption{240-cell Poiseuille PPC4 pressure, temperature, shear and streamwise heat flux: $557.746$ ms per seed and two seeds for every method. The forcing, time step and undivided collision cells are the same as in Fig.~\ref{fig:long2}; the particle count and sampling duration differ. Means use 32 FCT passes. Bands use unfiltered samples.}\label{fig:long4}
\end{figure}
\begin{figure}[!tbp]
\centering\includegraphics[width=\textwidth]{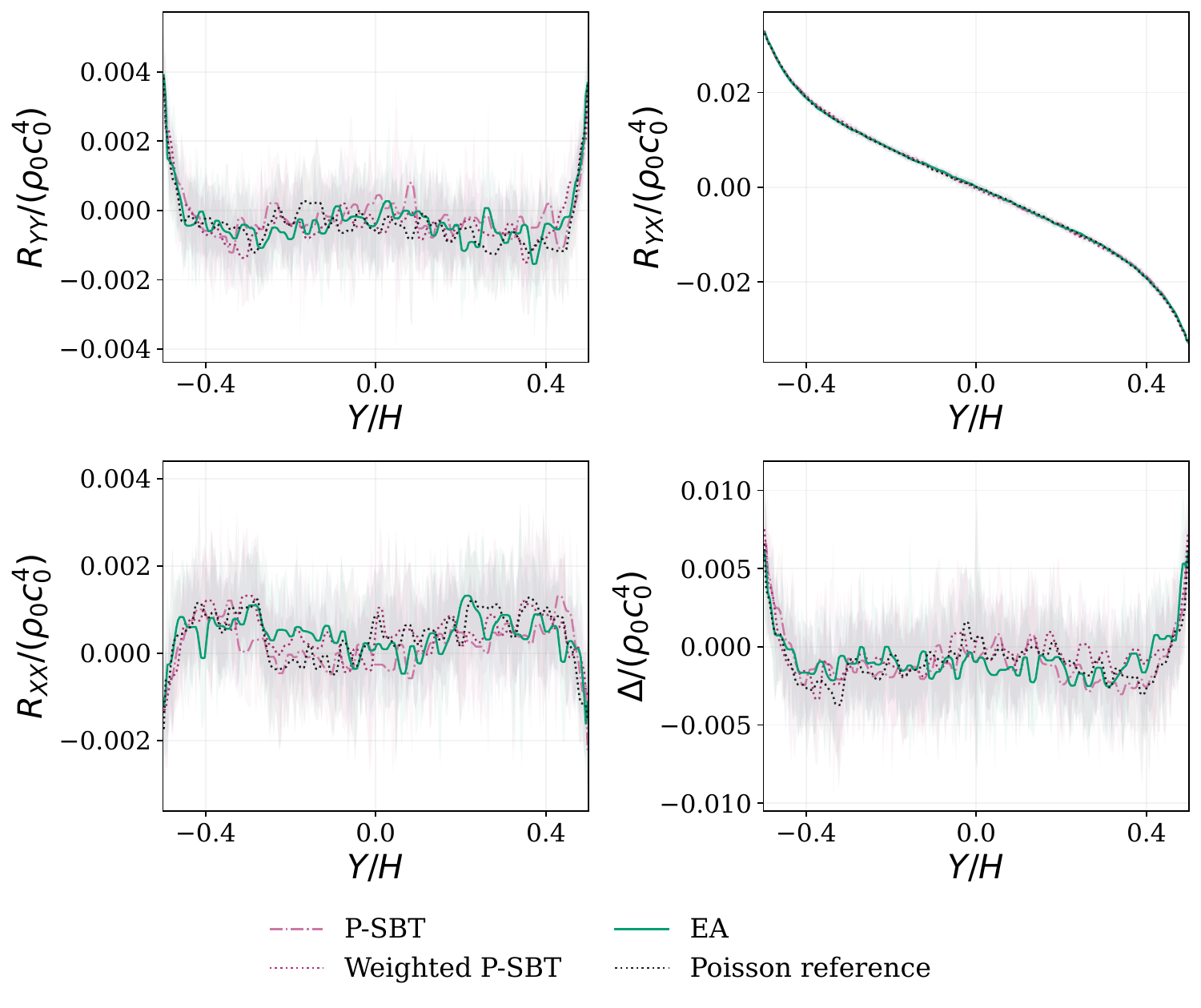}
\caption{Fourth-order contractions for the PPC4 comparison in Fig.~\ref{fig:long4}, using $557.746$ ms per independent trajectory. Means use 32 FCT passes. Bands use unfiltered samples.}\label{fig:long4mom}
\end{figure}

\begin{figure}[!tbp]
\centering\includegraphics[width=\textwidth]{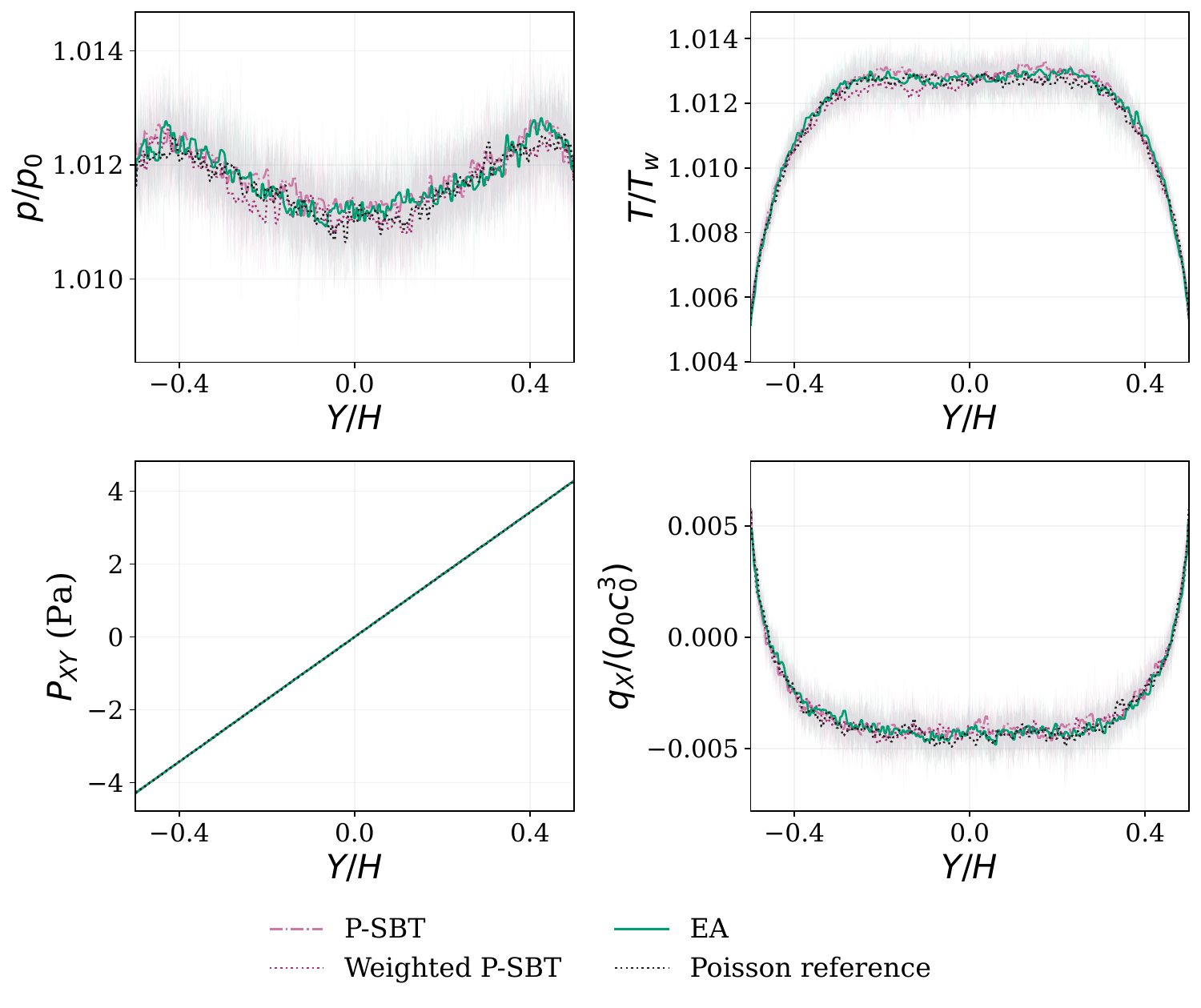}
\caption{720-cell Poiseuille PPC2 comparison at $\Delta t=0.336218$ ns and a common sampling interval: $185.915$ ms per seed and two seeds. The gas, forcing and undivided collision cells are the same as in Fig.~\ref{fig:long2}. Means use 32 FCT passes. Bands use unfiltered samples.}\label{fig:n720p2}
\end{figure}

\begin{figure}[!tbp]
\centering\includegraphics[width=\textwidth]{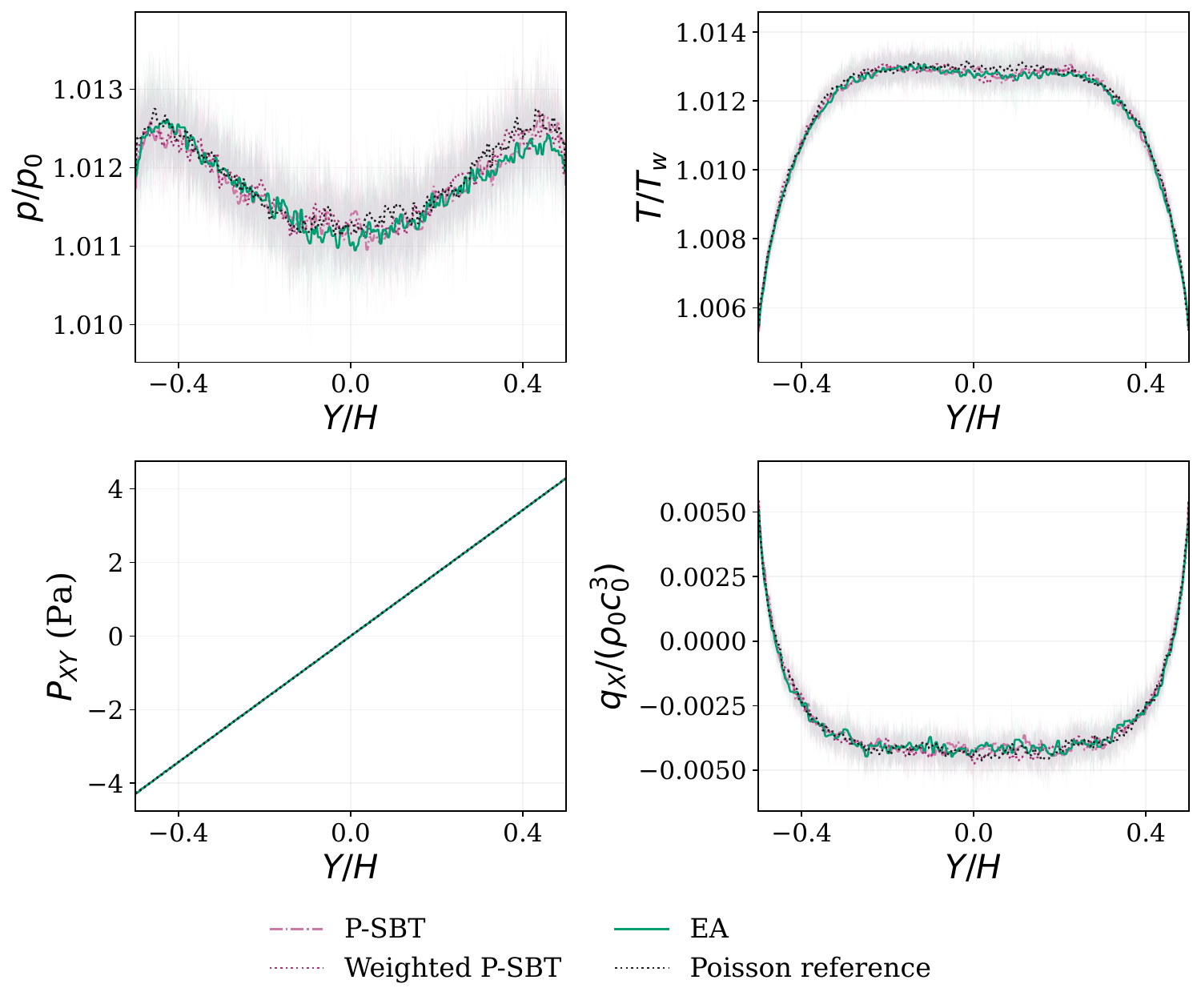}
\caption{720-cell PPC4 pressure, temperature, shear and streamwise heat flux at $\Delta t=0.336218$ ns, using the same $185.915$ ms common interval per seed and physical conditions as Fig.~\ref{fig:n720p2}. Means use 32 FCT passes. Bands use unfiltered samples.}\label{fig:n720p4}
\end{figure}

For the 720-cell comparisons, the common interval is 3072 production segments, or $185.915$ ms per seed, at both PPC2 and PPC4. Figures~\ref{fig:n720p2} and~\ref{fig:n720p4} use this interval and $\Delta t=0.336218$ ns. Cell width, time step and sampling duration all change between the 240- and 720-cell runs, so their comparison cannot isolate the spatial discretization error for Richardson extrapolation.

\FloatBarrier
\section{Fourier heat flux and spatial profiles}\label{sec:fourier-supp}
\begin{figure}[htbp]\centering
\includegraphics[width=.99\linewidth]{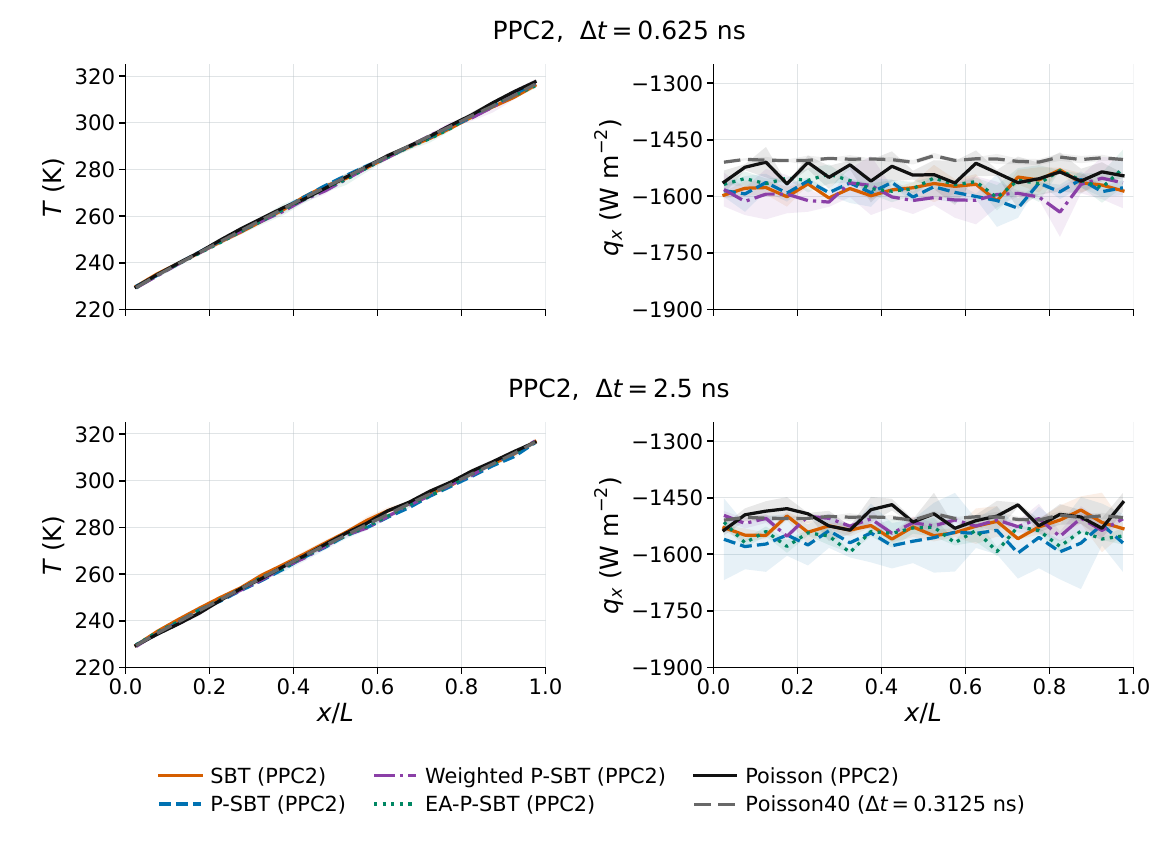}
\caption{Temperature and signed heat flux from the 223.15 K wall at $x/L=0$ to the 323.15 K wall at $x/L=1$, with $L=1$ mm. The low-population calculations use 800 cells and two simulator particles per cell. The upper and lower pairs of panels use $\Delta t=0.625$ and 2.5 ns, respectively. Both rows use Poisson40 at 0.3125 ns as the reference. Shaded ranges span the two individual seed estimates.}
\label{fig:fourier-profiles}\end{figure}

The 92 realizations represent 46 combinations of grid, time step, collision law, mean occupancy and transport ordering. Two independent seeds, 761501 and 761502, were used per combination. Each profile in Fig.~\ref{fig:fourier-profiles} covers 32 production blocks (0.8192 ms per seed) after a 0.2048 ms warm-up. We pool raw velocity moments from five adjacent bins before calculating each of the 20 plotted values. The collision grid has 800 cells in the figure. Each low-population calculation uses a collision--ballistic flight--collision (CBC) step, with a half collision step on each side of a full flight step. The no-time-counter calculation at forty particles per cell (NTC40) uses stochastic rounding. Whole-cell Poisson at the same occupancy (Poisson40) draws its candidate count from a Poisson distribution with the same mean.

A separate reproduction of the 400-cell, five-particle SBT result of \citet{Stefanov2019Fourier} uses conventional ballistic flight followed by collisions, $\Delta t=7.4120703$ ns and two different seeds. With 2,000 simulator particles in each trajectory, the seed numbers 761601 and 761602 specify independent velocity histories. The respective signed wall fluxes are $-1512.15$ and $-1501.54$ W\,m$^{-2}$, with temporal batch standard errors of 7.40 and 7.82 W\,m$^{-2}$. Each value averages the two walls over 22.2604 ms. A positive published wall-flux magnitude corresponds to the absolute value of this signed mean.

At this step the nominal SBT trial probability occasionally exceeds one. For this comparison we accept the proposed encounter whenever the nominal probability exceeds one and count these occurrences. We stop a run if the count exceeds 2\% of all accepted collisions. The two totals, including warm-up, are 57 out of 369,187,200 and 70 out of 369,164,089, or 0.0000154\% and 0.0000190\%. These counts characterize the 7.412 ns reproduction specifically. In Stefanov's Table~I \citep{Stefanov2019Fourier}, PPC5 SBT gives 1536~W\,m$^{-2}$. At half that step, PPC200 NTC gives 1512~W\,m$^{-2}$.

For the high-population case in Stefanov's Table~I, NTC carries the fractional part of the majorant candidate count into the next step. On 400 cells with PPC200, $\Delta t=3.7060$ ns and 1.8700 ms of sampling per seed, this NTC calculation and whole-cell Poisson give mean flux magnitudes of 1513.16 and 1511.86 W\,m$^{-2}$. They use the same geometry, population and time step. On 120,000 cells the SBT calculation at mean occupancy 0.1 (Stefanov's Table~II) gives 1510.03 W\,m$^{-2}$. For its two seeds the standard error of their mean is 13.28 W\,m$^{-2}$, and the larger error estimate obtained by temporal batching is 17.53 W\,m$^{-2}$. The published value of 1514~W\,m$^{-2}$ lies within both error estimates.

\paragraph{Reproduction of the fixed-particle-count test.}
In Fig.~15 of \citet{Taheri2017}, the total number of simulator particles was held at 24,240. As the collision cells became narrower, both the particles per cell and the time step decreased, with $\Delta x/(c_{\rm mp}\Delta t)=1.469$, where $c_{\rm mp}=\sqrt{2k_BT_0/m}$ is the most probable molecular speed at the reference temperature $T_0$. The published value of 1512~W\,m$^{-2}$ is the limiting heat-flux magnitude used for comparison. We checked the original Fortran SBT solver at two particles per cell (12,120 cells, $\Delta t=0.166523$~ns). Table~\ref{tab:fourier-native} compares its wall flux with two Python implementations over the same 0.80144~ms. The Python implementations differ only in the order in which particles enter a collision trial. Over the 1,202 blocks common to both Python runs, keeping the original order raised the flux magnitude by 31.1~W\,m$^{-2}$. With two seeds, the paired 95\% interval for this difference extends from $-248.7$ to $310.9$~W\,m$^{-2}$, leaving the size and sign of an ordering effect unresolved.

\begin{table}[htbp]
\caption{Fourier wall heat-flux magnitude averaged over both walls and the same 1,175 blocks (zero-based indices 325 through 1499). Positive flux denotes hot-to-cold transfer. The last column gives the original Fortran value minus the corresponding Python result, with one combined standard error accounting for independent trajectories and serial correlation. The temporal standard error for the single Fortran trajectory is 11.93~W\,m$^{-2}$.}
\label{tab:fourier-native}
\centering
\begin{tabular}{lrcc}
\hline
Implementation & Trajectories & Wall flux & Fortran minus Python \\
& & (W\,m$^{-2}$) & (W\,m$^{-2}$) \\
\hline
Original Fortran SBT & 1 & 1533.34 & --- \\
Python, original particle order & 2 & 1533.39 & $-0.05\pm19.77$ \\
Python, shuffled particle order & 2 & 1503.25 & $30.09\pm19.75$ \\
\hline
\end{tabular}
\end{table}

In a separate Poisson40 calculation on 800 cells at the same time step, eight independent trajectories each contributed 4,808 production blocks (3.27944~ms). The mean hot-to-cold wall flux was $1513.4\pm8.1$~W\,m$^{-2}$ (95\% interval). The hot-wall estimate exceeded the cold-wall estimate by $0.28\pm16.28$~W\,m$^{-2}$, and the mean flux in the second half exceeded that in the first by $0.53\pm16.29$~W\,m$^{-2}$. For the individual walls, the second-half changes were $1.27\pm20.05$ (cold) and $-0.22\pm25.67$ (hot), in the same units. All four intervals extend beyond the prescribed $\pm5$~W\,m$^{-2}$ tolerance, so these data leave stationarity at that precision unresolved.

\FloatBarrier
\section{Shock step size and higher moments}
\begin{figure}[!tbp]\centering\includegraphics[width=\textwidth]{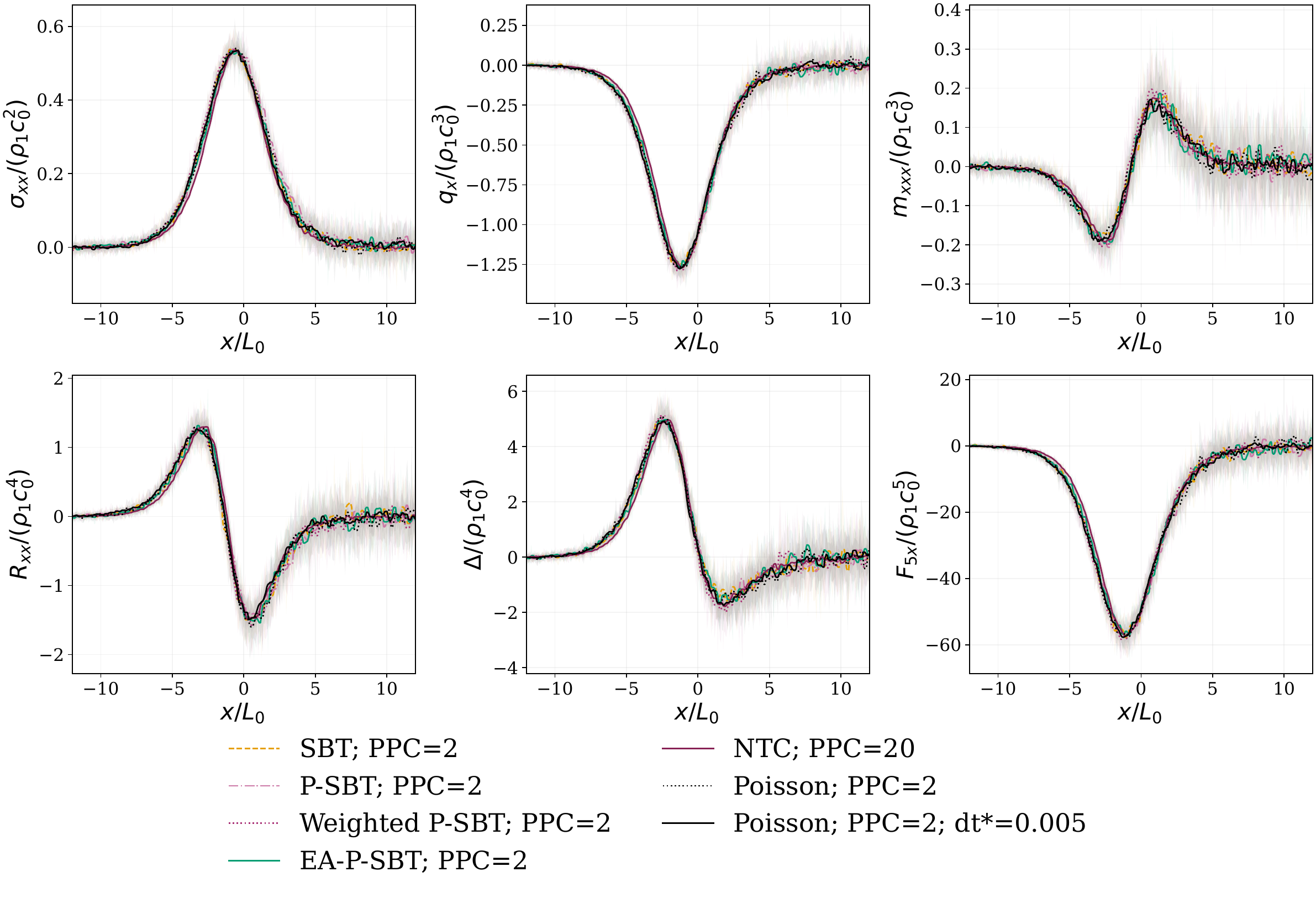}\caption{Mach 2. Larger-step stress, heat flux and higher moments. The dotted Poisson curve uses the same time step as the other samplers. The solid black curve uses the finer reference step. The mean profiles use eight FCT passes. The uncertainty estimates use unfiltered samples.}\label{fig:ma2-large}\end{figure}

\begin{figure}[!tbp]\centering\includegraphics[width=\textwidth]{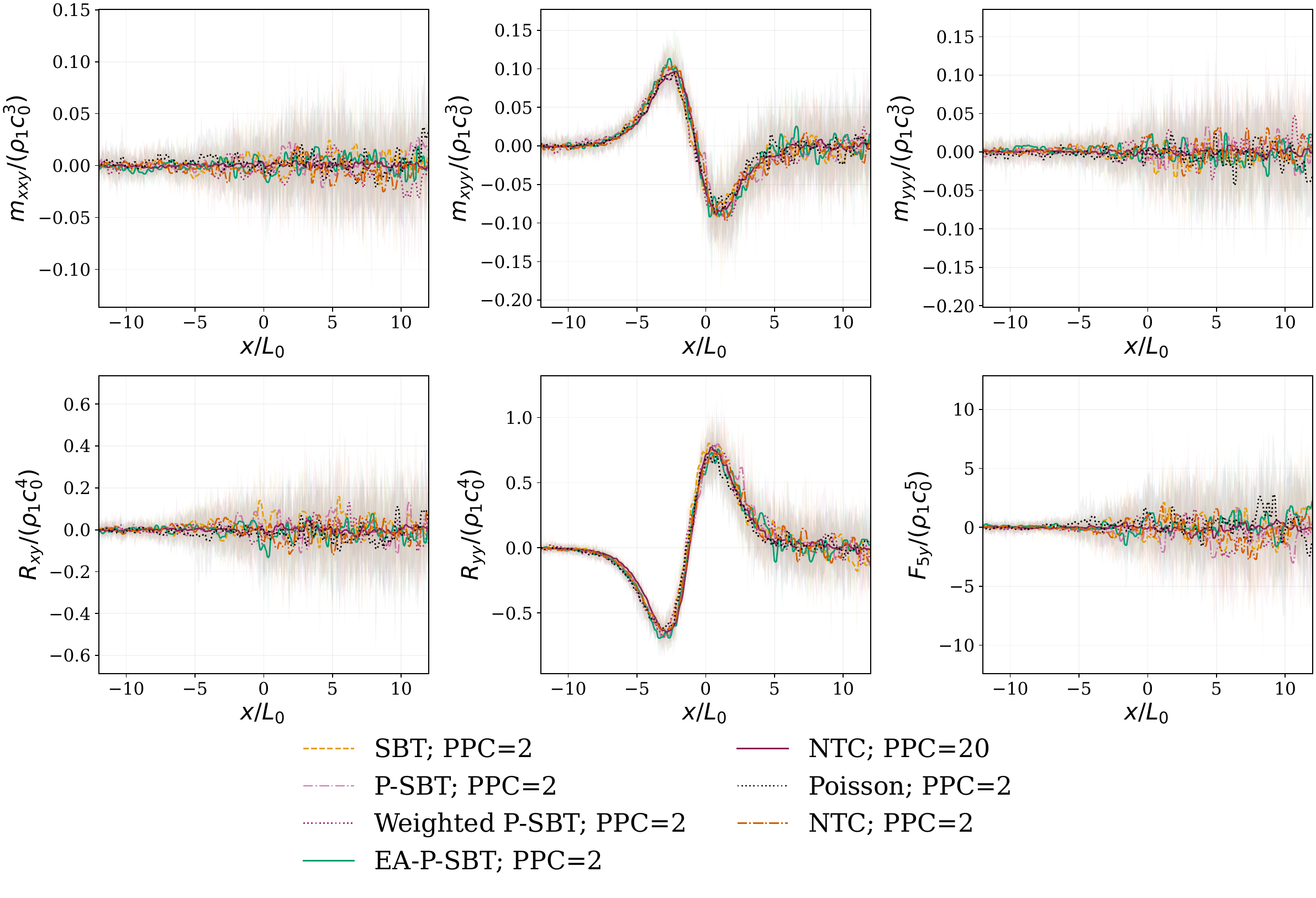}\caption{Mach 2. Additional components in reading order: $m_{xxy}$, $m_{xyy}$, $m_{yyy}$, $R_{xy}$, $R_{yy}$, $F_{5y}$. The mean profiles use eight FCT passes. The uncertainty estimates use unfiltered samples.}\label{fig:ma2-comp}\end{figure}

\begin{figure}[!tbp]\centering\includegraphics[width=\textwidth]{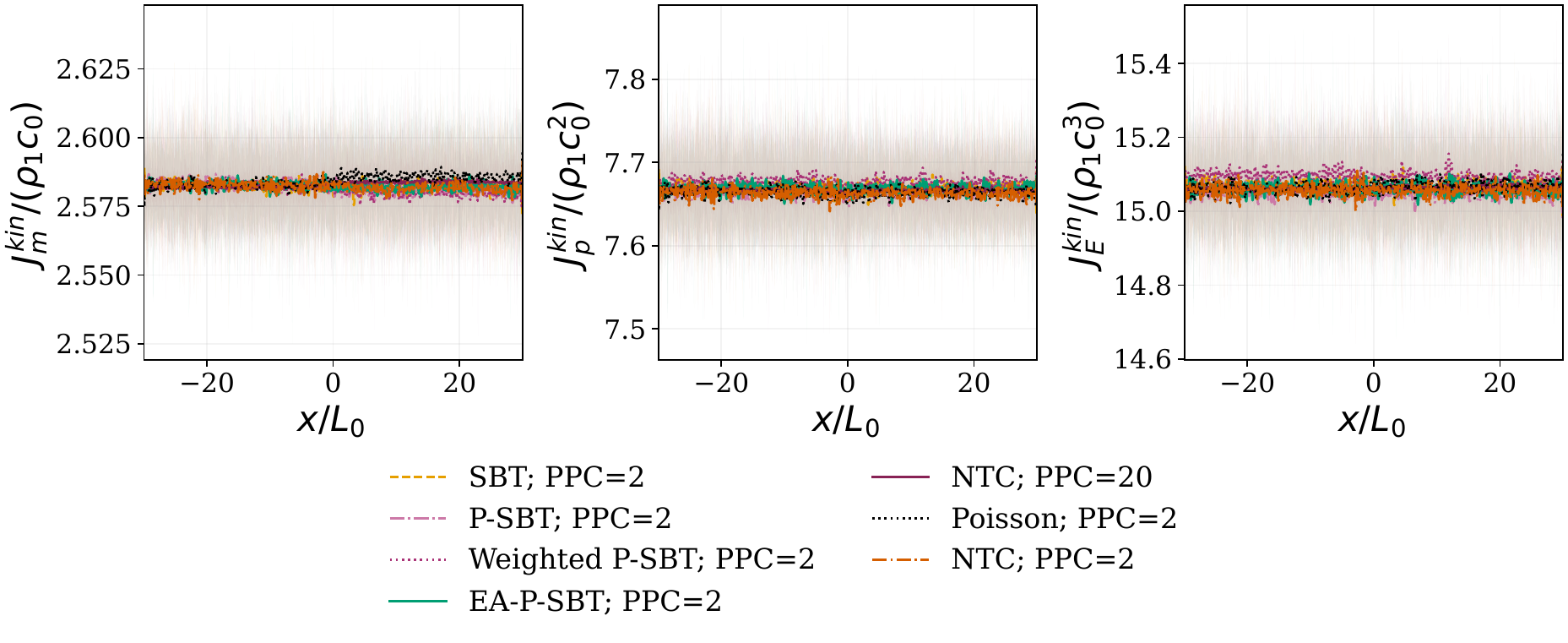}\caption{Mach 2. Full-domain kinetic mass, momentum and energy fluxes. The curves show transport by particle motion. The solver records collisional transfer and the population correction separately. The mean profiles use eight FCT passes. The uncertainty estimates use unfiltered samples.}\label{fig:ma2-flux}\end{figure}

\begin{figure}[!tbp]\centering\includegraphics[width=\textwidth]{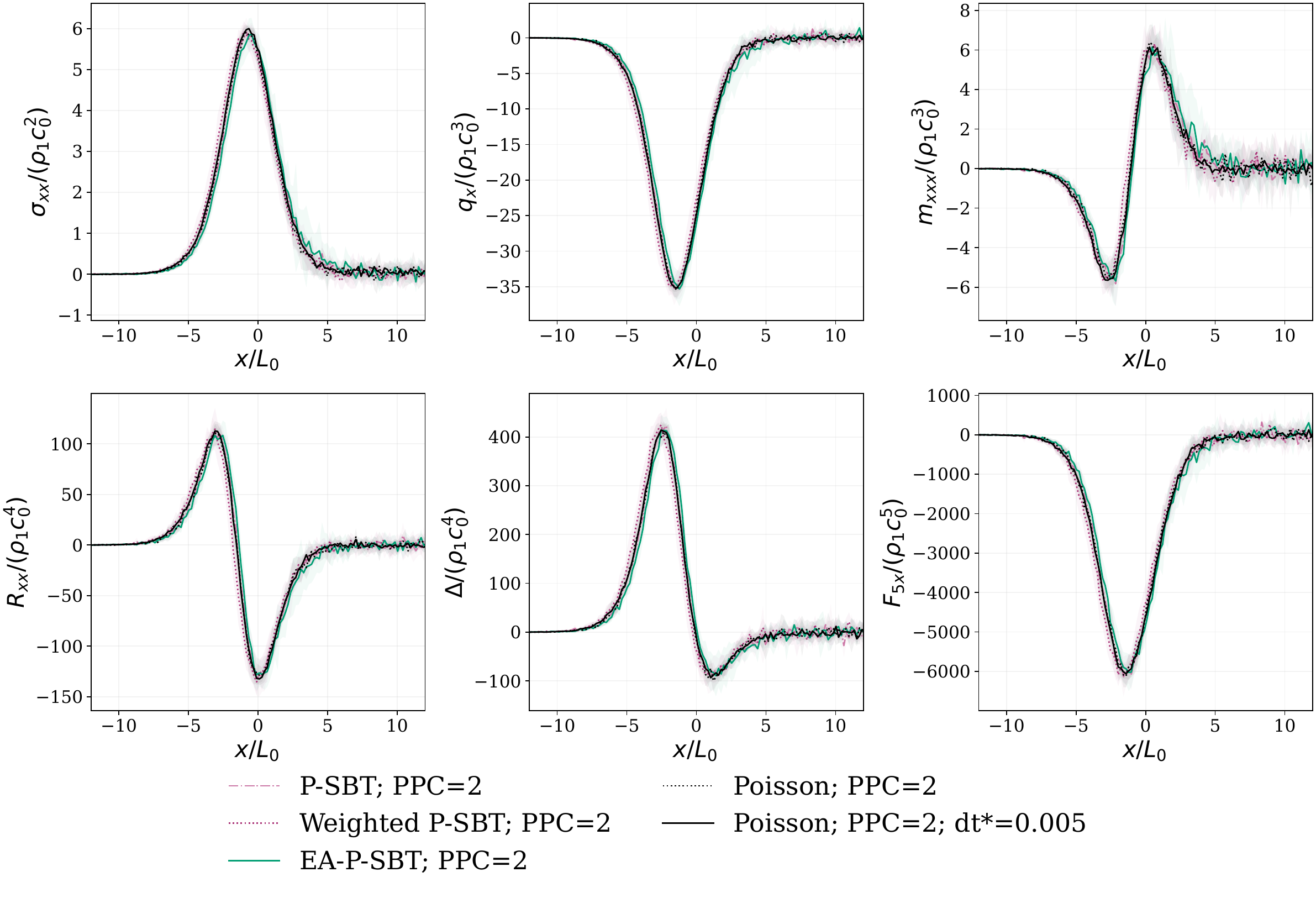}\caption{Mach 5. Larger-step stress, heat flux and higher moments. The dotted Poisson curve uses the same time step as the other samplers. The solid black curve uses the finer reference step. Mean profiles and uncertainty estimates use unfiltered samples.}\label{fig:ma5-large}\end{figure}

\begin{figure}[!tbp]\centering\includegraphics[width=\textwidth]{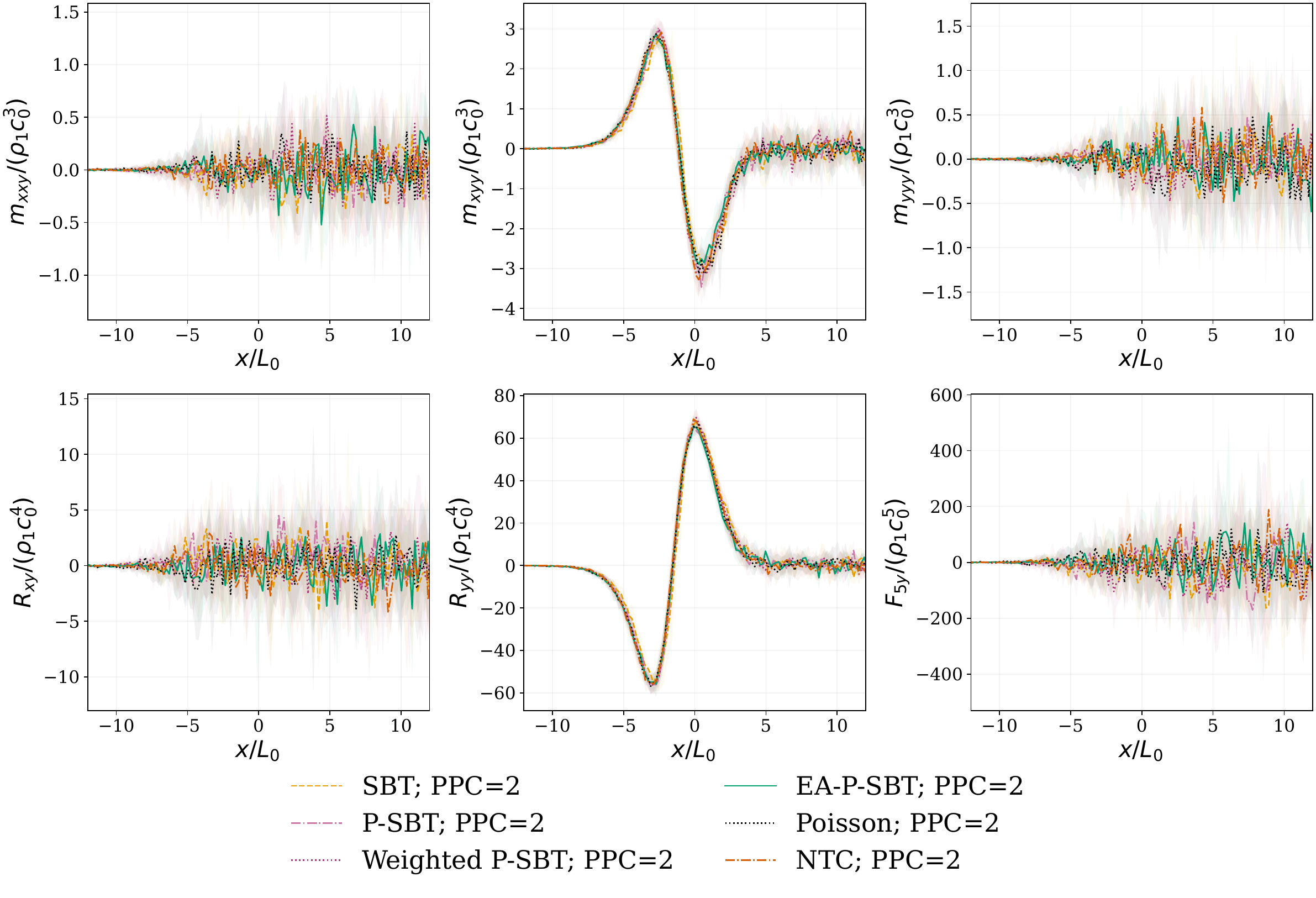}\caption{Mach 5. Additional components in reading order: $m_{xxy}$, $m_{xyy}$, $m_{yyy}$, $R_{xy}$, $R_{yy}$, $F_{5y}$. Mean profiles and uncertainty estimates use unfiltered samples.}\label{fig:ma5-comp}\end{figure}

\begin{figure}[!tbp]\centering\includegraphics[width=\textwidth]{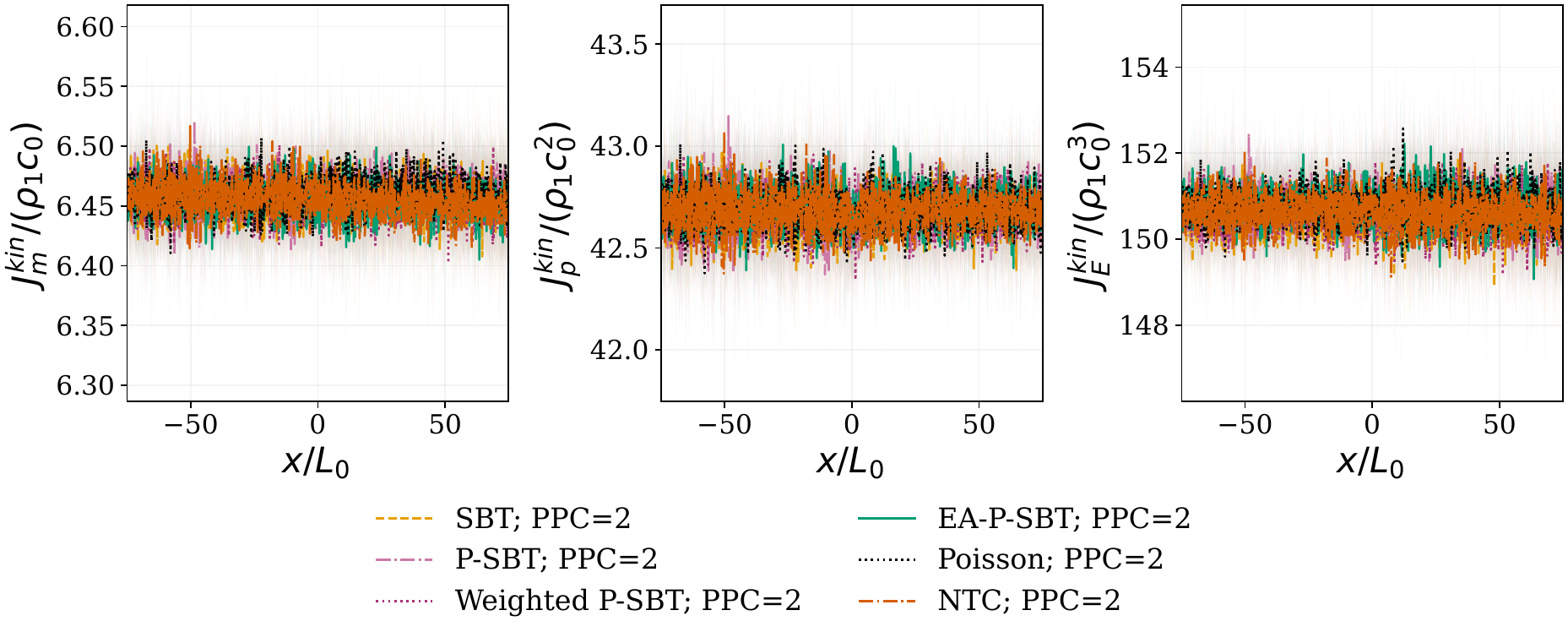}\caption{Mach 5. Full-domain kinetic mass, momentum and energy fluxes. The curves show transport by particle motion. The solver records collisional transfer and the population correction separately. Mean profiles and uncertainty estimates use unfiltered samples.}\label{fig:ma5-flux}\end{figure}

The larger-step profiles, additional moment components and fluxes appear for Mach 2 in Figs.~\ref{fig:ma2-large}, \ref{fig:ma2-comp} and~\ref{fig:ma2-flux}, and for Mach 5 in Figs.~\ref{fig:ma5-large}, \ref{fig:ma5-comp} and~\ref{fig:ma5-flux}. Table~\ref{tab:shockdiscrepancy} gives the relative profile differences.
The Mach-2 and Mach-5 shocks use the same hard-sphere model and sampling periods as in the main text. Larger-step figures compare methods at $\Delta t^*=0.015$ and the finer $0.005$ Poisson reference. With feedback every 1000 steps, the physical interval between corrections varies with the time step. The larger-step Mach-5 SBT case exceeded the Bernoulli probability bound. The additional moments and fluxes in Figs.~\ref{fig:ma2-comp}, \ref{fig:ma2-flux}, \ref{fig:ma5-comp} and~\ref{fig:ma5-flux} use $\Delta t/t_0=0.005$. For molecular velocity $\bv$, the kinetic mass, normal momentum and energy fluxes along $x$ are $J_m^{\rm kin}=\rho\langle v_x\rangle$, $J_p^{\rm kin}=\rho\langle v_x^2\rangle$ and $J_E^{\rm kin}=\tfrac12\rho\langle v_x|\bv|^2\rangle$, respectively.
\FloatBarrier
\begin{table}\centering\small
\begin{tabular}{rrlrrrrrr}\toprule
$M_1$ & $\Delta t^*$ & Sampler & PPC & $\sigma_{xx}$ & $q_x$ & $R_{xx}$ & $\Delta$ & $F_{5x}$\\\midrule
2 & 0.005 & SBT & 2 & 6.18 & 5.41 & 17.28 & 14.11 & 6.51\\
2 & 0.005 & P-SBT & 2 & 8.68 & 7.23 & 20.78 & 14.25 & 7.90\\
2 & 0.005 & Weighted P-SBT & 2 & 6.85 & 5.64 & 15.61 & 14.13 & 6.58\\
2 & 0.005 & EA-P-SBT & 2 & 6.91 & 5.48 & 15.30 & 13.98 & 6.32\\
2 & 0.005 & NTC & 20 & 8.78 & 8.65 & 16.71 & 14.54 & 8.72\\
2 & 0.005 & NTC & 2 & 6.86 & 6.12 & 17.92 & 13.71 & 6.85\\
2 & 0.015 & SBT & 2 & 7.88 & 6.71 & 21.18 & 17.42 & 8.40\\
2 & 0.015 & P-SBT & 2 & 10.72 & 7.93 & 22.61 & 16.78 & 9.07\\
2 & 0.015 & Weighted P-SBT & 2 & 7.91 & 6.97 & 21.04 & 18.13 & 8.54\\
2 & 0.015 & EA-P-SBT & 2 & 8.10 & 7.81 & 20.41 & 18.26 & 9.39\\
2 & 0.015 & NTC & 20 & 8.93 & 8.70 & 16.66 & 14.55 & 8.83\\
2 & 0.015 & Poisson & 2 & 8.38 & 7.60 & 20.59 & 17.35 & 9.17\\
5 & 0.005 & SBT & 2 & 5.84 & 5.59 & 11.09 & 9.12 & 5.51\\
5 & 0.005 & P-SBT & 2 & 4.05 & 2.92 & 5.94 & 4.36 & 3.22\\
5 & 0.005 & Weighted P-SBT & 2 & 2.94 & 2.81 & 4.79 & 4.48 & 3.17\\
5 & 0.005 & EA-P-SBT & 2 & 4.35 & 2.94 & 5.74 & 5.02 & 3.69\\
5 & 0.005 & NTC & 2 & 2.95 & 2.38 & 4.81 & 4.42 & 3.34\\
5 & 0.015 & P-SBT & 2 & 4.88 & 3.53 & 7.67 & 6.01 & 4.60\\
5 & 0.015 & Weighted P-SBT & 2 & 8.06 & 8.36 & 16.46 & 13.48 & 8.69\\
5 & 0.015 & EA-P-SBT & 2 & 9.06 & 8.32 & 16.12 & 12.35 & 8.39\\
5 & 0.015 & Poisson & 2 & 4.69 & 3.24 & 6.41 & 5.28 & 4.26\\
\bottomrule\end{tabular}
\caption{Relative $L_2$ differences (percent) between the unaligned profiles and Poisson at two particles per cell at $\Delta t^*=0.005$, using $100\|f-f_{\rm ref}\|_2/\|f_{\rm ref}\|_2$ over the original grid cells with $|x/L_0|\leq12$. Each curve averages two realizations over the same time interval. The differences include statistical variation in both the test and reference profiles.}\label{tab:shockdiscrepancy}\end{table}

\FloatBarrier

\begin{figure}[!tbp]
\centering\includegraphics[width=\linewidth]{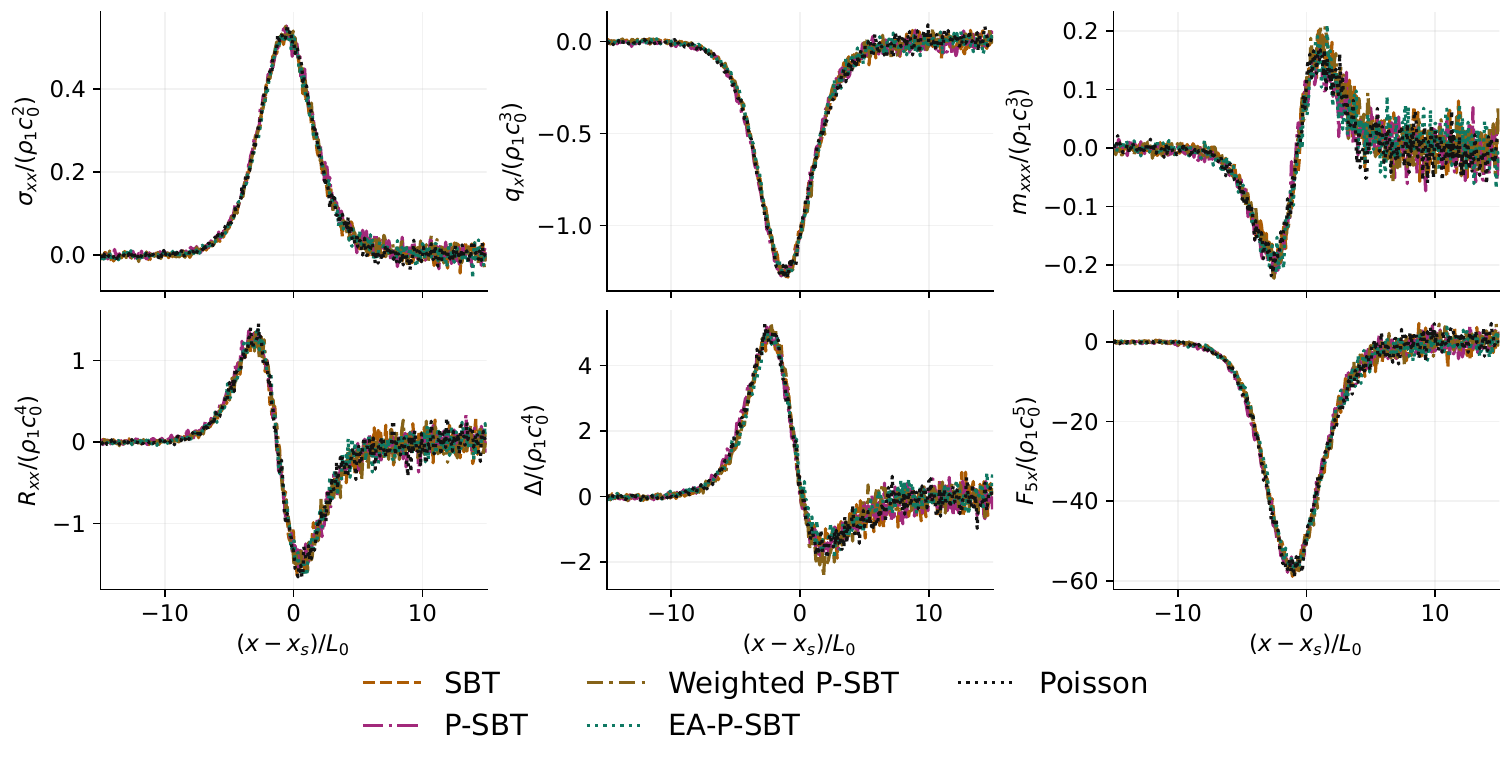}
\caption{Aligned Mach-2 core profiles at $\Delta t/t_0=0.015$ using 800 cells and two realizations. Lines and panels follow the Mach-2 figure in the main article.}
\label{fig:coarseMa2}
\end{figure}

\begin{figure}[!tbp]
\centering\includegraphics[width=\linewidth]{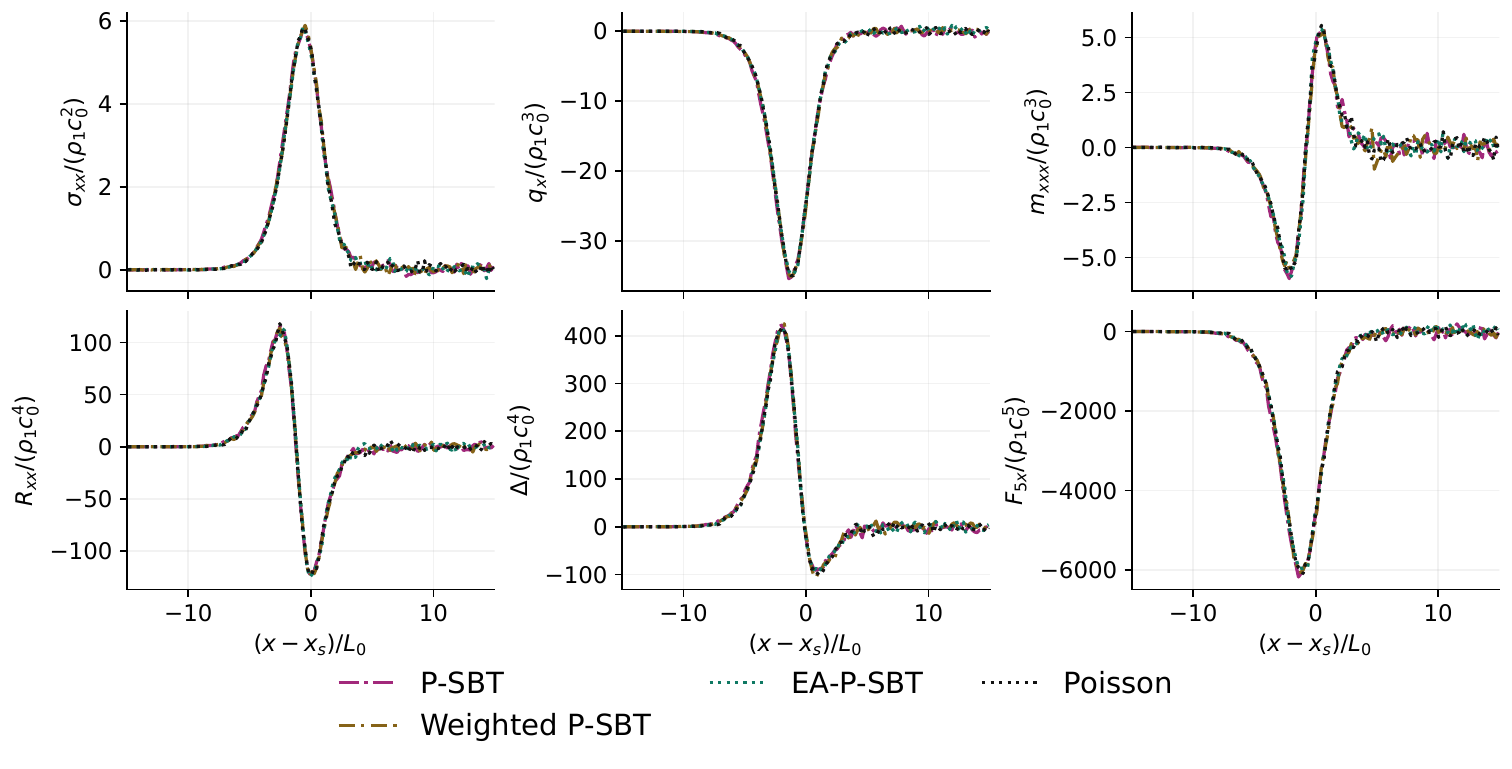}
\caption{Aligned Mach-5 core profiles at $\Delta t/t_0=0.015$ using 800 cells and two realizations. Alignment, averaging and panels follow the Mach-5 figure in the main article.}
\label{fig:coarseMa5}
\end{figure}

For all 54 trajectories, the accompanying data give the signed transport moments for each seed at both time steps, before and after shifting to the shock frame. Figures~\ref{fig:coarseMa2} and~\ref{fig:coarseMa5} show the aligned profiles at the larger step.

\FloatBarrier
\section{Reference coordinate at Mach 2}
Table~\ref{tab:source-width} compares the density and temperature errors before and after converting the published coordinates. We report the root-mean-square error (RMSE) for each profile.
The main article converts the horizontal coordinates of all 21 Ohwada--Xu data points using $X/\lambda_1=(\sqrt{\pi}/2)\xi$. This coordinate appears in the 1993 formulation \citep{OhwadaRIMS1993} and is specified for Mach 2 by \citet{XuHuang2011}. We compared marker positions in the 2004 vector figure and the 2011 raster image, resolving 19 density and 21 temperature symbols in the 2011 image. The ratios of the horizontal scales are 0.9987 and 0.9996, respectively. The root-mean-square (rms) differences in their plotted values are 0.0026 and 0.0025, below the value of one image pixel. This agreement supports using the documented coordinate for the vector data. All 21 vector points are retained in the physical comparison. The accompanying data include both source coordinates, the calibration and the marker-matching script.
\begin{table}[!tbp]\centering\small
\begin{tabular}{l@{\hspace{9pt}}l@{\hspace{9pt}}r@{\hspace{9pt}}r@{\hspace{12pt}}r@{\hspace{9pt}}r}\toprule
& & \multicolumn{2}{c}{RMSE before conversion} & \multicolumn{2}{c}{RMSE after conversion}\\
Method & Seed & Density & Temperature & Density & Temperature\\\midrule
Poisson40 & 202610160 & 0.02173 & 0.01934 & 0.00472 & 0.00242\\
Poisson40 & 202610161 & 0.02269 & 0.01982 & 0.00449 & 0.00236\\
NTC20 & 202610160 & 0.02180 & 0.01939 & 0.00491 & 0.00248\\
NTC20 & 202610161 & 0.02216 & 0.01814 & 0.00614 & 0.00267\\\bottomrule
\end{tabular}
\caption{Effect of the coordinate convention on the comparison using all published points. Density and temperature are normalized by their upstream values. Before conversion, the displayed coordinates are read as $X/\lambda_1$; the other columns use the published coordinate conversion. This separate check fits one translation shared by density and temperature for each seed and uses $210\le t/t_0\le3330$. The main table instead uses density-integral centring.}
\label{tab:source-width}\end{table}
The steepest secant between adjacent density points includes the point near $\xi=0.40$. The vertical separations on either side of that marker in the published figure are 14.437 and 8.827 typographic points. Omitting this point changes the secant width from 4.073 to 4.438 in printed-axis units. The reported widths use all the points. The accompanying files contain the digitized data, the four comparisons and the script for this test.

\FloatBarrier
\section{Fixed-coordinate Mach-2 profiles}
\begin{figure}[!tbp]\centering\includegraphics[width=\textwidth]{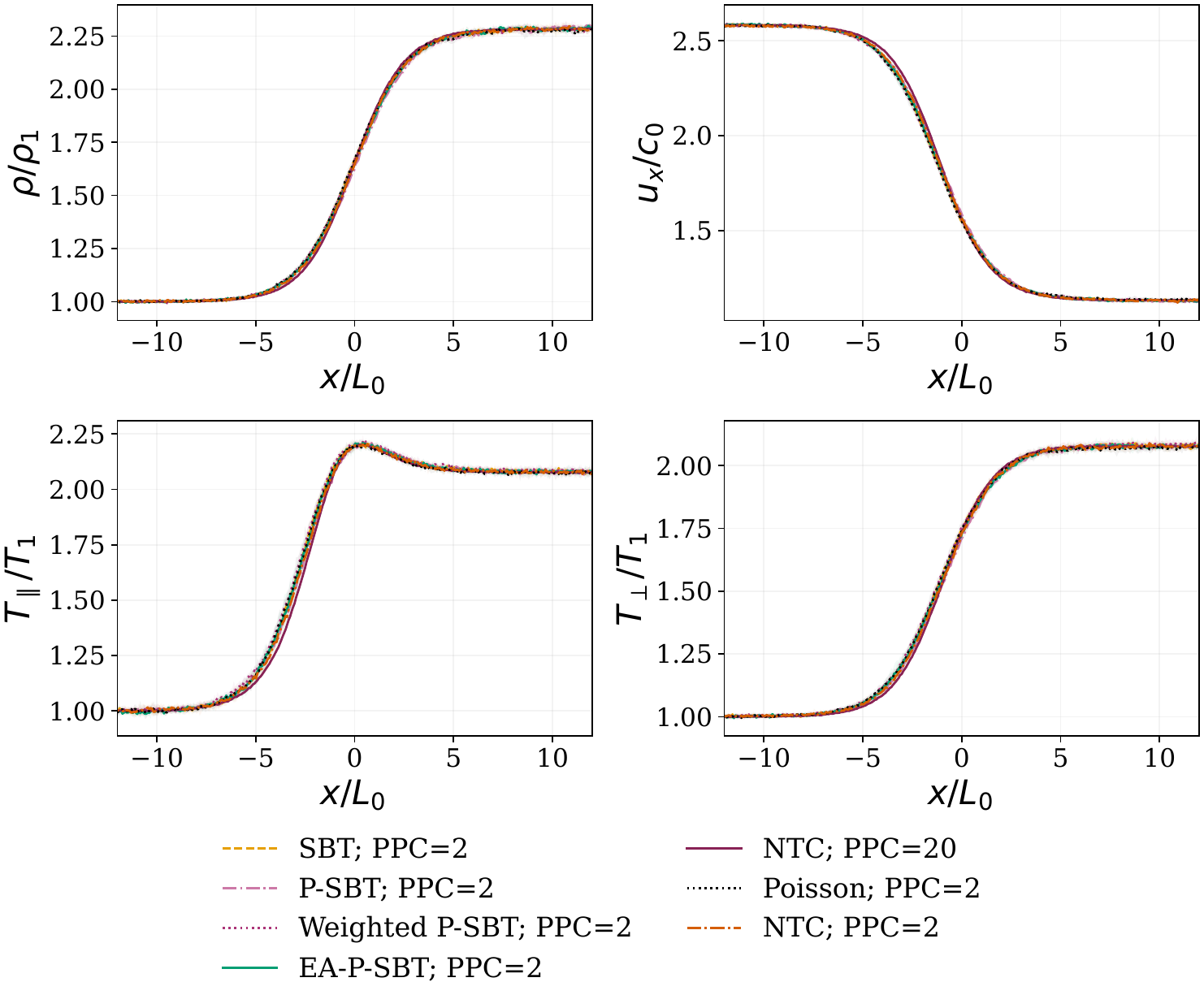}\caption{Unfiltered Mach-2 structure profiles corresponding to Fig.~\ref{fig:shock2structure}.}\label{fig:ma2-raw}\end{figure}

\begin{figure}[!tbp]\centering\includegraphics[width=\textwidth]{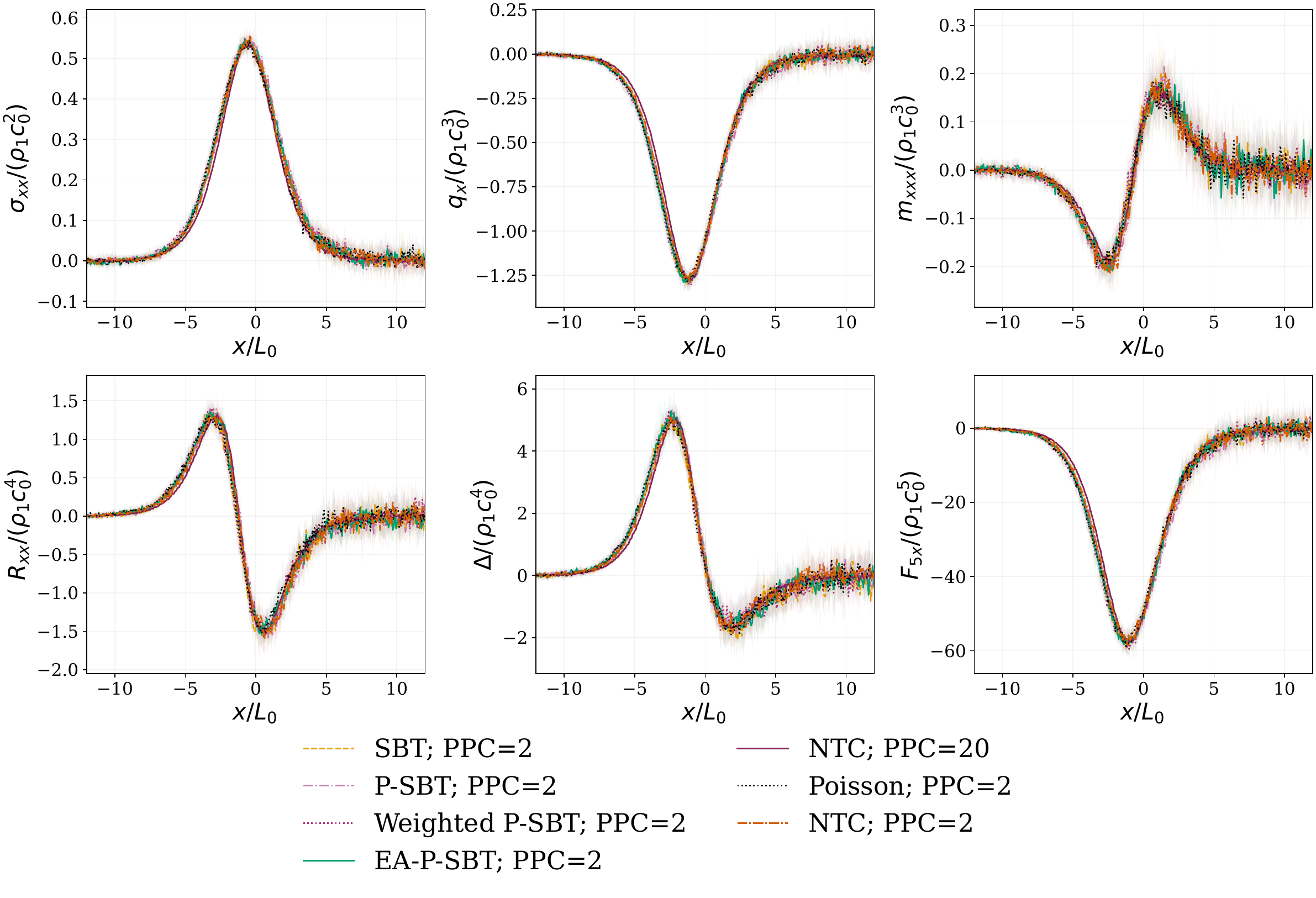}\caption{Mach-2 stress, heat flux and higher moments measured at fixed positions. The main article shows the same moments after alignment with the shock.}\label{fig:ma2-raw-mom}\end{figure}

Figures~\ref{fig:ma2-raw} and~\ref{fig:ma2-raw-mom} show unfiltered means at fixed positions for the same seeds, times and scales as the aligned profiles. Other Mach-2 mean curves use eight FCT passes with diffusion and antidiffusion coefficients $1/8$. Filtering preserves the profile sum and the two outermost cells at each end. Maximum pointwise and peak-amplitude changes are recorded in the source package; numerical comparisons use raw arrays.

\FloatBarrier
\section{Directional temperatures in Mach-2 and Mach-5 shocks}
\begin{figure}[!tbp]\centering\includegraphics[width=\textwidth]{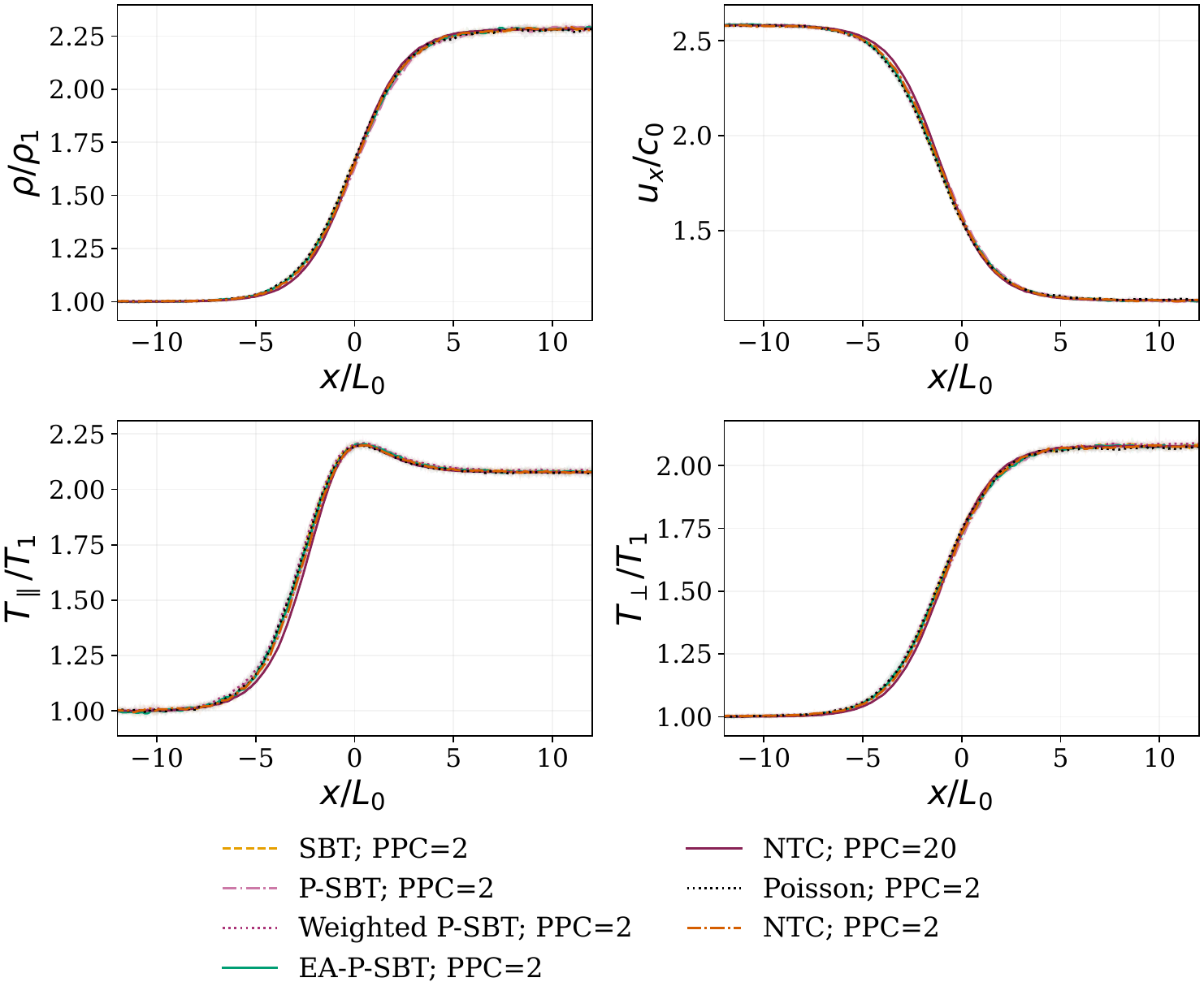}
\caption{Mach-2 density, streamwise velocity, parallel temperature and transverse temperature, in reading order, at $\Delta t^*=0.005$. All methods use the same hard-sphere gas, 800 full collision cells and two seeds over $133.098$ ms per seed. PPC denotes upstream mean occupancy. The mean lines use eight FCT passes. Bands span $\pm2$ pointwise standard errors from unfiltered samples. The cell width is $0.075L_0$. The panels show the shock core.}\label{fig:shock2structure}\end{figure}

\begin{figure}[!tbp]\centering\includegraphics[width=\textwidth]{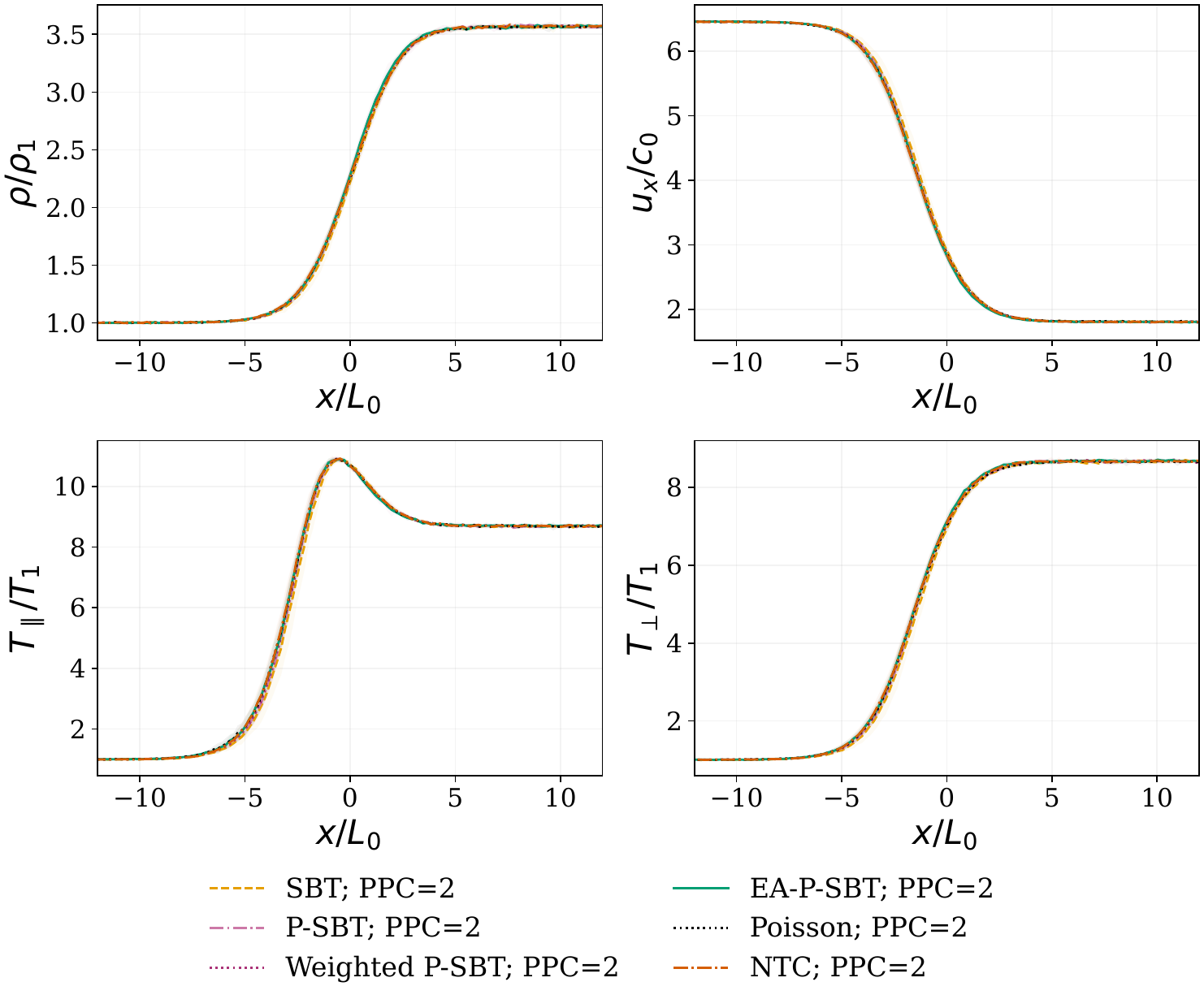}
\caption{Mach-5 shock structure at $\Delta t^*=0.005$, with the panels and gas parameters of figure~\ref{fig:shock2structure}; the cell width is $0.1875L_0$. Each curve uses two seeds and $133.098$ ms per seed. The mean lines and pointwise uncertainty bands use unfiltered samples.}\label{fig:shock5structure}\end{figure}

Figures~\ref{fig:shock2structure} and~\ref{fig:shock5structure} resolve the temperatures parallel and transverse to the shock normal. For a planar shock with normal along $x$, $T_\parallel=m\langle C_x^2\rangle/k_B$ and $T_\perp=m\langle C_y^2+C_z^2\rangle/(2k_B)$. The translational temperature is $(T_\parallel+2T_\perp)/3$.

\FloatBarrier
\section{Mach-5 population control}
\begin{figure}[!tbp]\centering
\includegraphics[width=\linewidth]{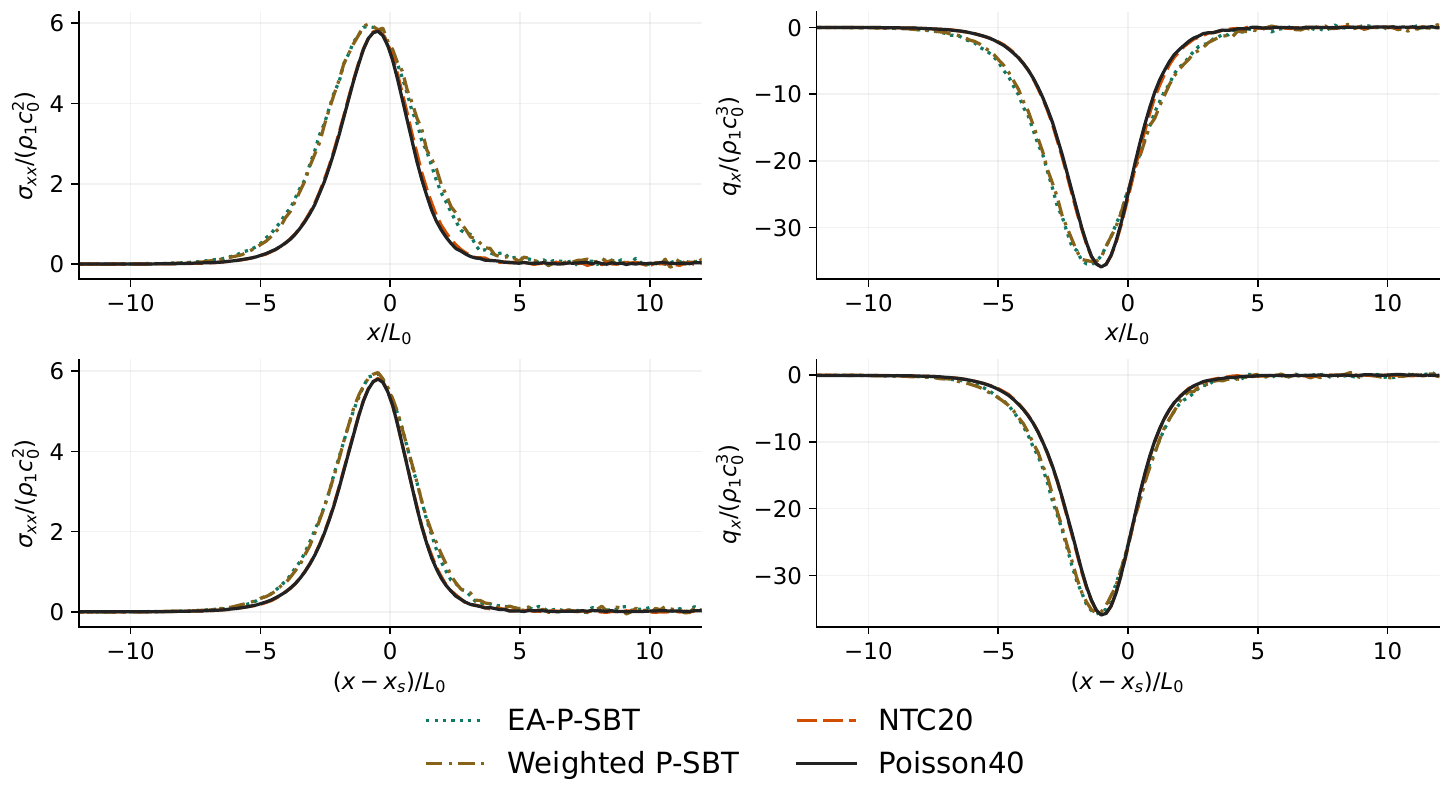}
\caption{Mach-5 stress and heat flux on 800 cells at $\Delta t/t_0=0.005$. The top row uses fixed coordinates. In the bottom row, the origin follows the front position estimated from the density integral. Each curve averages two realizations over the same 133.098 ms. EA-P-SBT and weighted P-SBT use upstream PPC2. The population controls use PPC20 and PPC40.}
\label{fig:mach5_particle_controls}
\end{figure}

We use the same alignment procedure as at Mach 10 to compare the finer-step Mach-5 profiles with the mean of two Poisson40 realizations (Fig.~\ref{fig:mach5_particle_controls}). After normalization by the Poisson40 profile norm over the same part of the shock, following the front changes the EA-P-SBT differences from 29.24\% to 15.32\% in stress and from 26.92\% to 13.60\% in heat flux. With weighted P-SBT they change from 30.30\% to 15.72\% and from 25.22\% to 12.23\%. The corresponding NTC20 differences decrease from 2.96\% to 0.94\% and from 2.10\% to 0.82\%. Front motion contributes to the differences in both shocks. Once the fronts are aligned, the PPC2 differences on 800 cells at Mach 5 remain larger than those of the selected Mach-10 runs, which use different grids and particle populations.

\FloatBarrier
\section{Mach-10 shock motion and sampling duration}\label{sec:shock10}
The superscript $\mathrm{cl}$ in the Mach-10 plots denotes the central moments defined in the article, $m_{xxx}^{\rm cl}=m_{xxx}$ and $R_{xx}^{\rm cl}=R_{xx}$.
At Mach 10 the domain spans $-150\leq x/L_0\leq150$ and the downstream-to-upstream density ratio is $400/103$. EA-P-SBT and weighted-partner P-SBT use upstream PPC2 on grids of 4,000 and 6,000 full collision cells. NTC20 and the Poisson40 numerical reference both use 2,400 cells. These configurations start with approximately 19,534, 29,301, 117,204 and 234,408 simulator particles for the 4,000-cell PPC2, 6,000-cell PPC2, NTC20 and Poisson40 cases, respectively. All use $\Delta t/t_0=0.000625$ ($26.66$ ns), with at most one threshold-triggered population correction per $5t_0$ output block. Two independent realizations per configuration contribute the same $86.8$ ms after warm-up ($210\leq t/t_0\leq2245$).

\begin{figure}[!tbp]
\centering
\includegraphics[page=2,width=\textwidth]{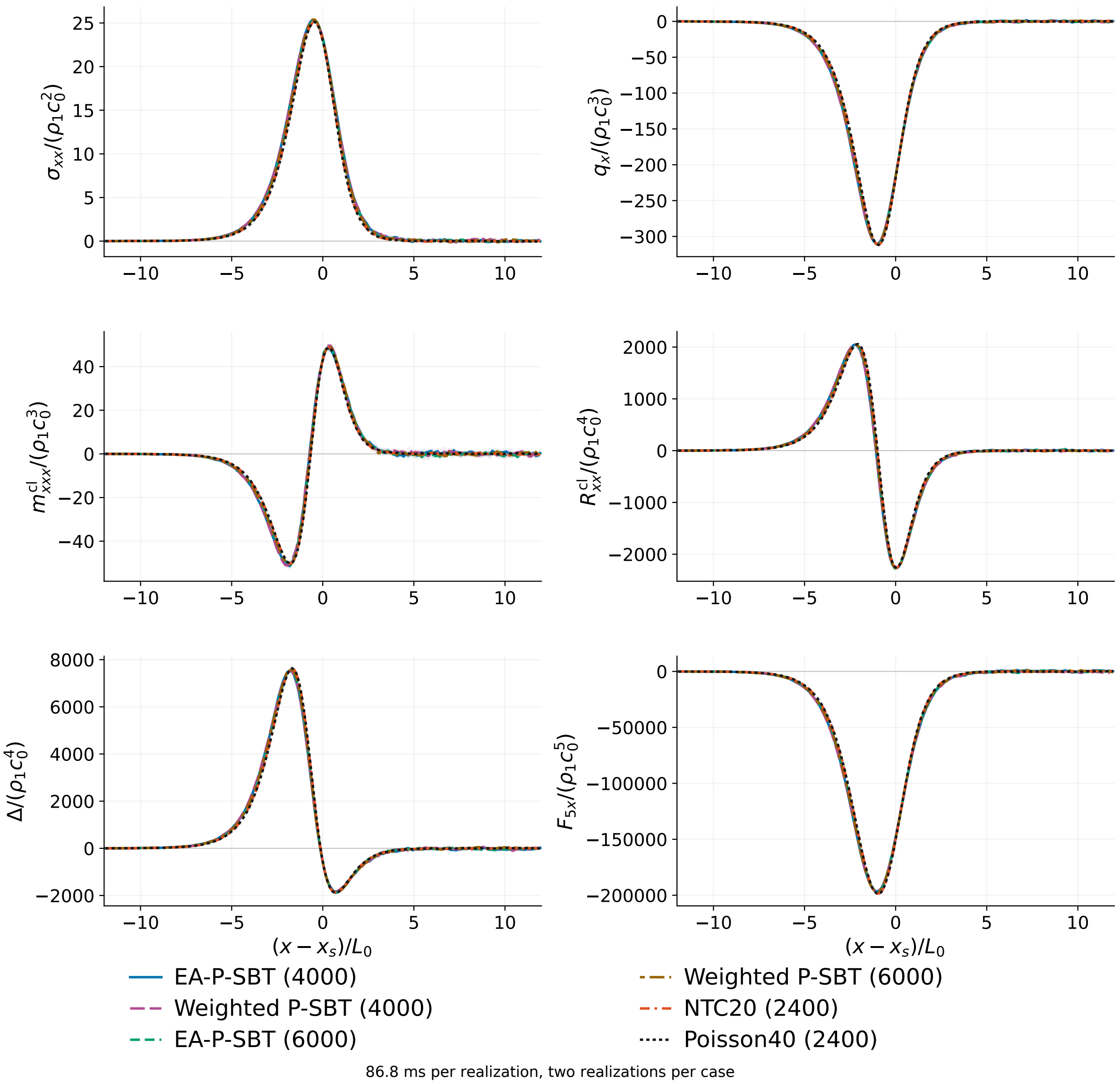}
\caption{Mach-10 normal stress, heat flux and four higher moments in the frame of the shock. The density integral gives the shock position. The legend gives the number of collision cells. EA-P-SBT and weighted-partner P-SBT use upstream PPC2. Each curve uses two realizations and $86.8$ ms per realization at $\Delta t=26.66$ ns. Shaded ranges are $\pm2$ estimated standard errors based on temporal batches and between-realization variation. Means use the remapped profiles.}
\label{fig:shock10moments}
\end{figure}

For each block we integrate the density on $|x/L_0|\leq15$ and locate the step between the upstream and downstream densities with the same integral. That location is $x_s$. We subtract $x_s$ and translate the raw moments onto cells of width $0.125L_0$, conserving their integrals. We then average in time and form the central moments. The comparison in the original spatial coordinates uses the same grid at fixed $x$. Figure~\ref{fig:shock10moments} shows stress, heat flux and higher moments after the translation.

\begin{table}
\centering
\small
\setlength{\tabcolsep}{4pt}
\begin{tabular}{lrrrrrrrr}
\toprule
Scheme & Cells & PPC & $\sigma_{xx}$ & $q_x$ & $m_{xxx}$ & $R_{xx}$ & $\Delta$ & $F_{5x}$ \\
\midrule
EA-P-SBT & 4000 & 2 & 6.6 & 5.1 & 9.3 & 10.1 & 7.3 & 4.4 \\
Weighted P-SBT & 4000 & 2 & 6.4 & 5.0 & 9.8 & 10.4 & 7.3 & 4.3 \\
EA-P-SBT & 6000 & 2 & 4.2 & 3.1 & 5.6 & 6.2 & 4.6 & 2.7 \\
Weighted P-SBT & 6000 & 2 & 4.3 & 3.3 & 5.9 & 6.2 & 4.9 & 3.1 \\
NTC20 & 2400 & 20 & 0.7 & 0.5 & 1.7 & 1.2 & 0.8 & 0.5 \\
\bottomrule
\end{tabular}
\caption{Relative $L_2$ profile differences (percent) from the 2,400-cell Poisson40 reference for Mach 10 after aligning the shock. Each entry uses the mean of two realizations over the same $86.8$ ms sampling interval. The norm covers $|(x-x_s)/L_0|\leq12$.}
\label{tab:shock10}
\end{table}

On 6,000 cells the two PPC2 stress profiles differ from Poisson40 by about 14 percent in relative $L_2$ at fixed coordinates. Following the front brings the differences down to 4.2--4.3 percent and gives heat-flux differences of 3.1--3.3 percent (Table~\ref{tab:shock10}). The 4,000-cell runs retain larger differences, up to 6.6 percent in stress and 10.4 percent in the higher moments. NTC20 remains closest to Poisson40. At fixed upstream PPC, refining the grid also adds particles, so this comparison measures both changes. The 6,000-cell profiles remain closer to Poisson40 when the sampling durations are chosen to give comparable numbers of particle observations. The aligned comparison uses 6,000 cells at PPC2 and 2,400 cells for NTC20 and Poisson40. Below we compare profiles at fixed positions and vary the sampling durations to match the nominal number of particle observations.

\subsection{Time-step sensitivity at fixed feedback interval}
At Mach 2 and 5, changing the time step also changes the time between population checks. A separate Mach-10 Poisson40 test holds the latter interval fixed. Both time steps use the same 1,600-cell grid, two realizations, $133.1$ ms after warm-up, and at most one threshold-triggered correction per $5t_0$ ($213.3\,\mu$s) output block. Between 106.65 and 53.32 ns, the profiles averaged over seeds at fixed positions differ by 1.14 percent in normal stress and 0.79 percent in heat flux in the relative $L_2$ norm on $|x/L_0|\leq12$, normalized by the respective 106.65 ns profile. Each profile is the mean of two realizations over $210\leq t/t_0\leq3330$. At each step, the two individual seed profiles differ by 1.20--1.80 percent in stress and 1.21--1.92 percent in heat flux, using the first seed as denominator. The between-step differences are thus comparable to sampling variation. Within-block samples are taken every 20 time steps, so their physical spacing also changes.

\FloatBarrier

\subsection{Profiles at fixed positions and at equal nominal observation counts}
\begin{figure}[!tbp]
\centering
\includegraphics[page=1,width=\textwidth]{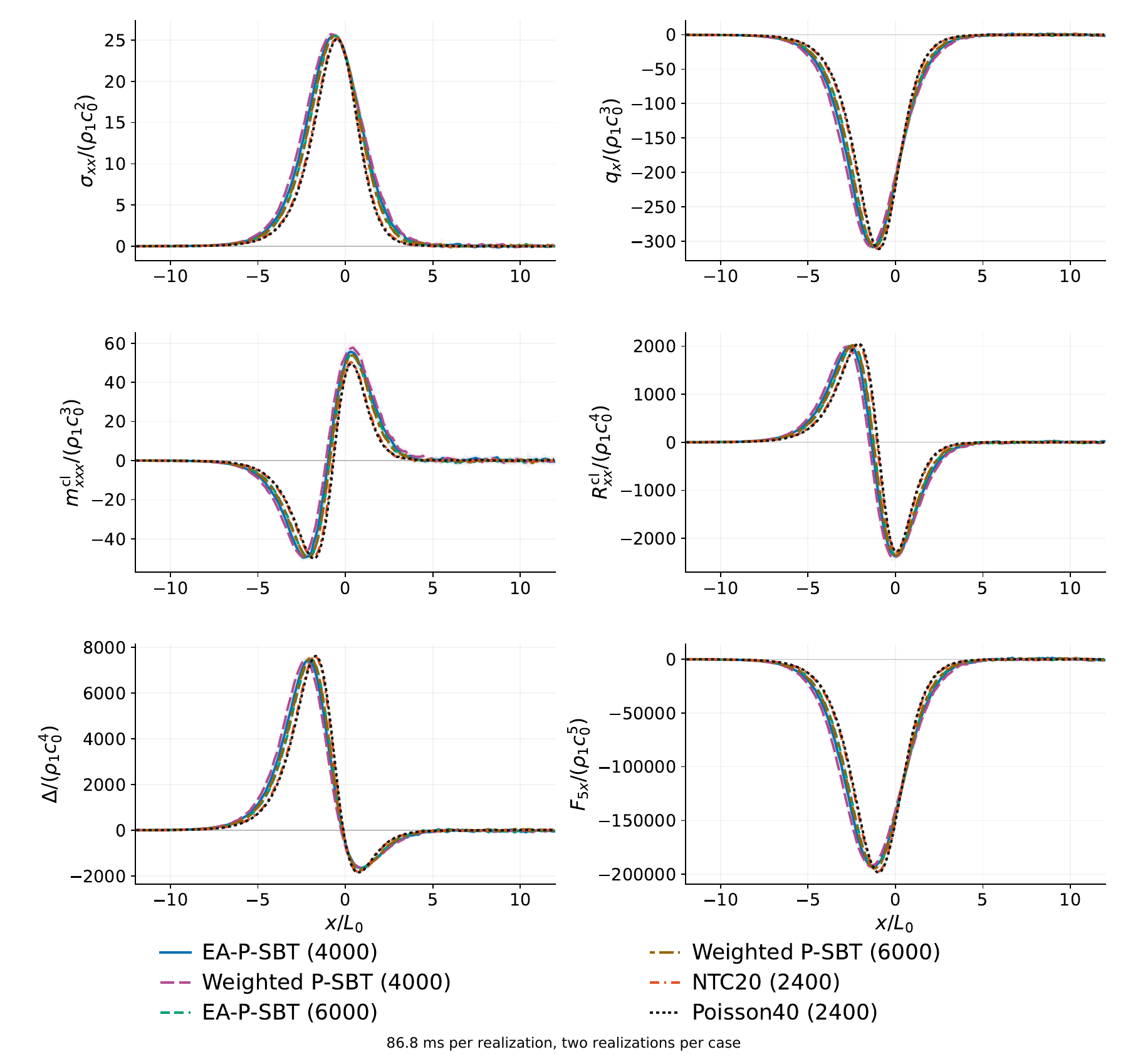}
\caption{Mach-10 profiles in fixed coordinates, sampled after warm-up for $86.8$ ms per realization. Grid sizes are in the legend. The panel order and methods are the same as in the aligned Mach-10 profiles above.}
\label{fig:shock10fixed}
\end{figure}

The fixed-coordinate profiles in Fig.~\ref{fig:shock10fixed} and the aligned profiles in Fig.~\ref{fig:shock10moments} use the same time blocks. Across the two realizations, the standard deviation of front position is approximately $0.89$--$1.08$ in units of $L_0$ on 4,000 cells at PPC2 and $0.71$--$0.78$ on 6,000 cells at PPC2. For Poisson40 it is $0.25$--$0.26$.

\begin{figure}[!tbp]
\centering
\includegraphics[page=3,width=\textwidth]{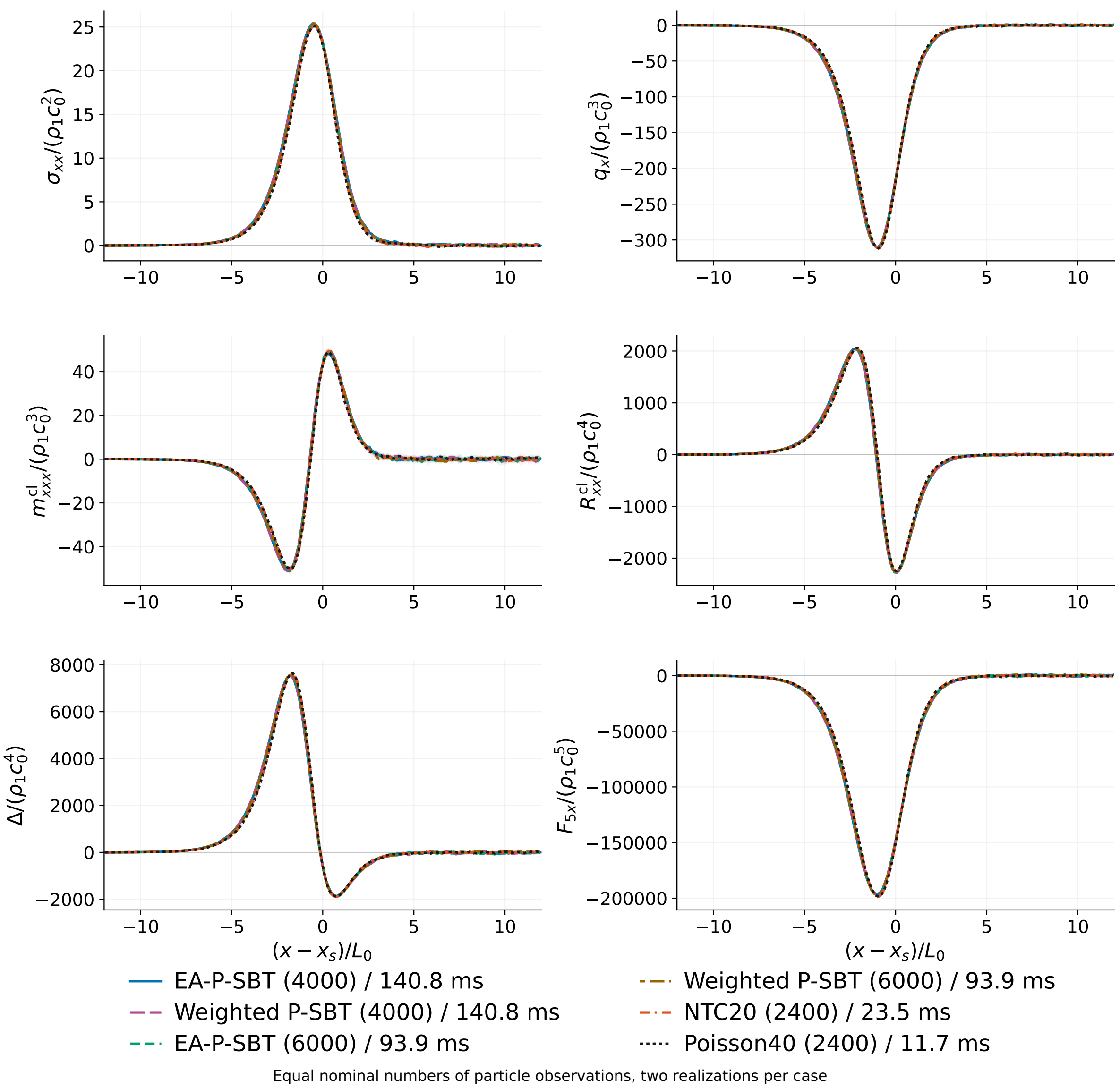}
\caption{Mach-10 profiles aligned with the front, with sampling durations chosen to match the nominal number of particle observations. The legend gives cell counts and sampling duration per realization. Shading includes variation between time blocks and between realizations.}
\label{fig:shock10matched}
\end{figure}

Figure~\ref{fig:shock10matched} uses post-warm-up durations of $140.8$, $93.9$, $23.5$ and $11.7$ ms for the 4,000-cell PPC2, 6,000-cell PPC2, NTC20 and Poisson40 configurations, respectively. Each duration is applied to two realizations. The product of the initial particle count and the number of sampling instants is approximately $5.16\times10^9$ per realization in each configuration. Temporal correlations can give different effective sample sizes. After alignment the relative stress differences are $6.3$--$6.5$ percent on 4,000 cells, $4.3$--$4.4$ percent on 6,000 cells and $1.2$ percent for NTC20.

\FloatBarrier
\section{Grid and particle-count checks for the shock profiles}\label{sec:shock-extra}
We repeated the Mach-5 high-occupancy comparison at 2,400 cells. For $210\le t/t_0\le1720$ (64.42~ms per seed) and $\Delta t/t_0=0.005$, NTC20 and Poisson40 differ by 0.96\% in aligned normal stress and 0.68\% in aligned heat flux, relative to Poisson40. The two Poisson seeds differ by 0.90\% and 0.55\% in these quantities. A separate comparison of 800 and 2,400 cells uses two seeds per method over $210\le t/t_0\le3330$. Conservative remapping precedes the calculation of central moments. Relative to the 800-cell profiles, Poisson40 stress and heat flux change by 0.99\% and 1.00\%. The NTC20 changes are 1.47\% and 1.33\%. At fixed upstream PPC, refining the grid increases the total particle population.

A further series pairs 16,000-cell PPC2 shocks with 800-cell Poisson40 at Mach 2 and 5, and pairs 32,000 cells with 1,600 at Mach 10. Table~\ref{tab:shock-extra} gives examples from the serial implementation. At each Mach number, the low-occupancy and reference profiles are averaged over the same time interval. The Mach-10 record here is much shorter than the moderate-grid comparison. The data file also gives fixed-coordinate values, the four higher moments and the remaining method and seed combinations. Four-thread records use separate random histories and are identified separately. Their runs are excluded from the serial comparisons in the table.
\begin{table}[!tbp]\centering\small
\begin{tabular}{@{}rrlrrrr@{}}\toprule
$M_1$ & PPC2 cells & Scheme & Seeds & Time (ms) & $\sigma_{xx}$ (\%) & $q_x$ (\%)\\\midrule
2 & 16,000 & Weighted P-SBT & 2 & 133.10 & 1.45 & 1.04\\
2 & 16,000 & EA-P-SBT & 2 & 133.10 & 1.56 & 0.97\\
5 & 16,000 & Weighted P-SBT & 2 & 50.34 & 2.25 & 1.43\\
5 & 16,000 & EA-P-SBT & 2 & 50.34 & 2.19 & 1.54\\
10 & 32,000 & Weighted P-SBT & 2 & 13.65 & 2.15 & 1.78\\
10 & 32,000 & EA-P-SBT & 1 & 13.65 & 2.32 & 1.63\\\bottomrule
\end{tabular}
\caption{Aligned relative $L_2$ differences from the corresponding Poisson40 profile, normalized by its norm. Sampling time is per trajectory. Only one serial Mach-10 EA trajectory had production samples on the common interval. Changes in grid, step and number of realizations limit comparisons across configurations.}
\label{tab:shock-extra}
\end{table}

\subsection{Mach-5 velocity marginals and the local fourth moment}\label{sec:sup-return}
\begin{figure}[!tbp]\centering
\includegraphics[width=\linewidth]{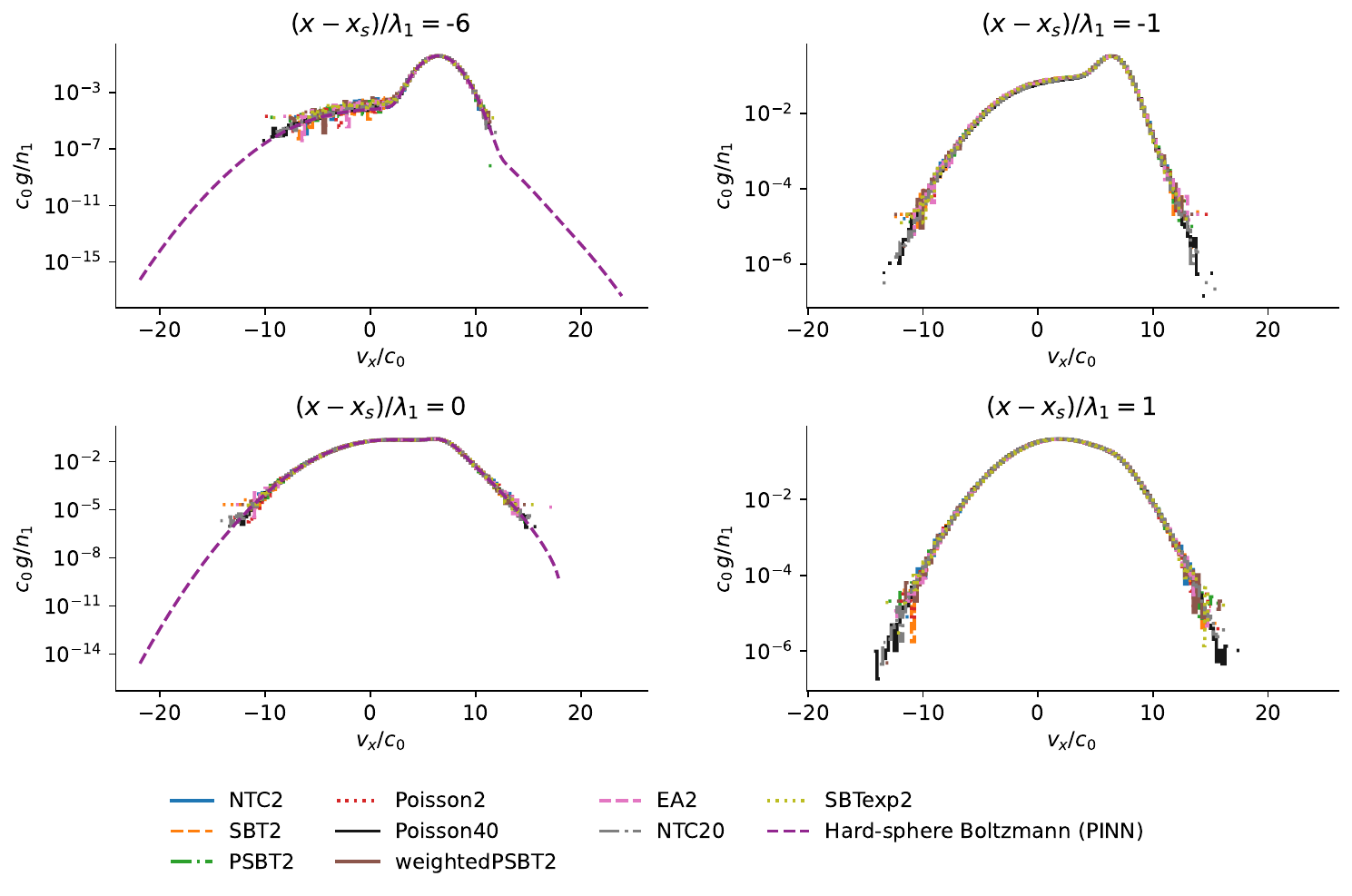}
\caption{Both signs of the axial velocity in the Mach-5 shock, at four distances from the aligned density midpoint. Step curves give the nine DSMC means over the same 470 blocks per seed. At the first and third positions, dashed curves give the PINN marginal. The vertical axis shows the density in each finite velocity bin on a logarithmic scale.}
\label{fig:sup-m5-marginal}
\end{figure}

Figure~\ref{fig:sup-m5-marginal} shows both signs of the axial velocity. We compare particle histograms with the velocity distribution from a physics-informed neural network (PINN) approximation to the hard-sphere Boltzmann equation. The scalar fourth moment integrates over all velocity directions; an occupied bin in the axial marginal alone cannot specify its contribution to this moment. The accompanying records include the contribution of each velocity direction to the moments and the probabilities below and above the histogram limits.

\begin{table}[!tbp]\centering\small
\caption{Mach-5 heat flux and contracted fourth moment on the upstream side. At fixed position $\xi=(x-x_s)/\lambda_1=-1$ the columns give $q_x/(\rho_1 c_0^3)$ and $R_{xx}/(\rho_1c_0^4)$. The last two columns give the mean crossing position and fourth moment evaluated at $n/n_1=1.44393$, with linear interpolation in each realization. Parentheses give half the difference between the two realizations, expressed in the last digits.}
\label{tab:sup-m5-return}\begin{tabular}{lrrrr}\toprule
Method & $q_x$ at $\xi=-1$ & $R_{xx}$ at $\xi=-1$ & $\xi$ at fixed $n$ & $R_{xx}$ at fixed $n$\\\midrule
Poisson40 & -35.08(3) & 39.97(9) & -1.000 & 39.97(25)\\
NTC20 & -35.21(4) & 39.66(109) & -1.007 & 41.14(54)\\
Poisson2 & -35.22(12) & 23.46(96) & -1.088 & 40.38(41)\\
NTC2 & -35.55(1) & 24.58(131) & -1.083 & 41.53(118)\\
SBT2 & -35.30(10) & 26.40(131) & -1.081 & 41.55(63)\\
SBT-exp2 & -35.54(19) & 22.65(87) & -1.102 & 41.04(105)\\
Representative P-SBT2 & -35.33(13) & 22.96(229) & -1.090 & 40.35(106)\\
Weighted P-SBT2 & -35.52(9) & 20.79(185) & -1.099 & 40.42(18)\\
EA2 & -35.53(9) & 21.72(181) & -1.097 & 40.81(160)\\
\bottomrule\end{tabular}\end{table}

On the upstream flank $-2\leq\xi\leq-0.5$, we compute $\|R_{xx}-R_{xx,\mathrm{P40}}\|_2/\|R_{xx,\mathrm{P40}}\|_2$ in the same reference coordinate $\xi$ for both comparisons. At fixed $\xi$ this relative difference ranges from 10.7 to 15.2\% for the seven PPC2 cases. Evaluating each seed at the local mean density of the Poisson40 profile before averaging reduces that range to 0.9--2.8\%. For NTC20 the corresponding differences are 1.1\% and 0.7\%.

\FloatBarrier
\subsection{Velocity contributions in the Mach-5 Boltzmann reference}
\begin{figure}[!htbp]\centering
\includegraphics[width=.94\linewidth]{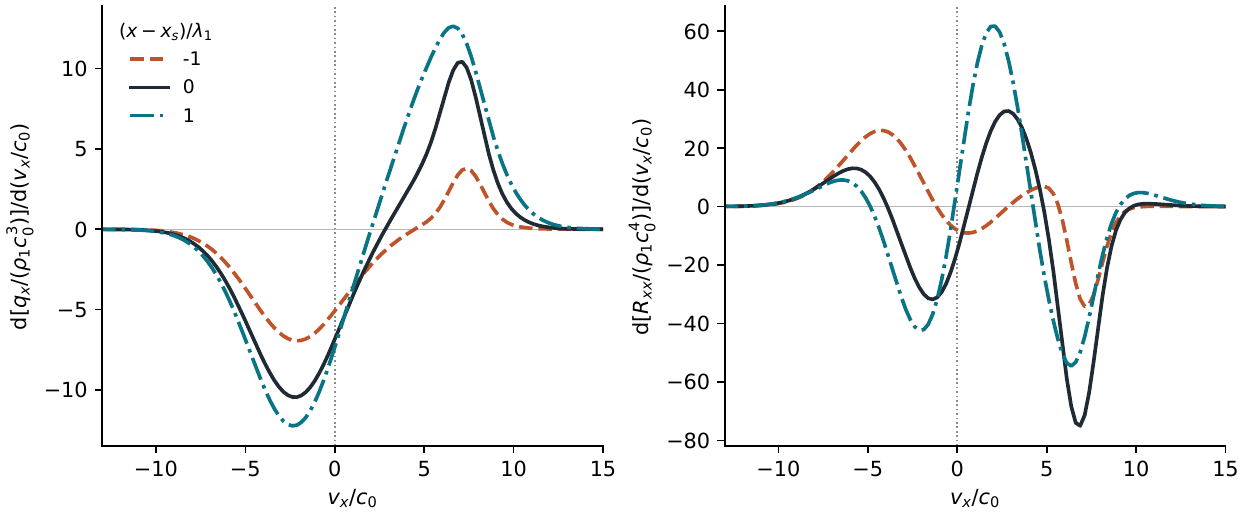}
\caption{Velocity spectra of heat flux and the contracted fourth moment from the independent Mach-5 Boltzmann PINN. Each curve represents a slab of width $0.3\lambda_1$ at the indicated position relative to the density midpoint. Values give the signed contribution per unit $v_x/c_0$ in bins of width 0.25. The line at $v_x=0$ separates the two directions in the shock frame. Summing the bin contributions, including the two open end bins, recovers the local moments.}
\label{fig:sup-m5-spectra}
\end{figure}

The full velocity distribution from the hard-sphere Boltzmann PINN \citep{Roohi2026KineticPartI} allows us to resolve the heat flux and fourth moment within each velocity sign. We bin the distribution by shock-frame $v_x$ and evaluate the central moments using the mean velocity of the whole slab. One mean free path upstream, velocities with $v_x<0$ contribute $115.14$ to $R_{xx}/(\rho_1c_0^4)$ and those with $v_x\geq0$ contribute $-72.47$, giving $42.67$ in total. At the shock midpoint the corresponding heat-flux contributions are $-53.56$ and $28.48$, giving $-25.08$. Figure~\ref{fig:sup-m5-spectra} shows where these opposing contributions lie in velocity space.

\FloatBarrier

\subsection{Heat flux and the fourth moment at equal local density}\label{sec:sup-flank-density}
\begin{figure}[!htbp]\centering
\includegraphics[width=.93\linewidth]{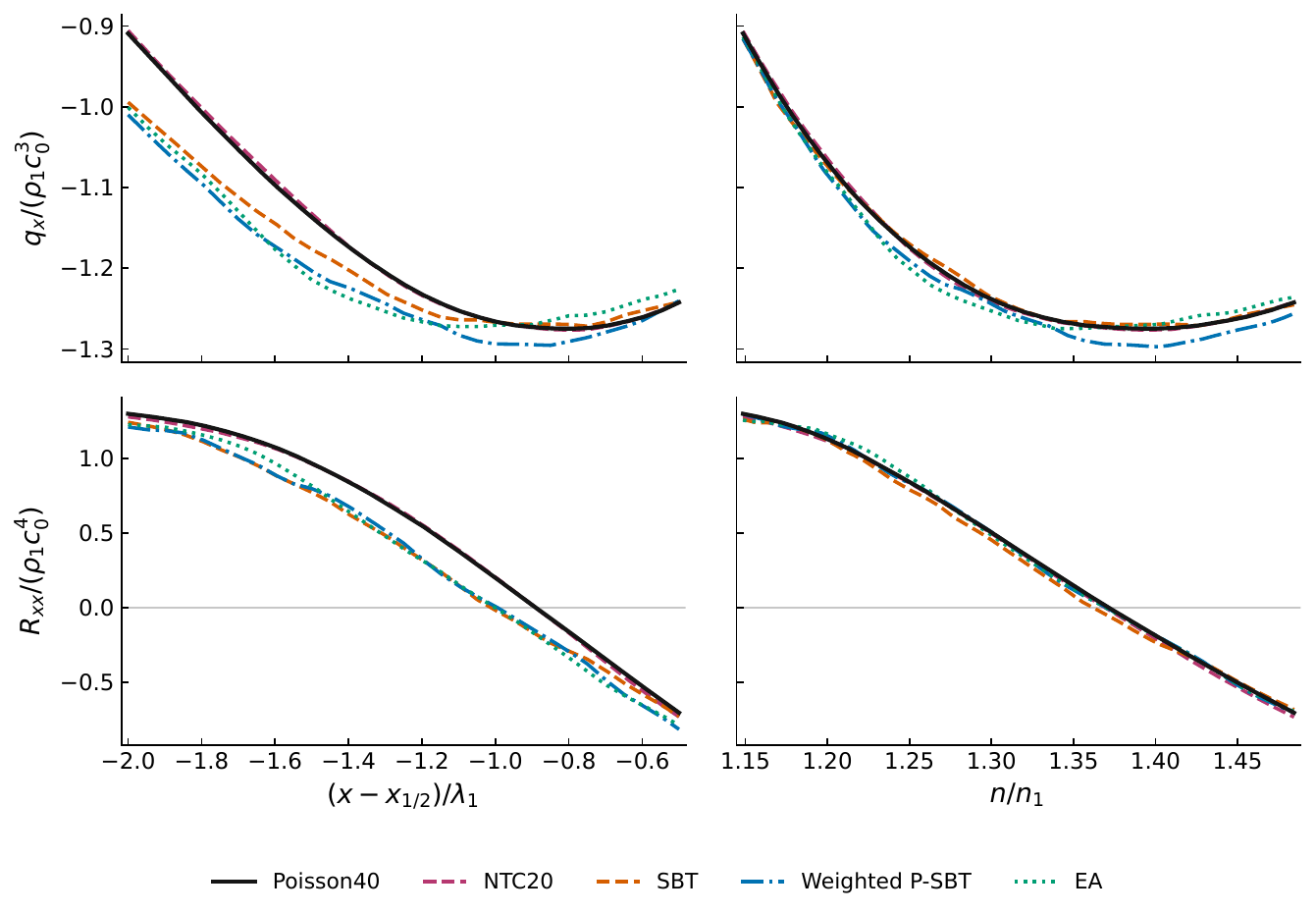}
\caption{Upstream flank of the Mach-2 shock. The upper and lower rows show heat flux and the contracted fourth moment. The left column uses distance from the density midpoint and the right column uses local density. Each curve averages two seeds. The gray band spans the two Poisson40 values. SBT, weighted P-SBT and EA have nominal PPC2.}
\label{fig:sup-flank-m2}
\end{figure}

Figure~\ref{fig:sup-flank-m2} compares the upstream flank of the Mach-2 shock on the 800-cell grid. Each of the two seeds contributes 624 blocks over $210\leq t/t_0\leq3330$ at $\Delta t/t_0=0.005$. We first translate each block by the shock displacement obtained from its density integral. We then pool its time-averaged raw moments over slabs of width $0.3\lambda_1$ before forming $q_x$ and $R_{xx}$. We measure distance from the density midpoint of each resulting profile. To compare at equal density, we find where each seed reaches the Poisson40 mean density at a given reference position and interpolate between the two bracketing measurements. The densities increase from $1.149n_1$ to $1.484n_1$ over the reference interval $-2\leq(x-x_{1/2})/\lambda_1\leq-0.5$. Each has a single rising crossing within this part of each profile.

For the six PPC2 schemes, using local density in place of distance reduces the relative $L_2$ difference in $R_{xx}$ from 10.1--20.7\% to 1.8--6.3\%. For $q_x$ the corresponding ranges are 2.6--4.8\% and 0.46--1.25\%. Both norms use the same quadrature in the Poisson40 spatial coordinate. NTC20 already follows the reference closely in the spatial comparison, and its fourth-moment difference changes from 1.95\% to 2.26\%. At equal density the two Poisson40 seeds differ by 0.15\% in heat flux and 0.80\% in the fourth moment, with both differences normalized by the norm of their mean. With two seeds, we use these separations as a measure of repeatability.

\begin{figure}[!htbp]\centering
\includegraphics[width=.93\linewidth]{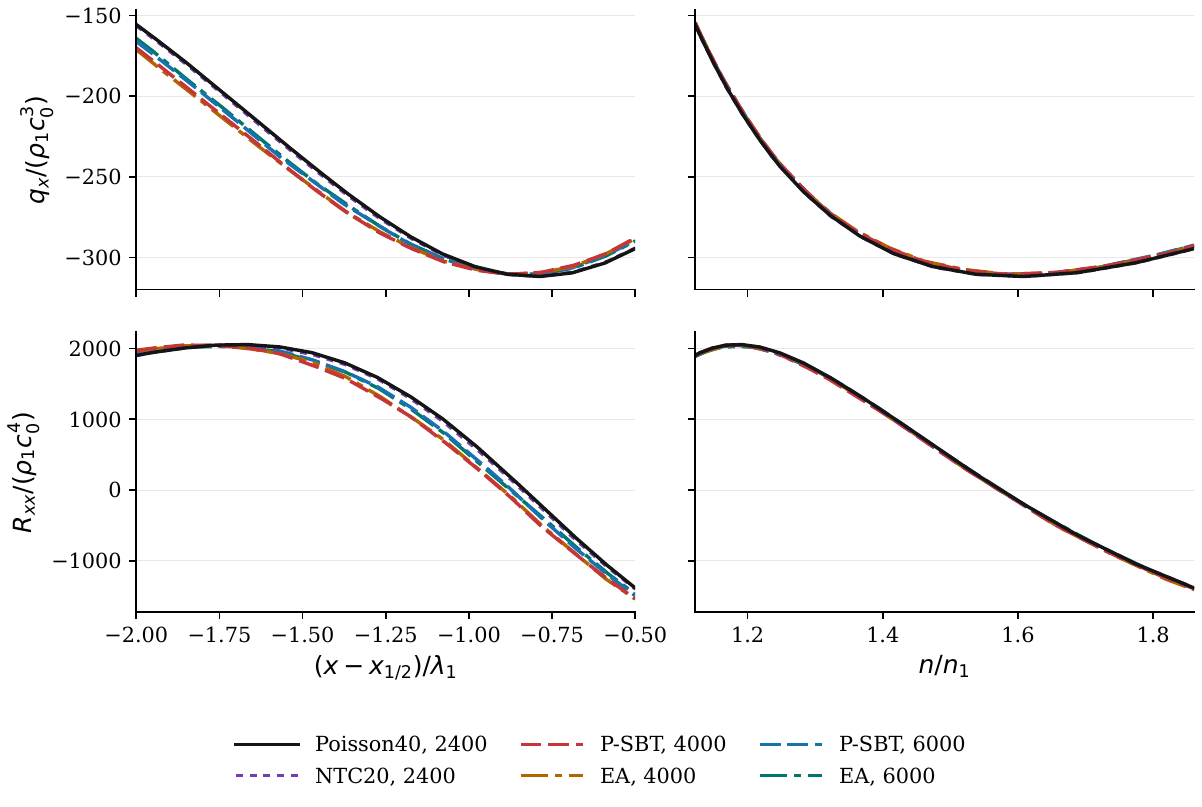}
\caption{Mach-10 heat flux and contracted fourth moment on the upstream flank, against distance from the density midpoint (left) and local density (right). The legend gives the collision scheme and the number of cells. P-SBT denotes weighted P-SBT. The gray band spans the two Poisson40 realizations.}
\label{fig:sup-flank-m10}
\end{figure}

\FloatBarrier
Figure~\ref{fig:sup-flank-m10} repeats the comparison at Mach 10. The weighted P-SBT and EA calculations have nominal PPC2 on 4000 and 6000 cells. Both controls use 2400 cells. At $\Delta t/t_0=0.000625$, all six cases share 407 production blocks (86.81~ms per seed). We translate each block by the density-integral displacement before pooling its raw moments on bins of width $0.125L_0$. The density midpoint of the resulting profile sets the spatial origin for each seed. On the upstream side of the shock, the reference interval covers densities from $1.122n_1$ to $1.861n_1$.

On 4000 cells the spatial differences in $R_{xx}$ are 14.16\% for weighted P-SBT and 13.49\% for EA. At equal density they fall to 1.53\% and 1.26\%. On 6000 cells the corresponding changes are 8.22\% to 0.72\% and 8.53\% to 1.20\%. Heat flux on the 6000-cell grid follows the same pattern, from 2.76\% to 0.40\% for weighted P-SBT and from 2.49\% to 0.51\% for EA. Over the full shock, both 6000-cell profiles have a 10--90\% density width of about $2.66\lambda_1$, compared with $2.54\lambda_1$ for Poisson40. The density comparison therefore answers a narrower physical question than the spatial comparison. It measures how the two moments evolve through the density rise while retaining a separate measure of the shock width.

\FloatBarrier
\subsection{Velocity-resolved transport in eight Mach-5 particle calculations}
We record the 35 Cartesian velocity moments through fourth order in each streamwise velocity bin. Within each slab we use the mean velocity and temperature of the whole particle population to form the signed contributions to $q_x$ and $R_{xx}$. Sixteen trajectories share blocks 1305--1464, or 34.13~ms and 8000 sampling instants per seed, on the 800-cell grid at $\Delta t/t_0=0.005$. The eight methods are SBT, representative-rate P-SBT, weighted P-SBT, EA, SBT-exp, NTC and Poisson at PPC2, and NTC at PPC20. We compare these spectra with the independent PINN approximation to the hard-sphere Boltzmann equation. The Poisson40 results in Table~\ref{tab:sup-m5-return} were averaged over the earlier interval.

\begin{figure}[!htbp]\centering
\includegraphics[width=.97\linewidth]{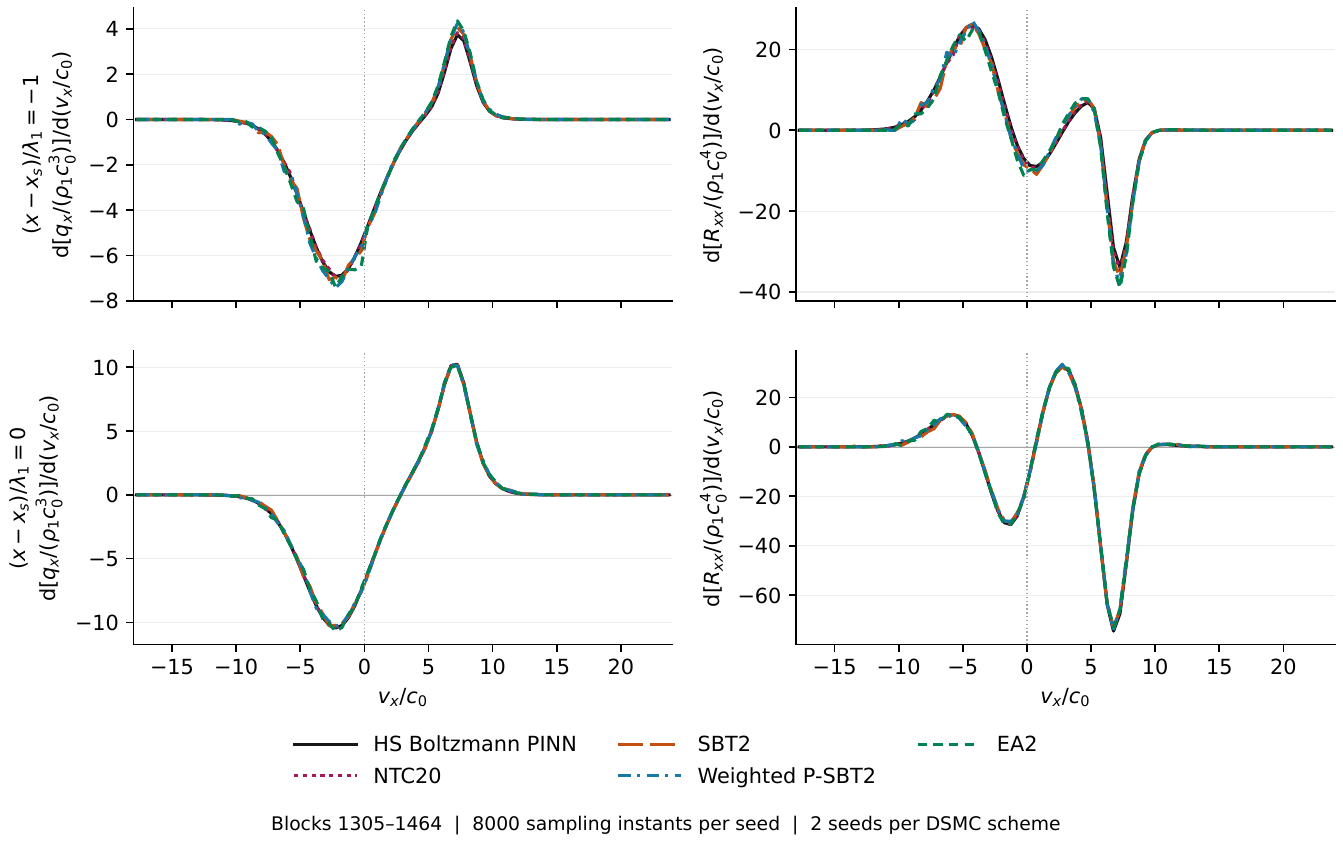}
\caption{Mach-5 velocity contributions to heat flux (left) and the contracted fourth moment (right), one upstream mean free path ahead of the density midpoint (top) and at the midpoint (bottom). Curves show two-realization means for the four particle schemes. The reference curve is the hard-sphere (HS) Boltzmann PINN. Each plotted value combines two adjacent $0.25c_0$ bins and gives the contribution per unit $v_x/c_0$. The shaded NTC20 band spans its two seeds. The accompanying data give both realizations of all eight schemes at five positions.}
\label{fig:sup-m5-sampler-spectra}
\end{figure}

At the density midpoint, all eight schemes and the independent reference give the largest positive contribution to $R_{xx}$ in the interval $2.5\leq v_x/c_0<3.0$, and the most negative in $6.5\leq v_x/c_0<7.0$, when evaluated in bins of width $0.5c_0$ (Fig.~\ref{fig:sup-m5-sampler-spectra}). The net heat flux ranges from $-24.74$ to $-24.01$ across the particle calculations, compared with $-25.08$ in the PINN. One mean free path upstream, the PPC2 heat fluxes range from $-35.77$ to $-34.82$, close to the reference value of $-35.38$. The fourth moment at the same position varies more across both methods and realizations. NTC20 gives $37.35$ and the PINN gives $42.67$. The two SBT realizations give $32.47$ and $16.72$. These seed differences limit the precision of the comparison among the PPC2 schemes.

\FloatBarrier
\subsection{Direct comparison with Poisson40}
\begin{figure}[!htbp]\centering
\includegraphics[width=.97\linewidth]{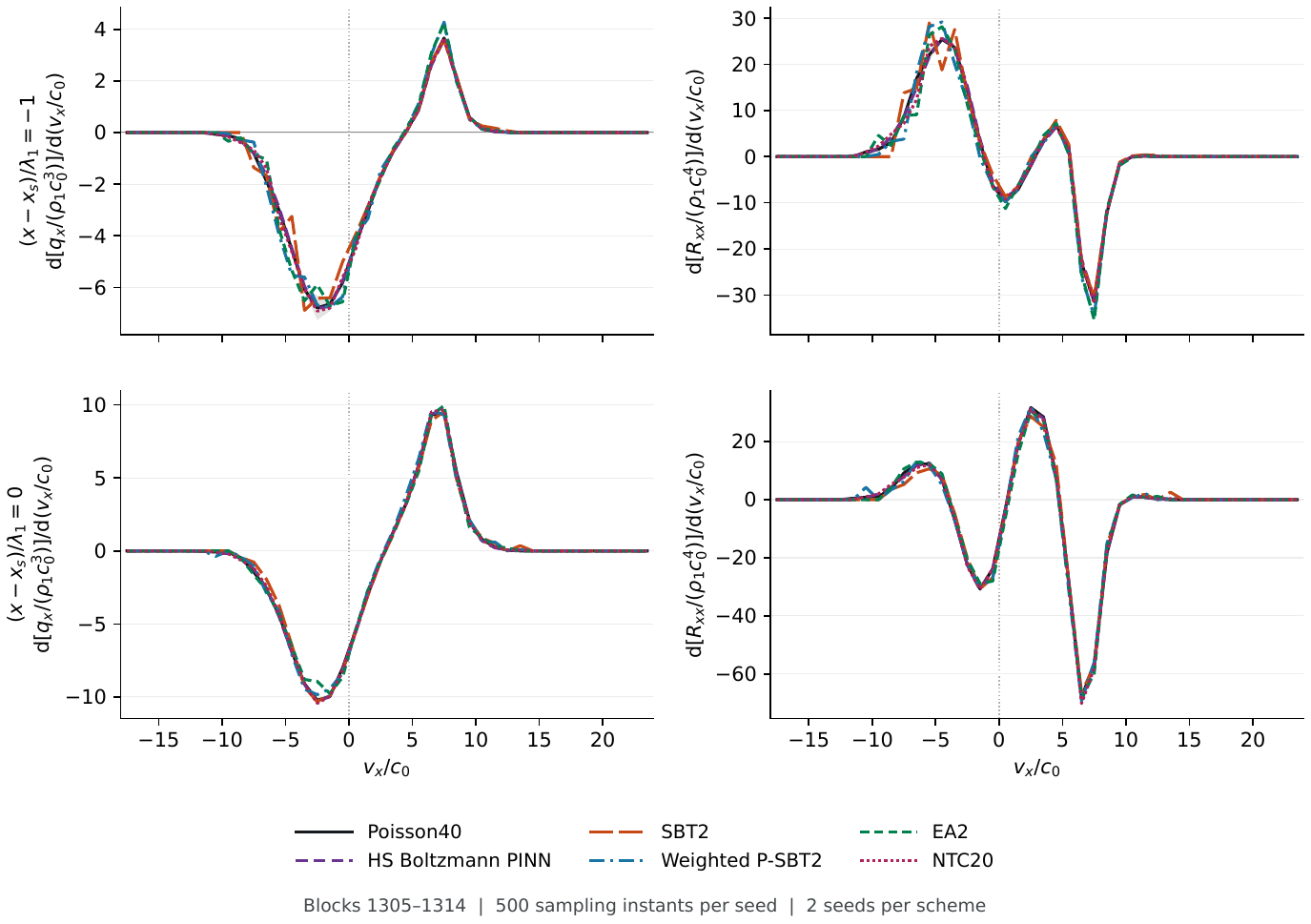}
\caption{Velocity contributions to Mach-5 heat flux and the contracted fourth moment, including the direct Poisson40 comparison on ten common blocks. Positions and normalization follow Fig.~\ref{fig:sup-m5-sampler-spectra}. For this shorter record we sum four adjacent native bins per plotted point, giving a width of $c_0$. The gray band spans the two Poisson40 seeds. The accompanying profiles at five positions include all nine particle schemes and the independent Boltzmann reference.}
\label{fig:sup-m5-p40-bins}
\end{figure}

The first ten blocks of the 160-block record (1305--1314) are available for both Poisson40 seeds as well as for the eight schemes above. This gives a common interval of 2.133~ms and 500 sampling instants per seed. At the density midpoint, the negative- and positive-velocity populations in Poisson40 contribute $-53.36$ and $+28.50$ to the normalized heat flux. The net flux, evaluated before rounding, is $-24.85$, compared with $-24.88$ for NTC20 on the same blocks and $-25.08$ in the independent Boltzmann reference. Figure~\ref{fig:sup-m5-p40-bins} shows the cancellation as a function of velocity. The longer record in Fig.~\ref{fig:sup-m5-sampler-spectra} resolves the spectra of the other eight methods over 16 times as many sampling instants.

\FloatBarrier
\section{Transient cavity centerlines and full fields}\label{sec:cavity-supp}
The gas consists of hard spheres with $m=6.63\times10^{-26}$~kg and $d=3.658\times10^{-10}$~m. Initially $n_0=1.68208\times10^{22}$~m$^{-3}$, $T_0=273.15$ K, and the gas is at rest. All four walls of the square ($L=1$~mm) reflect diffusely at $T_0$. At $t=0$ the upper wall starts moving in $+x$ at $U_{\rm lid}=153.95$~m\,s$^{-1}$. We use $c_m=\sqrt{2k_BT_0/m}$, $\rho_0=mn_0$ and $p_0=n_0k_BT_0=63.44$~Pa. The $64\times64$ uniform collision grid has one collision volume per cell. Each PPC2 trajectory contains 8192 simulator particles and each Poisson40 trajectory contains 163840. Ballistic transport precedes collisions.

At each of nine phases $t^*=4.5,4.5625,\ldots,5$, we pool the 35 Cartesian velocity monomials through degree four over all 21,000 trajectories of a method. We form central moments in each $16\times16$ reporting bin, then average the phases using trapezoidal weights $(1,2,2,2,2,2,2,2,1)/16$. Heat flux, third moments and fourth moments are scaled by $p_0c_m$, $\rho_0c_m^3$ and $\rho_0c_m^4$. Odd parity in the spanwise velocity gives zero ensemble expectation for the corresponding moments. The instantaneous fields at $t^*=5$ use the same bins. The archived snapshot gives the instantaneous fields on a finer $32\times32$ grid.

To estimate uncertainty, we delete the same 40 trajectory labels from all four methods, repeating the reduction for each of 525 groups. This retains correlations between phases, neighbouring bins and paired methods. The full-field comparison uses simultaneous 95\% intervals with a Bonferroni correction across three differences, 32 observables and 256 bins. In the nine-phase field, P-SBT has 32 resolved shear differences, of which 23 lie in the upper eighth of the cavity. EA differs from Poisson40 in 21 bins, including 16 in the upper eighth. SBT has no shear difference above this threshold. The five-phase fields recovered from the last five samples reproduce the earlier ensemble means and their jackknife standard errors.

\subsection*{Centerline profiles and near-lid shear}
\begin{figure}[htbp]\centering
\includegraphics[page=1,width=.96\linewidth]{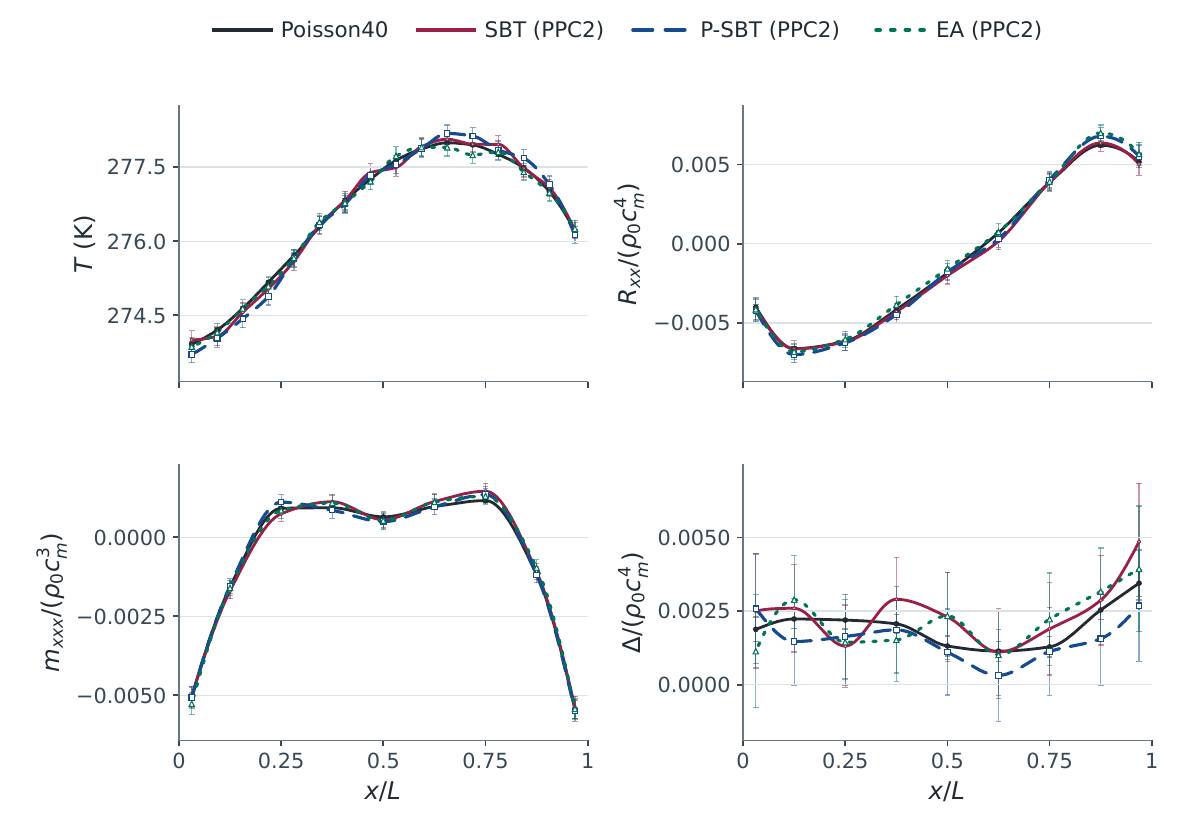}
\caption{Temperature and higher moments on the horizontal centerline. Each marker averages the nine phases $4.5\le t^*\le5$. The higher moments pool pairs of interior bins. The two end bins are shown separately.}\label{fig:cavity-line-horizontal}
\end{figure}

\begin{figure}[htbp]\centering
\includegraphics[page=2,width=.96\linewidth]{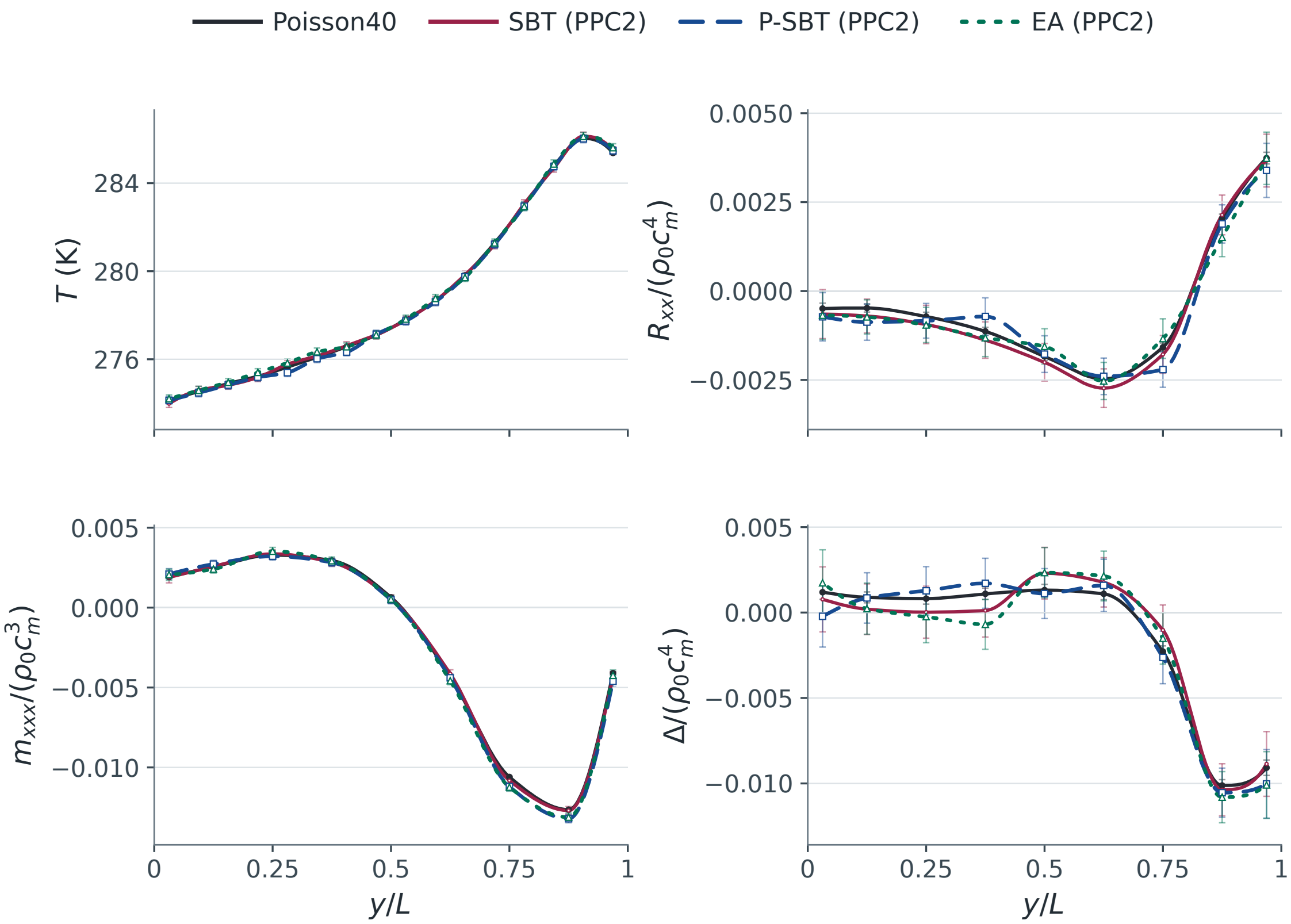}
\caption{Temperature and higher moments on the vertical centerline. The lid is at $y/L=1$. Whiskers span pointwise 95\% intervals at the sampled positions.}\label{fig:cavity-line-vertical}
\end{figure}

\begin{figure}[htbp]\centering
\includegraphics[page=3,width=.96\linewidth]{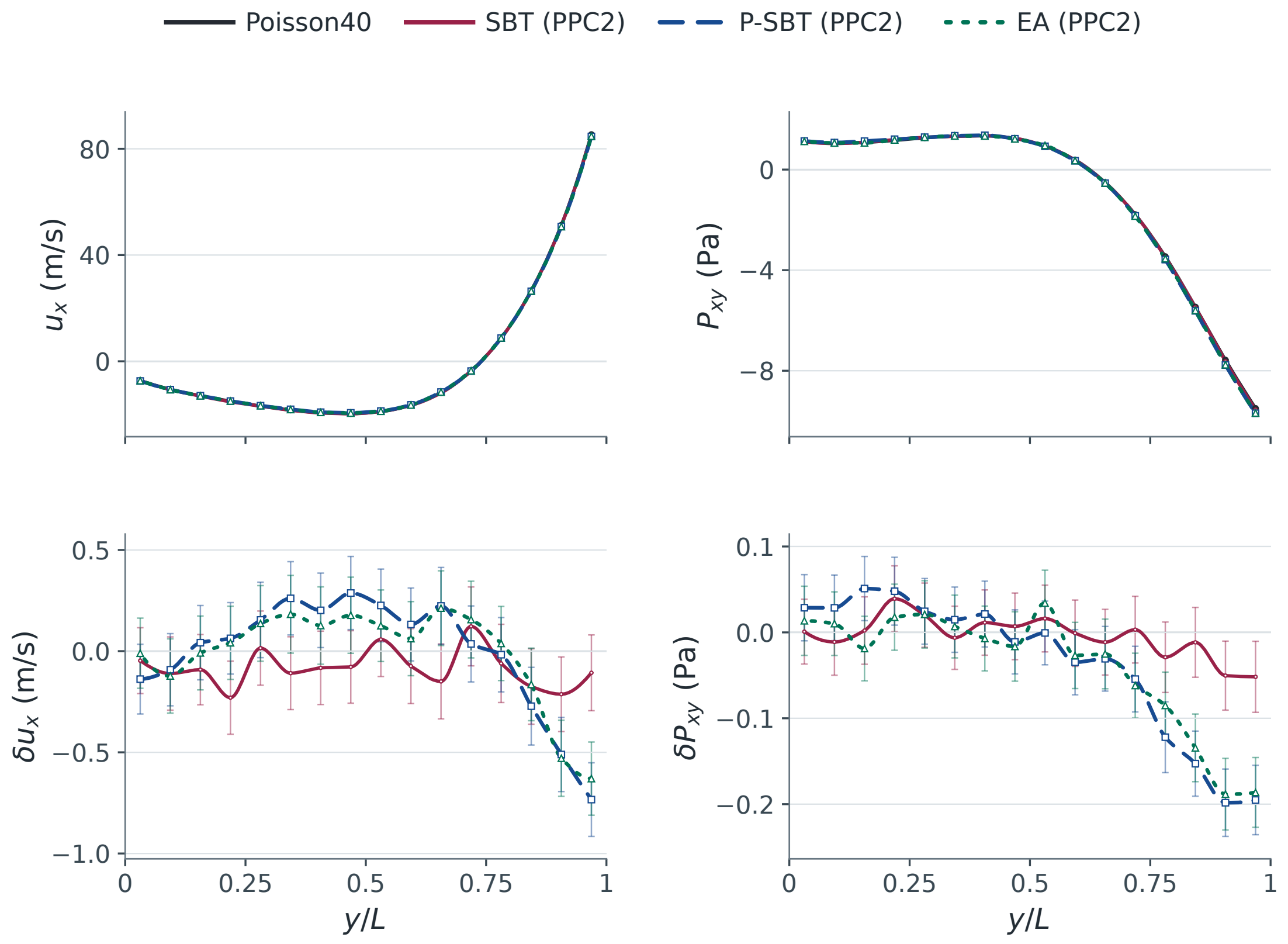}
\caption{Horizontal velocity and shear on the vertical centerline. The lower panels give the signed differences from Poisson40 with paired pointwise 95\% intervals. The nearest observation to the lid is at $y/L=0.96875$.}\label{fig:cavity-line-shear}
\end{figure}

The centerline is the mean of the two rows or columns bracketing $0.5$. Temperature, velocity and shear retain all 16 positions. For $R_{xx}$, $m_{xxx}$ and $\Delta$ we average adjacent interior bin means in seven fixed pairs. The first and last bins keep their native width. The resulting nine spatial positions resolve the near-wall rise with less scatter in the interior. All four methods use the same bin pairs. Markers give the measured averages, whiskers give pointwise 95\% jackknife intervals at those positions, and the curves interpolate between them. Figures~\ref{fig:cavity-line-horizontal} and~\ref{fig:cavity-line-vertical} show both centerlines. Figure~\ref{fig:cavity-line-shear} and Table~\ref{tab:cavity-local} give the velocity and shear close to the moving wall.

\begin{table}[htbp]
\caption{Nine-phase differences from Poisson40 at $\Delta t=11.58$~ns. The root-mean-square (rms) temperature difference uses both centerlines. Signed differences in velocity and shear are evaluated at $x/L=0.5$, $y/L=0.96875$, with one paired standard error. Values use the original reporting-bin observations.}
\label{tab:cavity-local}\centering
\begin{tabular}{lccc}\toprule
PPC2 method & Temperature rms (K) & $\delta u_x$ (m\,s$^{-1}$) & $\delta P_{xy}$ (Pa)\\\midrule
SBT & 0.091 & $-0.107\pm0.095$ & $-0.052\pm0.021$\\
P-SBT & 0.129 & $-0.734\pm0.093$ & $-0.195\pm0.021$\\
EA & 0.102 & $-0.630\pm0.092$ & $-0.186\pm0.021$\\
\bottomrule\end{tabular}
\end{table}
\clearpage
\subsection*{Contour comparisons}
\begin{figure}[htbp]\centering
\includegraphics[page=1,width=.97\linewidth,height=.76\textheight,keepaspectratio]{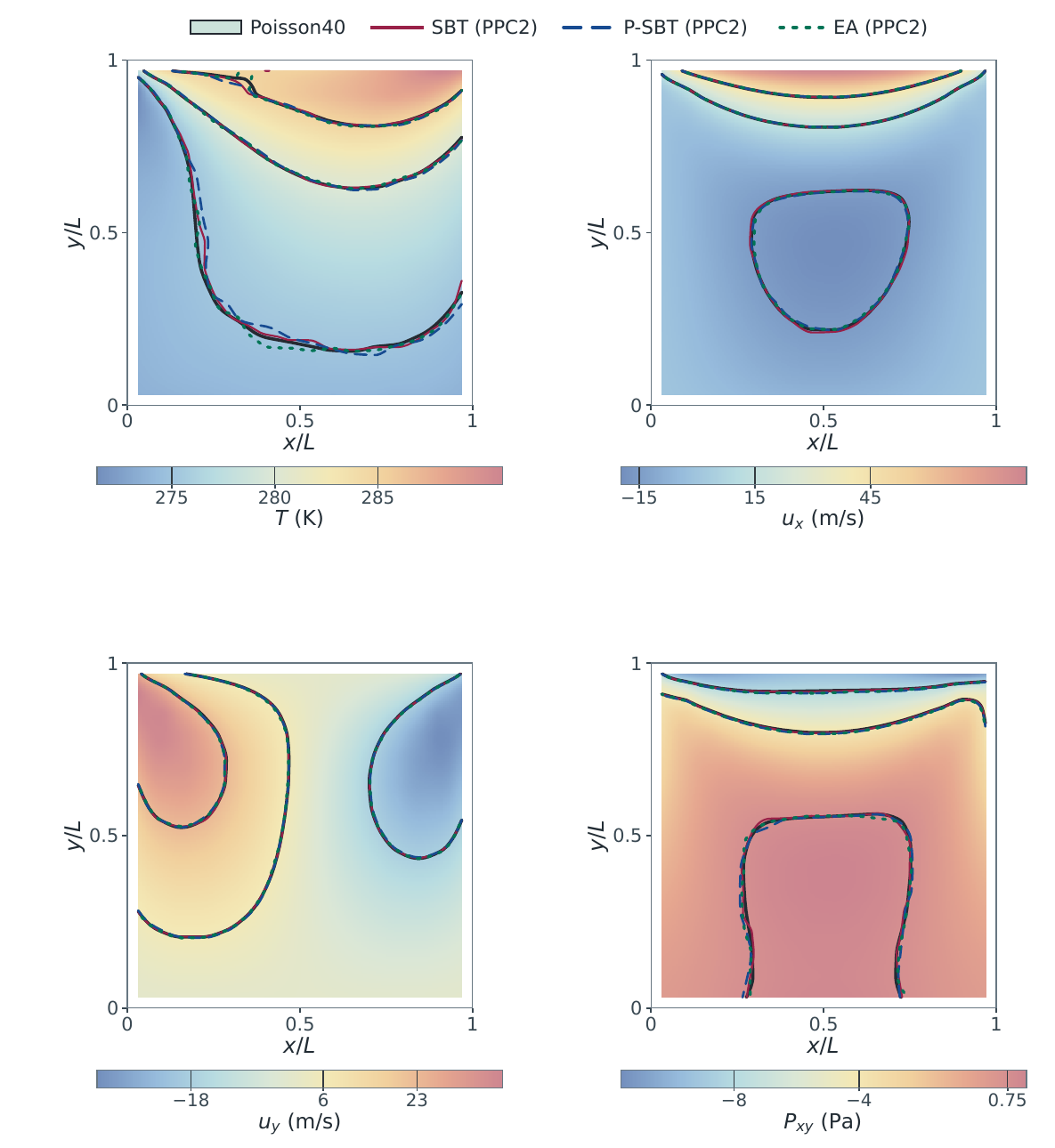}
\caption{$T$, $u_x$, $u_y$ and $P_{xy}$ in the cavity. Nine-phase mean, $4.5\le t^*\le5$. The color field shows Poisson40. The common line levels are marked on each colorbar.}\label{fig:cavity-atlas-1}
\end{figure}

Figures~\ref{fig:cavity-atlas-1}, \ref{fig:cavity-atlas-2}, \ref{fig:cavity-atlas-3}, \ref{fig:cavity-atlas-4}, \ref{fig:cavity-atlas-5} and \ref{fig:cavity-atlas-6} compare the methods over the whole cavity on the same $16\times16$ reporting grid. We show the same 14 mean fields and eight instantaneous fields for both averaging intervals. For this selection a field is shown only if its spatial 5th--95th percentile range in every method exceeds 25 times the median pointwise standard error and its median absolute departure from the four-neighbour average is at most 2.5\% of that range. The short marks on each colorbar locate its contour levels. Where the values lie close together, their labels are listed below the bar. A contour level with four or more closed loops shorter than three bin spacings in any method is omitted for every method. The companion files retain the original fields for all 32 observables.

For contours only, each interior bin takes up to half its value from the mean of its four neighbours. The four methods use the same local fraction. Its value limits any bin's change to one marginal standard error and any method-minus-reference change to half a paired standard error. The outermost bins retain their measured values. We also retain the measured values wherever averaging could change a resolved difference or whether its interval includes zero. A shape-preserving cubic interpolant joins the centers, averaged over the two possible orders of interpolation along $x$ and $y$. Tables, tests and numerical comparisons use the measured bin means. The centerline intervals come directly from the underlying observations.
\begin{figure}[htbp]\centering
\includegraphics[page=2,width=.97\linewidth,height=.76\textheight,keepaspectratio]{figures_cavity/cavity_full_field_atlas_21000.pdf}
\caption{$\rho$, $p$, $P_{xx}$ and $P_{yy}$ in the cavity. Nine-phase mean, $4.5\le t^*\le5$. The color field shows Poisson40. The common line levels are marked on each colorbar.}\label{fig:cavity-atlas-2}
\end{figure}
\begin{figure}[htbp]\centering
\includegraphics[page=3,width=.97\linewidth,height=.76\textheight,keepaspectratio]{figures_cavity/cavity_full_field_atlas_21000.pdf}
\caption{$P_{zz}$, $q_x$, $m_{xxx}$ and $m_{xxy}$ in the cavity. Nine-phase mean, $4.5\le t^*\le5$. The color field shows Poisson40. The common line levels are marked on each colorbar.}\label{fig:cavity-atlas-3}
\end{figure}
\begin{figure}[htbp]\centering
\includegraphics[page=4,width=.97\linewidth,height=.76\textheight,keepaspectratio]{figures_cavity/cavity_full_field_atlas_21000.pdf}
\caption{$m_{xyy}$ and $R_{xy}$ in the cavity. Nine-phase mean, $4.5\le t^*\le5$. The color field shows Poisson40. The common line levels are marked on each colorbar.}\label{fig:cavity-atlas-4}
\end{figure}
\begin{figure}[htbp]\centering
\includegraphics[page=5,width=.97\linewidth,height=.76\textheight,keepaspectratio]{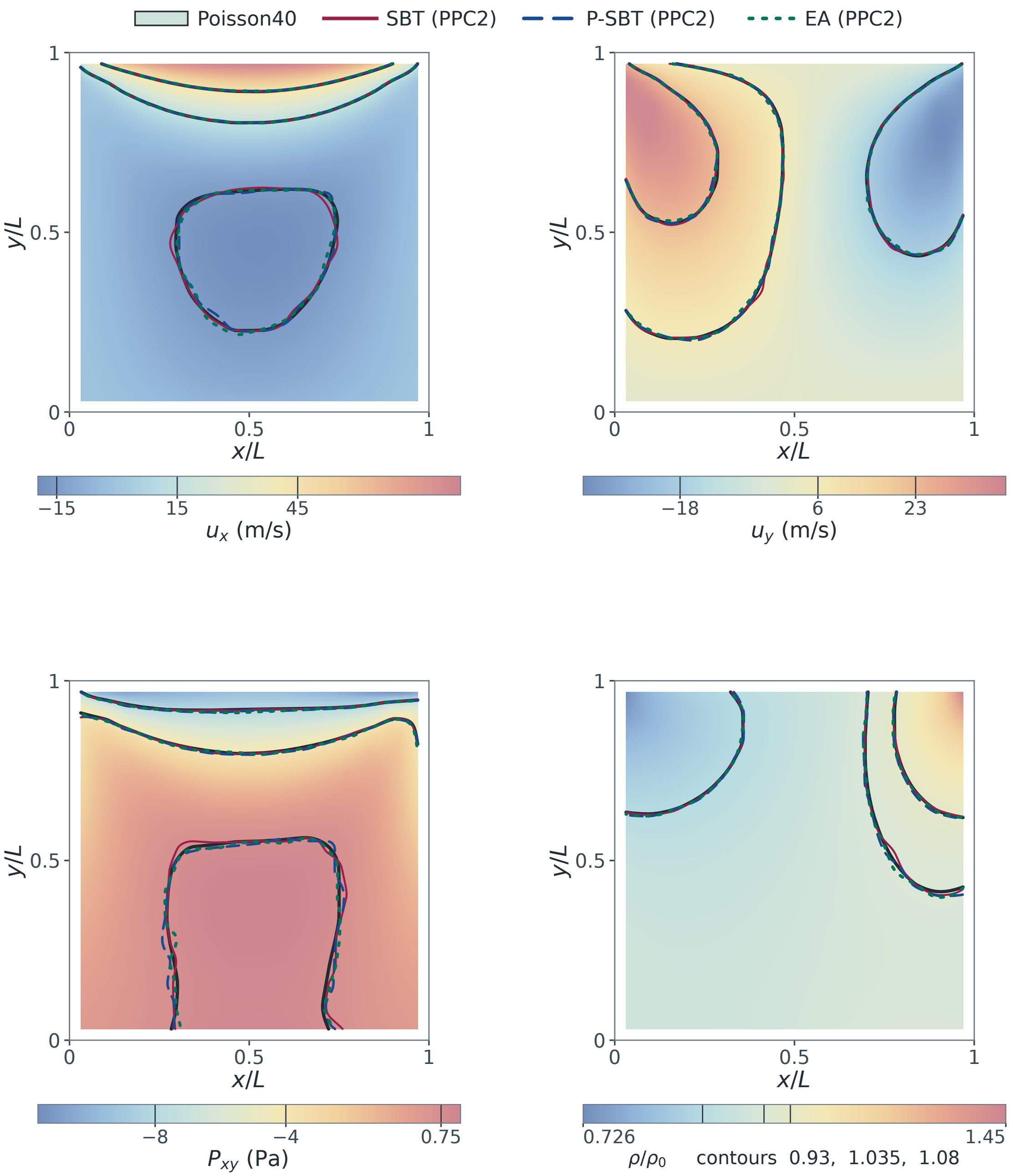}
\caption{$u_x$, $u_y$, $P_{xy}$ and $\rho$ in the cavity. Instantaneous ensemble at $t^*=5$. The color field shows Poisson40. The common line levels are marked on each colorbar.}\label{fig:cavity-atlas-5}
\end{figure}
\begin{figure}[htbp]\centering
\includegraphics[page=6,width=.97\linewidth,height=.76\textheight,keepaspectratio]{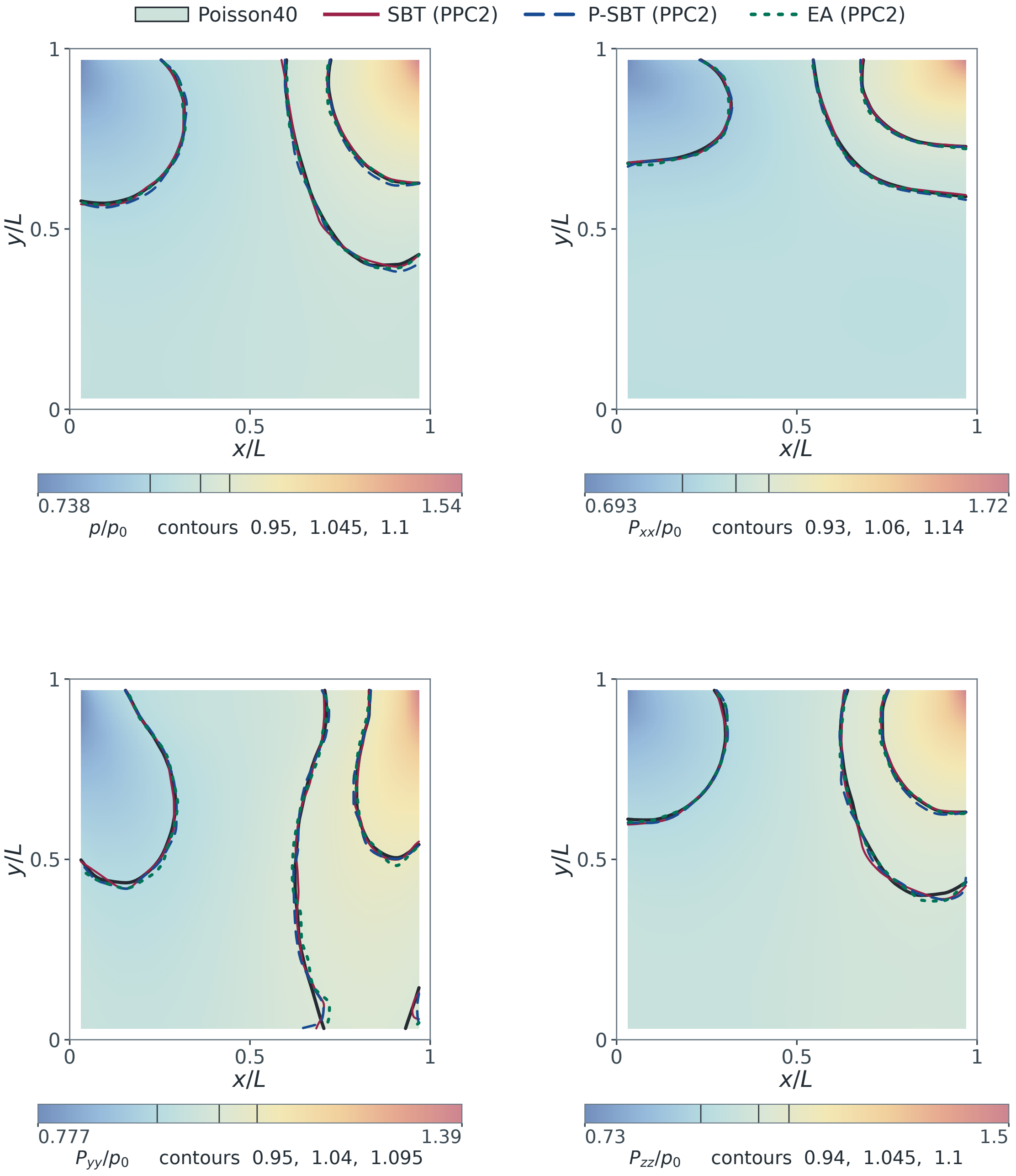}
\caption{$p$, $P_{xx}$, $P_{yy}$ and $P_{zz}$ in the cavity. Instantaneous ensemble at $t^*=5$. The color field shows Poisson40. The common line levels are marked on each colorbar.}\label{fig:cavity-atlas-6}
\end{figure}
\clearpage
\subsection*{Effect of the time step at equal nominal particle counts}
\begin{figure}[htbp]\centering
\includegraphics[width=.96\linewidth]{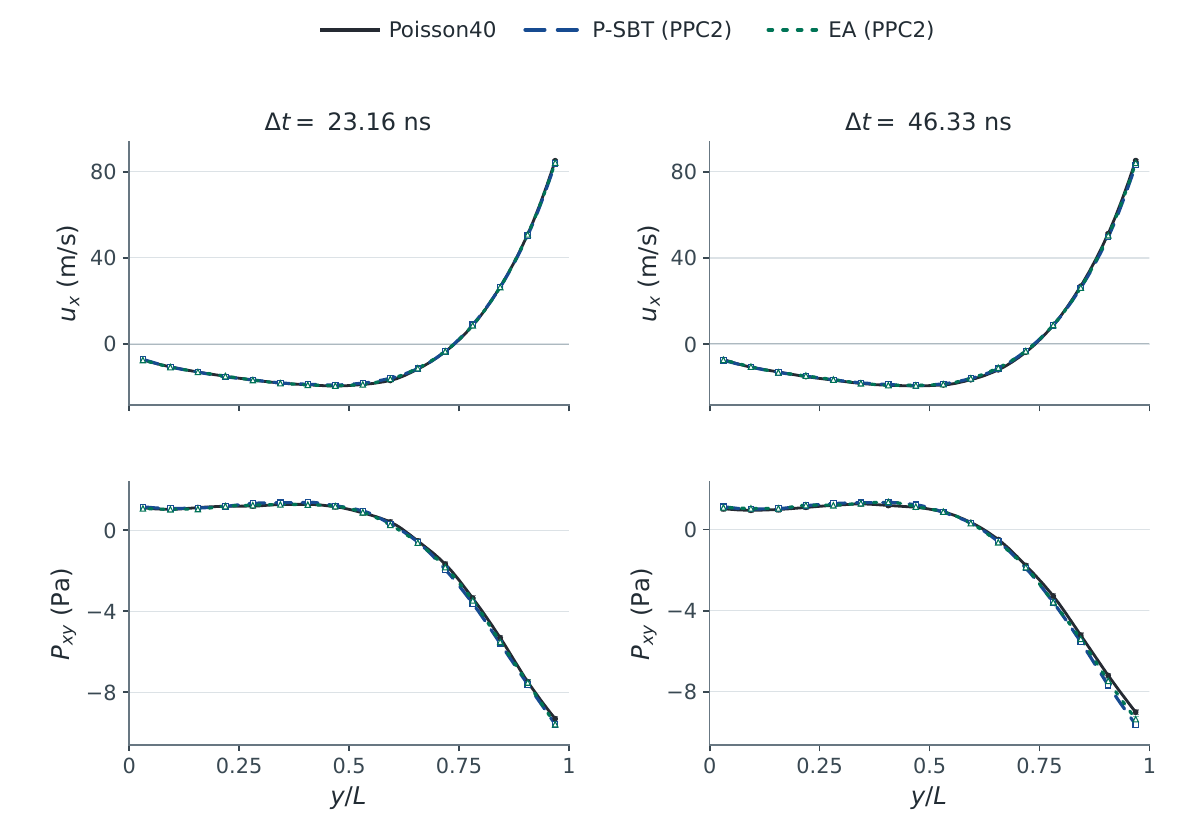}
\caption{Vertical-centerline horizontal velocity and shear averaged over $4.5\le t^*\le5$ at two time steps. Markers give the original 16 positions. The archived data give a marginal standard error at each observation time. Whiskers show $\pm\sum_j w_js_j$, a conservative estimate of the one-standard-error scale for the mean with positive time weights $w_j$ and per-phase errors $s_j$.}\label{fig:cavity-step-lines}
\end{figure}

At each phase, 5,120 trajectories at two particles per cell give 41,943,040 nominal particle observations, the same count as 256 trajectories at forty particles per cell. Each method uses the same $64\times64$ collision grid and the same 81 observation times from $t^*=0$ to 5. Individual particles may contribute at more than one time. Table~\ref{tab:cavity-step} gives the velocity and temperature differences at the final observation. Figure~\ref{fig:cavity-step-lines} compares P-SBT and EA with the Poisson40 reference at the same time step.
\begin{table}[htbp]
\caption{Instantaneous vertical-centerline rms differences from same-step Poisson40 at $t^*=5$. Velocities use $U_{\rm lid}=153.95$ m\,s$^{-1}$ and temperatures use $T_0=273.15$ K.}\label{tab:cavity-step}
\centering\begin{tabular}{lrrr}\toprule
Method & $\Delta t$ (ns) & $u_x$ rms (m\,s$^{-1}$) & $T$ rms (K)\\\midrule
P-SBT & 23.16 & 0.624 & 0.629\\
EA & 23.16 & 0.753 & 0.519\\
P-SBT & 46.33 & 1.011 & 0.522\\
EA & 46.33 & 0.848 & 0.624\\\bottomrule
\end{tabular}\end{table}

\FloatBarrier
\section{Eccentric bearing: mesh, occupancy and wall histories}\label{sec:bearing-supp}
\subsection{Sampling and the wall force}
The gas starts from a quiescent Maxwellian. We accumulate momentum at both circular walls in blocks of $2H/c_m=0.592966~\mu$s. The nominal number of particles per collision cell (PPC) is the total particle count divided by the number of collision cells. Cell areas vary around the eccentric gap and the instantaneous occupancy can differ from this mean. Table~\ref{tab:bearing-runs} gives the blocks used in each comparison. We discard the first 50 blocks for the equal-population averages. Because rotor force develops more slowly than torque, we also compare means at later times.

\begin{table}[!tbp]\centering\small
\begin{tabularx}{\linewidth}{@{}l >{\raggedright\arraybackslash}X l@{}}
\toprule
Comparison & Angular $\times$ radial grid and nominal occupancy & Blocks used\\\midrule
Grid and step & $576\times12$ and $864\times18$. PPC2 methods and higher-occupancy controls. & 51--150\\
Equal population & $2560\times54$, PPC2, uniform or rotor-graded. Reference $576\times12$, Poisson40. & 51--430\\
Fine-grid occupancy & $2560\times54$, Poisson2 and Poisson10. Same coarse Poisson40 reference. & 151--340\\
Four occupancies & $5120\times108$, regular Poisson at PPC2, 4, 10 and 20, and the repeated-pair control at PPC2. & 151--772\\
Repeated pairs & $5120\times108$, Poisson2 with and without the pair veto, and regular Poisson10. & 151--1190\\
Wall history & $576\times12$, Poisson2 and Poisson40, with histories tagged from block 151. & 153--250\\
\bottomrule
\end{tabularx}
\caption{Bearing calculations. The equal-population, fine-grid occupancy and wall-history runs use four seeds (911901--911904) and $\Delta t=3.706035$~ns. The graded SBT comparison uses three matched seeds. At $2560\times54$, PPC10 corresponds to 1,382,400 particles. The coarse-grid study also uses a doubled step, 7.41207~ns. Block ranges include both endpoints.}
\label{tab:bearing-runs}
\end{table}

For angular sector $j$ of width $\delta\theta$, each rotor impact contributes the gas-on-wall impulse $m(\boldsymbol v_{\rm in}-\boldsymbol v_{\rm out})$. With particle weight $F_{\rm num}$ and acquisition time $\Delta t_{\rm obs}$, the tangential stress is the sum of tangential impulses multiplied by $F_{\rm num}/(R\,\delta\theta\,\Delta t_{\rm obs})$ for unit span. We sum the Cartesian impulses before calculating the magnitude of the force. Pressure and shear both contribute to this load. If $\delta\tau(\theta)=\tau_{\rm method}-\tau_{\rm ref}$, define
\begin{equation}
 a_1=\frac1\pi\int_0^{2\pi}\delta\tau\cos\theta\,\mathrm d\theta,\quad
 b_1=\frac1\pi\int_0^{2\pi}\delta\tau\sin\theta\,\mathrm d\theta,\quad
 A_1=(a_1^2+b_1^2)^{1/2}.
\end{equation}
In the rotor frame, $\theta=0$ points from the stator centre towards the rotor centre. The shear contribution to the \emph{force difference} per unit length is $\pi R(-b_1,a_1)$ for a continuous traction field. The direct force measurement also includes the normal impulse and uses each particle's actual impact angle.

\subsection{Grid refinement and equal total population}
On the coarse grid, we also compare ensembles with the same nominal total number of particle observations. At 3.706~ns we use 80 realizations per PPC2 method, 16 at PPC10, eight at PPC20 and four for Poisson40. Using more trajectories reduces the sampling scatter, but a difference in the mean torque remains. At $576\times12$ the Poisson2 and Poisson40 torque magnitudes are 61.00 and 54.34~$\mu$N\,m per metre. At $864\times18$ they are 57.93 and 53.40. The same-grid gap drops from 12.3\% to 8.5\%. The change in the Poisson40 torque itself shows that the coarse reference depends on grid spacing. On the coarse grid the weighted and adaptive torque magnitudes fall from 61.05 and 61.12 at PPC2 to 55.03 and 54.72 at PPC20. At PPC2 the fractions of accepted within-step repeated-pair collisions are 3.71\% and 3.70\%; at 7.412~ns they rise to 7.18\% and 7.15\%. With the larger step the corresponding torque magnitudes are 62.04 and 62.02. In these ensembles the two primary schemes give very similar mean torques.

Table~\ref{tab:bearing-equal} compares the longer, equal-total-particle trajectories. To grade the mesh while keeping its cell count, we redistributed the six innermost radial layers in the area coordinate. The first occupied half its former area. The other 48 layers retained their edges. For a fine-grid run at PPC2, this redistribution keeps the same total population as the coarse Poisson40 reference. Filtering out the earlier blocks changes some means slightly. Over blocks 151--430, the three uniform-grid Poisson-family torque magnitudes differ from the coarse reference by 0.30--0.44\%, and the corresponding shear-profile differences are 4.28--4.60\%.

The uniform fine-grid trajectories also extend to block 1650. In the last 800 blocks, the torque differences from Poisson40 are $-0.10\pm0.17$, $-0.29\pm0.19$ and $-0.17\pm0.16~\mu$N\,m per metre for Poisson2, weighted P-SBT2 and EA-P-SBT2. The respective profile differences are 4.25\%, 4.45\% and 4.21\%. A longer average therefore preserves the distinction between agreement in torque and agreement in local stress.

\begin{table}[!tbp]\centering\small
\begin{tabular}{@{}llrrrr@{}}
\toprule
Mesh & Method & $\delta|\mathcal T|$ (\%) & $E_\tau$ (\%) & $A_1$ (Pa) & Load (N/m)\\\midrule
Coarse & Poisson40 & 0 & 0 & 0 & 0.3746\\
Uniform fine & Poisson2 & 0.44 & 4.24 & 0.47 & 0.3659\\
 & Weighted P-SBT2 & 0.53 & 4.45 & 0.52 & 0.3641\\
 & EA-P-SBT2 & 0.46 & 4.40 & 0.48 & 0.3657\\
Graded fine & Poisson2 & 0.18 & 4.46 & 0.50 & 0.3632\\
 & Weighted P-SBT2 & 0.89 & 4.69 & 0.53 & 0.3649\\
 & EA-P-SBT2 & 0.36 & 4.59 & 0.53 & 0.3649\\
 & SBT2 & $-5.33$ & 6.39 & 0.40 & 0.3565\\
 & NTC2 & $-5.63$ & 6.65 & 0.40 & 0.3548\\
\bottomrule
\end{tabular}
\caption{Equal-population bearing results over blocks 51--430. $\delta|\mathcal T|$ compares magnitudes of seed-mean torque with matched Poisson40. $E_\tau$ is the $L_2$ norm of the signed angular difference divided by the reference norm. $A_1$ is the amplitude of the first harmonic of that difference. Loads are means of the individual force magnitudes. The SBT2 row uses its three admissible trajectories and their corresponding reference seeds. Its fourth trajectory reached the prescribed probability limit.}
\label{tab:bearing-equal}
\end{table}

\begin{figure}[!tbp]\centering
\includegraphics[width=.99\linewidth]{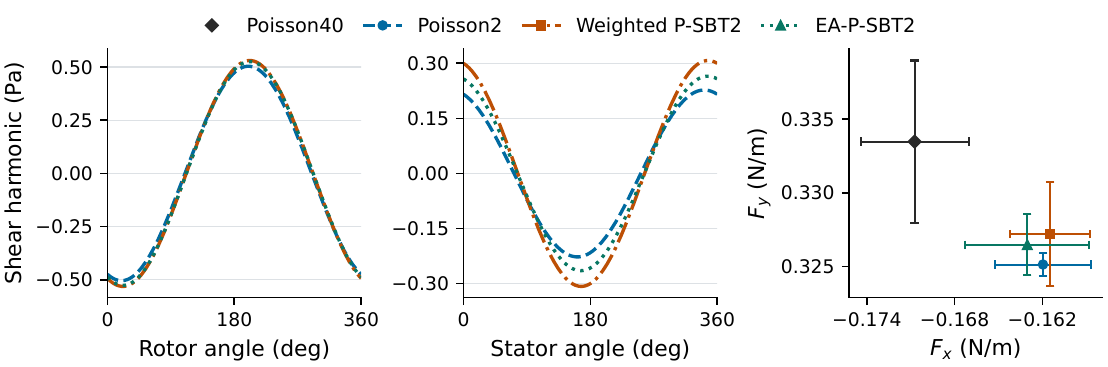}
\caption{First harmonic of the shear difference from coarse Poisson40 on each wall of the graded bearing, and the measured Cartesian gas force per unit span on the rotor. The wall angles refer to the respective cylinder centres. Coordinates in the right panel are those of the solver. Error bars are marginal 95\% intervals across four realizations. Blocks 51--430.}
\label{fig:bearing-load}
\end{figure}

For graded Poisson2 the rotor coefficients are $a_1=-0.474\pm0.124$~Pa and $b_1=-0.172\pm0.045$~Pa. Weighted and adaptive P-SBT give nearly the same phase. On the stator the three amplitudes are smaller, 0.23--0.31~Pa (Fig.~\ref{fig:bearing-load}). In the graded Poisson2 case the difference between the shear-force vectors is 16.9\% of the reference shear-force magnitude. Including the normal impulse gives a much smaller change in total load magnitude, $-3.04\pm1.19$\%. Using the force-angle convention in Eq.~(48) of \citet{Doi2025Bearing}, the coarse-reference and graded Poisson2 angles are $62.89^\circ$ and $63.52^\circ$. Our solver angle increases counterclockwise from the narrow-gap direction, so the reported Doi angle equals $180^\circ$ minus the solver angle.

\subsection{Increasing occupancy on the uniform fine grid}
\begin{figure}[!tbp]\centering
\includegraphics[width=.99\linewidth]{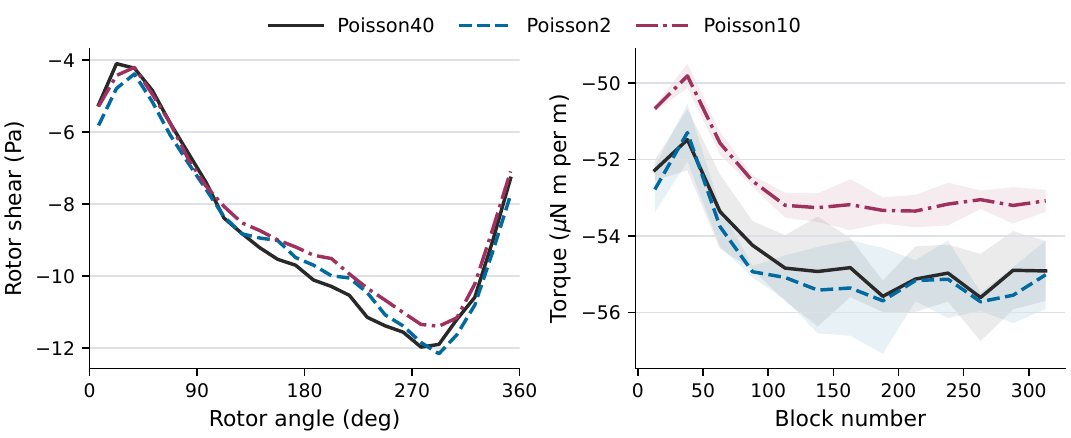}
\caption{Rotor shear for blocks 151--340 (left) and torque in successive complete groups of 25 blocks (right). The fine-grid Poisson runs differ fivefold in particle population. Poisson40 is on the coarser mesh. Shading denotes pointwise 95\% intervals over the four seed means. One block lasts $0.592966~\mu$s.}
\label{fig:bearing-occupancy}
\end{figure}

Poisson2 and Poisson10 use the same grid, collision algorithm, wall reflection law, step and initial distribution (Fig.~\ref{fig:bearing-occupancy}). For each occupancy we ran four independent trajectories, pairing them by seed across methods. The Poisson10 rotor load rose by about 15\% between blocks 51--75 and 76--100. Its means over blocks 151--200 and 201--250 were 0.3685 and 0.3678~N/m. For these later blocks, the paired 95\% intervals for changes in torque and in both first-harmonic coefficients include zero. We therefore also give averages starting at block 151 for comparison with the longer averages in Table~\ref{tab:bearing-equal}.

Over blocks 151--255, increasing PPC from 2 to 10 lowers the torque magnitude by $3.78\pm1.62$\%. The first-harmonic amplitude $A_1$ is 0.389 and 0.392~Pa for Poisson2 and Poisson10, respectively. Averaging through block 340 gives torques of $-55.38$, $-53.21$ and $-55.20~\mu$N\,m per metre for Poisson2, Poisson10 and coarse Poisson40. Their loads are 0.3766, 0.3680 and 0.3853~N/m. On the fine grid, the cosine and sine coefficients change by $+0.146\pm0.049$ and $-0.142\pm0.135$~Pa, respectively, from PPC2 to PPC10. The angular shear difference from the coarse reference therefore persists as the mean torque changes. These two averaging intervals overlap. The four occupancies in the next comparison were tested on a grid with twice as many cells in each direction.

\subsection{Occupancy at fixed mesh and time step}
\label{sec:bearing-occupancy-ladder}

On the $5120\times108$ grid, we kept the bearing geometry and time step fixed and changed the
particle weight to obtain nominal occupancies of 2, 4, 10 and 20. The four Poisson20
trajectories (seeds 911901--911904, in order) ended at blocks 853, 990, 880 and 772.
The planned endpoint was block 1190. For the comparison in
Table~\ref{tab:bearing-occupancy-ladder}, we discarded the first 150 blocks and
stopped every average at block 772. This gives 622 blocks, or $368.825~\mu$s,
per seed at each occupancy. The rotor radius is 1~mm. The angular average of
the shear and the torque per unit span are related by
$\mathcal T=2\pi R^2\langle\tau\rangle_\theta$.

\begin{table}[!tbp]
\centering\small
\begin{tabular}{@{}lrrrr@{}}
\toprule
& $\mathcal T$ & $\langle\tau\rangle_\theta$ & $\delta|\mathcal T|$ & $E_\tau$\\
Method & ($\mu$N\,m/m) & (Pa) & (\%) & (\%)\\
\midrule
Poisson2 & $-55.099\pm0.122$ & $-8.769\pm0.019$ & $4.25\pm0.21$ & 4.72\\
Poisson4 & $-53.829\pm0.057$ & $-8.567\pm0.009$ & $1.84\pm0.11$ & 2.04\\
Poisson10 & $-53.102\pm0.033$ & $-8.451\pm0.005$ & $0.47\pm0.05$ & 0.56\\
Poisson20 & $-52.854\pm0.023$ & $-8.412\pm0.004$ & \multicolumn{1}{c}{---} & \multicolumn{1}{c}{---}\\
Poisson2 with veto & $-54.997\pm0.112$ & $-8.753\pm0.018$ & $4.05\pm0.17$ & 4.48\\
\bottomrule
\end{tabular}
\caption{Bearing torque and signed rotor shear at fixed grid and step.
The torque and stress columns give means over the four trajectories.
For $\delta|\mathcal T|$, we formed the percentage change in torque magnitude
relative to Poisson20 separately for each seed, then averaged the four differences.
Each $\pm$ value is the half-width of a two-sided 95\% Student interval with
three degrees of freedom. $E_\tau$ compares the seed-averaged profiles in the
48 angular sectors with the Poisson20 profile using the relative $L_2$ norm.}
\label{tab:bearing-occupancy-ladder}
\end{table}

The two halves of the common Poisson20 record (blocks 151--461 and 462--772)
gave signed torques of $-52.8675$ and $-52.8400~\mu$N\,m/m. The paired
second-half minus first-half change was $+0.0275\pm0.0395~\mu$N\,m/m.
We also fitted the torque magnitude at PPC2, 4 and 10 to
$a+b/\mathrm{PPC}$ separately for each seed. With PPC20 withheld,
the predicted and measured magnitudes were $52.843\pm0.022$ and
$52.854\pm0.023~\mu$N\,m/m. The paired discrepancy (measured minus predicted)
was $0.021\pm0.053\%$ of the measured value. Fitting all four levels gave
$a=52.599\pm0.025$ and $b=4.986\pm0.273$, both in $\mu$N\,m/m.
At PPC4, the difference from a straight line in inverse occupancy through
PPC2 and PPC10 was $-0.022\pm0.044~\mu$N\,m/m. These estimates apply to
torque on this grid at $\Delta t=3.706035$~ns.

\subsection{Consecutive collisions with the same partner}
\label{sec:bearing-last-partner-veto}

To test whether a second collision of the same pair accounts for the torque
shift, we added a veto to Poisson2. If the ordinary acceptance test succeeds
and each particle was the other's partner in its most recent gas collision,
the veto rejects the event. We compared this control with ordinary Poisson2
on the same grid and step over blocks 151--1190 (1040 blocks per seed).
Of the 46,912,439,896 gas collisions accepted by ordinary Poisson2, 3.437\%
were repeats within the same time step and 0.336\% repeated a mutual partner
from an earlier step. In the veto runs, 46,922,291,809 proposals passed the
ordinary acceptance test. The veto rejected 1,771,759,691 (3.776\%) and
45,150,532,118 collisions were executed. The denominators thus count
accepted gas collisions in the ordinary runs and eligible proposals in the veto runs.
Suppressing mutual repeats changed the torque magnitude by
$-0.160\pm0.227\%$ relative to ordinary Poisson2. The full paired interval,
$[-0.388,\,0.067]\%$, lies within one percentage point of zero.

\FloatBarrier
\subsection{Which particles carry tangential momentum to the rotor?}
We first examine every rotor impact at each nominal occupancy. If a gas collision preceded the impact, we record how many of the two partners had reflected from the rotor since their previous gas collision. This identifies the most recent collision only. A particle reaching the rotor directly from a wall without a gas collision forms a separate group. We omit the first two blocks after recording these histories began. Blocks 153--250 span $58.11~\mu$s per seed. In each sector the group sums recover the independently recorded number of impacts and total signed impulse.

\begin{figure}[!tbp]\centering
\includegraphics[width=.86\linewidth]{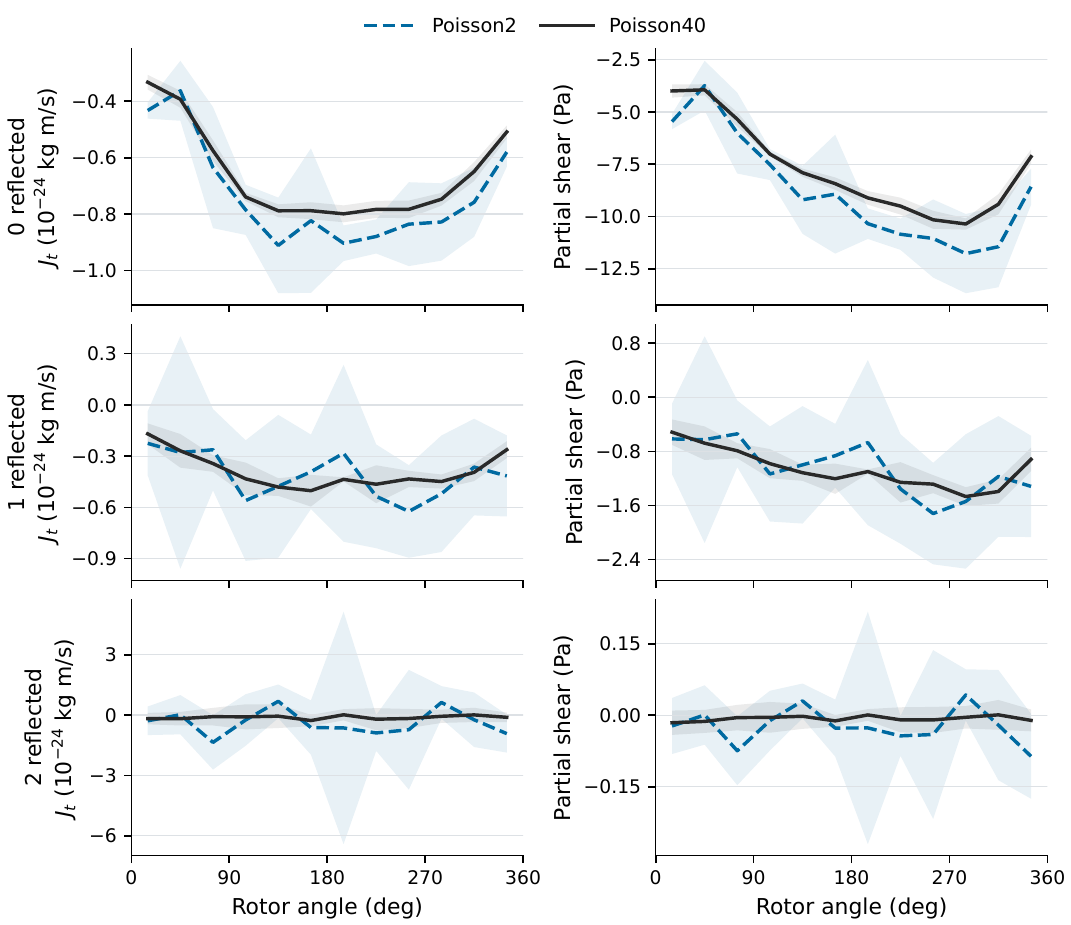}
\caption{Tangential impulse per rotor impact (left) and signed partial wall stress (right) on $576\times12$ cells. Rows give the number of rotor-reflected partners in the last gas collision. The angle is measured where the particle strikes the wall. Four adjacent 7.5-degree sectors were pooled by adding their counts and impulses. Bands give pointwise 95\% intervals across four seeds. Each row has its own ordinate scale.}
\label{fig:bearing-impulses}
\end{figure}

An infrequent type of impact can transfer considerable momentum per impact and still carry little mean shear. Figure~\ref{fig:bearing-impulses} compares these quantities. In each sector and trajectory, the per-impact mean is the sum of the tangential impulses per physical molecule divided by the number of impacts in the group. Its contribution to wall stress is the same sum multiplied by the simulator weight and divided by the wall arc length and sampling time. At PPC2, impacts following a collision between two reflected partners have large uncertainty in impulse per impact and contribute little to the shear difference. Where neither partner had recently reflected, the contribution to shear is more negative at PPC2 than at PPC40 in 11 of the 12 sectors. The sector intervals are wide. Table~\ref{tab:bearing-history} gives circumferential means for the four paired trajectories, retaining correlations among sectors.

\begin{table}[!tbp]\centering\small
\begin{tabular}{@{}lrrr@{}}\toprule
Last gas collision & Poisson2 (Pa) & Poisson40 (Pa) & Difference (Pa)\\\midrule
Neither partner reflected & $-8.738$ & $-7.689$ & $-1.049\pm0.451$\\
One reflected & $-1.046$ & $-1.059$ & $+0.013\pm0.198$\\
Both reflected & $-0.023$ & $-0.007$ & $-0.016\pm0.039$\\
Direct flight since wall & $-0.037$ & $-0.017$ & $-0.020\pm0.019$\\
All rotor impacts & $-9.844$ & $-8.772$ & $-1.072\pm0.291$\\\bottomrule
\end{tabular}
\caption{Contributions to circumferentially averaged gas-on-wall shear on the coarse grid over the same 98 blocks. Uncertainty is the half-width of the paired 95\% Student interval over four seeds. The difference for pairs with neither partner reflected separates into $-0.124\pm0.309$~Pa from the first layer and $-0.925\pm0.510$~Pa from the remaining gas layers.}
\label{tab:bearing-history}
\end{table}

Summing all three pair classes, last gas collisions outside the first layer account for $-0.978\pm0.561$~Pa of the $-1.072$~Pa difference in total shear. The corresponding fraction is a ratio of signed means. The uncertainty of the difference includes covariance among the classes.

\begin{figure}[!tbp]\centering
\includegraphics[width=.91\linewidth]{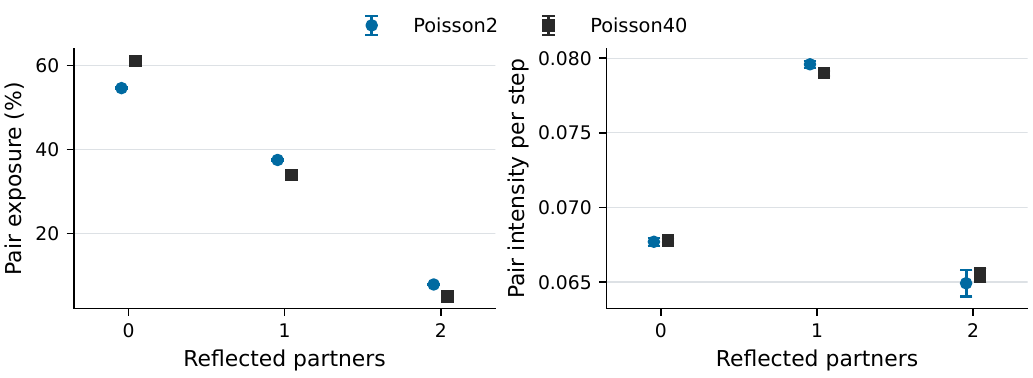}
\caption{Fraction of the available pair time by rotor reflection history (left) and mean dimensionless pair intensity per step at the PPC2 simulator weight (right). Both measurements use the first layer adjacent to the rotor on the common coarse grid. Bars give 95\% intervals across seeds.}
\label{fig:bearing-rate}
\end{figure}

For the second measurement we follow candidate pairs in the first gas layer and match them by angular collision cell. In Poisson2 we select cells containing exactly two particles. In Poisson40 we include all eligible pairs. Pair exposure is the time for which an unordered pair was available to collide, estimated by inverse proposal intensity. We compare pair intensities at the same physical location and simulator weight. Pairs with one or two rotor-reflected partners account for 45.4\% of pair exposure at PPC2 and 39.0\% at PPC40 (Fig.~\ref{fig:bearing-rate}). One-reflected-partner pairs have the highest mean intensity in both samples. The standardized Poisson2 intensity differs from Poisson40 by $+0.13\pm0.28$\%. This measurement concerns the mean rate per eligible pair in the selected layer. Shear weights the direction and amount of transferred momentum, as well as the number of encounters. A small difference in average pair intensity can therefore accompany a measurable difference in wall shear.

\FloatBarrier
\bibliographystyle{pof_linked}
\small
\ifdefined\bibsep\setlength{\bibsep}{1pt plus .25pt}\fi
\bibliography{references}